\documentclass[a4paper,makeidx,titlepage,openright,fleqn,11pt]{book}

\usepackage{these,graphicx,amsmath}

\makeindex

\def\etal{{\it et al.} }
\def\aj{{\it Astron.\ J.}}
\def\apj{{\it ApJ}}
\def\apjl{{\it ApJ Lett.}}
\def\apjs{{\it ApJ Suppl.}}
\def\aas{{\it A\&A Suppl.}}
\def\aa{{\it A\&A}}
\def\aal{{\it A\&A Lett.}}
\def\araa{{\it Ann.\ Rev.\ A\&A}}
\def\mnras{{\it MNRAS}}
\def\mnrasl{{\it MNRAS Lett.}}
\def\newa{{\it New Astr.}}
\def\nature{{\it Nature}}

\def\apss{{\it Ap\&SS}}
\def\astpart{{\it Astropart. Physics}}
\def\expas{{\it Experimental Astron.}}

\def\ar{{\it Astronomy Reports}}
\def\enprep{{\it en pr\'eparation}}
\def\mjysr{MJy/sr }
\def\inu{{I_{\nu}}}

\def\bnu{{B_{\nu}}}
\def\msol{{M$_{\odot}$}}
\def\mic{{{\mu}m}}
\def\uk{{{\mu}K}}
\def\ukc{{{\mu}K_{CMB}}}
\def\cm2{$cm^{-2}$}

\begin{document}


\thispagestyle{empty}

\vspace*{4em}

\begin{center}

{\bf \Large TH\`ESE}

\vspace*{2em}

pr\'esent\'ee au Centre d'\'Etude Spatiale des Rayonnements

\vspace*{1em}

par

\vspace*{1em}

{\bf \Large Xavier Dupac}

\vspace*{2em}

en vue de l'obtention du grade de

\vspace*{1em}

{\bf Docteur de l'Universit\'e Paul Sabatier - Toulouse III}

\vspace*{2em}

Sp\'ecialit\'e

\vspace*{1em}

{\bf ASTROPHYSIQUE}

\vspace*{1em}

\centerline{\underline{\hspace{5cm}}}
\vspace{0.5cm}

{\bf \huge Construction optimale}\\
\vspace{0.2cm}
{\bf \huge d'images bolom\'etriques}\\
\vspace{0.2cm}
{\bf \huge Contribution \`a l'\'etude du milieu}\\
\vspace{0.2cm}
{\bf \huge interstellaire et du rayonnement fossile}\\
\vspace{0.3cm}
\centerline{\underline{\hspace{5cm}}}

\vspace*{1em}

\end{center}

Soutenue le 18 septembre 2002 devant la commission d'examen compos\'ee de:

\vspace*{1em}

\begin{tabular}{ll}
{\large M. Dominique Le Qu\'eau}         & {\large Pr\'esident}\\
{\large M. Fran\c{c}ois Boulanger}       & {\large Rapporteur}\\
{\large M. Pierre L\'ena}	         & {\large Rapporteur}\\
{\large M. Alain Blanchard}	         & {\large Examinateur}\\
{\large M. Fran\c{c}ois-Xavier D\'esert} & {\large Examinateur}\\
{\large M. Martin Giard}	         & {\large Directeur de th\`ese}\\
{\large M. James Bartlett}               & {\large Invit\'e}\\
{\large M. Michel P\'erault}             & {\large Invit\'e}\\
\end{tabular}


\newpage
\thispagestyle{empty}
\vspace*{5em}

\newpage

\thispagestyle{empty}

\vspace*{4em}

\begin{center}

{\bf \Large TH\`ESE}

\vspace*{2em}

pr\'esent\'ee au Centre d'\'Etude Spatiale des Rayonnements

\vspace*{1em}

par

\vspace*{1em}

{\bf \Large Xavier Dupac}

\vspace*{2em}

en vue de l'obtention du grade de

\vspace*{1em}

{\bf Docteur de l'Universit\'e Paul Sabatier - Toulouse III}

\vspace*{2em}

Sp\'ecialit\'e

\vspace*{1em}

{\bf ASTROPHYSIQUE}

\vspace*{1em}

\centerline{\underline{\hspace{5cm}}}
\vspace{0.5cm}

{\bf \huge Construction optimale}\\
\vspace{0.2cm}
{\bf \huge d'images bolom\'etriques}\\
\vspace{0.2cm}
{\bf \huge Contribution \`a l'\'etude du milieu}\\
\vspace{0.2cm}
{\bf \huge interstellaire et du rayonnement fossile}\\
\vspace{0.3cm}
\centerline{\underline{\hspace{5cm}}}

\vspace*{1em}

\end{center}

Soutenue le 18 septembre 2002 devant la commission d'examen compos\'ee de:

\vspace*{1em}

\begin{tabular}{ll}
{\large M. Dominique Le Qu\'eau}         & {\large Pr\'esident}\\
{\large M. Fran\c{c}ois Boulanger}       & {\large Rapporteur}\\
{\large M. Pierre L\'ena}	         & {\large Rapporteur}\\
{\large M. Alain Blanchard}	         & {\large Examinateur}\\
{\large M. Fran\c{c}ois-Xavier D\'esert} & {\large Examinateur}\\
{\large M. Martin Giard}	         & {\large Directeur de th\`ese}\\
{\large M. James Bartlett}               & {\large Invit\'e}\\
{\large M. Michel P\'erault}             & {\large Invit\'e}\\
\end{tabular}



\newpage
\thispagestyle{empty}
\vspace*{5em}

\newpage

\chapter*{R\'esum\'e de la th\`ese}

Ce travail de th\`ese s'inscrit dans le d\'eveloppement de l'astrophysique des rayonnements infrarouge lointain et millim\'etrique.
Nous avons travaill\'e sur le traitement et l'analyse de donn\'ees concernant le milieu interstellaire galactique \`a travers l'\'emission thermique des poussi\`eres, et la cosmologie \`a travers l'observation des fluctuations du rayonnement fossile.
Nous nous sommes particuli\`erement int\'eress\'es \`a la construction de cartes optimales par des m\'ethodes d'inversion lin\'eaire.
Ceci nous a permis de d\'evelopper une nouvelle m\'ethode de construction de cartes pour l'exp\'erience ballon submillim\'etrique PRONAOS, fond\'ee sur une matrice d'inversion de Wiener, qui reconstruit la carte de fa\c{c}on globale.
L'analyse des cartes de PRONAOS dans les complexes de formation d'\'etoiles massives que sont Orion et M17 a ensuite permis de d\'ecouvrir les variations importantes des conditions physiques du milieu et des propri\'et\'es des poussi\`eres.
Nous avons notamment mis en \'evidence des condensations froides (T $\approx$ 10 - 20 K) \`a proximit\'e des centres actifs de formation d'\'etoiles.
Certains de ces nuages froids pourraient \^etre gravitationnellement instables.
Il appara\^\i t \'egalement une anticorr\'elation entre la temp\'erature et l'indice spectral submillim\'etrique.
Nos investigations concernant cet effet favorisent des causes li\'ees \`a la physique intrins\`eque des grains.
Nous avons \'egalement d\'evelopp\'e des m\'ethodes optimales de construction de cartes pour les exp\'eriences mesurant les fluctuations du rayonnement fossile.
Nous avons simul\'e diff\'erentes strat\'egies d'observation d'exp\'eriences ballon et satellite (tel que Planck), construit diff\'erents trains de donn\'ees et appliqu\'e nos m\'ethodes de construction de cartes \`a ces donn\'ees simul\'ees.
Les m\'ethodes it\'eratives d\'evelopp\'ees (COBE et Wiener) permettent de reconstruire la carte du ciel avec une grande pr\'ecision, malgr\'e la grande quantit\'e de bruit autocorr\'el\'e pr\'esent dans les trains de donn\'ees.
Nous avons \'egalement particip\'e au traitement et \`a l'analyse des donn\'ees de l'exp\'erience ballon Archeops.
En pr\'esence de bruit important et ayant des caract\'eristiques statistiques complexes, nous avons appliqu\'e des m\'ethodes simples de construction de cartes et d'estimation du spectre de puissance des fluctuations.

\vspace*{1em}

Mots clefs: construction de cartes - ballons - milieu interstellaire - poussi\`ere - cosmologie - rayonnement fossile

\vspace*{2em}

This work takes part of the development of far-infrared and millimeter astrophysics.
We have worked on the data processing and analysis in the fields of the Galactic interstellar medium,
through the dust thermal emission, and cosmology through the observation of the cosmic
microwave background fluctuations.
We have been particularly interested in optimal map-making by inverse linear methods.
We have developed a new map-making method for the balloon-borne submillimeter experiment PRONAOS, based on a Wiener inversion matrix, which allows to globally reconstruct the map.
The analysis of PRONAOS maps in massive star-forming complexes as Orion and M17 allowed us to discover the large variations of the physical conditions and the dust properties.
Notably, we showed cold condensations (T $\approx$ 10 - 20 K) near the active star-forming centers.
Some of these cold clouds could be gravitationally unstable.
Also, we showed an anticorrelation between the temperature and the submillimeter spectral index.
Our investigations concerning this effect favour causes related to the intrinsic physics of the grains.
We have also developed optimal map-making methods for the experiments aiming at measuring the cosmic microwave background fluctuations.
We simulated several different observation strategies for balloon-borne or satellite experiments (such as Planck), constructed timelines and applied our map-making methods to these simulated data.
The iterative methods that we have developed (COBE and Wiener) allow to reconstruct the sky maps very accurately, in spite of the large amount of self-correlated noise present in the timelines.
We have also worked on the data processing and analysis for the Archeops balloon-borne experiment.
In the presence of much noise having complex statistical properties, we have applied simple map-making and power spectrum estimation methods.

\vspace*{1em}

Key words: map-making - balloons - interstellar medium - dust - cosmology - cosmic microwave background

\chapter*{Remerciements}

Comme chacun sait, le travail de th\`ese se construit sur plus de trois ans,
pendant lesquels l'apprentissage et la pratique du m\'etier de chercheur
mettent en jeu des relations humaines \'etroites et irrempla\c{c}ables.
Dans les moments difficiles v\'ecus par les membres du d\'epartement Univers Froid du CESR
et par tous les Toulousains ces trois derni\`eres ann\'ees, cette affirmation prend
un sens encore plus grand.
De plus, la th\`ese est l'aboutissement de nombreuses ann\'ees d'\'etudes,
dont l'origine se perd dans la nuit des temps.
Il serait donc vain de pr\'etendre remercier de fa\c{c}on exhaustive toutes
les personnes qui m'ont aid\'e \`a un moment ou \`a un autre de cette longue
aventure estudiantine qui se termine aujourd'hui.
Je m'excuse donc par avance vis-\`a-vis de toutes celles et ceux que j'oublie.

Je voudrais tout d'abord remercier mon directeur de th\`ese Martin Giard,
qui m'a offert la possibilit\'e de travailler dans ce domaine de recherche au
CESR, et qui fut toujours prodigue en encouragements et conseils d\'ecisifs, en
toutes circonstances.
Je remercie les membres de mon jury de th\`ese, en particulier Dominique Le Qu\'eau
pour m'avoir accueilli au CESR et pour avoir accept\'e de pr\'esider le jury.
Je remercie chaleureusement mes rapporteurs: Fran\c{c}ois Boulanger et M.
l'Acad\'emicien Pierre
L\'ena, qui m'ont beaucoup aid\'e par leurs commentaires pr\'ecieux sur
le manuscrit.
Je remercie \'egalement James Bartlett, Alain Blanchard et Michel P\'erault pour
avoir accept\'e d'\^etre membres du jury.
Je souhaite remercier particuli\`erement Fran\c{c}ois-Xavier D\'esert en sa
qualit\'e de membre du jury, et pour l'aide pr\'ecieuse qu'il m'a apport\'ee dans mon
travail sur l'exp\'erience Archeops.

Je tiens \`a remercier chaleureusement Guy Serra, directeur du
D\'epartement Univers Froid, que j'ai eu la chance de conna\^\i tre,
malheureusement trop peu de temps, mais suffisamment pour l'appr\'ecier
profond\'ement.
Sa rigueur scientifique, ainsi que son enthousiasme et son optimisme
humanistes et r\'evolutionnaires m'ont beaucoup encourag\'e, et j'esp\`ere que son sens de l'humain restera toujours l'esprit dominant de
cette \'equipe de travail.
Je remercie l'ensemble des membres du d\'epartement Univers
Froid pour l'aide qu'ils ont pu m'apporter, en particulier Isabelle
Ristorcelli, Laurent Ravera, Claude M\'eny, Catherine Pech, Charlotte Vastel,
S\'ebastien Maret, Emmanuel Caux, et tous les autres.
Je remercie le personnel administratif, en particulier
Dorine Roma et Monique M\'eric, qui ont si souvent planch\'e sur des parcours
de mission complexes.
Je remercie Henri R\`eme et Gilbert Vedrenne, respectivement directeurs de
la formation doctorale APS et de l'\'ecole doctorale SDU2E, ainsi que la
secr\'etaire de l'\'ecole doctorale Marie-Claude Cathala.

Je remercie chaleureusement Alain Beno\^\i t, PI d'Archeops, dont
l'enthousiasme et l'\'energie furent indispensables \`a tout le monde,
notamment pendant les campagnes \`a Kiruna.
Je remercie toute la collaboration Archeops, en particulier Philippe
Filliatre, Sophie Henrot-Versill\'e, Olivier Dor\'e,
Fran\c{c}ois Bouchet et Jean-Christophe Hamilton.
Je remercie \'egalement l'ensemble de la collaboration PRONAOS, au CESR, \`a
l'IAS et ailleurs.

Remercions maintenant les personnes qui m'ont donn\'e le go\^ut de la
science, de la physique et de l'astronomie pendant mon jeune \^age, \`a
commencer par mes parents, ainsi que mes instituteurs et professeurs,
notamment M. Bobet et Mlle Pascaud.
Je remercie chaleureusement M. Louis Ousset, et mes camarades Sylvain
Genet, Micka\"el Mouton et Karine Soulat avec qui j'ai pu participer aux
Olympiades de Physique, en premi\`ere et terminale, ce qui m'a
merveilleusement initi\'e au monde complexe de l'exp\'erimentation.
Je remercie mes enseignants \`a l'Universit\'e, en particulier MM. Saulnier,
Abgrall et Parisot.

Je remercie enfin, et surtout, ma famille et mes amis, et tout d'abord mes
parents Annette et Mikel, dont l'aide morale et financi\`ere fut indispensable pendant mes
\'etudes.
Je dois tout \`a leur confiance, \`a leur int\'er\^et pour mon travail et \`a leur soutien sans faille.
Je remercie \'egalement mon fr\`ere R\'emi et ma s\oe ur Marie, ainsi que mes grands-parents,
qui m'ont toujours soutenu par leur confiance in\'ebranlable.
Je ne puis assez remercier Dominique Rougier et S\'ebastien Faure pour leur amiti\'e et le
partage de tant d'aventures en Terres du Milieu, dans l'eau sur (ou sous) des planches, ou dans la rue...
Merci \`a David Rousseau, Nelly Bosselut et Bruno Barrier pour leur amiti\'e et leur amour de la montagne (avec ou sans corde !)
Merci \`a Arnaud Jacotin, C\'esar Bertucci, Hanna Sabat, et tous les autres.
Merci enfin \`a B\'ereng\`ere pour ses pr\'ecieux commentaires sur le manuscrit, ses encouragements et son soutien de tous les instants...



\tableofcontents
\listoffigures
\listoftables

\chapter*{Introduction g\'en\'erale}
{\flushright\it \'Etudiez comme si vous deviez vivre toujours [...]\\}
{\flushright S$^t$ Isidore de S\'eville\\}

\vspace*{1em}

Il convient tout d'abord de pr\'eciser la nature de cette
th\`ese, et les comp\'etences auxquelles elle fait appel.
Cette th\`ese d'astrophysique comporte l'\'etude de deux sujets scientifiques
qui peuvent appara\^\i tre au premier abord assez \'eloign\'es l'un de l'autre, \`a savoir les
poussi\`eres galactiques froides et le rayonnement fossile, ce dernier \'etant
le plus important
en termes de temps et de travail.
Le premier est reli\'e au milieu interstellaire de notre galaxie, le second
\`a la cosmologie de notre univers.
Cependant, comme ce sera d\'evelopp\'e plus loin, les instruments de mesure de
ces deux objets sont semblables, parfois les m\^emes, et de ce fait les cha\^\i
nes de traitement des donn\'ees concernant ces deux domaines scientifiques
sont fortement corr\'el\'ees, y compris en ce qui concerne le probl\`eme de la
construction de cartes.

Le sujet phare de cette th\`ese est bien la construction d'images, qui est un
probl\`eme math\'ematique relativement simple \`a poser mais moins \'evident
\`a r\'esoudre et \`a appliquer dans la pratique.
Environ un tiers de notre temps de travail concernant l'analyse des donn\'ees de PRONAOS fut
occup\'e par la construction d'images proprement dite, ainsi que l'essentiel
de notre temps sur l'analyse des donn\'ees d'Archeops.
Il s'agit d'un probl\`eme de math\'ematiques appliqu\'ees qui fait appel \`a
de l'analyse num\'erique, de l'algorithmique, et bien s\^ur \`a
la programmation associ\'ee.
Cependant, la parfaite connaissance du ciel que l'on observe est
indispensable, car l'objectif de la construction de cartes est bien moins de
faire de belles cartes que de sauvegarder toute l'information astrophysique
utile, tout en r\'eduisant le bruit et la masse d'information.
Il s'agit d'une \'etape essentielle dans l'analyse de donn\'ees
astrophysiques, en particulier en ce qui concerne le rayonnement fossile.
Nous avons pu contribuer \`a la cr\'eation, au d\'eveloppement et aux
applications de nouvelles m\'ethodes de construction de cartes, \`a la fois
pour PRONAOS, exp\'erience concern\'ee par l'\'etude du milieu interstellaire,
et pour les exp\'eriences concern\'ees par l'\'etude du rayonnement fossile.

Bien entendu, le travail scientifique ne s'arr\^ete pas lorsque l'on a construit
les cartes.
Nous avons contribu\'e \`a l'\'etude scientifique du milieu interstellaire
dense, principalement dans les r\'egions de formation d'\'etoiles massives
Orion et Messier 17, \`a partir de donn\'ees PRONAOS.
Cette \'etude fait appel \`a des connaissances g\'en\'erales physico-chimiques
et astrophysiques, ainsi qu'\`a des notions de statistique.
Nous avons aussi particip\'e \`a l'\'etude des cartes d'Archeops, en ce qui
concerne l'estimation du spectre de puissance des fluctuations du rayonnement
fossile.

Ce m\'emoire est structur\'e comme suit: le premier chapitre introduit le contexte scientifique de cette th\`ese, le chapitre \ref{chapconst} est consacr\'e au probl\`eme de la construction de cartes dans sa g\'en\'eralit\'e, le chapitre \ref{pous} expose notre travail sur la construction de cartes pour PRONAOS et l'\'etude des poussi\`eres dans la Galaxie, et le chapitre \ref{rf} expose notre travail concernant le rayonnement fossile.

Une grande partie du travail fourni pendant ces trois ann\'ees a pu donner lieu \`a des publications
(liste fournie en section \ref{publis}), mais celles-ci ne sauraient en aucun
cas repr\'esenter la totalit\'e du travail de th\`ese, comme d'ailleurs ne le
repr\'esente pas parfaitement non plus ce m\'emoire.
Car en d\'efinitive, comment rendre compte de l'essentiel, \`a savoir que la
recherche reste avant tout une aventure humaine
collective, mais aussi une qu\^ete personnelle ?

\chapter{L'Univers froid, un domaine \`a explorer}
{\flushright\it Suivant la lumi\`ere du soleil, nous quitt\^ames le Vieux Monde.\\}
{\flushright Christophe Colomb\\}

\section{ Introduction}
En l'an 1800, Sir Frederick William Herschel (1738-1822)
d\'ecouvrit ce qu'il appela les "rayons calorifiques" (voir figure \ref{herschel}).
L'exp\'erience qu'il r\'ealisa mit en effet en \'evidence le rayonnement infrarouge
gr\^ace \`a l'effet de chauffage qu'il produisit.
D\`es sa d\'ecouverte premi\`ere, le rayonnement infrarouge fut donc associ\'e
\`a la notion de temp\'erature.
Mais, tandis que dans le sens commun ce rayonnement est associ\'e \`a
l'id\'ee de chaleur, la connaissance du spectre \'electromagn\'etique en
de\c{c}\`a du rouge s'\'etendit jusqu'\`a des longueurs d'onde qui
caract\'erisent des temp\'eratures tr\`es
froides.

\begin{figure}
\begin{center}
\caption[Sir Herschel en 1800]{Sir Herschel d\'ecouvrant le rayonnement infrarouge (1800).
Image tir\'ee du site de l'IPAC ({\it http:// www.ipac.caltech.edu)}.}
\label{herschel}
\end{center}
\end{figure}

Aujourd'hui, nous savons que le rayonnement infrarouge lointain qui
caract\'erise l'Univers froid repr\'esente
la majorit\'e de la puissance de radiation qui nous vient du
ciel.
Cependant, paradoxalement, cette partie du spectre \'electromagn\'etique fut
largement ignor\'ee jusqu'aux trente derni\`eres ann\'ees, alors m\^eme que
l'astronomie radio s'\'etait largement d\'evelopp\'ee, et que bien entendu
l'astronomie optique existait depuis l'Antiquit\'e, faisant de cette
science l'une des plus anciennes au monde.
Il est vrai que le rayonnement infrarouge lointain provenant de
l'ext\'erieur de notre plan\`ete subit l'effet de
l'absorption par l'atmosph\`ere, et que de ce fait l'observation au sol est
limit\'ee \`a quelques fen\^etres de transmission et/ou \`a des sites
d'observation exceptionnels et \`a haute altitude.
Malgr\'e des progr\`es fulgurants r\'ealis\'es \`a la fin du vingti\`eme
si\`ecle pour l'observation de ce rayonnement infrarouge lointain,
c'est bien en dehors de l'atmosph\`ere terrestre que son observation est la
plus prometteuse.
Par la suite, nous ne parlerons essentiellement plus que de ce rayonnement
infrarouge lointain, autrement appel\'e submillim\'etrique et millim\'etrique:
dont la longueur d'onde est comprise entre environ 100 $\mic$ et 3
mm, ce qui correspond au domaine de fr\'equence compris entre 100 et 3000 GHz.
Encore aujourd'hui, de nombreuses longueurs d'onde dans ce domaine sont peu
explor\'ees, en particulier les cartographies compl\`etes du ciel sont
limit\'ees \`a des r\'esolutions angulaires tr\`es larges (7 degr\'es pour le satellite
COBE, COsmic Background Explorer, dans le millim\'etrique).
Nous nous sommes int\'eress\'es pour cette th\`ese \`a l'observation de ce
rayonnement en dehors de l'atmosph\`ere terrestre, essentiellement gr\^ace \`a des instruments
embarqu\'es en ballon stratosph\'erique.
Nous pr\'esentons dans les sections suivantes une introduction aux domaines
scientifiques reli\'es \`a ces observations de l'Univers froid, et nous
pr\'esentons succintement les modes d'observations et d\'etecteurs concern\'es
en section \ref{instr}.

\section{ Nature des rayonnements observ\'es en infrarouge lointain}
Le rayonnement infrarouge lointain en astronomie caract\'erise principalement
deux natures radiatives distinctes: le rayonnement thermique continuum des solides
froids, et les raies de rotation du gaz mol\'eculaire.
D'autres rayonnements sont aussi pr\'esents de fa\c{c}on marginale
(bremsstrahlung des \'electrons par exemple).
Le rayonnement thermique caract\'erise un corps (solide, liquide ou gaz) en
\'equilibre thermodynamique.
Les particules composant ce corps ont une \'energie cin\'etique moyenne
caract\'eris\'ee par la temp\'erature de ce corps.
La densit\'e d'\'etats d'\'energie du corps \'emetteur d\'efinit la forme de l'\'emission: continuum dans le cas d'un solide, discr\`ete (raies) dans le cas d'un gaz.
Pour les solides (grains), connaissant la temp\'erature,
la loi d'\'emission est approximativement celle du corps noir, c'est-\`a-dire la loi de Planck.
L'intensit\'e \'emise par unit\'e de fr\'equence (ou de longueur d'onde) est
proportionnelle \`a:

\begin{equation}
\bnu(\nu,T) = {2 h \nu^3 \over c^2} {1 \over e^{{h \nu \over k T}} - 1}
\label{eqplanck}
\end{equation}

o\`u $\bnu$ est appel\'ee la fonction de Planck (W/m$^2$/Hz/sr), $\nu$ est la
fr\'equence (Hertz), T
est la temp\'erature (Kelvin), h la constante de Planck, k la constante de Boltzmann et
c la vitesse de la lumi\`ere.
L'\'emission d'un solide en \'equilibre thermodynamique est donc parfaitement
d\'ecrite (\`a un facteur d'\'echelle pr\`es) par un seul param\`etre, la
temp\'erature.
Ce rayonnement thermique des grains est un continuum, comme l'indique l'\'equation de
Planck (\ref{eqplanck}), il pourra donc \^etre observ\'e dans des bandes de
fr\'equence larges.

L'autre rayonnement important en astronomie dans le domaine
millim\'etrique - submillim\'etrique est l'\'emission (thermique \'egalement) des raies
du gaz (mol\'ecules, atomes ou ions).
Dans ce domaine de longueur d'onde, ces raies sont principalement dues \`a la quantification
des niveaux d'\'energie de rotation des mol\'ecules, mais il existe aussi des raies de
structure fine des atomes et des ions.
Ainsi, la relaxation d'un niveau d'\'energie plus \'elev\'e vers un niveau plus
bas permet d'\'emettre un photon de longueur d'onde discr\`ete.
De ce fait, ces rayonnements pourront \^etre observ\'es dans des bandes de
fr\'equence \'etroites ou par des spectrom\`etres.

Ces rayonnements que l'on observe dans le domaine infrarouge
lointain proviennent de diff\'erentes distances dans l'Univers.

\section{ Dans le syst\`eme solaire}
Le rayonnement infrarouge lointain dans le syst\`eme solaire correspond \`a
l'\'emission thermique des plan\`etes et autres objets froids, c'est-\`a-dire
ceux situ\'es \`a une grande distance du Soleil.
En particulier, les plan\`etes g\'eantes (Jupiter, Saturne, Uranus, Neptune)
sont une source importante de radiation infrarouge lointaine.
Ces objets ont en effet une temp\'erature de surface de $\approx$ 50 K pour
Neptune et $\approx$ 120 K pour Jupiter.
Nous n'avons pas \'etudi\'e l'\'emission thermique de ces objets, mais ils
peuvent \^etre utilis\'es pour \'etalonner un instrument destin\'e \`a
observer le rayonnement infrarouge lointain plus faible provenant de la
Galaxie ou du rayonnement fossile.
En particulier, un instrument poss\'edant un lobe relativement grand (quelques
minutes d'arc) peut utiliser les plan\`etes g\'eantes comme sources
ponctuelles puissantes, afin de cartographier les lobes de l'instrument.

\section{ Dans la Voie Lact\'ee}
Notre galaxie, la Voie Lact\'ee, tire son nom du rayonnement visible diffus des \'etoiles.
Elle doit l'existence de cette lumi\`ere stellaire \`a la pr\'esence d'un milieu interstellaire gazeux, dans lequel
naissent les \'etoiles et les syst\`emes stellaires.
Cependant, si la pr\'esence de nombreuses \'etoiles dans la Voie Lact\'ee saute aux yeux
depuis l'Antiquit\'e, celle de mati\`ere interstellaire se r\'esume seulement
\`a de l'intuition.
Des objets diffus appel\'es ``n\'ebuleuses'' ont \'et\'e observ\'es depuis la Renaissance, gr\^ace \`a l'invention des instruments optiques (lunettes, t\'elescopes).
La pr\'esence de mati\`ere diffuse a donc \'et\'e imagin\'ee, mais la confusion r\'egnait entre les n\'ebuleuses effectivement pr\'esentes dans la Galaxie (mati\`ere interstellaire gazeuse) et celles lui \'etant ext\'erieures (galaxies).

Ce n'est que pendant la deuxi\`eme moiti\'e du XIX$^e$ si\`ecle que fut mise en
\'evidence de fa\c{c}on irr\'efutable la pr\'esence de mati\`ere
interstellaire dans la Galaxie par l'observation de raies d'\'emission dans le visible.
En particulier, les travaux de Sir William Huggins et d'autres, principalement sur la Grande
N\'ebuleuse d'Orion (par exemple Huggins 1865 et Huggins 1882), permirent de mettre en \'evidence cette mati\`ere.
La d\'ecouverte de nuages interstellaires tels que la N\'ebuleuse
d'Orion permit \'egalement d'expliquer l'att\'enuation de la lumi\`ere
stellaire dans la voie Lact\'ee (taches sombres telles que le Sac \`a Charbon
pr\`es de la Croix du Sud).

Apr\`es les r\'egions ionis\'ees mises en \'evidence dans le visible, le
d\'eveloppement de la radioastronomie pendant la premi\`ere moiti\'e du XX$^e$
si\`ecle permit d'identifier dans les ann\'ees cinquante l'hydrog\`ene atomique par sa raie de structure
hyperfine \`a 21 cm (par exemple van de Hulst \etal 1951, Lilley 1955).
Ce constituant majoritaire du milieu interstellaire permit de tracer la grande
structure (spirale) de la Voie Lact\'ee.
Les raies de rotation mol\'eculaires furent les cibles suivantes de la
radioastronomie, qui d\'etecta les mol\'ecules OH, CO, H$_2$O, NH$_3$,
H$_2$CO, CN, et de nombreuses autres, pour arriver \`a environ 120 aujourd'hui
\cite{snyder97}.
Depuis la fin des ann\'ees 1970, l'astronomie infrarouge et millim\'etrique de la Galaxie
s'est d\'evelopp\'ee en particulier gr\^ace \`a l'observation spatiale, notamment par les
satellites IRAS (InfraRed Astronomical Satellite, 1983-84, voir la carte du ciel en figure \ref{figgal}), COBE (COsmic
Background Explorer, 1992) et ISO (Infrared Space Observatory, 1995-98).

Le milieu interstellaire gazeux se pr\'esente sous trois \'etats: mol\'eculaire, atomique et ionis\'e (voir McKee \& Ostriker 1977).
Sa structure fractale fait appara\^\i tre des contrastes de densit\'e et de
temp\'erature tr\`es importants, la densit\'e pouvant varier de moins de
10$^{-3}$ H.cm$^{-3}$ dans les bulles d'expansion des supernov{\ae}, jusqu'\`a
plus de 10$^8$ H.cm$^{-3}$ dans les nuages mol\'eculaires denses.
Inversement, la temp\'erature peut varier de moins de 10 K dans les nuages mol\'eculaires
denses et/ou \'eloign\'es des sources de rayonnement des \'etoiles, jusqu'\`a
des millions de degr\'es dans les r\'esidus de supernov{\ae}.
La masse totale de milieu interstellaire dans la Galaxie repr\'esente seulement
4 \% environ de la masse totale de la Galaxie, pour l'essentiel due aux
\'etoiles.
Dans le milieu interstellaire, l'essentiel de la masse est contenue dans les
nuages mol\'eculaires (H$_2$, $\approx$ 40 \%),
les nuages atomiques (H I, $\approx$ 40 \%), et les r\'egions diffuses atomiques inter-nuages (H I, $\approx$ 20 \%, voir par exemple Bakes 1997).
Bien que les bulles de supernov{\ae} repr\'esentent environ 50 \% du volume du
milieu interstellaire galactique, elles ne contribuent pas significativement
\`a sa masse.
Le cycle de la vie stellaire dans la Galaxie est donc comme suit: les
\'etoiles naissent par effondrement gravitationnel dans les coeurs denses des nuages mol\'eculaires, vivent
leur vie d'\'etoile (fusion thermonucl\'eaire de l'hydrog\`ene, puis de
l'h\'elium, du carbone, de l'oxyg\`ene, etc, pour les \'etoiles massives),
puis meurent en laissant une bulle de mati\`ere diffuse en expansion
(n\'ebuleuses "plan\'etaires" ou r\'esidus de supernova suivant la masse de
l'\'etoile).
Une partie de la masse de l'\'etoile reste sous forme de r\'esidu stellaire
(naine blanche, \'etoile \`a neutrons ou trou noir (?) suivant la masse
stellaire initiale), le reste retournant au milieu qui s'en trouve enrichi en
\'el\'ements lourds ("m\'etallicit\'e"), i.e. plus lourds que l'h\'elium.
Cette mati\`ere interstellaire \'evolue pour former de nouveaux nuages denses,
qui formeront la g\'en\'eration stellaire suivante.
Ce cycle r\'ep\'et\'e a pour cons\'equence de faire \'evoluer la composition
chimique de la Galaxie vers toujours plus d'\'el\'ements lourds, modifiant
ainsi la chimie \'egalement \`a l'int\'erieur des \'etoiles.

On voit donc que l'\'etude des nuages mol\'eculaires est particuli\`erement int\'eressante
pour caract\'eriser la formation stellaire, ainsi que la structure de la
Galaxie.
Outre le gaz (H$_2$ et He essentiellement), de la poussi\`ere
est pr\'esente dans les nuages mol\'eculaires.
Elle repr\'esente 1 \% environ de la masse du milieu, et consiste en
diff\'erentes esp\`eces carbon\'ees et silicat\'ees.
L'\'emission thermique des gros grains est observ\'ee dans le domaine
submillim\'etrique (longueur d'onde entre 100 $\mic$ et 1 mm).
Cependant, cette \'emission n'est pas un corps noir parfait, et la
temp\'erature ne d\'efinit pas compl\`etement la poussi\`ere galactique.
Nous pr\'esentons les caract\'eristiques des poussi\`eres galactiques, ainsi que notre \'etude effectu\'ee
gr\^ace aux donn\'ees PRONAOS, au chapitre \ref{pous}.

En conclusion de cette section, rappelons que les connaissances
actuelles sur la structure de la Galaxie, la formation stellaire, ou
l'\'evolution physico-chimique du milieu interstellaire, sont tr\`es
sommaires, surtout si l'on compare \`a d'autres domaines de l'astrophysique
tels que la structure et l'\'evolution stellaire.
En effet, par exemple, le r\^ole probablement pr\'epond\'erant des champs
magn\'etiques dans la structure galactique, ou la formation des mol\'ecules et
des grains dans le milieu, sont mal connus et peu mod\'elis\'es. 
La m\^eme chose peut \^etre dite pour les premiers stades de la formation stellaire.
L'Univers froid galactique reste donc largement un domaine \`a explorer.

\begin{figure}
\begin{center}
\caption[Vue de la Galaxie par le satellite IRAS]{Vue de la Galaxie par le satellite IRAS.
Les trois couleurs utilis\'ees pour cette image sont le bleu (donn\'ees IRAS \`a 12 $\mic$), le vert (60 $\mic$) et le rouge (100 $\mic$).
Image tir\'ee du site web de l'IPAC: {\it http:// www.ipac.caltech.edu}.}
\label{figgal}
\end{center}
\end{figure}

\section{ Dans l'Univers}
Aujourd'hui, la tr\`es grande majorit\'e des cosmologistes adh\`ere \`a la
th\'eorie du Big Bang, dans le contexte de la Relativit\'e G\'en\'erale d'Einstein.

\subsection{La th\'eorie du Big Bang\label{bigbang}}
Les postulats de la Relativit\'e G\'en\'erale impliquent que l'espace et le temps forment une
vari\'et\'e riemannienne \`a quatre dimensions, \'eventuellement dot\'ee d'une
courbure.
La g\'eom\'etrie et la courbure de l'espace-temps sont reli\'ees de fa\c{c}on simple \`a la
densit\'e d'\'energie et de masse.
Si l'on a une distribution simple de masse-\'energie, il est possible
d'exprimer analytiquement la g\'eom\'etrie de l'espace-temps, par sa m\'etrique.
Dans le cadre de la cosmologie, une hypoth\`ese extr\^emement simple, mais qui
s'av\`ere valable \`a grande \'echelle, est de consid\'erer que la densit\'e
de masse-\'energie dans l'Univers est homog\`ene et isotrope \`a toutes les
\'epoques (c'est le principe cosmologique).
La g\'eom\'etrie de l'Univers s'exprime alors par la m\'etrique de
Friedmann-Robertson-Walker:

\begin{equation}
ds^2 = c^2 dt^2 - R^2(t) [{dr^2 \over 1-kr^2} + r^2 (d\theta^2 + sin^2\theta
d\phi^2)]
\label{eqmetric}
\end{equation}

o\`u r, $\theta$ et $\phi$ sont les coordonn\'ees sph\'eriques, c la vitesse de la
lumi\`ere, k le param\`etre de courbure de l'espace, valant -1, 0 ou 1, et R(t) le facteur
d'\'echelle.
Deux grandeurs d\'efinissent donc un univers donn\'e qui respecte le principe
cosmologique: la courbure de l'espace (ferm\'e, plat ou ouvert), et
l'\'evolution temporelle du facteur d'\'echelle.
Cette derni\`ere est d\'ecrite par les \'equations de Friedmann, qui
d\'erivent des \'equations d'Einstein appliqu\'ees \`a la m\'etrique de Friedmann-Robertson-Walker.

Les observations cosmologiques, \`a la lumi\`ere des \'equations de Friedmann,
ont mis en \'evidence que l'Univers \'etait en
expansion depuis un temps z\'ero o\`u l'Univers \'etait infiniment dense:
c'est la th\'eorie du Big Bang.
Les deux preuves observationnelles majeures de cette th\'eorie sont, d'une part, la r\'ecession des
galaxies, mise en \'evidence par le d\'ecalage spectral vers le rouge (z)
syst\'ematique des raies observ\'ees dans les galaxies \'eloign\'ees,
\'etablissant que la vitesse de r\'ecession est (sans tenir compte des
mouvements locaux) proportionnelle \`a la distance entre deux galaxies, tels
deux points sur la surface d'un ballon de baudruche.
Le coefficient de proportionnalit\'e est la constante de Hubble H$_0$ \cite{hubble31}.
D'autre part, la deuxi\`eme preuve majeure de la th\'eorie du Big Bang est la d\'ecouverte (fortuite)
du rayonnement fossile ou CMB (Cosmic Microwave Background) en 1965 par
Penzias et Wilson, ing\'enieurs au Bell Telephone Laboratories Incorporated.

Notons que nous utiliserons le sigle CMB bien qu'il s'agisse d'un acronyme anglais, car il s'agit de l'acronyme le plus couramment utilis\'e pour d\'esigner le rayonnement fossile, et qu'il n'en existe pas de r\'epandu en fran\c{c}ais.

\begin{figure}
\begin{center}
\caption[R\'esum\'e de l'Histoire de l'Univers]{Un r\'esum\'e de l'Histoire de l'Univers d'apr\`es la th\'eorie du Big Bang et th\'eories modernes attenant.
Figure tir\'ee du site {\it http:// www.damtp.cam.ac.uk}.}
\label{histuniv}
\end{center}
\end{figure}

L'Histoire de l'Univers est aujourd'hui suppos\'ee avoir eu lieu comme sch\'ematis\'e en figure \ref{histuniv}.
Une th\'eorie int\'eressante pour l'Histoire de l'Univers est celle de l'inflation.
Celle-ci pr\'evoit que l'Univers est pass\'e par une phase d'expansion acc\'el\'er\'ee exponentielle (et non plus seulement de d\'ec\'el\'eration comme aujourd'hui), ce qui signifierait que des r\'egions aujourd'hui causalement ind\'ependantes (car il existe un horizon des \'ev\`enements et donc de l'Univers observable) ont \'et\'e causalement li\'ees avant l'inflation, et que leurs faibles anisotropies primordiales se sont retrouv\'ees apr\`es l'inflation.
Ceci expliquerait pourquoi l'on observe aujourd'hui une si grande isotropie dans le CMB, malgr\'e la non-causalit\'e aux grandes \'echelles.

\subsection{Param\`etres cosmologiques\label{params}}

Les \'equations de Friedmann introduisent des param\`etres cosmologiques,
qui d\'ependent de R(t), de ses d\'eriv\'ees, de la courbure k et
des densit\'es de mati\`ere et d'\'energie.
D\'ecrivons rapidement les plus importants:

\vspace*{1em}

- $\Omega$, le param\`etre de densit\'e totale de l'Univers, aujourd'hui (z = 0).
M\^eme si nous omettons l'indice 0 pour les param\`etres par commodit\'e, il s'agit bien des valeurs \`a l'\'epoque actuelle.
La valeur 1 est la densit\'e critique:
$\Omega$ = 1 signifie un univers plat, \`a g\'eom\'etrie euclidienne (k = 0), $\Omega < 1$ signifie un univers ouvert (l'Univers est infini dans ces deux cas), et $\Omega > 1$ un univers ferm\'e, \`a g\'eom\'etrie sph\'erique, donc fini.
Les exp\'eriences r\'ecentes concernant les fluctuations du rayonnement fossile favorisent clairement $\Omega \approx 1$ (voir par exemple de Bernardis \etal 2000).
Le param\`etre de densit\'e totale se d\'ecompose en plusieurs composantes d'\'energie:

\begin{equation}
\Omega = \Omega_m + \Omega_\Lambda = \Omega_{CDM} + \Omega_{bar} + \Omega_\Lambda
\end{equation}

\vspace*{1em}

- $\Omega_m$, le param\`etre de densit\'e effective.
Il s'agit de la fraction de la densit\'e critique sous forme de mati\`ere et d'\'energie.
La mati\`ere domine largement aujourd'hui.
Les exp\'eriences r\'ecentes concernant le rayonnement fossile et les supernov{\ae} favorisent $\Omega_m \approx 0.3$ (voir notamment Netterfield \etal 2002).

\vspace*{1em}

- $\Omega_{CDM}$, le param\`etre de densit\'e de mati\`ere noire froide ({\it Cold Dark Matter} soit CDM).
Ce terme de mati\`ere noire signifie ici: non baryonique.
Il s'agit de la fraction de la densit\'e critique sous forme de mati\`ere noire froide.
La mati\`ere habituelle est en effet constitu\'ee de baryons (protons, neutrons) et de leptons (\'electrons essentiellement).
Les baryons sont les particules qui dominent la masse de la mati\`ere traditionnelle, aussi qualifie-t-on simplement cette mati\`ere de ``baryonique''.
L'existence de mati\`ere invisible est invoqu\'ee pour expliquer le d\'esaccord entre les mesures de masse effectu\'ees d'apr\`es les objets que nous voyons dans l'Univers, et les mesures cosmologiques qui montrent des masses de mati\`ere bien sup\'erieures.
Des objets baryoniques invisibles peuvent \^etre invoqu\'es pour expliquer ce d\'esaccord (naines brunes et autres MACHOs, mais aussi agglom\'erats de quarks exotiques...), mais les mesures cosmologiques contraignent \'egalement la densit\'e baryonique totale, pas seulement celle qui est visible, et montrent qu'une autre mati\`ere doit exister pour expliquer les observations.
Cette mati\`ere noire pourrait \^etre compos\'ee de mati\`ere noire chaude ou HDM (neutrinos...), mais surtout de mati\`ere sombre froide (CDM) form\'ee de particules exotiques, extensivement imagin\'ees par les th\'eoriciens, mais jamais d\'etect\'ees (particules SUSY, photinos, axions...).
La mati\`ere sombre froide repr\'esente la grande majorit\'e de la densit\'e effective.

\vspace*{1em}

- $\Omega_{bar}$, le param\`etre de densit\'e de mati\`ere baryonique.
Il s'agit de la mati\`ere nucl\'eaire habituelle (protons, neutrons, antiprotons, atomes, mol\'ecules, \'etoiles, etc) mais aussi d'une \'eventuelle mati\`ere noire baryonique exotique (agglom\'erats de quarks...).
$\Omega_{bar}$ est la fraction de la densit\'e critique sous forme de mati\`ere baryonique.
Les exp\'eriences r\'ecentes favorisent $\Omega_{bar}$ de l'ordre de 0.03 \`a  0.1 \cite{netterfield02}.

\vspace*{1em}

- $\Lambda$, la constante cosmologique, est un postulat qui appara\^\i t naturellement dans les \'equations d'Einstein.
Son histoire est assez mouvement\'ee; en effet elle fut originellement introduite par Einstein dans les \'equations de la Relativit\'e G\'en\'erale, plus par n\'ecessit\'e philosophico-mystique d'avoir un Univers statique que par souci de g\'en\'eralit\'e math\'ematique,
puis elle fut oubli\'ee et jet\'ee aux orties lorsque les preuves de l'expansion de l'Univers arriv\`erent, pour \^etre ressortie du placard afin d'expliquer le d\'esaccord entre la densit\'e totale de l'Univers et la densit\'e de mati\`ere-\'energie.
Les exp\'eriences r\'ecentes concernant le rayonnement fossile et les supernov{\ae} favorisent $\Omega_{\Lambda} \approx 0.7$ (voir par exemple Netterfield \etal 2002).

\vspace*{1em}

- $H_0$, la constante de Hubble, est d\'efinie observationnellement par le rapport entre la vitesse de r\'ecession des galaxies et leur distance \cite{hubble31}, et th\'eoriquement par le rapport R'(t)/R(t).
Ce param\`etre qui s'exprime en unit\'e exotique de km/s/Mpc est reli\'e assez simplement \`a l'\^age de l'Univers (en premi\`ere approximation, l'\^age de l'Univers est l'inverse de $H_0$).
$H_0$ a une valeur mal connue entre environ 45 et 90 km/s/Mpc, ce qui correspond \`a un \^age de l'Univers d'environ 15 milliards d'ann\'ees.

\vspace*{1em}

- $q_0$, le param\`etre de d\'ec\'el\'eration de l'expansion de l'Univers \`a z=0.
q(t) est d\'efini par:

\begin{equation}
q(t) = - {R''(t) \: R(t) \over R'(t)^2}
\end{equation}


\vspace*{1em}

- n, l'indice spectral du spectre de puissance des fluctuations de densit\'e (voir section \ref{rfint}).

Ces param\`etres, avec quelques autres, sont souvent appel\'es les ``12 nombres''.
Leur d\'etermination pr\'ecise est l'un des objectifs essentiels de la cosmologie observationnelle du d\'ebut du XXI$^e$ si\`ecle, afin de pouvoir contraindre les th\'eories cosmologiques.

\subsection{Le rayonnement fossile\label{rfint}}
Pr\'edit en 1948 par Gamow (Gamow 1948{\it a} et Gamow 1948{\it b}), ce rayonnement est \'emis environ 300000 ans
apr\`es le Big Bang (z $\approx$ 1100).
Avec l'expansion, le fluide
cosmique se dilate et la temp\'erature du milieu diminue.
Lorsque l'\'energie des photons a diminu\'e jusqu'\`a une temp\'erature d'environ 4000 K, ceux-ci cessent d'\^etre ionisants et permettent que les protons et les \'electrons libres de l'Univers primordial se combinent pour former des atomes d'hydrog\`ene.
L'Univers devient alors transparent aux ondes \'electromagn\'etiques.
Le premier rayonnement libre \'emis dans l'Univers est donc le CMB, image de
la surface de derni\`ere diffusion lorsque la temp\'erature dans l'Univers
\'etait de l'ordre de 4000 K.
Auparavant, d'autres fossiles pourraient th\'eoriquement \^etre observ\'es
(rayonnement fossile de neutrinos et des ondes gravitationnelles) mais ne sont pas d\'etect\'es
aujourd'hui.
Le CMB est donc l'information la plus ancienne qu'il nous est donn\'e
d'observer de nos jours, et en tout cas la plus ancienne lumi\`ere observable
existante.
Ce rayonnement originellement chaud (environ 4000 K) a \'et\'e refroidi par l'expansion de l'Univers, et aujourd'hui il appara\^\i t comme un corps noir parfait de temp\'erature 2.728 $\pm$ 0.004
K, mesure effectu\'ee gr\^ace au satellite COBE (COsmic Background Explorer,
Boggess \etal 1992, voir figure \ref{cncmb}).
Cette temp\'erature correspond \`a une \'emission maximale \`a la longueur
d'onde de 1.87 mm.
L'Univers \'etant extr\^emement homog\`ene \`a l'\'epoque de la surface de derni\`ere diffusion,
le CMB est lui-m\^eme tr\`es isotrope: on l'observe dans toutes les directions
avec la m\^eme intensit\'e et les m\^emes caract\'eristiques spectrales, \`a
ceci pr\`es que le mouvement local de l'observateur (la Terre) par rapport au
fluide cosmique introduit un dip\^ole (le dip\^ole cosmologique), et que de
tr\`es faibles fluctuations (10$^{-4}$) sont observables.
Nous pr\'esentons les caract\'eristiques de ces fluctuations, ainsi que notre
analyse des donn\'ees Archeops, au chapitre \ref{rf}.
Le CMB est, par d\'efinition, \`a l'arri\`ere-plan de tout autre rayonnement
\'electromagn\'etique observable,
aussi son observation est-elle sujette \`a toutes sortes de pollutions
astrophysiques, aussi honorables que la poussi\`ere galactique.
Il est donc clair que les observations du continuum galactique et du CMB vont de
pair dans nombre d'exp\'eriences astronomiques.

\begin{figure}
\begin{center}
\caption[Le spectre de corps noir du CMB]{Le spectre de corps noir du CMB comme vu par le satellite COBE (principalement).
Image tir\'ee du site {\it http:// aether.lbl.gov/www/projects/cobe}.}
\label{cncmb}
\end{center}
\end{figure}

Pour conclure cette section, nous pouvons dire que l'\`ere de la cosmologie de pr\'ecision n'est
pas encore entr\'ee dans son \^age d'or.
La mesure des param\`etres cosmologiques est encore entach\'ee d'erreurs
larges, d'incertitudes syst\'ematiques et de d\'eg\'en\'erescences
fondamentales.
L\`a encore, l'exploration de l'Univers froid ne fait que commencer.

\section{ Instrumentation du continuum (sub-)millim\'etrique\label{instr}}
Mesurer ces rayonnements continuum caract\'erisant l'Univers froid
est particuli\`erement d\'elicat.
En effet, l'absorption du rayonnement millim\'etrique par l'atmosph\`ere ne
permet d'observer efficacement au sol que dans quelques fen\^etres de
transmission (voir figure \ref{figtatm}).
L'observation de rayonnements continuum est donc assez limit\'ee depuis le
sol, c'est pourquoi l'observation stratosph\'erique et spatiale s'est
particuli\`erement d\'evelopp\'ee dans ce domaine.
L'observation spatiale gr\^ace aux satellites permet de s'affranchir compl\`etement des
contaminations atmosph\'eriques, et plus ou moins efficacement de toute
contamination provenant de la Terre, suivant les caract\'eristiques de
l'orbite du satellite.
L'observation en ballon stratosph\'erique permet de s'affranchir efficacement
des contaminations atmosph\'eriques pour un co\^ut et un temps de
d\'eveloppement du projet tr\`es inf\'erieurs.
En revanche, le temps d'observation en ballon est habituellement tr\`es court
(quelques heures ou quelques jours) en comparaison d'une exp\'erience
satellite, et les exp\'eriences ballon sont souvent con\c{c}ues
pour d\'efricher le terrain inconnu d'un projet satellite.
Mais de ce fait, les exp\'eriences ballons jouent un r\^ole pionnier dans
l'exploration du ciel (sub-)millim\'etrique et ont permis des d\'ecouvertes
majeures dans ce domaine.

\begin{figure}
\begin{center}
\caption[Transmission atmosph\'erique]{Transmission atmosph\'erique des ondes \'electromagn\'etiques.
Image tir\'ee du site {\it http://violet.pha.jhu.edu/.wpb/spectroscopy}.}
\label{figtatm}
\end{center}
\end{figure}

Les instruments permettant d'observer le rayonnement
submillim\'etrique/millim\'etrique ont une optique proche de celle permettant
d'observer dans le visible: ce sont des t\'elescopes, c'est-\`a-dire des
syst\`emes optiques poss\'edant un miroir primaire qui focalise le rayonnement
re\c{c}u vers un miroir secondaire plus petit, lui-m\^eme focalisant le
rayonnement vers le plan focal o\`u sont install\'es les d\'etecteurs.
Cependant, les instruments modernes poss\`edent souvent des optiques plus complexes utilisant des miroirs suppl\'ementaires.
Les d\'etecteurs permettant d'observer efficacement un continuum infrarouge lointain et
millim\'etrique sont les bolom\`etres (voir figure \ref{figbol}).
Ces d\'etecteurs fonctionnent suivant le principe suivant: le rayonnement
chauffe un \'el\'ement absorbeur reli\'e \`a un thermom\`etre dont la
r\'esistance \'electrique varie avec la temp\'erature.
Ces variations de r\'esistance sont proportionnelles aux variations de la tension
\'electrique aux bornes du bolom\`etre lorsqu'on le soumet \`a un courant.
Les variations de tension ainsi mesur\'ees sont proportionnelles (dans le
r\'egime utilis\'e pour observer) au flux re\c{c}u.
La d\'etection r\'ealis\'ee avec des bolom\`etres est tr\`es sensible mais a tendance g\'en\'eralement \`a produire du bruit corr\'el\'e dans le temps, en plus du bruit blanc (sans corr\'elations).
Ce bruit corr\'el\'e, aussi appel\'e bruit rouge ou bruit en 1/f car il a tendance \`a augmenter quand la fr\'equence diminue, en suivant une loi de puissance, est un probl\`eme pour le traitement des donn\'ees car il peut cr\'eer des stries et autres art\'efacts sur les cartes si l'on ne prend pas en compte les propri\'et\'es statistiques de ce bruit dans le processus de construction de cartes.
Aussi ce processus est-il une \'etape d\'ecisive et d\'elicate du traitement des donn\'ees d'exp\'eriences bolom\'etriques.

\begin{figure}
\begin{center}
\caption[Bolom\`etres]{Bolom\`etres install\'es avec leurs cornets sur le plan focal de l'exp\'erience Archeops (voir section \ref{archeops}).}
\label{figbol}
\end{center}
\end{figure}

\chapter{La construction de cartes\label{chapconst}}
{\flushright\it Wanderer hei\ss t mich die Welt\\
weit wandert' ich schon\\
auf der Erde R\"ucken\\
r\"uhrt' ich mich viel\\
\vspace*{1em}
Le Voyageur m'appelle le monde\\
Beaucoup ai-je d\'ej\`a voyag\'e\\
Sur la face de la Terre\\
J'ai fait mon chemin\\}
\vspace*{1em}
{\flushright Richard Wagner, {\it Siegfried}\\}

\section{ Introduction}
Nous nous sommes particuli\`erement int\'eress\'es \`a la construction de
cartes, qui est un
probl\`eme math\'ematique relativement simple \`a poser mais moins \'evident
\`a r\'esoudre et \`a appliquer dans la pratique.

Comme pour notre bonne vieille Terre, la connaissance pr\'ecise de notre ciel
n\'ecessite d'abord de conna\^\i tre sa g\'eographie.
Lorsque l'on observe le ciel, nous sommes au centre d'une sph\`ere que nous
observons de l'int\'erieur.
De ce fait, la cartographie c\'eleste est tr\`es similaire \`a la cartographie
terrestre, \`a ceci pr\`es que les points cardinaux est et ouest sont
invers\'es, et que bien s\^ur il n'y a pas un axe des p\^oles \'evident, qui
s'impose dans toute cartographie.
C'est pourquoi plusieurs topologies de la sph\`ere c\'eleste sont utilis\'ees,
en fonction de quels points du ciel sont fix\'es pour \^etre les p\^oles: coordonn\'ees locales,
\'equatoriales, \'ecliptiques ou galactiques.
Ayant donc choisi un syst\`eme de coordonn\'ees, il faudra encore choisir une
projection si l'on veut visualiser une carte sur une surface plane.
En ce qui concerne les cartes d'une petite portion du ciel, la projection
tangente suffira, pour les cartes de tout le ciel la projection Mollweide (par
exemple) sera n\'ecessaire.

La sph\`ere c\'eleste a \'et\'e observ\'ee depuis la plus haute Antiquit\'e dans le
domaine visible, gr\^ace \`a l'{\oe}il humain tout d'abord, puis \`a divers
instruments.
Parall\`element, la volont\'e de cartographier le ciel a toujours \'et\'e
extr\^emement forte, de la m\^eme fa\c{c}on que pour la surface de la Terre,
mais pas pour les m\^emes motivations.
Il suffira de jeter un coup d'oeil \`a une carte des constellations, et d'observer le nombre
d'\'etoiles nomm\'ees par les anciens (Arabes) pour se convaincre de la
volont\'e intemporelle de cartographier la vo\^ute c\'eleste.
Les progr\`es de l'astronomie moderne ont conduit \`a une connaissance
multi-longueur d'onde du ciel, qui traduit la diversit\'e in\'epuisable des
observations astronomiques.
Cependant, la complexification de l'instrumentation introduit aussi une
distance plus grande entre l'observation et la compr\'ehension imm\'ediate par
l'Homme, c'est-\`a-dire, en g\'en\'eral, la carte.
Les observations modernes, notamment dans le domaine millim\'etrique, peuvent
faire appel \`a des proc\'edures d'observation qui demandent un traitement
pouss\'e des donn\'ees pour obtenir une carte satisfaisante.
En particulier, l'observation en balayage du ciel n\'ecessite de reconstruire
la carte \`a partir de donn\'ees temporelles \`a une dimension, contenant du
bruit dont les caract\'eristiques s'expriment \`a une dimension, etc.
Passer des donn\'ees temporelles \`a la carte en traitant le bruit de fa\c{c}on
optimale, ou au moins correcte, est un probl\`eme d'inversion que nous traitons de mani\`ere lin\'eaire en faisant appel aux caract\'eristiques de l'instrument et aux
propri\'et\'es statistiques du bruit et du signal observ\'e.
Nous pr\'esentons la th\'eorie de la construction de cartes en section suivante.

\section{ Th\'eorie de l'inversion lin\'eaire\label{invlin}}

\subsection{Position du probl\`eme}

L'observation d'un signal sur le ciel par un instrument peut \^etre
mod\'elis\'ee en supposant que la r\'eponse instrumentale est lin\'eaire, ce
qui est le cas pour les instruments qui nous int\'eressent, et qui est une propri\'et\'e g\'en\'eralement requise.
Soit {\bf x} un vecteur repr\'esentant le vrai ciel que l'on observe: dans le cas
d'une carte d'une partie du ciel (ou de tout le ciel), ce peut \^etre un
vecteur repr\'esentant une grandeur \`a deux dimensions.
Soit {\bf A} la matrice repr\'esentant la convolution effectu\'ee par l'instrument
sur le ciel: il s'agit pour nous du passage de 2-D \`a 1-D (autrement dit: le
pointage sur le ciel), la convolution par la g\'eom\'etrie du faisceau,
\'eventuellement l'observation diff\'erentielle du ciel, etc.
Ces effets sont mod\'elisables par une matrice connue.
Ce n'est pas le cas du bruit, que l'on mod\'elise par un vecteur {\bf n}
additif.
Les donn\'ees instrumentales ordonn\'ees dans le temps sont alors repr\'esent\'ees par un vecteur {\bf y}:

\begin{equation}
{\bf y} = {\bf A \: x} + {\bf n}
\label{obs}
\end{equation}

L'inversion lin\'eaire consiste \`a rechercher une estimation {\bf \~x} du vrai ciel {\bf x} comme d\'erivant lin\'eairement des observations {\bf y}:

\begin{equation}
{\bf \Tilde x} = {\bf W \: y}
\label{const}
\end{equation}

o\`u {\bf W} est la matrice de reconstruction.
Notons que ce formalisme est tout \`a fait universel et qu'il peut s'appliquer
\`a tout autre chose que la construction de cartes.
En fait, tout effet instrumental lin\'eaire peut \^etre mod\'elis\'e de cette fa\c{c}on.

Comment estimer la matrice {\bf W} optimale ?
Pour cela, il faut chercher \`a minimiser l'\'ecart entre le ciel reconstruit et le vrai ciel.
Ceci peut \^etre r\'ealis\'e directement sur le ciel, on minimise alors l'erreur de reconstruction $<|\epsilon|^2>$:

\begin{equation}
\epsilon = {\bf \Tilde x} - {\bf x} = [{\bf W \: A} - {\bf I}] \: {\bf x} + {\bf W \: n}
\label{epsilon}
\end{equation}

Mais cela peut aussi \^etre effectu\'e sur les mesures en minimisant le $\chi^2$:

\begin{equation}
\chi^2 = ({\bf y} - {\bf A \: \Tilde x})^t \: {\bf N}^{-1} \: ({\bf y} - {\bf
  A \: \Tilde x}) = ({\bf y} - {\bf A \: W \: y})^t \: {\bf N}^{-1} \: ({\bf
  y} - {\bf A \: W \: y})
\label{chi2}
\end{equation}

o\`u les signes $<$ $>$ d\'enotent une moyenne statistique, {\bf I} est la
matrice identit\'e et {\bf N} est la matrice de covariance du bruit dans les donn\'ees
temporelles:

\begin{equation}
{\bf N} = <{\bf n \: n}^t>
\end{equation}

\subsection{La m\'ethode COBE}
La minimisation du $\chi^2$ par rapport \`a la matrice {\bf W} (d\'eriver
l'\'equation \ref{chi2} par rapport \`a {\bf W} puis \'egaler \`a z\'ero) conduit
imm\'ediatement \`a:

\begin{equation}
{\bf W} = [{\bf A}^t \: {\bf N}^{-1} \: {\bf A}]^{-1} \: {\bf A}^t \: {\bf N}^{-1}
\label{cob}
\end{equation}

Il s'agit de la m\'ethode de construction de cartes dite COBE, car elle a \'et\'e
utilis\'ee (entre autres) pour la construction des cartes du CMB de cette exp\'erience \cite{janssen92}.
Autres propri\'et\'es int\'eressantes de cette m\'ethode: elle minimise
$<|\epsilon|^2>$ sous la condition que ${\bf W \: A} = {\bf I}$, et elle
correspond au maximum de vraisemblance pour {\bf x} d'une analyse bay\'esienne
si {\bf n} est une variable al\'eatoire gaussienne.
Cette m\'ethode est la meilleure si l'on n'a pas d'{\it a priori} sur la
statistique du signal.
Il faut signaler aussi qu'aucune hypoth\`ese n'a \'et\'e faite sur une
\'eventuelle corr\'elation entre le bruit et le signal.

Notons enfin que dans le cas o\`u la matrice {\bf A} est inversible (donc au moins carr\'ee, ce qui veut dire que l'on ne r\'eduit pas la quantit\'e d'information en passant des TOI \`a la carte), la matrice {\bf W} se r\'eduit \`a {\bf A}$^{-1}$, ce qui signifie que l'information statistique sur le bruit est inutilisable.

\subsection{La m\'ethode de Wiener\label{introwie}}
La minimisation de l'erreur de reconstruction $<|\epsilon|^2>$ conduit
imm\'ediatement \`a:

\begin{equation}
{\bf W} = <{\bf x \: y}^t> \: <{\bf y \: y}^t>^{-1}
\end{equation}

c'est-\`a-dire, en faisant au passage l'hypoth\`ese de l'ind\'ependance entre
le bruit et le signal:

\begin{equation}
{\bf W} = {\bf S \: A}^t \: [{\bf A \: S \: A}^t + {\bf N}]^{-1}
\label{wie1}
\end{equation}

o\`u {\bf S} est la matrice de covariance du signal dans le domaine de la carte:

\begin{equation}
{\bf S} = <{\bf x \: x}^t>
\end{equation}

L'\'equation \ref{wie1} est la forme la plus courante du filtre de Wiener \cite{wiener49}.
Nous l'appellerons m\'ethode Wiener 1, suivant ainsi la notation de Tegmark (1997).
Cette m\'ethode a une autre forme matricielle que l'on obtient \`a partir de
l'\'equation \ref{wie1} de la fa\c{c}on suivante:

\begin{equation}
{\bf W} = {\bf S \: A}^t \: [{\bf I} + {\bf N}^{-1} \: {\bf A \: S \:
  A}^t]^{-1} \: {\bf N}^{-1}
\end{equation}

L'inversion entre crochets peut \^etre d\'evelopp\'ee en sa s\'erie
de Taylor:

\begin{equation}
{\bf W} = {\bf S \: A}^t \: [{\bf I} - {\bf N}^{-1} \: {\bf A \: S \:
  A}^t + {\bf N}^{-1} \: {\bf A \: S \: A}^t {\bf N}^{-1} \: {\bf A \: S \:
  A}^t - \: ...] \: {\bf N}^{-1}
\end{equation}

En d\'eveloppant ${\bf S \: A}^t$ \`a gauche et en le factorisant \`a droite, on
tombe sur une autre s\'erie qui donne:

\begin{equation}
{\bf W} = [{\bf I} + {\bf S \: A}^t \: {\bf N}^{-1} \: {\bf A}]^{-1} \: {\bf S
  \: A}^t \: {\bf N}^{-1}
\end{equation}

Soit encore:
\begin{equation}
{\bf W} = [{\bf S}^{-1} + {\bf A}^t \: {\bf N}^{-1} \: {\bf A}]^{-1} \: {\bf
  A}^t \: {\bf N}^{-1}
\label{wie2}
\end{equation}

Cette forme matricielle, bien qu'identique \`a celle de l'\'equation
\ref{wie1}, a une forme plus proche de la m\'ethode COBE (\'eq. \ref{cob}).
Nous l'appellerons m\'ethode Wiener 2.

Comme nous l'avons vu, la m\'ethode de Wiener minimise l'erreur de
reconstruction (voir aussi le travail en harmoniques sph\'eriques par Bunn
\etal 1994).
Elle correspond \'egalement au maximum de vraisemblance pour {\bf x} d'une
analyse bay\'esienne si {\bf n} et {\bf x} sont des variables al\'eatoires
gaussiennes \cite{zaroubi95}.
Autrement dit, la m\'ethode de construction de cartes de Wiener est la
meilleure solution si l'on sait {\it a priori} que le signal est gaussien, et
que le bruit est gaussien et ind\'ependant du signal.

\subsection{La m\'ethode du moyennage par pixel\label{intpixavg}}
Si l'on ne souhaite pas (ou que l'on ne peut pas) traiter le bruit pendant la construction de cartes,
la m\'ethode du moyennage par pixel peut \^etre utilis\'ee.
Il s'agit d'additionner les observations dans chaque pixel et de diviser par
le nombre d'observations dans le pixel.
Le bruit dans les pixels est ainsi r\'eduit de la racine du nombre
d'observations par pixel (i.e. le poids ou la redondance), mais les corr\'elations \'eventuelles du bruit ne
sont pas utilis\'ees pour r\'eduire ce bruit efficacement.
Cependant, cette m\'ethode peut \^etre efficace si les corr\'elations du bruit
sont faibles ou que la redondance est grande.
Le moyennage par pixel s'\'ecrit:

\begin{equation}
{\bf W} = [{\bf A}^t \: {\bf A}]^{-1} \: {\bf A}^t
\label{pixelavg}
\end{equation}

Cette forme correspond \`a la solution optimale du probl\`eme de construction de cartes en pr\'esence d'un bruit blanc (non autocorr\'el\'e) uniforme.
En effet, elle peut \^etre d\'eriv\'ee de l'\'equation \ref{cob} en consid\'erant que la matrice {\bf N} est proportionnelle \`a l'identit\'e.
Si l'on applique strictement cette \'equation, on moyenne par pixel en
d\'econvoluant des effets que l'on a inclus dans {\bf A} (observation
diff\'erentielle par exemple).
Si l'on n'inclut dans {\bf A} que le pointage sur le ciel, on ne fait que construire une carte
en moyennant les observations dans les pixels.
Si l'on a un pixel pour un \'echantillon de donn\'ees temporelles, et que l'on
n'inclut que le pointage dans {\bf A}, on ne fait
que passer des donn\'ees \`a la carte en affectant chaque point \`a un pixel (${\bf W} = {\bf A}^{-1}$).

\subsection{M\'ethodes lin\'eaires avanc\'ees\label{metadv}}
En g\'en\'eralisant les m\'ethodes pr\'ec\'edemment pr\'esent\'ees, il est possible de choisir l'influence de la corr\'elation du signal avec la m\'ethode suivante:

\begin{equation}
{\bf W} = [\eta \: {\bf S}^{-1} + {\bf A}^t \: {\bf N}^{-1} \: {\bf A}]^{-1} \: {\bf
  A}^t \: {\bf N}^{-1}
\label{sask}
\end{equation}

qui s'\'ecrit aussi:

\begin{equation}
{\bf W} = {\bf S \: A}^t \: [{\bf A \: S \: A}^t + \eta \: {\bf N}]^{-1}
\label{sask1}
\end{equation}

o\`u $\eta$ est un coefficient scalaire.
Cette m\'ethode a \'et\'e utilis\'ee pour la construction des cartes de l'exp\'erience CMB Saskatoon \cite{tegmarketal97}, nous l'appellerons donc m\'ethode Saskatoon (\'eq. \ref{sask}).
Le cas $\eta$ = 0 correspond \`a la m\'ethode optimale sans {\it a priori} sur le signal (m\'ethode COBE, \'eq. \ref{cob}), tandis que le cas $\eta$ = 1 correspond \`a la m\'ethode optimale avec {\it a priori} gaussien sur le signal, c'est-\`a-dire au filtre de Wiener (\'eq. \ref{wie2}).
En fait, la m\'ethode de Wiener produit des cartes en g\'en\'eral moins bruit\'ees que la m\'ethode COBE, en r\'eduisant la puissance dans les pixels de fa\c{c}on in\'egale.
Ceci induit un rapport signal sur bruit donn\'e dans la carte reconstruite, qui est optimal par rapport \`a la minimisation de l'erreur de reconstruction, dans le cas de la m\'ethode de Wiener.
Puisque le bruit est plus important aux petites \'echelles (par d\'efinition, les grandes \'echelles b\'en\'eficient de plus d'observations donc ont un bruit plus r\'eduit), la m\'ethode de Wiener lisse la carte de fa\c{c}on ad\'equate.
Cependant, il est int\'eressant de pouvoir supposer un rapport signal sur bruit diff\'erent afin par exemple de faire ressortir des petites \'echelles.
La m\'ethode Saskatoon est adapt\'ee \`a ce genre d'investigations (pour $\eta$ positif inf\'erieur \`a 1).

Puisque la m\'ethode de Wiener r\'eduit la puissance in\'egalement dans les pixels, il a \'et\'e propos\'e notamment par Tegmark \& Efstathiou (1996) la m\'ethode suivante, afin de r\'e-\'echelonner chaque pixel:

\begin{equation}
{\bf W} = {\bf \Lambda \: S \: A}^t \: [{\bf A \: S \: A}^t + {\bf N}]^{-1}
\label{rechwie1}
\end{equation}

soit aussi:

\begin{equation}
{\bf W} = {\bf \Lambda} \: [{\bf S}^{-1} + {\bf A}^t \: {\bf N}^{-1} \: {\bf A}]^{-1} \: {\bf
  A}^t \: {\bf N}^{-1}
\label{rechwie2}
\end{equation}

Ces deux \'equations correspondent aux deux formes de la m\'ethode de Wiener (\ref{wie1} et \ref{wie2}) auxquelles ont met en facteur une matrice diagonale ${\bf \Lambda}$ qui r\'e-\'echelonne chaque pixel de la carte de sorte que ${\bf W \: A} = {\bf I}$.
Ceci permet d'\'eviter le biais de la m\'ethode de Wiener qui enl\`eve de la puissance dans les pixels bruit\'es.
Cette m\'ethode peut \^etre d\'eriv\'ee en minimisant l'erreur de reconstruction sous la condition que ${\bf W \: A} = {\bf I}$ \cite{tegmark96}.
Les coefficients de la diagonale de ${\bf \Lambda}$ sont:

\begin{equation}
{\bf \Lambda}_{ii} = {1 \over ({\bf A}^t \: [{\bf A \: S \: A}^t + {\bf N}]^{-1} \: {\bf A})}_{ii}
\end{equation}

Bien s\^ur, il est aussi possible de combiner les m\'ethodes Saskatoon (\'eq. \ref{sask}, \'eq. \ref{sask1}) et Wiener r\'e-\'echelonn\'e (\'eq. \ref{rechwie1}, \'eq. \ref{rechwie2}).

Pour conclure, nous voyons qu'appliquer l'une de ces m\'ethodes \'elabor\'ees (COBE: \'eq. \ref{cob}, Wiener: \'eq. \ref{wie1}, \ref{wie2}, Saskatoon et Wiener r\'e-\'echelonn\'e) n\'ecessite de conna\^\i tre la corr\'elation du bruit dans les donn\'ees ({\bf N}) et \'eventuellement celle du signal dans la carte ({\bf S}).
Estimer correctement la matrice de covariance du bruit n'est pas forc\'ement imm\'ediat, de m\^eme qu'appliquer des inversions matricielles \`a des trains de donn\'ees de grande taille, comme c'est le cas pour les exp\'eriences actuelles sur le rayonnement fossile.
Chaque instrument, chaque objet astrophysique, a diff\'erentes caract\'eristiques qui font que l'application des m\'ethodes optimales de construction de cartes est un probl\`eme en lui-m\^eme pour chaque exp\'erience.


\chapter{La poussi\`ere galactique\label{pous}}
{\flushright\it All that is gold does not glitter.\\
Tout ce qui est or ne brille pas. \\}
{\flushright J.R.R. Tolkien, {\it The Lord of the Rings}\\}
\vspace*{1em}

\section{ Introduction sur la poussi\`ere galactique\label{secpous}}

\subsection{G\'en\'eralit\'es\label{pousgen}}

\subsubsection{Introduction}

Composante minoritaire du milieu interstellaire (environ 1 \% en masse pour 99 \% de gaz, atomique ou mol\'eculaire), la poussi\`ere comprend une grande vari\'et\'e de petits corps solides, depuis les grosses mol\'ecules (quelques nanom\`etres) jusqu'aux gros grains (quelques microm\`etres).
Dans le bestiaire intersid\'eral, la poussi\`ere se place donc entre les objets microscopiques (particules, atomes, mol\'ecules) et les objets macroscopiques (m\'et\'eorites, ast\'ero\"\i des, plan\`etes, \'etoiles).
Faisant le lien entre ces deux mondes, les poussi\`eres ou grains sont des corps de nature chimique tr\`es diverse, qui ont la propri\'et\'e d'absorber efficacement les rayonnements ultraviolet et visible.
L'absorption des poussi\`eres dans le visible donne aux nuages mol\'eculaires leur aspect sombre qui se d\'etache sur l'\'emission brillante des r\'egions ionis\'ees qui les entourent et des \'etoiles situ\'ees en arri\`ere-plan.
Les propri\'et\'es d'absorption de ces grains sont expliqu\'ees en premi\`ere approximation par l'optique physique gr\^ace \`a la th\'eorie de Mie.
L'\'emission des grains d\'epend de leur taille et de leur nature chimique.
Plus les grains sont gros, plus les ph\'enom\`enes d'interaction quantique mati\`ere - rayonnement sont remplac\'es par l'\'equilibre thermodynamique et une \'emission continue caract\'eris\'ee par une temp\'erature.

\subsubsection{Consid\'erations th\'eoriques}

Suivant les diff\'erents mod\`eles (Mathis \etal 1977, Draine \& Lee 1984, Draine \& Anderson 1985, D\'esert \etal 1990, Li \& Greenberg 1997, Dweck \etal 1997, Draine \& Li 2001...), les grains sont class\'es en plusieurs cat\'egories en fonction de leurs tailles et de leurs caract\'eristiques chimiques.
Les observations de l'\'emission infrarouge et de l'extinction ultraviolet - optique du milieu interstellaire contraignent les mod\`eles de grains, de sorte qu'aujourd'hui l'on consid\`ere de fa\c{c}on assez consensuelle que les grains sont classables en trois composantes (ou plus), qui permettent d'expliquer \`a la fois l'extinction et l'\'emission: de tr\`es petits grains carbon\'es porteurs des bandes aromatiques, d'autres petits grains carbon\'es, et de gros grains silicat\'es et poss\'edant \'eventuellement des manteaux de glaces.
De grandes incertitudes subsistent sur la nature majoritaire des grains interstellaires, notamment concernant la structure des petits grains carbon\'es (le graphite est favoris\'e pour expliquer la bosse d'extinction \`a 2175 \AA, mais \'etant difficile \`a former dans le milieu interstellaire, d'autres compos\'es carbon\'es sont \'evoqu\'es tels que le carbone amorphe), et les porteurs des bandes aromatiques (PAH ou charbons).

Si l'on r\'esume les choses en s'en tenant au mod\`ele embl\'ematique de D\'esert \etal (1990), les grains se classent en trois cat\'egories: les PAH, les tr\`es petits grains et les gros grains.
Les PAH (Polycyclic Aromatic Hydrocarbons ou hydrocarbures aromatiques polycycliques) sont de grosses mol\'ecules carbon\'ees compos\'ees de noyaux benz\'eniques, attach\'es les uns aux autres pour former une structure plane de diam\`etre de l'ordre du nanom\`etre.
Ils repr\'esentent environ 6 \% de la masse de grains dans le mod\`ele de D\'esert \etal (1990).
Les tr\`es petits grains sont graphitiques, sph\'eriques, d'un diam\`etre d'environ 10 nanom\`etres, et \'emettent dans l'infrarouge moyen entre environ 20 et 80 $\mic$.
Ils sont \'egalement responsables de la bosse d'extinction \`a 2175 \AA, et repr\'esentent environ 6 \% de la masse de grains.
Les PAH et les tr\`es petits grains se comportent de fa\c{c}on quantique: leur \'energie fluctue en fonction des photons qu'ils absorbent et \'emettent.

Ce n'est pas le cas des gros grains, qui, d'apr\`es le mod\`ele de D\'esert \etal (1990), sont des particules silicat\'ees et carbon\'ees d'environ 100 nm de diam\`etre.
Leur taille macroscopique leur permet d'avoir une densit\'e d'\'etat d'\'energie continue.
Leur loi d'\'emission est habituellement mod\'elis\'ee par un corps noir modifi\'e de la mani\`ere suivante (loi de Planck modifi\'ee):

\begin{equation}
\inu(\lambda,T,\beta)=C \: \bnu(\lambda,T) \: \lambda^{-\beta}
\label{cnm}
\end{equation}

o\`u $\inu$ est l'intensit\'e spectrale, $\bnu$ la loi de Planck, $\lambda$ la longueur d'onde, C une constante, T la temp\'erature (Kelvin) et $\beta$ l'indice spectral des grains (sans unit\'e).
Les gros grains sont responsables de l'\'emission continuum du ciel dans l'infrarouge lointain et le submillim\'etrique entre environ 80 et 900 $\mic$ (au-del\`a de 900 $\mic$ c'est le rayonnement fossile qui domine dans la majeure partie du ciel).
La composition des gros grains habituellement invoqu\'ee est essentiellement silicat\'ee (MgSiO$_3$, SiO$_2$, MgFeSiO$_4$...), avec aussi du carbone.
Des manteaux de glace volatils (H$_2$O, NH$_3$, H$_2$CO, CO$_2$...) peuvent se former sur les gros grains lorsque les conditions physiques du milieu le permettent (temp\'eratures basses...).
Le mod\`ele de D\'esert \etal (1990) ne prend pas en compte les manteaux de glace \'eventuels des grains, mais certains mod\`eles les prennent en compte, tel celui d'Ossenkopf \& Henning (1994) concernant les c{\oe}urs protostellaires.
D'autres mod\`eles plus exotiques existent, y compris celui de Hoyle \& Wickramasinghe (1988) portant sur des bact\'eries congel\'ees (!).


\subsubsection{Observations de la poussi\`ere galactique\label{obspousgal}}

Notre connaissance de l'\'emission continuum des grains a \'et\'e marqu\'ee par les observations du satellite IRAS (InfraRed Astronomical Satellite, 1983-84, {\it http:// www.ipac.caltech.edu/ipac/ iras/iras.html}), qui a cartographi\'e la quasi-totalit\'e du ciel dans quatre bandes infrarouges larges autour de 12, 25, 60 et 100 $\mic$ de longueurs d'onde.
Cette mission a marqu\'e un progr\`es d\'ecisif pour toute l'astronomie.
Par exemple, elle a permis d'accro\^\i tre de 70 \% le nombre de sources astronomiques connues en en d\'etectant 350000.
Ses cartes dans les longueurs d'onde infrarouges font r\'ef\'erence depuis pr\`es de vingt ans.
Le satellite COBE (COsmic Background Explorer, Boggess \etal 1992), autre observation de tout le ciel, con\c{c}u principalement pour l'observation du rayonnement fossile (voir chapitre \ref{rf}), a \'egalement permis la premi\`ere exploration \`a grande \'echelle de l'\'emission des grains, gr\^ace aux instruments DIRBE et FIRAS.
Ce satellite poss\'edait en effet trois instruments: DIRBE, Diffuse InfraRed Background Experiment, d\'etectant depuis l'infrarouge proche (1.2 $\mic$) jusqu'au submillim\'etrique (300 $\mic$) avec 40 minutes d'arc de r\'esolution angulaire, FIRAS, Far InfraRed Absolute Spectrophotometer, de 100 $\mic$ \`a 1 cm, avec 7 $^o$ de r\'esolution angulaire, et DMR, Differential Microwave Radiometer, de 3.3 mm \`a 9.5 mm avec 7$^o$ de r\'esolution.
Aucune carte \`a grande \'echelle n'existait avant COBE dans les longueurs d'onde au-del\`a de 100 $\mic$, aussi ce satellite marqua-t-il un progr\`es consid\'erable de l'astrophysique, outre la tr\`es importante d\'ecouverte des fluctuations du rayonnement fossile (voir section \ref{rfobs}).

Les projets Herschel ({\it http:// www.sci.esa.int/ home/herschel}) et Planck ({\it http:// www.sci.esa.int/ home/planck} ou Tauber 2000), en cours de pr\'eparation, fourniront une nouvelle g\'en\'eration de connaissances \`a l'astrophysique de la poussi\`ere galactique.

Entretemps, l'observation au sol et en ballons stratosph\'eriques tels qu'AGLAE \cite{caux86} et AROME
\cite{giard88} a permis de mieux conna\^\i tre le ciel infrarouge et les propri\'et\'es des grains.
L'exp\'erience AGLAE fournit en effet la premi\`ere cartographie quasi-compl\`ete du plan galactique dans le domaine submillim\'etrique (380 $\mic$).
Auparavant, de nombreuses observations infrarouge \`a grande \'echelle avaient n\'eanmoins eu lieu (voir notamment la revue d'Okuda 1981, Hayakawa \etal 1981 et Hauser \etal 1984).
L'exp\'erience AROME d\'etecta l'\'emission \`a 3.3 $\mic$ dans le milieu diffus galactique.

Le lecteur d\'esireux de parfaire sa connaissance des grains pourra se reporter par exemple \`a Stepnik (2001).

L'exp\'erience PRONAOS, que nous d\'ecrivons en section \ref{pronaos}, est en train d'apporter une contribution d\'ecisive \`a la connaissance des propri\'et\'es physiques des grains et de la structure du milieu interstellaire dense, en ayant cartographi\'e dans quatre bandes spectrales submillim\'etriques de nombreuses zones du milieu interstellaire.
Nous nous sommes particuli\`erement int\'eress\'es aux complexes de formation d'\'etoiles massives que sont Orion et M17.
Ces r\'egions contiennent une grande vari\'et\'e de milieux, depuis les zones ionis\'ees jusqu'aux c{\oe}urs denses protostellaires.
Elles sont donc particuli\`erement complexes, et font appara\^\i tre des contrastes d'intensit\'e tr\`es importants, ce qui induit des difficult\'es pour la reconstruction des images.
Nous pr\'esentons succintement dans les sections suivantes l'\'etat des connaissances sur ces r\'egions avant l'apport de PRONAOS.


\subsection{Le complexe mol\'eculaire d'Orion\label{introorion}}
Situ\'ee \`a une distance d'environ 470 parsecs (pc), la N\'ebuleuse d'Orion Messier 42, historiquement appel\'ee Grande N\'ebuleuse d'Orion, est une r\'egion ionis\'ee (H II) associ\'ee \`a des nuages mol\'eculaires.
Elle fait partie du complexe mol\'eculaire g\'eant d'Orion, qui a \'et\'e r\'ev\'el\'e par des cartes d'\'emission des raies de rotation de la mol\'ecule $^{12}$CO \cite{kutner77}.
La carte CO de Maddalena \etal (1986) a montr\'e la r\'epartition \`a grande \'echelle du gaz mol\'eculaire dans le complexe g\'eant Orion-Licorne.
Cette r\'egion de milieu interstellaire dense s'\'etend sur approximativement 30 degr\'es du sud-est au nord-ouest.
Elle est form\'ee de plusieurs zones distinctes, la plus au nord \'etant la r\'egion gazeuse r\'epartie autour de l'\'etoile $\lambda$ Orionis.
Au sud, les nuages mol\'eculaires g\'eants d'Orion A (voir figure \ref{figorionge}) et B s'\'etendent sur environ 15 degr\'es du nord au sud.
Au sud-est sont situ\'es les complexes mol\'eculaires de la Licorne et du Filament Sud.
Cet ensemble de nuages mol\'eculaires g\'eants situ\'e \`a 150 pc sous le plan galactique pourrait avoir \'et\'e form\'e par la collision d'un nuage g\'eant tombant sur le disque galactique depuis l'h\'emisph\`ere sud galactique \cite{franco88}.
Ces nuages mol\'eculaires g\'eants sont associ\'es \`a de grandes r\'egions H II, les associations d'\'etoiles OB (jeunes, chaudes et massives) nomm\'ees Orion Ia, Ib, Ic, qui sont situ\'ees sur une cha\^\i ne de 15 degr\'es de long du c\^ot\'e du complexe oppos\'e au plan galactique.

\begin{figure}
\begin{center}
\caption[Carte CO d'Orion]{Carte en \'emission mol\'eculaire du CO du complexe d'Orion A \`a grande \'echelle.
L'ISF (N\'ebuleuse M42) est bien visible \`a la droite de l'image.
Figure tir\'ee de Nagahama \etal (1998).}
\label{figorionge}
\end{center}
\end{figure}

La N\'ebuleuse d'Orion (M42) est une partie du complexe mol\'eculaire g\'eant d'Orion A, et en est \'egalement la r\'egion la mieux \'etudi\'ee.
Derri\`ere cette r\'egion ionis\'ee se trouve le nuage mol\'eculaire OMC-1 (OMC pour Orion Molecular Cloud),
qui correspond \`a un pic d'\'emission en CO.
Ce nuage est chauff\'e par le rayonnement ultraviolet des \'etoiles OB de l'amas ouvert du Trap\`eze (voir par exemple Hillenbrand 1997).
Deux sources infrarouges intenses sont pr\'esentes \`a l'int\'erieur d'OMC-1: il s'agit de l'objet ponctuel de Becklin-Neugebauer (BN, Becklin \& Neugebauer 1967) et de la N\'ebuleuse Kleinmann-Low (KL, Kleinmann \& Low 1967), situ\'ee 10$''$ vers le sud, qui a un diam\`etre angulaire d'environ 30$''$.
On pense que BN est une jeune \'etoile massive ($\approx$ 25 \msol) ayant un taux de perte de masse important.
La premi\`ere observation de M42 dans l'infrarouge a \'et\'e r\'ealis\'ee par Low \& Aumann (1970) dans le domaine spectral entre 30 et 1000 $\mic$.
Harper (1974) a ensuite cartographi\'e M42 \`a 90 $\mic$, et a montr\'e que l'\'emission infrarouge s'\'etendait sur une r\'egion plus large.
Cette r\'egion a depuis \'et\'e intensivement \'etudi\'ee, depuis les longueurs d'onde optiques jusqu'au domaine radio (voir \`a ce sujet par exemple la revue de Genzel \& Stutzki 1989).

On peut observer au nord d'OMC-1 un grand filament \'etendu (voir figure \ref{figisf}) appel\'e ISF (Integral-Shaped Filament) \`a cause de sa forme en S.
Deux r\'egions de densit\'e plus importante sont pr\'esentes dans l'ISF: les nuages mol\'eculaires OMC-2 et OMC-3 (voir Bally \etal 1987).
Ces condensations ont \'et\'e \'etudi\'ees en \'emission de raies mol\'eculaires; voir notamment les travaux r\'ecents de Dutrey \etal (1993), Castets \& Langer (1995) et Nagahama \etal (1998).
Elles ont aussi \'et\'e observ\'ees en continuum; voir les travaux r\'ecents de Chini \etal (1997), Lis \etal (1998) et Johnstone \& Bally (1999).
Les cartes \`a 1300 $\mic$ de OMC-2 et OMC-3 de Chini \etal (1997) ont montr\'e pr\'ecis\'ement la structure filamentaire de ces condensations et ont permis d'\'etudier plusieurs sources \`a l'int\'erieur.
Sur la cha\^\i ne dense qui relie OMC-1, OMC-2 et OMC-3, les trois condensations sont le si\`ege d'une formation d'\'etoiles active et semblent avoir \'et\'e form\'ees \`a peu pr\`es en m\^eme temps.
Cependant, il existe d'importantes diff\'erences entre les trois condensations OMC-1, OMC-2 et OMC-3.
OMC-1 est le si\`ege de formation d'\'etoiles massives et de fortes interactions entre le milieu gazeux et les sources infrarouges protostellaires, qui accroissent la temp\'erature du milieu jusqu'\`a environ 70 K, tandis qu'OMC-2 est moins brillant, contient de jeunes \'etoiles de plus faible masse, et a moins de pertes \'energ\'etiques qu'OMC-1.
On pense qu'OMC-3 a une masse comparable \`a OMC-2 d'environ 100 \msol, mais en ayant moins de pertes d'\'energie et une temperature du gaz plus faible, ce qui sugg\`ere qu'OMC-3 est moins \'evolu\'e qu'OMC-2, lui-m\^eme \'etant moins avanc\'e qu'OMC-1 \cite{castets95}.
La cartographie r\'ecente en $^{13}$CO de Nagahama \etal (1998) pour la totalit\'e du nuage mol\'eculaire d'Orion A a permis de pr\'eciser la structure de ce complexe mol\'eculaire.
Les travaux r\'ecents en continuum submillim\'etrique (Lis \etal 1998, Johnstone \& Bally 1999) ont une haute r\'esolution angulaire et ont donc montr\'e la structure complexe, filamentaire et fractale, de l'ISF d'Orion.
Ces auteurs ont \'egalement d\'ecouvert de nombreuses condensations de petite taille.
Cependant, les observations continuum au sol souffrent des contaminations atmosph\'eriques et n'ont donc pas la sensibilit\'e n\'ecessaire pour cartographier l'\'emission \'etendue de faible intensit\'e provenant de la poussi\`ere situ\'ee en dehors de la cha\^\i ne dense OMC-1 - OMC-3 (ISF).
Une observation ballon r\'ecente dans l'infrarouge lointain \cite{mookerjea00} a d\'eduit une r\'epartition des temp\'eratures dans Orion, la source la plus froide qu'ils aient observ\'ee ayant une temp\'erature de 15 K.
Nous d\'ecrivons l'apport d\'ecisif de l'exp\'erience ballon PRONAOS \`a la connaissance du milieu interstellaire dans Orion en section \ref{orion}.

\begin{figure}
\begin{center}
\caption[Carte continuum d'Orion]{Carte en continuum submillim\'etrique (350 $\mic$) de l'ISF dans Orion.
Figure tir\'ee de Lis \etal (1998).}
\label{figisf}
\end{center}
\end{figure}

\subsection{Le complexe mol\'eculaire Messier 17\label{introm17}}
La N\'ebuleuse Messier 17, aussi appel\'ee N\'ebuleuse Omega, du Cygne ou du Fer \`a Cheval, est une r\'egion ionis\'ee associ\'ee \`a un nuage mol\'eculaire g\'eant.
Cette r\'egion est le si\`ege de formation d'\'etoiles massives, comme le complexe d'Orion.
Elle est situ\'ee \`a environ 2200 pc de nous \cite{chini80} dans la constellation du Sagittaire, et poss\`ede le plus haut taux d'ionisation connu dans la Galaxie pour une r\'egion de formation d'\'etoiles (voir par exemple Glushkov 1998).
Ce taux d'ionisation tr\`es haut est d\^u \`a l'excitation du gaz par de jeunes \'etoiles de type O et B (Chini \etal 1980, Lemke \& Harris 1981).

Le complexe mol\'eculaire M17 (voir figure \ref{figm17}) a \'et\'e cartographi\'e en \'emission de rotation mol\'eculaire du monoxyde de carbone par Lada (1976), qui a montr\'e les deux condensations les plus intenses dans ce complexe, que l'on appelle habituellement M17 SW (sud-ouest) et M17 N (nord).
Ce nuage mol\'eculaire Messier 17 fait partie du complexe mol\'eculaire g\'eant qui s'\'etend sur 170 pc vers le sud-ouest le long du bras spiral du Sagittaire (\'emission CO montr\'ee par Elmegreen \etal 1979).
L'interaction de ce nuage mol\'eculaire g\'eant avec les r\'egions ionis\'ees est particuli\`erement visible dans M17, o\`u un front de choc en expansion interagit avec les nuages de gaz, et semble avoir fragment\'e le nuage mol\'eculaire d'origine \cite{rainey87}.
La r\'egion M17 SW est la mieux \'etudi\'ee, particuli\`erement pour la r\'egion domin\'ee par les photons (PDR) situ\'ee pr\`es de la fronti\`ere de la r\'egion H II au nord-est (voir par exemple Meixner \etal 1992 et Greaves \etal 1992).

\begin{figure}
\begin{center}
\caption[Cartes CO de M17]{Cartes CO du complexe Messier 17 tir\'ees de Wilson \etal (1999).}
\label{figm17}
\end{center}
\end{figure}

Les premi\`eres observations en infrarouge lointain de M17 ont \'et\'e r\'ealis\'ees par Low \& Aumann (1970) et Harper \& Low (1971).
M17 SW a ensuite \'et\'e cartographi\'e dans l'infrarouge moyen et lointain par Harper \etal (1976) et Gatley \etal (1979).
Wilson \etal (1979) ont cartographi\'e \`a 69 $\mic$ la totalit\'e du nuage M17 (i.e. avec M17 N).
Le satellite IRAS ({\it http:// www.ipac.caltech.edu/ipac/ iras/iras.html}) a produit des cartes infrarouges compl\`etes du ciel, qui ont montr\'e la r\'epartition de la poussi\`ere dans le complexe M17 sans cependant donner beaucoup d'information sur les zones les moins intenses de ce complexe.
Les mesures r\'ecentes de Wilson \etal (1999) ont produit des cartes CO pr\'ecises du nuage mol\'eculaire M17, ainsi que le travail de Sekimoto \etal (1999) sur une r\'egion plus large.
Dans ce contexte, nous d\'ecrivons l'apport essentiel de PRONAOS \`a la connaissance de cette r\'egion en section \ref{m17}.

\section{ L'exp\'erience PRONAOS\label{pronaos}}

\subsection{Pourquoi PRONAOS ?}
Sortir de l'atmosph\`ere terrestre pour observer les rayonnements submillim\'etriques appara\^\i t une n\'ecessit\'e au cours des ann\'ees 1980, alors que le ciel infrarouge lointain est tr\`es peu explor\'e au-del\`a de 100 $\mic$ de longueur d'onde (IRAS).
Le satellite COBE (1992) fournit en effet des cartes tr\`es limit\'ees en r\'esolution angulaire (40$'$ - 7$^o$), et les mesures au sol (t\'elescopes CSO: Caltech Submillimeter Observatory, {\it http:// www.submm.caltech.edu/cso}; JCMT: James Clerk Maxwell Telescope, {\it http:// www.ast.cam.ac.uk:81/ JACpublic/JCMT}; SEST: Swedish-ESO Submillimetre Telescope, {\it http:// www.ls.eso.org/lasilla/telescope/ sest/sest.html}) sont limit\'ees \`a quelques fen\^etres spectrales.
Quant aux observations effectu\'ees \`a bord d'avions (KAO: Kuiper Airborne Observatory, {\it http:// jean-luc.arc.nasa.gov/ kao/home/kao.html} et SOFIA: Stratospheric Observatory For Infrared Astronomy, {\it http:// sofia.arc.nasa.gov}), leur altitude relativement basse ($\approx$ 10 km) ne leur permet pas de s'affranchir totalement des contaminations atmosph\'eriques, et donc d'avoir la sensibilit\'e n\'ecessaire pour d\'etecter l'\'emission \'etendue de la poussi\`ere.
La n\'ecessit\'e d'une exp\'erience en ballon stratosph\'erique se fait donc sentir.
Une bonne r\'esolution angulaire sera n\'ecessaire pour d\'ecrire pr\'ecis\'ement les structures des r\'egions cartographi\'ees, ainsi que plusieurs bandes spectrales submillim\'etriques.

Le domaine des longueurs d'onde submillim\'etriques est en effet particuli\`erement int\'eressant pour caract\'eriser les propri\'et\'es des poussi\`eres dans le milieu interstellaire.
L'\'emission de la poussi\`ere dans ce domaine de longueurs d'onde est d\^u aux gros grains \`a l'\'equilibre thermique (voir section \ref{pousgen} de ce m\'emoire), dont l'\'emission est mod\'elis\'ee par la loi de corps noir modifi\'e (\'equation \ref{cnm}).
La temp\'erature d'un nuage mol\'eculaire est un param\`etre clef qui contr\^ole (avec d'autres) la structure et l'\'evolution des fragments (nuages denses pr\'ecurseurs d'un amas de jeunes \'etoiles, par opposition au terme c{\oe}ur qui d\'esigne un nuage pr\'ecurseur d'une seule \'etoile).
Pouvoir \'etudier efficacement la poussi\`ere froide galactique est donc un atout essentiel pour la compr\'ehension de la structure du milieu interstellaire et des processus de formation stellaire.
De plus, les travaux sur le r\'esidu de corr\'elation 60 - 100 $\mic$ dans les donn\'ees IRAS ont montr\'e qu'une grande partie de la masse pouvait \^etre contenue dans une composante froide (voir notamment Laureijs \etal 1988).
Dans ce contexte, l'imagerie spectrale submillim\'etrique de nuages mol\'eculaires a \'et\'e r\'ealis\'ee par PRONAOS.

La science galactique \'etait l'une des deux motivations principales de cette exp\'erience, avec aussi l'\'etude de l'effet Sunyaev-Zeldovich positif (voir section \ref{rfsec}).
N'oublions pas \'egalement la motivation technologique de pr\'eparer les futures missions satellites dans les domaines de longueurs d'onde infrarouge lointain et millim\'etrique (notamment Herschel et Planck), que ce soit au niveau de l'optique, des d\'etecteurs bolom\'etriques (voir section \ref{instr}), des syst\`emes \'electroniques de mesure, ou de l'\'etalonnage.

\subsection{Qu'est-ce que PRONAOS ?\label{quepron}}
PRONAOS (PROgramme NAtional d'Observations Submillim\'etriques ou autres sigles \'equivalents) est le projet fran\c{c}ais d'astronomie submillim\'etrique ballon des ann\'ees 1990.
Ce projet, dont le PI \'etait Guy Serra, directeur de recherche au CESR \`a Toulouse, a \'et\'e con\c{c}u en 1985 et men\'e \`a bien gr\^ace au CNES (Centre National des \'Etudes Spatiales) avec la collaboration des laboratoires CESR, IAS (Institut d'Astrophysique Spatiale \`a Orsay), Service d'A\'eronomie du CNRS \`a Verri\`ere-le-Buisson, IAP (Institut d'Astrophysique de Paris) et Observatoire de Gene\`eve.

\begin{figure}
\begin{center}
\caption[La nacelle PRONAOS]{Vue d'artiste de la nacelle PRONAOS en vol. Il manque n\'eanmoins la cha\^\i ne de vol qui la relie au ballon.}
\label{figpronaos}
\end{center}
\end{figure}

La nacelle PRONAOS (voir fig. \ref{figpronaos}) repr\'esente une r\'ealisation consid\'erable pour les exp\'eriences ballon, aussi bien par la taille que par les technologies mises en {\oe}uvre: il s'agit d'une nacelle de 8 m de c\^ot\'e, pesant pr\`es de 3 tonnes.
Le t\'elescope n\'ecessairement large pour la sensibilit\'e et la r\'esolution angulaire, mesure 2 m\`etres de diam\`etre.
Il a \'et\'e construit de fa\c{c}on \`a minimiser le plus possible sa masse pour pouvoir embarquer cette nacelle sous un ballon stratosph\'erique.
Ce t\'elescope de 2 m\`etres de type Cassegrain p\`ese au total 248 kg.
Le miroir primaire en carbone est form\'e de six p\'etales r\'ealis\'es en nid d'abeille et recouverts d'une couche d'or.
Les p\'etales sont orientables avec une pr\'ecision de quelques $\mic$ de mani\`ere \`a conserver en permanence la forme d\'esir\'ee.
Le t\'elescope est d\'ecrit pr\'ecis\'ement dans Buisson \& Duran (1990).

L'instrument focal SPM (Syst\`eme Photom\'etrique Multibande ou Spectro-Photom\`etre Multibande, voir Lamarre \etal 1994) est compos\'e d'un miroir vibrant qui permet au faisceau de se d\'eplacer sur le ciel avec une amplitude d'environ 6 minutes d'arc \`a une fr\'equence de 19.5 Hz, et de quatre bolom\`etres refroidis \`a 0.3 K.
Les moyens cryog\'eniques utilisent de l'h\'elium 4 liquide et un syst\`eme de r\'efrig\'eration \`a l'h\'elium 3.
Les quatre bolom\`etres mesurent le flux submillim\'etrique dans les domaines spectraux 180-240, 240-340, 340-540 et 540-1200 $\mic$, avec une sensibilit\'e aux faibles gradients de flux d'environ 1 \mjysr dans la voie 4 (1 Jansky ou Jy = 10$^{-26}$ \: Watt/m$^2$/Hz).
Les longueurs d'onde effectives correspondant \`a ces bandes spectrales, adopt\'ees par la collaboration PRONAOS, sont 200, 260, 360 et 580 $\mic$.
Gr\^ace \`a un arrangement de filtres dichro\"\i ques, la mesure est r\'ealis\'ee simultan\'ement dans les quatre voies au m\^eme endroit du ciel.
Un senseur stellaire (petit t\'elescope optique permettant de rep\'erer la position des \'etoiles), install\'e hors de l'axe principal du t\'elescope PRONAOS, et un gyroscope \`a trois axes, permettent de conna\^\i tre et de contr\^oler le pointage sur le ciel.
Avant chaque vol, le t\'elescope et l'ensemble du dispositif exp\'erimental sont align\'es et contr\^ol\'es.
Deux corps noirs internes sont \'etalonn\'es avec une r\'ef\'erence absolue, et permettent d'\'etalonner l'instrument en vol avec une pr\'ecision relative entre bandes de 5 \%, ce qui a \'et\'e v\'erifi\'e en observant la plan\`ete Saturne (source ponctuelle brillante).
L'\'etalonnage absolu de l'instrument a une pr\'ecision d'environ 8 \% (Pajot {\it et al.}, \enprep).
Les mesures sur Saturne permettent de d\'eriver la forme angulaire du faisceau jusqu'\`a un rayon de 6$'$ hors axe.
Les largeurs \`a mi-hauteur du faisceau sont 2$'$ dans les voies 1 et 2, 2.5$'$ dans la voie 3 et 3.5$'$ dans la voie 4.
On utilise les int\'egrales de ces faisceaux pour comparer la mesure sur Saturne \`a l'\'etalonnage au sol, qui est r\'ealis\'e sur des corps noirs \'etendus qui remplissent le faisceau.
Les efficacit\'es de faisceau (soit le rapport du faisceau principal au faisceau total) varient de 0.78 dans la voie 1 \`a 0.96 dans la voie 4.
Ceci revient \`a dire que l'erreur moyenne sur le front d'onde vaut environ 15 $\mic$.
Les principales caract\'eristiques de l'instrument PRONAOS sont d\'ecrites en table \ref{tabpp}.

\begin{table}
\begin{center}
\caption[Caract\'eristiques de PRONAOS]{Caract\'eristiques principales de la nacelle PRONAOS \'equip\'ee de l'instrument SPM.}
\begin{tabular}{|l|cccc|}
\hline
Diam\`etre du miroir primaire & \multicolumn{2}{c}{2045 mm, f/10}&  &  \\
Miroir vibrant & \multicolumn{2}{c}{f=19.5 Hz, $\delta _{hauteur}=6'$}&  & \\
Pr\'ecision de pointage & \multicolumn{2}{c}{$20''$ absolue, $5''$ relative}& & \\
Bande& 1 & 2 & 3 & 4 \\
($\mic$) &  $180-240$ & $240-340$ & $340-540$ & $540-1200$ \\
R\'esolution angulaire (arcmin)& 2 &  2  &  2.5  & 3.5 \\
NEB ($MJy sr^{-1}Hz^{-1/2}$)& 22 & 26 & 8.5 & 5.3 \\
\hline
\end{tabular}
\label{tabpp}
\end{center}
\end{table}

\subsection{Quand et comment observe PRONAOS ?\label{compron}}

La strat\'egie d'observation consiste \`a balayer le ciel en bandes altazimutales, c'est-\`a-dire \`a \'el\'evation constante.
Le miroir vibrant de l'instrument SPM permet en fait d'observer un gradient de flux, ce qui a pour avantage d'\'eliminer une grande partie de l'\'emission atmosph\'erique r\'esiduelle \`a 40 km d'altitude (\'el\'evation constante), ainsi que l'\'emission thermique de l'instrument (fr\'equence rapide de vibration du miroir de 19.5 Hz).
La mesure ainsi effectu\'ee est celle du gradient de flux astrophysique sur le ciel, qu'il faut donc transformer en flux absolu pendant la proc\'edure de traitement des donn\'ees.
De plus, les rotations du t\'elescope autour de la direction d'observation d\'eforment la carte par rapport \`a une grille rectangulaire.
Nous d\'ecrivons la proc\'edure de traitement des donn\'ees ainsi que notre m\'ethode de construction de cartes pour PRONAOS en section \ref{cartespron}.

L'\'equipe PRONAOS a effectu\'e trois campagnes de lancement couronn\'ees de succ\`es, en utilisant le turn-around, c'est-\`a-dire une p\'eriode de temps assez courte (\`a peu pr\`es une semaine) pendant laquelle les vents stratosph\'eriques baissent d'intensit\'e, permettant au ballon d'\'evoluer lentement en cercle au-dessus de la base de lancement.
Celle-ci est situ\'ee \`a Fort-Sumner, dans l'\'etat du Nouveau-Mexique, aux \'Etats-Unis.
Le premier vol a eu lieu le 17 septembre 1994.
Des probl\`emes techniques au niveau du senseur stellaire et du ballon ont consid\'erablement diminu\'e l'int\'er\^et scientifique de ce premier vol, mais ce fut un succ\`es technologique total pour le t\'elescope et les d\'etecteurs.
La seule observation scientifique exploitable a n\'eanmoins permis une d\'ecouverte importante dans Orion (voir sections suivantes).

Le deuxi\`eme vol, en septembre 1996, a \'et\'e le succ\`es majeur de PRONAOS, avec pr\`es de 13 heures de donn\'ees scientifiques exploitables portant sur une quinzaine d'objets (voir table \ref{tabobspron}).
Le troisi\`eme vol, en septembre 1999, a \'et\'e d\'ecevant \`a cause d'un bruit parasite important et de probl\`emes de trajectoire du ballon.
Six objets ont pu \^etre observ\'es de fa\c{c}on exploitable, mais pour seulement 1 heure environ de donn\'ees.
J'ai pu participer \`a cette campagne de septembre 1999, au CNES \`a Toulouse (voir section \ref{campagnes}).

\subsection{Qu'observe PRONAOS ?}
PRONAOS a essentiellement observ\'e des r\'egions du milieu interstellaire de la Galaxie, et des amas de galaxies en effet SZ.
Les observations de l'effet Sunyaev-Zeldovich (voir section \ref{rfsec}) ont \'et\'e effectu\'ees sur trois amas: Abell 2142, Abell 2163 et Abell 478.
Les observations du milieu interstellaire ont \'et\'e effectu\'ees sur des objets tr\`es divers caract\'erisant la diversit\'e du milieu interstellaire de notre galaxie, depuis les complexes mol\'eculaires g\'eants si\`eges de formation d'\'etoiles massives (Orion, M17) jusqu'au milieu plus diffus des cirrus \`a haute latitude galactique (Polaris par exemple).
La liste compl\`ete des observations PRONAOS exploitables scientifiquement est donn\'ee en table \ref{tabobspron}.

\begin{table}
\begin{center}
\caption[Observations de PRONAOS]{Liste des objets observ\'es de mani\`ere scientifiquement utile par PRONAOS.
Les lignes horizontales s\'eparent les trois vols de 1994, 1996 et 1999.}
\begin{tabular}{llll}
\hline
Source & Taille des cartes & Temps d'obs. (mn) & Publication (journal) \\
\hline
Orion M42 & 50$'$ x 10$'$ & 50 & Ristorcelli \etal 1998, \apj \\
\hline
M82 & 26$'$ x 11$'$ & 22 & \\
A 2142 (SZ) & --- & 68 & \\
$\rho$ Ophiuci & 50$'$ x 35$'$ & 44 & Ristorcelli {\it et al.}, \enprep \\
NCS & 20$'$ x 15$'$ & 14 & Bernard {\it et al.}, soumis \\
Serpentis & 50$'$ x 24$'$ & 50 & \\
M17 & 50$'$ x 40$'$ & 67 & Dupac \etal 2002, \aa \\
G34 & 20$'$ x 15$'$ & 12 & \\
VGA & 80$'$ x 3.5$'$ & 13 & \\
VGB & 80$'$ x 3.5$'$ & 13 & \\
A 2163 (SZ) & --- & 105 & Lamarre \etal 1998, \apjl \\
NGC 891 & 26$'$ x 10$'$ & 32 & \\
Cygnus B & 50$'$ x 32$'$ & 30 & M\'eny {\it et al.}, \enprep \\
A 478 (SZ) & --- & 71 & \\
Polaris & 50$'$ x 3.5$'$ & 39 & Bernard \etal 1999, \aa \\
Orion M42 & 80$'$ x 50$'$ & 131 & Dupac \etal 2001, \apj \\
Taurus & 50$'$ x 3.5$'$ & 40 & Stepnik \etal 2002, sous pr. dans \aa \\
\hline
M82 & 20$'$ x 11$'$ & 5 & \\
L134S & 32$'$ x 28$'$ & 13 & \\
L1689 & 34$'$ x 34$'$ & 17 & \\
Cygnus B & 39$'$ x 36$'$ & 20 & M\'eny {\it et al.}, \enprep \\
IC5146 & 39$'$ x 20$'$ & 11 & \\
L507A & 34$'$ x 14$'$ & 7 & \\
\label{tabobspron}
\end{tabular}
\end{center}
\end{table}

Nous nous sommes particuli\`erement int\'eress\'es aux complexes de formation d'\'etoiles massives que sont Orion et M17.
Ces r\'egions contiennent une grande vari\'et\'e de milieux et de conditions physiques (temp\'erature, densit\'e...), depuis les zones ionis\'ees jusqu'aux c{\oe}urs denses protostellaires.
Les mesures de PRONAOS permettent de mieux conna\^\i tre la r\'epartition de la mati\`ere interstellaire dans ces complexes de formation d'\'etoiles massives, et de contraindre ses propri\'et\'es physiques ainsi que les premiers stades de la formation stellaire.
Ces r\'egions sont particuli\`erement int\'eressantes et complexes, et font appara\^\i tre des contrastes d'intensit\'e tr\`es importants, ce qui induit des difficult\'es pour la reconstruction des images.

\section{ Construction de cartes pour PRONAOS-SPM\label{cartespron}}

\subsection{Introduction}
Comme expliqu\'e en section \ref{invlin}, l'observation du ciel est mod\'elis\'ee par un probl\`eme lin\'eaire \`a inverser.
Si nous reprenons les notations de l'\'equation \ref{obs}, nous voyons que nous devons mod\'eliser dans le cas de PRONAOS la matrice {\bf A} et le bruit {\bf n}.
Les donn\'ees temporelles ({\bf y}) de PRONAOS se pr\'esentent sous la forme d'un vecteur ordonn\'e dans le temps, dans lequel chaque \'echantillon est la diff\'erence de flux mesur\'ee entre les deux directions de faisceau du miroir vibrant.
Ainsi, les donn\'ees temporelles sont un gradient de flux, convolu\'e par la forme du faisceau de l'instrument, lobe de forme et de taille d\'etermin\'ees par les mesures sur une source ponctuelle (Saturne).

La proc\'edure que nous avons utilis\'ee pour les donn\'ees PRONAOS que nous avons trait\'ees (Orion et M17) consiste tout d'abord \`a appliquer une r\'eduction basique des donn\'ees brutes: d\'eglitchage (suppression des pics tr\`es intenses d\^us \`a des impacts de rayons cosmiques sur un d\'etecteur) et filtrage des bruits \'electroniques \`a haute fr\'equence.
Puis, nous corrigeons de la d\'eformation de la grille de la carte en prenant en compte le pointage du t\'elescope, y compris les erreurs fines de pointage dues aux balancements de la nacelle.

Une fois ces premi\`eres \'etapes du traitement effectu\'ees, nous pouvons construire la carte de flux absolu en d\'econvoluant le signal.

\subsection{La m\'ethode de construction de cartes Wiener-Fourier}
Habituellement, les donn\'ees PRONAOS \'etaient d\'econvolu\'ees par une m\'ethode de Wiener traitant les donn\'ees ligne par ligne.
L'op\'eration de d\'econvolution est effectu\'ee dans l'espace de Fourier, ce qui consiste simplement en une multiplication par le bon filtre, en l'occurrence un filtre de Wiener calcul\'e par minimisation de l'erreur de reconstruction.
Ce genre de m\'ethodes est appel\'e en g\'en\'eral EKH \cite{emerson79}, et cette m\'ethode de Wiener-Fourier est d\'ecrite dans Sales \etal (1991).
Son avantage principal r\'eside dans le fait qu'elle r\'ealise une d\'econvolution ligne par ligne: cela signifie qu'on ne manipule pas de grosses matrices et que donc les calculs informatiques sont rapides.
Mais dans cette m\^eme caract\'eristique se trouve son principal d\'efaut.
En effet, chaque ligne des donn\'ees temporelles n'est en r\'ealit\'e pas ind\'ependante des autres, car il existe des corr\'elations du bruit et du signal.
Les cartes reconstruites par cette m\'ethode font appara\^\i tre des art\'efacts tels que visibles en figure \ref{figwief}.
Aussi avons-nous d\'evelopp\'e une nouvelle m\'ethode de reconstruction.

\begin{figure}
\begin{center}
\caption[Carte par la m\'ethode Wiener-Fourier]{Image de la premi\`ere s\'equence de la N\'ebuleuse d'Orion M42 vue par la voie 1 (200 $\mic$) de PRONAOS-SPM.
Cette carte a \'et\'e construite en utilisant la m\'ethode de Wiener-Fourier.
L'\'echelle de couleurs est logarithmique, et fait appara\^\i tre l'intensit\'e positive reconstruite jusqu'\`a 0.5 en log, puis le bleu clair et le bleu fonc\'e montrent des art\'efacts de valeurs n\'egatives, clairement visibles le long de la d\'eclinaison d'OMC-1 sous forme d'oscillations.
Les lignes noires dans la barre de couleur montrent les niveaux des contours.
}
\label{figwief}
\end{center}
\end{figure}

\subsection{La m\'ethode de construction de cartes Wiener globale\label{wieglob}}

\subsubsection{Fondements}
Nous utilisons la m\'ethode Wiener 2 (\'eq. \ref{wie2}) pour construire la carte.
La matrice r\'eponse de l'instrument {\bf A} est tout d'abord construite en consid\'erant la transformation de l'image du ciel r\'eel par l'instrument.
Nous introduisons donc dans la matrice {\bf A} l'effet du miroir vibrant (mesure d'un gradient de flux), l'effet de dilution par le lobe (voir figure \ref{figlobmat}), et bien s\^ur le pointage.
La matrice est donc construite en consid\'erant que chaque point des donn\'ees temporelles, c'est-\`a-dire chaque ligne de la matrice {\bf A}, est cr\'e\'e par l'ensemble des points du ciel situ\'es \`a l'int\'erieur d'un disque de rayon correspondant \`a la taille du lobe, autour du pointage consid\'er\'e.
La contribution de chaque pixel du ciel concern\'e d\'ecro\^\i t avec le rayon autour du point central, avec la forme indiqu\'ee en figure \ref{figlobmat}, et on consid\`ere que cette contribution s'annule au-del\`a d'une certaine limite de rayon (S. Roques, communication priv\'ee).
Nous introduisons cette op\'eration de convolution par le lobe pour les deux faisceaux du miroir vibrant, l'un en positif, l'autre en n\'egatif.
La diff\'erence des deux contributions permet de construire la matrice {\bf A}.

\begin{figure}
\begin{center}

\caption[Lobe et matrice]{Lobes de l'instrument PRONAOS dans les quatre voies (haut) et une ligne de la matrice {\bf A} (bas).
Les diff\'erents pics dans cette ligne sont d\^us au passage du faisceau dans les lignes successives de la carte: en effet une ligne de la matrice {\bf A} a la dimension de la carte totale.
On voit donc que la matrice {\bf A} tient compte de la taille du faisceau dans les deux directions.
La pr\'esence de deux blocs de pics (n\'egatif et positif) est due au miroir vibrant.
}
\label{figlobmat}
\end{center}
\end{figure}

Pour construire la carte, plut\^ot que d'appliquer brutalement la matrice Wiener 2 (\'eq. \ref{wie2}), nous transformons le probl\`eme d'inversion en un syst\`eme lin\'eaire:

\begin{equation}
[{\bf S}^{-1} + {\bf A}^t \: {\bf N}^{-1} \: {\bf A}] \: {\bf \Tilde x} = {\bf A}^t \: {\bf N}^{-1} \: {\bf y}
\label{wie2pron}
\end{equation}

Il faut alors calculer ${\bf A}^t \: {\bf N}^{-1} \: {\bf A}$.
Nous le faisons de fa\c{c}on brutale, ce qui n\'ecessite, m\^eme pour des donn\'ees temporelles aussi courtes que celles de PRONAOS (quelques milliers de points de donn\'ees par objet), environ 15 mn de temps sur un processeur Pentium III \`a 500 MHz.
Nous calculons la solution de l'\'equation \ref{wie2pron} en utilisant une m\'ethode it\'erative standard des gradients biconjugu\'es (routine IDL fond\'ee sur un algorithme de {\it Numerical Recipes}, publi\'e chez {\it Cambridge University Press}).
La solution trouv\'ee par convergence de la m\'ethode apr\`es environ 50/100 it\'erations est par d\'efinition la solution optimale du probl\`eme de reconstruction, \`a condition d'estimer correctement les matrices {\bf N} et {\bf S}.
L'image ainsi reconstruite prend en compte l'inversion du lobe instrumental, ce qui fait ressortir de l'information cach\'ee dans la tache de faisceau, mais cela n'a pas une grande signification astronomique.
Aussi, nous lissons le r\'esultat avec le lobe de l'instrument pour obtenir une carte d'aspect parfaitement satisfaisant et homog\`ene.
Nous it\'erons ce processus (gradients conjugu\'es it\'eratif + lissage) afin d'affiner l'estimation de {\bf S}, dont nous ne consid\'erons que les \'el\'ements diagonaux.
\`A chaque it\'eration, nous injectons dans la diagonale de {\bf S} le carr\'e de la carte de signal reconstruite.

Pour la matrice de covariance du bruit {\bf N}, nous avons \'egalement consid\'er\'e uniquement les \'el\'ements diagonaux, c'est-\`a-dire que l'on n\'eglige les corr\'elations temporelles du bruit dans les donn\'ees.
Ceci se justifie dans la mesure o\`u le miroir vibrant permet d'\'eliminer les \'eventuelles d\'erives basse fr\'equence que peuvent cr\'eer les d\'etecteurs bolom\'etriques.

\subsubsection{M\'ethode du bruit uniforme}
Une m\'ethode simple consiste alors \`a supposer que le bruit est ind\'ependant du signal et uniforme, c'est-\`a-dire qu'il a le m\^eme \'ecart quadratique moyen pour tous les points de donn\'ees.
Ayant suppos\'e que le bruit n'\'etait pas autocorr\'el\'e, la matrice de covariance du bruit {\bf N} se r\'eduit alors aux \'el\'ements diagonaux uniformes.
Il faut alors estimer le niveau de bruit dans les donn\'ees.
On teste donc la m\'ethode avec diff\'erents niveaux de bruit, et l'on se rend compte que plus le niveau de bruit suppos\'e est \'elev\'e, plus la carte reconstruite est libre d'art\'efacts de reconstruction (bruit granuleux et/ou oscillations le long des lignes d'observation intenses), mais plus les zones de faible intensit\'e sont lav\'ees de leur signal comme de leur bruit.
C'est une propri\'et\'e naturelle des m\'ethodes de Wiener que de supprimer de fa\c{c}on in\'egale la puissance dans les pixels en fonction du rapport signal sur bruit, et de construire ainsi la carte optimale du point de vue de l'erreur global de reconstruction.
Le choix de diff\'erents rapports signal sur bruit suppos\'es permet ainsi de reconstruire de fa\c{c}on diff\'erente la carte, soit en essayant de faire ressortir des zones de faible rapport signal sur bruit au prix de cr\'eer des art\'efacts dans les zones intenses, soit en \'eliminant les art\'efacts (le bruit) au prix de l'\'elimination d'une partie du signal, essentiellement dans les zones de faible intensit\'e.
Comme largement expliqu\'e dans Dupac (1999), nous avons reconstruit nos cartes en prenant le niveau de bruit permettant d'\'eliminer les art\'efacts en \'eliminant le moins possible de signal.

Afin de valider cette m\'ethode de d\'etermination du niveau de bruit, nous avons r\'ealis\'e des simulations d'observations (bruit\'ees), faites par PRONAOS-SPM dans la voie 1, sur une source \'etendue ayant une forme voisine de celle du filament ISF dans Orion (voir figure \ref{figsimuspron}).
La simulation d'observation est r\'ealis\'ee en utilisant la matrice {\bf A} telle que d\'ecrite plus haut, et en rajoutant un bruit blanc (non autocorr\'el\'e) uniforme d'\'ecart quadratique moyen arbitraire que l'on fixe \`a 10 USPM (unit\'es SPM).
Cette valeur du niveau de bruit simul\'e est tr\`es sup\'erieure au niveau du bruit dans les donn\'ees.
Apr\`es simulation de l'observation PRONAOS, nous reconstruisons la carte en supposant diff\'erents niveaux de bruit.
On observe que la carte reconstruite en prenant un niveau de bruit dans la m\'ethode de construction de cartes \'egal au niveau de bruit r\'eel, se caract\'erise par l'absence d'art\'efacts.
On observe que plus le bruit suppos\'e sous-estime le bruit r\'eel, plus la carte reconstruite fait appara\^\i tre des art\'efacts.
En cons\'equence, le niveau de bruit que l'on d\'etermine en minimisant les bruits r\'esiduels dans la carte reconstruite correspond, d'apr\`es les r\'esultats des simulations, au bruit r\'eel dans les donn\'ees.
La carte reconstruite en supposant ce niveau de bruit est donc bien la solution optimale du probl\`eme de reconstruction, ce qui valide la m\'ethode utilis\'ee.

\begin{figure}
\begin{center}
\caption[Construction de cartes sur simulation]{Carte simul\'ee d'une source \'etendue ayant une forme voisine de celle du filament ISF dans Orion (en bas \`a droite), telle que visible par la voie 1 de PRONAOS-SPM.
Apr\`es simulation de l'observation PRONAOS, nous reconstruisons la carte en supposant diff\'erents niveaux de bruit: le bruit r\'eel (en bas \`a gauche), 1.5 fois moins que le bruit r\'eel (en haut \`a droite), et 2 fois moins que le bruit r\'eel (en haut \`a gauche).
}
\label{figsimuspron}
\end{center}
\end{figure}

Gr\^ace \`a cette nouvelle m\'ethode, nous avons reconstruit efficacement les cartes de la N\'ebuleuse d'Orion: comparer les r\'esultats en figures \ref{figorion12} et \ref{figorion34} par rapport \`a la m\'ethode Wiener-Fourier en figure \ref{figwief}.
Cette m\'ethode de construction de cartes a fait l'objet d'une section de l'article sur Orion \cite{dupac01}.

\subsubsection{M\'ethode du bruit non uniforme}
Nous avons \'egalement cherch\'e \`a am\'eliorer encore cette m\'ethode, en supposant cette fois que le bruit n'\'etait pas ind\'ependant du signal.
Nous consid\'erons alors qu'il existe deux composantes de bruit: un bruit uniforme comme celui de la m\'ethode pr\'ec\'edente (le bruit r\'esiduel), et un bruit proportionnel au signal (bruit non stationnaire, voir section \ref{propcorr}).
Dans ce cas, l'optimisation de la m\'ethode se fait de la m\^eme fa\c{c}on que pr\'ec\'edemment, sauf que les deux param\`etres ajustables sont le niveau de bruit uniforme et le coefficient de proportionnalit\'e d\'efinissant le bruit corr\'el\'e au signal.
La matrice de covariance du bruit {\bf N} est alors toujours diagonale mais les \'el\'ements diagonaux ne sont plus uniformes.

Ainsi, il est possible de reconstruire tr\`es efficacement les zones de faible intensit\'e en supposant un bruit r\'esiduel faible, hypoth\`ese soutenue par les mesures de bruit instrumental, tout en \'eliminant efficacement les art\'efacts de reconstruction dans les zones intenses gr\^ace \`a la superposition du bruit proportionnel au signal.
La pr\'esence d'un tel bruit se justifie car la r\'eponse du bolom\`etre n'est pas forc\'ement parfaite lorsque l'on observe une zone tr\`es intense, ce qui est le cas des zones centrales de M42 et M17.
Nous avons test\'e cette m\'ethode sur les donn\'ees d'Orion et de M17, et nous avons montr\'e que les r\'esultats \'etaient consid\'erablement am\'elior\'es en ce qui concerne M17 (voir le r\'esultat de la m\'ethode du bruit uniforme en figure \ref{figcompuni}, \`a comparer aux figures \ref{figm1712} et \ref{figm1734}).
Pour Orion, le fait d'avoir deux s\'equences de donn\'ees (zone centrale et nuages de faible intensit\'e) que l'on traite s\'epar\'ement permet d\'ej\`a une reconstruction tr\`es satisfaisante, sans que la m\'ethode du bruit proportionnel apporte un mieux significatif.
Cette m\'ethode de construction de cartes qui tient compte du bruit corr\'el\'e au signal a fait l'objet d'une section de l'article sur M17 \cite{dupacetal02}.

\begin{figure}
\begin{center}
\caption[Carte de M17 par la m\'ethode du bruit uniforme]{Carte de M17 dans la voie 1 (200 $\mic$) de PRONAOS-SPM, construite par la m\'ethode du bruit uniforme.
L'\'echelle de couleurs est logarithmique, et fait appara\^\i tre l'intensit\'e positive reconstruite jusqu'\`a -0.5 en log, puis le bleu fonc\'e montre le bruit n\'egatif.
Des zones de faible intensit\'e \`a l'ouest sont perdues par rapport \`a la m\'ethode du bruit non uniforme (voir figures \ref{figm1712} et \ref{figm1734}).}
\label{figcompuni}
\end{center}
\end{figure}

\subsubsection{Incertitudes de reconstruction}
\`A partir des simulations d\'ecrites pr\'ec\'edemment, nous pr\'esentons en figure \ref{errreconstpron} un graphe montrant l'erreur de reconstruction constat\'ee sur la carte reconstruite par rapport au ciel simul\'e, pixel par pixel, en fonction du flux dans chaque pixel.

\begin{figure}
\begin{center}
\caption[Erreurs de reconstruction]{Erreur absolue de reconstruction constat\'ee sur la carte reconstruite par rapport au ciel simul\'e, pixel par pixel, en fonction du flux dans chaque pixel.
La droite horizontale marque le niveau de bruit dans les donn\'ees (3 $\sigma$).}
\label{errreconstpron}
\end{center}
\end{figure}

Ce graphique montre que l'erreur absolue de reconstruction constat\'ee est du m\^eme ordre que le niveau de bruit dans les simulations (10 USPM \`a 1 $\sigma$).
L'\'ecart quadratique moyen de l'erreur de reconstruction sur la carte est de 31 USPM, soit trois fois plus que le niveau dans les donn\'ees.
Cependant, cette valeur relativement \'elev\'ee est due \`a quelques points, principalement des pixels aux flux importants pour lesquels l'erreur absolue est certes \'elev\'ee mais l'erreur relative quand m\^eme tr\`es faible.
Ces erreurs absolues plus grandes pour les pixels ayant un flux important peuvent s'expliquer par le fait que le faisceau de l'instrument dilue le vrai signal sur plusieurs pixels, et que la reconstruction peut difficilement rem\'edier \`a cet effet.
Si l'on enl\`eve les 32 pixels les plus bruit\'es (sur 266), l'\'ecart quadratique moyen des pixels restant tombe \`a 17 USPM.
Ceci montre la robustesse de cette nouvelle m\'ethode de construction de cartes pour PRONAOS.
Les incertitudes photom\'etriques dues \`a la reconstruction, mises en \'evidence en figure \ref{errreconstpron}, sont difficilement mod\'elisables, et nous ne les avons donc pas prises en compte pour l'analyse.
Dans la pratique, nous avons \'elimin\'e de l'analyse les zones trop bruit\'ees des cartes, comme expliqu\'e dans les sections suivantes.

En conclusion de cette section, nous pouvons dire que ces investigations ont permis d'am\'eliorer consid\'erablement la construction de cartes pour PRONAOS.
L'\'etude des deux nuages froids les plus faibles mis en \'evidence dans Orion (nuages 3 et 4, voir section \ref{orion}) n'aurait pu \^etre effectu\'ee sans cette nouvelle m\'ethode, et l'\'etude des zones froides \`a l'ouest de M17 (voir section \ref{m17}) non plus sans la m\'ethode du bruit en partie corr\'el\'e au signal.
Les tr\`es importants contrastes d'intensit\'e (jusqu'\`a 10000) de ces r\'egions, reconstruits gr\^ace aux m\'ethodes de construction de cartes que nous avons d\'evelopp\'ees, sont le t\'emoin de la structure complexe du milieu interstellaire dans ces nuages mol\'eculaires g\'eants en interaction avec des r\'egions ionis\'ees.
Outre la construction de cartes, nous avons travaill\'e longuement sur l'analyse des donn\'ees PRONAOS des complexes de formation d'\'etoiles massives que sont Orion A et Messier 17.

\section{ Analyse des cartes de PRONAOS de la N\'ebuleuse d'Orion M42\label{orion}}

PRONAOS a observ\'e la N\'ebuleuse d'Orion Messier 42, partie la plus active du complexe mol\'eculaire g\'eant d'Orion (voir section \ref{introorion}), lors du deuxi\`eme vol de l'exp\'erience en septembre 1996.
Avec plus de deux heures d'int\'egration, cette observation est la plus longue jamais r\'ealis\'ee par PRONAOS.
Elle a \'et\'e r\'ealis\'ee en deux s\'equences, la premi\`ere concernant l'ISF et la deuxi\`eme ce qui s'av\`erera \^etre les nuages 2, 3 et 4.
L'utilisation de notre m\'ethode de construction de cartes d\'ecrite en section \ref{wieglob} a permis de construire les cartes que nous pr\'esentons et \'etudions dans les sections suivantes.

 \subsection{\'Etude des cartes d'intensit\'e}
Nous pr\'esentons en figure \ref{figorion12} les cartes obtenues dans les bandes 1 et 2 de PRONAOS-SPM, et en figure \ref{figorion34} les cartes des voies 3 et 4.
Nous pr\'esentons \'egalement en table \ref{fluxorion} les positions centrales des objets observ\'es et les flux dans les quatre voies int\'egr\'es sur un diam\`etre indiqu\'e dans la table.
Ce diam\`etre repr\'esente approximativement la taille de la r\'egion centrale, la plus brillante, du nuage.

\begin{figure}
\begin{center}
\caption[Cartes PRONAOS d'Orion \`a 200-260 $\mic$]{Cartes de la N\'ebuleuse d'Orion M42 dans les deux premi\`eres voies de PRONAOS-SPM: 200 $\mic$ (haut) et 260 $\mic$ (bas).
Nous avons rassembl\'e les deux s\'equences, reconstruites s\'epar\'ement, en une seule image.
L'\'echelle de couleur est logarithmique, et montre l'intensit\'e reconstruite positive jusqu'\`a -1.5 en log, puis le bleu fonc\'e et le violet montrent le bruit n\'egatif.
Le niveau de bruit est d'environ 4 \mjysr rms dans la voie 1.
Cependant, l'incertitude d'\'etalonnage fait que la pr\'ecision sur l'intensit\'e est de 5 \% (1 $\sigma$) en relatif entre bandes (8 \% en absolu).
Le cadre noir qui appara\^\i t sur la carte \`a 260 $\mic$ est la zone cartographi\'ee lors du premier vol de PRONAOS \cite{ristorcelli98}.
}
\label{figorion12}
\end{center}
\end{figure}

\begin{figure}
\begin{center}
\caption[Cartes PRONAOS d'Orion \`a 360-580 $\mic$]{Cartes de la N\'ebuleuse d'Orion M42 dans les voies 3 et 4 de PRONAOS-SPM: 360 $\mic$ (haut) et 580 $\mic$ (bas).
Nous avons rassembl\'e les deux s\'equences, reconstruites s\'epar\'ement, en une seule image.
L'\'echelle de couleur est logarithmique, et montre l'intensit\'e reconstruite positive jusqu'\`a -1.5 en log, puis le bleu fonc\'e et le violet montrent le bruit n\'egatif.
Le niveau de bruit est d'environ 0.8 \mjysr dans la voie 4.
Cependant, l'incertitude d'\'etalonnage fait que la pr\'ecision sur l'intensit\'e est de 5 \% (1 $\sigma$) en relatif entre bandes (8 \% en absolu).
}
\label{figorion34}
\end{center}
\end{figure}

\begin{table}
\begin{center}
\caption[Flux int\'egr\'es pour les donn\'ees PRONAOS]{
Flux int\'egr\'es pour les donn\'ees PRONAOS.
Les diam\`etres approximatifs des r\'egions centrales \'etudi\'ees sont donn\'es en minutes d'arc.}
\begin{tabular}{lllllll}

\hline
 & $\alpha_{1950}$ & $\delta_{1950}$ & F$_{\nu}$(Jy) & F$_{\nu}$(Jy) & F$_{\nu}$(Jy) & F$_{\nu}$(Jy) \\
 & (h,mn,sec) & ($^o$,$'$) & 200 $\mic$ & 260 $\mic$ & 360 $\mic$ & 580 $\mic$ \\
\hline

OMC-1(3.6$'$) & 5h 32mn 38& -5$^o$ 25$'$& 28500 $\pm $ 1400 & 14120 $\pm $ 710 & 6010 $\pm $ 300 & 1413 $\pm $ 71\\
\hline

ISF & & & & & &\\
(sud) 5.4$'$ & 5h 32mn 52& -5$^o$ 13$'$& 5130 $\pm $ 260 & 3110 $\pm $ 160 & 1489 $\pm $ 74 & 404 $\pm $ 20\\
\hline

ISF & & & & & &\\
(nord) 4.2$'$  & 5h 32mn 38& -4$^o$ 58$'$& 2490 $\pm $ 120 & 1449 $\pm $ 72 & 699 $\pm $ 35 & 182.9 $\pm $ 9.1\\
\hline

Nuage 1 (3.5$'$) & 5h 31mn 30& -5$^o$ 25$'$& 461 $\pm $ 23 & 308 $\pm $ 15 & 209 $\pm $ 10 & 51.0 $\pm $ 2.6\\
\hline

Nuage 2 (3.5$'$) & 5h 31mn 16& -5$^o$ 08$'$& 37.9 $\pm $ 1.9 & 41.0 $\pm $ 2.0 & 27.0 $\pm $ 1.3 & 9.95 $\pm $ 0.50\\
\hline

Nuage 3 (3.5$'$) & 5h 30mn 15& -5$^o$ 01$'$& 25.0 $\pm $ 1.3 & 25.8 $\pm $ 1.3 & 15.26 $\pm $ 0.76 & 5.82 $\pm $ 0.29\\
\hline

Nuage 4 (5.2$'$) & 5h 29mn 11& -4$^o$ 57$'$& 63.2 $\pm $ 3.2 & 59.5 $\pm $ 3.0 & 21.5 $\pm $ 1.1 & 12.54 $\pm $ 0.63\\
\hline

\end{tabular}
\label{fluxorion}
\end{center}
\end{table}

La r\'esolution angulaire de ces cartes est 2 minutes d'arc dans les voies 1 et 2 (longueurs d'onde effectives de 200 et 260 $\mic$), 2.5$'$ dans la troisi\`eme voie (360 $\mic$) et 3.5$'$ dans la quatri\`eme (580 $\mic$).
Cette r\'esolution angulaire correspond \`a une r\'esolution spatiale d'environ 0.4 parsec \`a la distance de M42 de 470 pc.
Le niveau du bruit dans les cartes se situe \`a environ 1 MJy/sr.
Cependant, les incertitudes d'\'etalonnage font que la pr\'ecision sur les intensit\'es est de 5 \% en relatif entre les bandes SPM (intercalibration), et de 8 \% en absolu (voir section \ref{quepron} de ce m\'emoire et Pajot {\it et al.}, \enprep).

Nous observons dans les figures \ref{figorion12} et \ref{figorion34} que la zone la plus brillante est le nuage mol\'eculaire OMC-1 (voir l'introduction sur Orion en section \ref{introorion} et les noms des sources identifi\'ees en figure \ref{figorion12}).
Il appara\^\i t en effet sur nos cartes comme une r\'egion tr\`es intense d'\'emission thermique de la poussi\`ere.
En particulier dans les environs de BN/KL, l'\'emission est tr\`es \'elev\'ee, et il s'agit du maximum d'\'emission pour toute la carte: 49000 \mjysr dans la voie 1 (200 $\mic$).

Au nord de ce c{\oe}ur central, on observe tr\`es clairement un grand nuage mol\'eculaire qui s'\'etend sur environ 3 pc, il s'agit de l'ISF (Integral-Shaped Filament ou filament en forme d'int\'egrale).
Il pr\'esente une intensit\'e spectrale moyenne d'environ 2300 \mjysr \`a 200 $\mic$ de longueur d'onde.

On voit \`a l'ouest du c{\oe}ur central OMC-1 une condensation de plus faible intensit\'e, que nous appelons Nuage 1, et qui poss\`ede une intensit\'e de 570 \mjysr dans la voie 1.
Une \'emission plus faible est visible autour de cette condensation, qui est reli\'ee au nuage 2 au nord.

Les nuages 2, 3 et 4 apparaissent au nord-ouest dans les cartes en figures \ref{figorion12} et \ref{figorion34}, d'est en ouest.
Le nuage 2 est une condensation de faible intensit\'e d\'ecouverte lors du premier vol de PRONAOS.
L'observation d'OMC-1 et de ce nuage est d\'ecrite dans Ristorcelli \etal (1998).
Dans cet article, le nuage 2 est appel\'e NorthWest Condensation (source 4).
Dans nos observations du deuxi\`eme vol de PRONAOS, ce nuage appara\^\i t avec une intensit\'e spectrale maximale dans la carte de 47 \mjysr \`a 200 $\mic$, mais son maximum spectral est plus proche de 260 $\mic$, dans la voie 2 dans laquelle l'intensit\'e atteint 50 \mjysr.

Les nuages 3 et 4 sont deux condensations d'intensit\'e extr\^emement faible.
L'intensit\'e \`a 260 $\mic$ du nuage 3 est 32 \mjysr, et l'intensit\'e \`a 200 $\mic$ du nuage 4 est 35 \mjysr (respectivement les bandes dans lesquelles les intensit\'es sont maximales).
Il faut noter que ces deux nuages froids, ainsi que le nuage 2, sont invisibles sur les cartes en infrarouge lointain d'IRAS.
Il s'agit, \`a notre connaissance, de la premi\`ere d\'etection en continuum de ces objets (nuages 3 et 4).

 \subsection{Donn\'ees compl\'ementaires}

Nous avons utilis\'e un certain nombre de donn\'ees compl\'ementaires pour analyser nos mesures submillim\'etriques sur Orion.

 \subsubsection{Observations avec DIABOLO}

Nous avons utilis\'e des observations r\'ealis\'ees avec DIABOLO pour mieux contraindre le spectre du nuage 2.
Celles-ci ont \'et\'e effectu\'ees les 6 et 7 d\'ecembre 1996 avec le photom\`etre DIABOLO au foyer du t\'elescope millim\'etrique de 30 m de l'IRAM, situ\'e en Sierra Nevada, Espagne, pr\`es du Pic Veleta.
DIABOLO est un photom\`etre \`a deux canaux millim\'etriques, 1200 et 2100 $\mic$, d\'edi\'e \`a l'\'etude de l'effet Sunyaev-Zeldovich.
Cet instrument utilise des bolom\`etres refroidis \`a 0.1 K.
Une description d\'etaill\'ee du dispositif instrumental se trouve dans Beno\^\i t \etal (2000).

La configuration utilis\'ee pour l'observation du nuage 2 a \'et\'e la suivante.
L'instrument utilise un bolom\`etre par canal, avec une taille de faisceau de 30 secondes d'arc.
Un miroir secondaire vibrant est utilis\'e, afin de s'affranchir des contaminations de fond provenant du ciel et du t\'elescope lui-m\^eme, comme expliqu\'e pour PRONAOS en section \ref{compron}.
L'amplitude de vibration de ce miroir secondaire \'etait de 120$''$ \`a une fr\'equence de 0.8 Hz.
Une carte de 6$'$ par 6$'$ a \'et\'e produite, centr\'ee sur le pic d'intensit\'e du nuage, en balayant le faisceau du t\'elescope en azimut et hauteur.
La vitesse de balayage est de 6$''$/s, et le pas d'\'el\'evation entre deux lignes est de 15$''$.
Le temps total d'observation pour le nuage 2 est d'environ une heure.
Les cartes sont trait\'ees de fa\c{c}on \`a enlever les glitches (impacts de rayons cosmiques sur un d\'etecteur) en filtrant le signal de chaque ligne avec un filtre temporel.
Apr\`es avoir soustrait une ligne de base et r\'e-\'echantillonn\'e, les cartes de gradient (m\^eme syst\`eme que PRONAOS, section \ref{compron}) sont d\'econvolu\'ees en utilisant des programmes IDL maison suivant la m\'ethode EKH \cite{emerson79}.

 \subsubsection{Donn\'ees compl\'ementaires dans la litt\'erature}

Les autres donn\'ees compl\'ementaires utilis\'ees sont, d'une part, celles de l'observation compl\`ete du ciel r\'ealis\'ee par le satellite IRAS.
En effet, l'intensit\'e \`a 100 $\mic$ d'IRAS (voir figure \ref{iras100orion}) permet de mieux contraindre la temp\'erature des poussi\`eres (les gros grains dominent encore largement l'\'emission \`a 100 $\mic$, alors que ce n'est plus forc\'ement le cas \`a 60 $\mic$).
Cependant, les donn\'ees IRAS souffrent d'une sensibilit\'e nettement plus faible que PRONAOS, et de ce fait, les nuages faibles observ\'es par PRONAOS (nuages 2, 3 et 4) ne sont pas visibles dans ces donn\'ees IRAS.
On observe dans les donn\'ees \`a 100 $\mic$ que le halo de l'ISF et d'OMC-1 s'\'etend jusqu'aux zones des nuages froids.
On peut donc consid\'erer que dans ce cas pr\'ecis (complexe chaud proche de zones de poussi\`ere froide), ces zones contiennent deux composantes de gros grains, une chaude, le halo, observ\'ee par IRAS, et une froide, les nuages plus ou moins pr\'ecurseurs de formation d'\'etoiles, observ\'ee par PRONAOS.
N\'eanmoins, au vu de la carte IRAS \`a 100 $\mic$, le manque de sensibilit\'e semble \^etre une raison au moins aussi importante de ne pas voir les nuages froids qu'une suppos\'ee composante chaude \'etendue qui \'ecraserait toute autre \'emission \`a 100 $\mic$ mais serait invisible \`a 200 $\mic$ dans les donn\'ees PRONAOS.

\begin{figure}
\begin{center}
\caption[Carte IRAS d'Orion]{Carte IRAS \`a 100 $\mic$ de la N\'ebuleuse d'Orion M42 sur le champ d'observation de PRONAOS.
L'\'echelle de couleur est logarithmique, et montre l'intensit\'e reconstruite positive jusqu'\`a -1 en log, puis le bruit n\'egatif.
La zone blanche centrale correspond approximativement \`a la r\'egion dans laquelle les donn\'ees IRAS sont satur\'ees.
}
\label{iras100orion}
\end{center}
\end{figure}

Dans OMC-1, les donn\'ees IRAS sont satur\'ees, aussi l'intensit\'e IRAS \`a 100 $\mic$ n'est-elle dans cette zone qu'une limite inf\'erieure sans int\'er\^et majeur.
En revanche, les donn\'ees IRAS sont tout \`a fait utilisables dans l'ISF (zones d'OMC-2 et OMC-3).
Nous les utilisons pour l'ajustement en consid\'erant une erreur d'intercalibration de 25 \% entre IRAS et PRONAOS.

Nous avons \'egalement utilis\'e pour l'ajustement des donn\'ees spectrales d'OMC-1 provenant de donn\'ees compl\'ementaires de la litt\'erature.
Ce travail, r\'ealis\'e par Ristorcelli \etal (1998), a permis d'inclure les donn\'ees \`a 90 $\mic$ d'Harper (1974), en remplacement des donn\'ees IRAS satur\'ees, ainsi que les donn\'ees millim\'etriques de Chini \etal (1984) \`a 1 mm et de Mezger \etal (1990) \`a 1.3 mm.
Ces donn\'ees ont \'et\'e moyenn\'ees sur l'\'etendue de faisceau de la quatri\`eme voie de PRONAOS (3.5$'$).

Les flux int\'egr\'es de toutes ces donn\'ees compl\'ementaires sont indiqu\'es en table \ref{fluxcomporion}.

\begin{table}
\begin{center}
\caption[Flux int\'egr\'es autres que PRONAOS]{
Flux int\'egr\'es provenant de donn\'ees autres que PRONAOS.
Les diam\`etres approximatifs des r\'egions centrales \'etudi\'ees sont donn\'es en minutes d'arc.}
\begin{tabular}{lllllll}
\hline

& F$_{\nu}$(Jy) & F$_{\nu}$(Jy) & F$_{\nu}$(Jy) & F$_{\nu}$(Jy) &
F$_{\nu}$(Jy) & F$_{\nu}$(Jy) \\
& \tiny{90 $\mic$ Harper} & \tiny{100 $\mic$ IRAS} & \tiny{1000 $\mic$ Chini} & \tiny{1200 $\mic$ DIABOLO}
& \tiny{1300 $\mic$ Mezger} & \tiny{2100 $\mic$ DIABOLO} \\
\hline

OMC-1(3.6$'$) & \tiny{258000 $\pm $ 77000} & --- & \small{480 $\pm$ 140} & --- &
\small{187 $\pm$ 56} & --- \\
\hline

ISF &&&&&& \\
(sud) 5.4$'$ & --- & \small{10600 $\pm $ 2600} & --- & --- & --- & --- \\
\hline

ISF &&&&&& \\
(nord) 4.2$'$ & --- & \small{4000 $\pm $ 1000} & --- & --- & --- & --- \\
\hline

Nuage 2 (3.5$'$) & --- & --- & --- & \small{0.77 $\pm $ 0.13} & --- & \small{0.120 $\pm $ 0.030} \\
\hline

\end{tabular}
\label{fluxcomporion}
\end{center}
\end{table}

 \subsection{D\'erivation de la temp\'erature et de l'indice spectral\label{fitorion}}

L'originalit\'e des donn\'ees PRONAOS par rapport aux mesures submillim\'etriques au sol est que tout le domaine submillim\'etrique est couvert par les observations.
Ceci nous permet de contraindre efficacement \`a la fois la temp\'erature et l'indice spectral (\'equation \ref{cnm}) des gros grains responsables de l'\'emission continue submillim\'etrique (voir section \ref{pousgen}).
Nous ajustons donc les spectres obtenus avec PRONAOS et les donn\'ees compl\'ementaires avec le mod\`ele du corps noir modifi\'e.

Pour faire cela, il est n\'ecessaire auparavant de d\'egrader la r\'esolution de chaque carte des voies 1, 2 et 3 de PRONAOS-SPM \`a celle de la voie 4 (3.5$'$), afin de comparer des intensit\'es dilu\'ees dans la m\^eme taille et la m\^eme forme de faisceau.
Sinon, il est bien clair que les spectres sont biais\'es.
Nous avons donc liss\'e chacune des cartes PRONAOS obtenues dans les trois premi\`eres voies avec un profil calcul\'e pour obtenir la m\^eme forme de faisceau pour les quatre voies.

Nous r\'ealisons l'ajustement des spectres ainsi obtenus gr\^ace \`a la m\'ethode des moindres carr\'es, que nous appliquons pour d\'eterminer les meilleurs param\`etres C, T, et $\beta$.
Nous faisons un ajustement par pixel de la carte pour l'ensemble des donn\'ees spectrales disponibles.
Nous supposons de ce fait que l'\'emission dominante des gros grains peut \^etre caract\'eris\'ee par une seule temp\'erature.
Cette approximation peut ne pas \^etre satisfaisante dans le cas o\`u l'on observe de la poussi\`ere chaude majoritairement, mais qu'une composante de poussi\`ere froide est aussi pr\'esente sur la ligne de vis\'ee ou dans le faisceau.
Nous discutons de cette limitation en section \ref{anticor}.

En supposant que l'\'emission est optiquement mince, ce qui se justifie parfaitement dans le cas de l'\'emission submillim\'etrique de la poussi\`ere, qui est tr\`es peu absorb\'ee, et ce qui peut \^etre d\'emontr\'e {\it a posteriori} sur les r\'esultats obtenus, nous pouvons alors calculer la profondeur optique:

\begin{equation}
\tau_{\nu} = {\inu \over \bnu} = C \: \lambda^{-\beta}
\label{tau}
\end{equation}

Nous pr\'esentons en figure \ref{spectresorion} les spectres obtenus pour les sources identifi\'ees.
Les param\`etres temp\'erature et indice spectral d\'eriv\'es par l'ajustement et leurs barres d'erreur sont pr\'esent\'es en table \ref{paramorion}, ainsi que la profondeur optique dans les bandes extr\^emes de PRONAOS-SPM.
Cette table pr\'esente les r\'esultats pour les pics d'intensit\'e des sources pr\'esent\'ees, mais nous pr\'esentons \'egalement en figure \ref{cartestborion} les cartes compl\`etes de temp\'erature et d'indice spectral.
Celles-ci ont \'et\'e r\'ealis\'ees en utilisant les donn\'ees IRAS \`a 100 $\mic$ dans la majeure partie de la carte, mais pas dans M17 SW (satur\'ees) ni dans les nuages froids pour lesquels les donn\'ees IRAS ne d\'etectent pas de flux, comme bien visible en figure \ref{iras100orion}.
Sur l'ensemble des pixels des cartes pour lesquels l'ajustement est possible (l\`a o\`u il y a du signal), la pr\'ecision relative obtenue sur les deux param\`etres T et $\beta$ est meilleure que 50 \% dans 88 \% des cas, et meilleure que 20 \% dans 55 \% des cas.
La r\'egion d'OMC-1 souffre de l'absence des donn\'ees IRAS (satur\'ees), ce qui se traduit par des param\`etres T et $\beta$ assez mal d\'efinis sur les cartes.
En revanche, gr\^ace aux donn\'ees compl\'ementaires utilis\'ees, l'ajustement du pic d'intensit\'e d'OMC-1 donne des r\'esultats satisfaisants (voir table \ref{paramorion}).

\begin{figure}
\begin{center}
\caption[Spectres en intensit\'e de la N\'ebuleuse d'Orion]{Spectres en intensit\'e de la N\'ebuleuse d'Orion M42.
Les erreurs trac\'ees sont \`a 3 $\sigma$.
Les croix marquent les points PRONAOS, les triangles les points DIABOLO (Nuage 2), et les losanges les autres donn\'ees.
Pour la clart\'e de la figure, les points PRONAOS du nuage 4 sont visualis\'es par des carr\'es.
Les barres d'erreur des donn\'ees PRONAOS correspondent \`a l'incertitude relative d'\'etalonnage entre les bandes.
Les lignes pleines sont le r\'esultat des ajustements par le corps noir modifi\'e.}
\label{spectresorion}
\end{center}
\end{figure}

\begin{table}
\begin{center}
\caption[Temp\'erature et indice spectral dans Orion]{
Temp\'erature et indice spectral d'apr\`es l'ajustement des spectres des r\'egions centrales, profondeur optique.
Les barres d'erreur sont donn\'ees pour l'intervalle de confiance \`a 68 \%.
Les diam\`etres approximatifs des r\'egions centrales sont donn\'es en minutes d'arc.
}
\begin{tabular}{lllll}
\hline
   & T (K) & $\beta$ & ${\tau_{\nu}}\over{10^{-3}}$ & ${\tau_{\nu}}\over{10^{-3}}$\\
   & & & 200 $\mic$ & 580 $\mic$\\

\hline

OMC-1 (3.6$'$) & 66.1 $^{+10.2} _{-9.4}$ &
1.13 $^{+0.06} _{-0.09}$ & 47 $\pm $ 6 & 10.4 $\pm $ 0.8\\
\hline

ISF & & & & \\
(sud) 5.4$'$ & 22.4 $^{+2.1} _{-2.0}$ & 1.71 $^{+0.12} _{-0.19}$ & 5.3 $\pm $ 1.0 & 1.3 $\pm $ 0.1\\
\hline

ISF & & & & \\
(nord) 4.2$'$  & 25.2 $^{+2.5} _{-2.5}$ & 1.68 $^{+0.13}
_{-0.17}$ & 15 $\pm $ 3 & 1.9 $\pm $ 0.2\\
\hline

Nuage 1 (3.5$'$) & 17.0 $^{+3.4} _{-2.5}$ & 2.21 $^{+0.23}
_{-0.48}$ & 4.0 $\pm $ 3.2 & 0.76 $\pm $ 0.23\\
\hline

Nuage 2 (3.5$'$) & 11.8 $^{+0.6} _{-0.7}$ & 2.20
$^{+0.15} _{-0.18}$ & 4.2 $\pm $ 1.5 & 0.43 $\pm $ 0.07 \\ 
\hline

Nuage 3 (3.5$'$) & 13.3 $^{+2.6} _{-2.1}$ &
1.98 $^{+0.39} _{-0.79}$ & 1.4 $<4.2 (3 \sigma)$ & 0.19 $\pm $ 0.07\\
\hline

Nuage 4 (5.2$'$) & 16.9 $^{+7.3} _{-4.1}$ & 1.91 $^{+0.00}
_{-1.29}$ & 0.5 $<2.1 (3 \sigma)$ & 0.12 $\pm $ 0.06\\
\hline

\end{tabular}
\label{paramorion}
\end{center}
\end{table}

\begin{figure}
\begin{center}
\caption[Cartes de temp\'erature et d'indice d'Orion]{Cartes de temp\'erature en Kelvin (haut) et d'indice spectral (bas) de la N\'ebuleuse d'Orion M42.
Ces cartes ont \'et\'e r\'ealis\'ees en utilisant les donn\'ees IRAS \`a 100 $\mic$ dans la majeure partie de la carte, mais pas dans les environs de BN/KL (satur\'ees) ni dans les nuages froids pour lesquels les donn\'ees IRAS ne d\'etectent pas de flux, comme bien visible en figure \ref{iras100orion}.
La r\'esolution angulaire des deux cartes est de 3.5$'$.
Le contour ext\'erieur marque la limite de la zone observ\'ee.
Les zones en blanc \`a l'int\'erieur de ce contour correspondent aux zones pour lesquelles le rapport signal sur bruit est trop faible pour obtenir les param\`etres T et $\beta$.
Sur l'ensemble des pixels des cartes pour lesquels l'ajustement est possible (l\`a o\`u il y a du signal), la pr\'ecision relative obtenue sur les deux param\`etres T et $\beta$ est meilleure que 50 \% dans 88 \% des cas, et meilleure que 20 \% dans 55 \% des cas.
La r\'egion d'OMC-1 souffre de l'absence des donn\'ees IRAS (satur\'ees), ce qui se traduit par des param\`etres T et $\beta$ assez mal d\'efinis sur les cartes.
}
\label{cartestborion}
\end{center}
\end{figure}

 \subsection{Variations de la temp\'erature et de l'indice spectral\label{vartborion}}
La temp\'erature varie tr\`es fortement \`a l'int\'erieur et aux environs du complexe M42 observ\'e.
Elle varie entre environ 10 K et 100 K, et l'indice spectral change aussi beaucoup suivant les r\'egions, entre 1 et 2.5 environ.
Les nuages s\'epar\'es des r\'egions les plus denses (nuages 3 et 4) sont froids et ont des indices \'elev\'es, mais il y a \'egalement des nuages froids pr\`es de la r\'egion centrale active (OMC-1).
Ainsi, les nuages 1 et 2 (17 K et 11.8 K) sont situ\'es \`a environ 1.5 pc d'OMC-1.
En ce qui concerne les sources en commun (OMC-1 et le nuage 2), les param\`etres d\'eriv\'es dans ce travail sont en bon accord avec ceux de Ristorcelli \etal (1998, donn\'ees du premier vol de PRONAOS).

L'indice spectral d\'eriv\'e pour OMC-1 est 1.13 avec des barres d'erreur faibles ($^{+0.06} _{-0.09}$), gr\^ace au grand nombre de points de donn\'ees que nous avons dans la queue Rayleigh-Jeans du spectre, \'etant donn\'e la temp\'erature autour de 65-70 K.
Nous renvoyons \`a Ristorcelli \etal (1998) pour une discussion approfondie de l'objet OMC-1 (BN/KL).
Pour une r\'egion aussi complexe et active qu'OMC-1, il est probable que le spectre que nous observons entre 90 $\mic$ et 1300 $\mic$ traduise un m\'elange de diff\'erentes composantes de poussi\`eres \`a des temp\'eratures diff\'erentes, soit l'une \`a c\^ot\'e de l'autre dans le faisceau, soit sur la ligne de vis\'ee.
Notamment, le fait d'observer un indice spectral faible dans cette r\'egion pourrait s'expliquer par une composante de poussi\`eres froides qui a son maximum d'\'emission dans le domaine submillim\'etrique, augmentant ainsi l'\'emission totale mesur\'ee dans ce domaine, et r\'eduisant ainsi la pente de la queue Rayleigh-Jeans du spectre observ\'e.
Ceci se traduirait en effet par un indice spectral plus faible par rapport \`a une valeur suppos\'ee constante pour chaque composante de grains, disons $\approx$ 2.
Cependant, les investigations d\'ej\`a r\'ealis\'ees par Ristorcelli \etal (1998) dans OMC-1 ont montr\'e que la colonne densit\'e de poussi\`eres froides requise pour expliquer les mesures de cette mani\`ere devait \^etre 40 fois plus massive que la composante chaude.
Ceci est difficilement imaginable dans une r\'egion dense et active comme OMC-1, aussi devons-nous conclure que l'indice spectral de la composante dominante des poussi\`eres se trouve effectivement autour de 1.

L'\'emission \'etendue du filament ISF fait appara\^\i tre une temp\'erature assez froide (22 K au sud et 25 K au nord), avec des indices spectraux d'environ 1.7.
Dans ce domaine de temp\'erature, les points de donn\'ees PRONAOS combin\'es aux donn\'ees \`a 100 $\mic$ donnent des barres d'erreur faibles aussi bien sur la temp\'erature que sur l'indice spectral.
Afin d'en savoir plus sur la r\'ealit\'e des indices spectraux que l'on observe dans l'ISF, nous avons simul\'e une distribution de temp\'erature avec $\beta$ = 2, en utilisant une fonction cloche centr\'ee sur 17 K et de largeur \`a mi-hauteur de 9 K environ.
Ceci conduit \`a ajuster le spectre observ\'e par PRONAOS par un corps noir modifi\'e de temp\'erature un peu plus \'elev\'ee que la moyenne de la simulation (21 K) pour un indice spectral de 1.8.
Nous voyons donc que des temp\'eratures observ\'ees de l'ordre de 20 K avec des indices un peu plus faibles que 2 peuvent \^etre expliqu\'ees par une distribution de temp\'erature centr\'ee sur une valeur plus froide, avec un indice de 2.
Cependant, cet effet est faible et n'explique pas les \'ecarts d'indice spectral importants que nous observons dans Orion.

Les nuages 1 et 2 ont l'indice spectral le plus \'elev\'e observ\'e (2.2), le nuage 2 \'etant le plus froid (11.8 K), avec des barres d'erreur consid\'erablement r\'eduites par rapport \`a la mesure de Ristorcelli \etal (1998) gr\^ace aux mesures millim\'etriques de DIABOLO.
Le nuage 3 est un autre nuage froid (13.3 K) avec un indice spectral proche de 2 (1.98), cependant, l'erreur sur l'indice spectral est importante car nous ne disposons pas de donn\'ees millim\'etriques pour ce nuage.
Le nuage 4 a des param\`etres temp\'erature et indice spectral mal d\'efinis, mais semble \'egalement \^etre froid ($\approx$ 17 K).

R\'ecemment, des cartes de l'ISF \`a haute r\'esolution angulaire dans le domaine submillim\'etrique ont \'et\'e produites gr\^ace \`a des observations au sol, au CSO (Caltech Submillimeter Observatory), \`a 350 $\mic$, par Lis \etal (1998), et au JCMT (James Clerk Maxwell Telescope), \`a 450 et 850 $\mic$, par Johnstone \& Bally (1999).
Cependant, ces cartes n'incluent pas les r\'egions de faible intensit\'e montr\'ees dans ce travail (nuages 1 \`a 4), ni l'\'emission diffuse autour de l'ISF, que nous montrons en figures \ref{figorion12} et \ref{figorion34}.
Si l'on compare leurs cartes \`a celles de PRONAOS, il est visible que l'\'emission \'etendue autour des sources les plus brillantes n'est pas d\'etect\'ee par ces observations au sol.
En revanche, leur r\'esolution angulaire est bien sup\'erieure \`a la n\^otre, et permet ainsi de mieux d\'ecrire la structure fractale des nuages mol\'eculaires composant l'ISF.
Nos cartes apportent donc une information diff\'erente et compl\'ementaire des observations au sol.

En combinant leur carte \`a 350 $\mic$ avec la carte de Chini \etal (1997) \`a 1300 $\mic$ (IRAM), Lis \etal (1998) ont d\'eduit une carte d'indice spectral de la poussi\`ere sur la r\'egion brillante entourant OMC-1, y compris Orion-S et la Barre d'Orion.
En supposant une temp\'erature uniforme (55 K) dans cette r\'egion, ils ont d\'eriv\'e des indices spectraux entre 1.3 et 2.7, les valeurs les plus faibles \'etant observ\'ees dans la Barre d'Orion (environ 1.7) et BN/KL (1.8), tandis que les valeurs les plus \'elev\'ees (environ 2.3) sont d\'eriv\'ees dans le nord d'OMC-1.
Les valeurs qu'ils d\'erivent dans la zone la plus brillante (BN/KL et Orion-S) sont dans le domaine 1.8-2.0 quand on les moyenne sur la taille de faisceau de nos observations.
M\^eme si l'on tient compte du fait que la temp\'erature qu'ils supposent (55 K) est trop basse par rapport \`a celle que l'on mesure (66 K), ces valeurs restent clairement incompatibles avec notre mesure de l'indice spectral (1.13).
Cette diff\'erence pourrait \^etre expliqu\'ee par le fait qu'ils utilisent deux jeux de donn\'ees diff\'erentes avec d'\'eventuelles erreurs syst\'ematiques.

 \subsection{Densit\'es de colonne et masses\label{colorion}}

 \subsubsection{D\'erivation de la densit\'e de colonne}

Nous nous int\'eressons \`a l'estimation des densit\'es de colonne, c'est-\`a-dire la masse de gaz et de poussi\`eres le long de la ligne de vis\'ee, et des masses des r\'egions \'etudi\'ees.
Nous souhaitons mod\'eliser l'\'emission des poussi\`eres de fa\c{c}on \`a la fois simple et auto-coh\'erente avec nos observations et notre mode d'ajustement.
Pour cela, il nous para\^\i t n\'ecessaire d'utiliser l'indice spectral que nous mesurons pour caract\'eriser l'\'emission des grains, afin d'en d\'eduire une estimation de la densit\'e de colonne.

Cependant, l'indice spectral nous donne la pente de l'\'emissivit\'e, mais pas la valeur absolue (opacit\'e).
Nous consid\'erons uniquement l'\'emission thermique des gros grains qui dominent tr\`es largement dans ce domaine de longueur d'onde.
Il nous faut alors uniquement supposer une valeur de l'opacit\'e des gros grains, \`a une longueur d'onde donn\'ee.
L'opacit\'e est une grandeur qui traduit quelle quantit\'e d'\'energie est absorb\'ee et r\'e-\'emise par les grains par unit\'e de densit\'e de gaz et de poussi\`eres.
Sa valeur \`a 100 $\mic$ dans le mod\`ele de D\'esert \etal (1990) est bien contrainte par les donn\'ees IRAS, nous la choisissons donc.
La variation de l'opacit\'e dans le domaine submillim\'etrique observ\'e par PRONAOS (200 - 600 $\mic$) est alors d\'ecrite par l'indice spectral observ\'e.
De ce fait, nous supposons implicitement que la pente de la courbe d'\'emissivit\'e ne change pas dans le domaine de longueur d'onde observ\'e par PRONAOS, comme nous l'avions d'ailleurs d\'ej\`a suppos\'e lors de l'ajustement.
Cette supposition est largement acceptable dans l'intervalle de longueur d'onde observ\'e.
En revanche, nous prenons en compte la variabilit\'e spatiale de l'indice spectral dans les r\'egions observ\'ees.

Ceci nous donne un mod\`ele simple auto-coh\'erent qui nous permet de calculer la colonne densit\'e N$_H$ en fonction de l'intensit\'e spectrale et de l'indice spectral.
La valeur de l'opacit\'e adopt\'ee \cite{desert90} est:

\begin{equation}
\kappa_{100 \: \mic} = 0.361 \: cm^{2}/g
\end{equation}

par gramme de milieu total: gaz et poussi\`ere, \`a une longueur d'onde de 100 $\mic$.
Cette valeur est calcul\'ee d'apr\`es les courbes d'extinction donn\'ees dans D\'esert \etal (1990).
L'opacit\'e $\kappa$ est d\'efinie de la mani\`ere suivante:

\begin{equation}
\kappa = {\tau \over N_H \: m_H}
\end{equation}

La relation entre la densit\'e de colonne de gaz et les param\`etres de l'ajustement est alors:

\begin{equation}
N_{H} = {C \: \lambda^{-\beta}\over\kappa \: m_H}
\end{equation}

o\`u $m_H$ est la masse du proton et C et $\beta$ deux param\`etres de l'ajustement par la loi de corps noir modifi\'ee d\'ecrite en \'equation \ref{cnm}.
Pour la valeur de D\'esert \etal (1990) de $\kappa_{100 \mic}$, on a alors:

\begin{equation}
N_{H} = 1.67 \: 10^{24} \: C \: (100  \mic)^{-\beta}
\label{coldesert}
\end{equation}

avec N$_H$ exprim\'ee en protons/cm$^2$ et C en $\mic^\beta$.

Nous pouvons \'egalement adopter une valeur diff\'erente de l'opacit\'e \`a 100 $\mic$, afin de prendre en compte des milieux interstellaires froids et denses.
Le mod\`ele d'Ossenkopf \& Henning (1994) - et Ossenkopf (1993), sp\'ecifique aux c{\oe}urs protostellaires,
prend en compte la coagulation des grains dans des nuages froids et denses prot\'eg\'es du champ de rayonnement ultraviolet interstellaire.
L'opacit\'e est alors:

\begin{equation}
\kappa_{100 \: \mic} \approx 1 \: cm^{2}/g
\end{equation}

La relation entre la densit\'e de colonne et les param\`etres de l'ajustement est alors:

\begin{equation}
N_{H} = 6.02 \: 10^{23} \: C \: (100  \mic)^{-\beta}
\label{coloss}
\end{equation}

avec N$_H$ en en protons/cm$^2$ et C en $\mic^\beta$.

\subsubsection{Analyse}

Nous comparons les densit\'es de colonne estim\'ees de la mani\`ere d\'ecrite ci-dessus, avec celles que nous d\'erivons d'observations de la transition de rotation mol\'eculaire J = 1-0 du monoxyde de carbone $^{13}$CO.
Nous utilisons les donn\'ees de Nagahama \etal (1998), gracieusement fournies par les auteurs.
En adoptant l'\'equilibre thermique local et en supposant que l'\'emission est optiquement mince, la densit\'e de colonne de $^{13}$CO est:

\begin{equation}
N_{^{13}CO} = 2.6 \: 10^{14} \: {W_{^{13}CO} \over (1-e^{-5.3/T})}
\end{equation}

o\`u $W_{^{13}CO}$ est l'intensit\'e int\'egr\'ee dans la raie (voir par exemple Rohlfs \etal 2000).
Alors, en supposant un rapport en nombre ${H_2 \over ^{13}CO}$ \'egal \`a 4.6 $10^5$ \cite{rohlfs00}, un poids mol\'eculaire moyen de 2.36 m$_H$ (voir par exemple Elmegreen \etal 1979), et une temp\'erature du gaz de 30 K, nous d\'erivons un rapport densit\'e de colonne sur intensit\'e int\'egr\'ee dans la raie du $^{13}$CO de 17 $10^{20}$ protons $cm^{-2} (K.km/s)^{-1}$.
Ce coefficient nous permet de calculer des densit\'es de colonne que nous pr\'esentons dans la table \ref{tabmass}, afin de comparer aux estimations PRONAOS.

\begin{table}
\begin{center}
\caption[Densit\'es de colonne et masses dans Orion]{Densit\'es de colonne estim\'ees d'apr\`es les donn\'ees PRONAOS en utilisant les opacit\'es de D\'esert \etal (1990), Ossenkopf \& Henning (1994), et d'apr\`es les donn\'ees $^{13}$CO de Nagahama \etal (1998).
Masses, masses de Jeans, densit\'es.}
\begin{tabular}{lllllllll}
\hline

&N$_{H}$ \tiny{D\'esert} & N$_{H}$ \tiny{Ossenkopf} & N$_{H}$ \tiny{$^{13}$CO}
& \tiny{Masse D\'es.} & \tiny{Masse Oss.} & \tiny{M Jeans} &
\tiny{Densit\'e D\'es.} & \tiny{Densit\'e Oss.} \\

&$10^{20} cm^{-2}$ &$10^{20} cm^{-2}$ & $10^{20} cm^{-2}$ & (\msol)
& (\msol) & (\msol) & \tiny{(protons/cm${^3}$)} & \tiny{(protons/cm${^3}$)} \\
\hline

OMC-1(3.6$'$) & 1400 & ------ & 920 & 212 & --- & --- &  138000 & --- \\
\hline

ISF & & & & & & & \\
(sud) 5.4$'$ & 245 & ------ & 460 & 83 & --- & --- &  16100 & --- \\
\hline

ISF & & & & & & & \\
(nord) 4.2$'$  & 525 & ------ & 580 & 108 & --- & --- &  44400 & --- \\
\hline

Nuage 1 (3.5$'$) & 320 & 115 & 300 & 42 & 15 & 12.6 & 33800 & 12200 \\
\hline

Nuage 2 (3.5$'$)  & 320 & 120 & 170 & 46 & 17 & 8.7 & 32900 & 11900 \\ 
\hline

Nuage 3 (3.5$'$)  & 95 & 34 & 34 & 9.9 & 3.6 & 8.4 & 11300 & 4070 \\
\hline

Nuage 4 (5.2$'$)  & 33 & 12 & 9 & 10.3 & 3.7 & 18.5 & 2220 & 802 \\
\hline

\end{tabular}
\label{tabmass}
\end{center}
\end{table}

Comme nous pouvons l'observer dans cette table, la correspondance est plut\^ot bonne entre les densit\'es de colonne d'OMC-1 et de l'ISF estim\'ees d'apr\`es les mesures PRONAOS et le mod\`ele de D\'esert \etal (1990), et l'estimation d'apr\`es le CO.
En effet, le rapport PRONAOS/CO des densit\'es de colonne trouv\'ees est de 1.52 pour OMC-1, 0.53 pour le sud de l'ISF et 0.91 pour le nord.
Le nuage 1 fait appara\^\i tre un tr\`es bon accord avec l'estimation d'apr\`es D\'esert \etal 1990 (rapport 1.1).
Les nuages froids 2, 3 et 4 ont des estimations CO en tr\`es bon accord avec l'estimation d'apr\`es Ossenkopf \& Henning (rapports 0.71, 1 et 1.33).
Ces r\'esultats montrent que les valeurs trouv\'ees sont assez robustes.
Le bon accord entre les valeurs estim\'ees par le CO pour les nuages froids et le mod\`ele d'Ossenkopf \& Henning (1994) peut \^etre expliqu\'e par le fait que les condensations froides peuvent \^etre le si\`ege de formation de manteaux de glaces mol\'eculaires sur les grains, et de coagulation des grains.
Ces processus sont pris en compte par le mod\`ele d'Ossenkopf \& Henning (1994).

Nous montrons \'egalement en table \ref{tabmass} la masse de chaque r\'egion \'etudi\'ee.
Elle est obtenue simplement en multipliant la densit\'e de colonne de gaz par la surface approximative du nuage, en supposant une distance de 470 pc.
Bien entendu, l'erreur sur ces masses est importante et difficile \`a estimer correctement, car les sources d'incertitudes sont nombreuses.
\`A l'erreur sur la profondeur optique s'ajoute l'incertitude d'avoir pris une valeur de l'opacit\'e issue d'un mod\`ele et celle d'avoir pris une pente constante pour l'indice spectral.
N\'eanmoins, c'est une estimation int\'eressante qui permet d'avoir une mesure des masses ind\'ependante des raies mol\'eculaires, qui par ailleurs ne sont pas forc\'ement plus pr\'ecises.

Pour les nuages mol\'eculaires froids plus ou moins en dehors du complexe de formation d'\'etoiles de M42 (c'est-\`a-dire les nuages 1 \`a 4), nous pouvons raisonnablement nous demander s'ils pourraient s'effondrer sous l'effet de leur masse pour devenir des objets pr\'e-stellaires.
Comme indicateur de la stabilit\'e gravitationnelle des nuages \'etudi\'es, nous pr\'esentons en table \ref{tabmass} une estimation de leur masse de Jeans.
Nous utilisons l'expression d\'eriv\'ee de l'\'egalit\'e des \'energies gravitationnelle et thermique \cite{larson69}:

\begin{equation}
M_J = 10^{-18} \: R_{cm} \: T
\label{eqjeans}
\end{equation}

o\`u $M_J$ est exprim\'ee en masses solaires, $R_{cm}$ est le rayon, exprim\'e en centim\`etres, du nuage suppos\'e sph\'erique, et T la temp\'erature du nuage en Kelvin.
La temp\'erature que nous avons utilis\'ee est bien s\^ur celle des poussi\`eres que nous mesurons, alors que dans l'expression \ref{eqjeans}, il s'agit de la temp\'erature du gaz mol\'eculaire.
Cependant, dans le milieu interstellaire dense, les deux temp\'eratures sont peu diff\'erentes (voir par exemple Hollenbach 1989 et Tielens \& Hollenbach 1985).
En effet, diff\'erents m\'ecanismes permettent aux grains de chauffer efficacement le gaz (collisions gaz-grains, rayonnement infrarouge des grains absorb\'e par le gaz, etc).

La comparaison de la masse du nuage \`a sa masse de Jeans doit \^etre prise comme un indicateur de l'instabilit\'e du nuage.
Il y a bien s\^ur de larges incertitudes \`a la fois sur la masse du nuage mesur\'ee et sur les sources non thermiques d'\'energie interne qui peuvent contre-balancer l'effondrement gravitationnel, telles que le champ magn\'etique (voir par exemple Dudorov \& Stepanov 1997) et la turbulence du gaz (voir par exemple Gazol \etal 2001).
La pression ext\'erieure peut, au contraire, mener \`a la stabilisation de clumps qui sinon s'\'evaporeraient.
Malgr\'e ces restrictions, on voit en table \ref{tabmass} que dans notre \'echantillon de quatre nuages froids, il y a une tendance \`a ce que les clumps les plus proches de la r\'egion active (nuages 1 et 2) soient gravitationnellement les plus instables.

 \subsection{Conclusion sur Orion}

Tout d'abord, notre travail sur la N\'ebuleuse d'Orion montre l'\'emission submillim\'etrique des poussi\`eres \`a l'int\'erieur et autour du complexe mol\'eculaire, avec une grande sensibilit\'e et une m\'ethode de construction de cartes avanc\'ee qui permettent de mettre en \'evidence des nuages de tr\`es faible intensit\'e \`a proximit\'e de r\'egions tr\`es brillantes.
Nous montrons une large distribution de temp\'eratures et d'indices spectraux: la temp\'erature varie de 10 \`a 100 K, et l'indice spectral de 1 \`a 2.5 environ.
La d\'ecouverte de deux nuages froids suppl\'ementaires dans cette r\'egion (nuages 3 et 4), en plus de celle du nuage 2 par Ristorcelli \etal (1998), confirme que l'existence de telles condensations froides ne sont pas rares dans ces r\'egions.
Cependant, nous voyons \'egalement que les nuages froids \'etendus se trouvent en dehors de la zone active de formation d'\'etoiles.
Ils pourraient \^etre les futurs emplacements de l'activit\'e de formation d'\'etoiles.

Nous avons estim\'e les densit\'es de colonne et les masses des r\'egions observ\'ees en mod\'elisant de fa\c{c}on simple l'\'emission thermique des gros grains, et en utilisant une valeur de l'opacit\'e \`a 100 $\mic$ tir\'ee des mod\`eles (D\'esert \etal 1990, Ossenkopf \& Henning 1994).
Il y a un bon accord entre les densit\'es de colonne estim\'ees d'apr\`es les donn\'ees $^{13}$CO de Nagahama \etal (1998) et celles que nous d\'erivons de nos mesures du continuum submillim\'etrique.
Ceci montre que les valeurs d'opacit\'e dans les mod\`eles de grains sont correctes, et que l'imagerie spectrale submillim\'etrique en bandes larges est un moyen efficace de d\'eriver des masses dans le milieu interstellaire.

Finalement, nous observons une tendance \`a ce que plus les nuages froids se trouvent pr\`es du complexe de formation stellaire, plus ils sont instables gravitationnellement.
Or, l'histoire de la formation stellaire autour d'OMC-1 montre qu'il y a d\'ej\`a eu trois \`a quatre flamb\'ees successives de formation d'\'etoiles dans cette r\'egion, l'amas correspondant \`a l'objet BN/KL \'etant le dernier.
Les nuages froids que nous mettons en \'evidence pr\`es de la r\'egion active peuvent donc \^etre les pr\'ecurseurs d'une prochaine g\'en\'eration d'\'etoiles.
Ceci est bien s\^ur assez sp\'eculatif, et pourrait \^etre pr\'ecis\'e par d'autres observations de ces r\'egions, en particulier des possibles objets proto-stellaires enfouis.

L'ensemble de cette analyse des donn\'ees PRONAOS du vol 2 sur la N\'ebuleuse d'Orion M42 a \'et\'e publi\'ee dans Dupac \etal (2001) et a fait l'objet d'articles d'actes de conf\'erences (Dupac \etal 2001 {\it coll. a}, Dupac \etal 2002 {\it coll. c}).

\section{ Analyse des cartes de PRONAOS du complexe mol\'eculaire M17\label{m17}}

PRONAOS a observ\'e le complexe mol\'eculaire Messier 17 (voir l'introduction sur cet objet en section \ref{introm17}) lors du deuxi\`eme vol de l'exp\'erience en septembre 1996.
Cette observation de plus d'une heure, la plus longue du milieu interstellaire apr\`es Orion, r\'ealis\'ee lors du deuxi\`eme vol de PRONAOS, permet d'observer un autre complexe proche de formation d'\'etoiles massives, situ\'e n\'eanmoins \`a une distance plus importante qu'Orion (2200 pc).
L'utilisation de notre m\'ethode de construction de cartes avec bruit non uniforme d\'ecrite en section \ref{wieglob} a permis de construire les cartes que nous pr\'esentons et \'etudions dans les sections suivantes, et de faire appara\^\i tre des nuages de tr\`es faible intensit\'e.

 \subsection{\'Etude des cartes d'intensit\'e de PRONAOS}
Nous pr\'esentons en figure \ref{figm1712} les cartes obtenues dans les bandes 1 et 2 de PRONAOS-SPM, et en figure \ref{figm1734} les cartes des voies 3 et 4.
Nous pr\'esentons \'egalement en table \ref{fluxm17} les positions centrales des objets observ\'es et les flux dans les quatre voies int\'egr\'es sur un diam\`etre de 3.5$'$.

\begin{figure}
\begin{center}
\caption[Cartes de M17 \`a 200-260 $\mic$]{Cartes d'intensit\'e de PRONAOS dans les voies 1 (200 $\mic$, en haut) et 2 (260 $\mic$, en bas).
La r\'esolution angulaire est de 2$'$ pour les deux cartes.
L'\'echelle de couleur est logarithmique, et montre l'intensit\'e positive reconstruite jusqu'\`a -1 en log, puis le bruit n\'egatif.
Les zones non observ\'ees apparaissent en blanc.
}
\label{figm1712}
\end{center}
\end{figure}

\begin{figure}
\begin{center}
\caption[Cartes de M17 \`a 360-580 $\mic$]{Cartes d'intensit\'e de PRONAOS dans les voies 3 et 4 (360 $\mic$, en haut, et 580 $\mic$, en bas).
La r\'esolution angulaire est de 2.5$'$ pour la voie 3 et de 3.5$'$ pour la voie 4.
L'\'echelle de couleur est logarithmique, et montre l'intensit\'e positive reconstruite jusqu'\`a -1 en log, puis le bruit n\'egatif.
Les zones non observ\'ees apparaissent en blanc.
}
\label{figm1734}
\end{center}
\end{figure}

\begin{table}
\begin{center}
\caption[Flux int\'egr\'es dans M17]{
Coordonn\'ees \'equatoriales des pics d'intensit\'e, et flux (Jy) int\'egr\'es sur 3.5$'$ autour des pics d'intensit\'e des r\'egions identifi\'ees.
Les erreurs absolues sur les flux sont de 8 \% (1 $\sigma$).
Ces valeurs prennent en compte les corrections fines de couleur dues aux bandes larges de l'instrument SPM.
}
\begin{tabular}{lllllll}

\hline
 & $\alpha_{1950}$ & $\delta_{1950}$ & F$_{\nu}$(Jy) & F$_{\nu}$(Jy) & F$_{\nu}$(Jy) & F$_{\nu}$(Jy) \\
 & (h,mn,sec) & ($^o$,$'$) & 200 $\mic$ & 260 $\mic$ & 360 $\mic$ & 580 $\mic$ \\
\hline

M17 SW & 18 17 35 & -16 15 & 28000 & 15000 & 6600 & 1500 \\
\hline

M17 N & 18 17 45 & -16 03 & 4800 & 2500 & 1200 & 290 \\
\hline

Nuage A & 18 16 22 & -16 12 & 340 & 210 & 120 & 25 \\
\hline

Nuage B & 18 16 16 & -16 20 & 280 & 210 & 130 & 36 \\
\hline

Nuage C & 18 16 29 & -16 28 & 260 & 160 & 98 & 31 \\
\hline

Nuage D & 18 15 34 & -16 11 & 240 & 210 & 140 & 44 \\
\hline

\end{tabular}
\label{fluxm17}
\end{center}
\end{table}

La r\'esolution angulaire de ces cartes est 2 minutes d'arc dans les voies 1 et 2 (longueurs d'onde effectives de 200 et 260 $\mic$), 2.5$'$ dans la troisi\`eme voie (360 $\mic$) et 3.5$'$ dans la quatri\`eme (580 $\mic$).
Cette r\'esolution angulaire correspond \`a une r\'esolution spatiale d'environ 2 parsecs \`a la distance de M17 de 2200 pc \cite{chini80}.
Le niveau du bruit dans les cartes se situe comme pour Orion \`a environ 1 MJy/sr.
Les incertitudes d'\'etalonnage font que la pr\'ecision sur les intensit\'es est de 5 \% en relatif entre les bandes SPM (intercalibration), et de 8 \% en absolu (voir section \ref{quepron} de ce m\'emoire et Pajot {\it et al.}, \enprep).

Les cartes en figures \ref{figm1712} et \ref{figm1734} mettent en \'evidence les tr\`es importants contrastes d'intensit\'e existant dans ces complexes actifs.
Les zones de M17 Sud-Ouest (SW) et Nord (N) apparaissent sur nos cartes comme les r\'egions d'\'emission la plus intense.
La zone M17 SW atteint une intensit\'e maximale de 46000 \mjysr dans la voie 1 (200 $\mic$), au niveau de la source M17a (voir Lada 1976, Wilson \etal 1979 et Gatley \etal 1979).
Une petite condensation intense est visible au nord-est de la zone la plus intense.
Elle appara\^\i t clairement sur la figure \ref{figm1712} avec le deuxi\`eme contour le plus \'elev\'e sur l'image \`a 200 $\mic$, et poss\`ede un maximum d'intensit\'e \`a 200 $\mic$ de 9600 MJy/sr.
Cette condensation correspond \`a la source M17b.
Quant \`a la r\'egion M17 N, aussi appel\'ee M17c, elle atteint un pic d'intensit\'e de 7800 \mjysr \`a 200 $\mic$.
La source M17d \cite{wilson79} est elle bien visible \`a l'est de M17b, et son intensit\'e maximale \`a 200 $\mic$ est 2000 \mjysr environ.

Nous avons \'egalement cartographi\'e la r\'egion \`a l'ouest du complexe M17, dans laquelle nous mettons en \'evidence quatre condensations avec de faibles intensit\'es.
Le nuage A (voir figure \ref{figm1712} pour les noms) est situ\'e \`a l'ouest de M17 SW et appara\^\i t comme une condensation \'etendue atteignant une intensit\'e maximale de 630 \mjysr \`a 200 $\mic$.
Le nuage B se trouve au sud du nuage A et lui est reli\'e par un pont mol\'eculaire bien visible sur les cartes.
Le nuage B atteint une intensit\'e maximale \`a 200 $\mic$ de 440 \mjysr.
Une autre condensation \'etendue est visible au sud du nuage B, que nous appelons Nuage C.
Il a une intensit\'e maximale de 470 \mjysr, et les cartes \`a grande longueur d'onde montrent qu'il s'\'etend plus au sud, au-del\`a de la zone cartographi\'ee.
Le nuage D est une condensation situ\'ee au nord-ouest des cartes, reli\'ee au nuage B par un filament \`a l'int\'erieur duquel on peut observer une petite condensation (2$'$).
Le nuage D poss\`ede un pic d'intensit\'e \`a 200 $\mic$ de 470 \mjysr.

Il faut mentionner que ces nuages de faible intensit\'e sont difficilement perceptibles sur les cartes IRAS.
Comme nous le voyons en figures \ref{figm1712} et \ref{figm1734}, la carte \`a 580 $\mic$ appara\^\i t plus lisse que celle \`a 200 $\mic$, faisant moins ressortir les contrastes d'intensit\'e entre r\'egions de la carte.
Comme nous le discuterons plus tard, ceci est d\^u \`a une tendance des zones intenses \`a avoir des temp\'eratures plus chaudes que les zones de faible intensit\'e.
Ceci induit une intensit\'e relative plus forte des zones peu intenses dans les cartes \`a grandes longueurs d'onde, r\'eduisant de ce fait les contrastes entre r\'egions dans les cartes \`a grandes longueurs d'onde.

\subsection{Donn\'ees IRAS\label{irasm17}}

Nous utilisons les donn\'ees IRAS \`a 100 $\mic$ (voir figure \ref{iras100m17}) dans la plus grande partie de la carte pour mieux contraindre la temp\'erature des poussi\`eres.
Cependant, les donn\'ees IRAS apparaissent satur\'ees dans la zone de M17 SW, et ne sont donc qu'une limite inf\'erieure sans int\'er\^et particulier.
Nous n'utilisons pas non plus ces donn\'ees dans la zone \`a l'ouest de M17 (nuages A \`a D) car l'information donn\'ee par la carte IRAS \`a 100 $\mic$ (voir figure \ref{iras100m17}) nous semble trop bruit\'ee.
En revanche, dans la quasi-totalit\'e de M17 (sauf la zone centrale de M17 SW), nous utilisons les donn\'ees IRAS \`a 100 $\mic$, en consid\'erant pour cela que l'erreur d'intercalibration entre IRAS et PRONAOS est de 25~\%.
Dans les quelques zones o\`u l'ajustement est difficile m\^eme avec les donn\'ees 100 $\mic$ du fait des hautes temp\'eratures estim\'ees ($>$ 70 K), nous incluons \'egalement les donn\'ees \`a 60 $\mic$ d'IRAS.
Ceci est justifi\'e par le fait que, pour des temp\'eratures \'elev\'ees, les gros grains dominent encore le continuum d'\'emission \`a cette longueur d'onde, ce qui n'est plus le cas pour des temp\'eratures froides, pour lesquelles l'\'emission \`a 60 $\mic$ est domin\'ee par les tr\`es petits grains.

\begin{figure}
\begin{center}
\caption[Carte IRAS \`a 100 $\mic$ de M17]{Carte IRAS \`a 100 $\mic$ de M17 sur le champ d'observation de PRONAOS.
L'\'echelle de couleur est logarithmique, et montre l'intensit\'e reconstruite positive jusqu'\`a -1 en log, puis le bruit n\'egatif.
La zone blanche centrale correspond approximativement \`a la r\'egion dans laquelle les donn\'ees IRAS sont satur\'ees.
}
\label{iras100m17}
\end{center}
\end{figure}

 \subsection{D\'erivation de la temp\'erature et de l'indice spectral}

L'ajustement des donn\'ees pour d\'eriver la temp\'erature et l'indice spectral est expos\'e en section \ref{fitorion}.
Nous prenons en compte it\'erativement les corrections fines de couleur dues aux grandes largeurs de bande de SPM.
Nous pr\'esentons en figure \ref{spectresm17} les spectres obtenus pour certaines sources principales (pics d'intensit\'e).
Les param\`etres temp\'erature et indice spectral d\'eriv\'es par l'ajustement et leurs barres d'erreur sont pr\'esent\'es en table \ref{paramm17}, ainsi que la profondeur optique dans les bandes extr\^emes de PRONAOS-SPM.
La profondeur optique est donn\'ee par l'\'equation \ref{tau}, et l'erreur sur la profondeur optique est alors:

\begin{equation}
{\Delta \tau_{\nu}\over\tau_{\nu}} = {\Delta I_{\nu}\over I_{\nu}} + {h \nu \over kT^2} \: \Delta T
\end{equation}

La table \ref{paramm17} pr\'esente les r\'esultats pour les pics d'intensit\'e des sources pr\'esent\'ees, mais nous pr\'esentons aussi en figure \ref{cartestbm17} les cartes compl\`etes de temp\'erature et d'indice spectral.
Nous avons utilis\'e les donn\'ees IRAS tel qu'expos\'e en section \ref{irasm17}.
Avec ce processus, l'ajustement est r\'ealis\'e pour les pixels dans lesquels la carte fait appara\^\i tre du signal reconstruit.
Cependant, il existe bien \'evidemment quelques zones trop bruit\'ees pour obtenir une estimation de la temp\'erature et de l'indice spectral.
Ces zones apparaissent en blanc sur les cartes en figure \ref{cartestbm17}.

\begin{figure}
\begin{center}
\caption[Spectres en intensit\'e de M17]{Spectres en intensit\'e du complexe M17.
Les erreurs trac\'ees sont \`a 3 $\sigma$.
Les barres d'erreur des donn\'ees PRONAOS correspondent \`a l'incertitude relative d'\'etalonnage entre les bandes.
Les lignes pleines sont le r\'esultat des ajustements par le corps noir modifi\'e.}
\label{spectresm17}
\end{center}
\end{figure}

\begin{table}
\begin{center}
\caption[Temp\'erature et indice spectral de M17]{Temp\'erature, indice spectral et profondeur optique des pics d'intensit\'e observ\'es dans M17.
Les barres d'erreur sont donn\'ees pour l'intervalle de confiance de 68 \% (1 $\sigma$).}
\begin{tabular}{lllll}
\hline
   & T (K) & $\beta$ & $\tau_{\nu} \centerdot 10^3$ & $\tau_{\nu} \centerdot 10^3$\\
   & & & 200 $\mic$ & 580 $\mic$\\

\hline

M17 SW & 29 $\pm$ 8 & 1.7 $\pm$ 0.3 & 80 $\pm$ 60 & 12 $\pm$ 3 \\
\hline

M17 N & 28 $\pm$ 3 & 1.6 $\pm$ 0.2 & 15 $\pm$ 6 & 2.6 $\pm$ 0.4 \\
\hline

Nuage A & 17 $\pm$ 3 & 2.3 $\pm$ 0.3 & 6 $\pm$ 4 & 0.5 $\pm$ 0.1 \\
\hline

Nuage B & 17 $\pm$ 3 & 1.7 $\pm$ 0.3 & 5 $\pm$ 3 & 0.7 $\pm$ 0.2 \\
\hline

Nuage C & 26 $\pm$ 4 & 1.2 $\pm$ 0.2 & 1 $\pm$ 0.5 & 0.3 $\pm$ 0.07 \\
\hline

Nuage D & 14 $\pm$ 2 & 1.9 $\pm$ 0.3 & 9 $\pm$ 6 & 1.3 $\pm$ 0.3 \\
\hline

\end{tabular}
\label{paramm17}
\end{center}
\end{table}

\begin{figure}
\begin{center}
\caption[Cartes de temp\'erature et d'indice dans M17]{Cartes de temp\'erature en Kelvin (haut) et d'indice spectral (bas) du complexe M17.
La r\'esolution angulaire est de 3.5$'$ pour chaque carte.
Le contour ext\'erieur marque la limite de la r\'egion observ\'ee.
Les r\'egions en blanc \`a l'int\'erieur de ce contour n'ont pas un rapport signal sur bruit suffisant pour pouvoir d\'eriver la temp\'erature et l'indice spectral.
}
\label{cartestbm17}
\end{center}
\end{figure}

 \subsection{Variations de la temp\'erature et de l'indice spectral\label{vartbm17}}

Les cartes en figure \ref{cartestbm17} montrent tout d'abord une anticorr\'elation entre la temp\'erature et l'indice spectral.
Nous d\'ecrivons cet effet en section \ref{anticor}.
Dans la plus grande partie des cartes, nous observons que la temp\'erature varie entre 10 K et 80 K, et que l'indice spectral varie aussi significativement entre 1 et plus de 2.5.
Des temp\'eratures sup\'erieures \`a 80 K sont \'egalement observ\'ees, autour de 100 K ou au-dessus, dans quelques pixels situ\'es pr\`es du front d'ionisation.
Des variations \`a grande \'echelle de la temp\'erature et de l'indice spectral sont largement visibles sur ces cartes, ce qui montre, soit dit en passant, leur int\'er\^et scientifique.
En effet, m\^eme si la r\'esolution angulaire de l'instrument est relativement modeste compar\'ee aux instruments au sol, les variations de temp\'erature et d'indice mises en \'evidence ici sont significatives.
La temp\'erature augmente globalement d'ouest (nuages froids) en est (pr\`es du front d'ionization).
Les zones de faible intensit\'e \`a l'ouest des cartes sont nettement plus froides que la zone intense de M17.

La table \ref{paramm17} montre la temp\'erature, l'indice spectral et la profondeur optique d\'eriv\'es, ainsi que leurs barres d'erreur \`a 1 $\sigma$.
La r\'egion de M17 SW (sud-ouest) fait appara\^\i tre une temp\'erature de 29 K ($\pm$ 8) au pic d'intensit\'e de M17a.
Pour cette r\'egion, nous d\'erivons un indice spectral de 1.7 ($\pm$ 0.3).
Cependant, les donn\'ees IRAS sont satur\'ees dans cette zone, et de ce fait, inutilisables.
L'ajustement est donc assez incertain, et la temp\'erature pourrait \^etre plus \'elev\'ee.
De plus, dans une r\'egion aussi active que M17 SW, d'importants m\'elanges de composantes de poussi\`eres de temp\'eratures diff\'erentes pourraient avoir lieu sur la ligne de vis\'ee et dans le faisceau.
Nous en reparlerons en d\'etails en section \ref{anticor}.
La condensation M17 N (nord) poss\`ede une temp\'erature au pic d'intensit\'e de 28 K, et un indice spectral de 1.6.
Les barres d'erreur sont faibles gr\^ace aux donn\'ees IRAS \`a 100 $\mic$ qui permettent de contraindre efficacement la temp\'erature.
Des barres d'erreur comparables sont d\'eriv\'ees pour la majorit\'e des pixels de M17, sauf en ce qui concerne les nuages de faible intensit\'e \`a l'ouest.
La temp\'erature du complexe M17 est d'environ 30 \`a 50 K \`a l'ext\'erieur des condensations M17 SW et M17 N, et l'indice spectral varie entre 1 et 1.5.
Sur les bords du complexe, on peut observer des zones plus chaudes pr\`es des r\'egions ionis\'ees \`a l'est et au nord-ouest.
La temp\'erature atteint 80 K ou plus dans ces r\'egions, et l'indice spectral observ\'e varie entre 0.7 et 1.1.
Notons que pour ces hautes temp\'eratures, la temp\'erature est efficacement contrainte par les donn\'ees IRAS \`a 60 $\mic$, et l'indice spectral par les donn\'ees PRONAOS.

Les nuages A, B et D, dans la partie ouest des cartes, montrent des temp\'eratures basses (14-17 K) et des indices spectraux \'elev\'es, entre 1.7 et 2.3.
Certaines zones de ces condensations atteignent des temp\'eratures tr\`es basses, jusqu'\`a 10 K.

 \subsection{Densit\'es de colonne et masses\label{colm17}}

La m\'ethode d'estimation des densit\'es de colonne \`a partir des donn\'ees continuum est similaire \`a celle utilis\'ee pour l'analyse des donn\'ees d'Orion (section \ref{colorion}).
Nous pr\'esentons les r\'esultats sur les densit\'es de colonne et les masses en table \ref{massm17}.
L'erreur relative sur $N_{H}$ est, au premier ordre:

\begin{equation}
{\Delta N_H \over N_H} = {\Delta \tau \over \tau} + ln({\lambda \over 100 \mic}) \: \Delta\beta
\end{equation}

\begin{table}
\begin{center}
\caption[Densit\'es de colonne et masses dans M17]{Densit\'es de colonne estim\'ees d'apr\`es les donn\'ees PRONAOS + opacit\'e de D\'esert \etal (1990), et d'apr\`es les donn\'ees $^{13}$CO de Wilson \etal (1999).
Masses PRONAOS + D\'esert {\it et al.}, masses PRONAOS + Ossenkopf \& Henning, masses de Jeans.
Les barres d'erreur sur les densit\'es de colonne estim\'ees d'apr\`es PRONAOS sont donn\'ees pour l'intervalle de confiance \`a 68 \%.
}
\begin{tabular}{llllll}
\hline
&N$_{H}$ \tiny{PRON.+D\'es.} & N$_{H}$ \tiny{$^{13}$CO} & Masse
\tiny{PRON.+D\'es.} & Masse \tiny{PRON.+Oss.} & M Jeans \\

&$10^{20} H cm^{-2}$ & $10^{20} H cm^{-2}$ & \msol & \msol & \msol \\
\hline

M17 SW & 4100 $\pm 3600$ & 1200 & 16000 & --- & --- \\
\hline

M17 N & 730 $\pm 340$ & 460 & 4600 & --- & --- \\
\hline

Nuage A & 450 $\pm 380$ & --- & 2000 & 730 & 70\\
\hline

Nuage B & 250 $\pm 210$ & --- & 2400 & 870 & 100\\
\hline

Nuage C & 38 $\pm 25$ & --- & 350 & --- & 150 \\
\hline

Nuage D & 630 $\pm 530$ & --- & 2100 & 760 & 100\\
\hline

\end{tabular}
\label{massm17}
\end{center}
\end{table}

Afin de d\'eriver une estimation ind\'ependante des densit\'es de colonne, nous utilisons les observations r\'ealis\'ees par Wilson \etal (1999) dans la raie de rotation J = 1-0 du $^{13}CO$.
Nous remercions chaleureusement les auteurs de nous avoir transmis leurs donn\'ees sous forme num\'erique.
Leur carte couvre le complexe M17 dans sa totalit\'e, mais pas les zones de faible intensit\'e \`a l'ouest (nuages A \`a D).
Nous estimons les densit\'es de colonne d'apr\`es les donn\'ees CO de la m\^eme fa\c{c}on que pour Orion.

Il n'y a pas un tr\`es bon accord dans M17 SW entre l'estimation PRONAOS + mod\`ele de D\'esert \etal (1990) et le CO, bien que ce d\'esaccord soit coh\'erent avec notre barre d'erreur.
Il est possible que cela soit d\^u \`a une \'eventuelle \'emission optiquement \'epaisse du $^{13}$CO dans la transition 1-0.
La r\'egion M17 N fait appara\^\i tre un bien meilleur accord.
Si l'on fait confiance \`a notre mesure PRONAOS + D\'esert, ce qui semble d'ailleurs assez sp\'eculatif au vu des barres d'erreur, il est possible de contraindre le rapport densit\'e de colonne sur intensit\'e int\'egr\'ee $^{13}$CO.
Ce rapport varie entre environ 15 $10^{20}$ et 100 $10^{20} \: cm^{-2} (K.km/s)^{-1}$ en fonction des r\'egions de M17.
Puisque le rapport correspondant \`a l'\'emission CO optiquement mince est de 17 $10^{20}$, et que les zones les plus intenses font appara\^\i tre les valeurs les plus \'elev\'ees de ce coefficient, il est possible de conclure que le d\'esaccord entre les estimations continuum et CO est d\^u (au moins en partie) \`a une \'emission $^{13}$CO 1-0 optiquement \'epaisse.
On peut aussi invoquer des incertitudes dans le rapport ${H_2 \over ^{13}CO}$ quand les densit\'es de colonne sont grandes.
De plus, la temp\'erature d'excitation peut \^etre sup\'erieure \`a 30 K dans M17 SW, bien que nous la mesurions autour de 30 K.
Si nous supposons une temp\'erature d'excitation de 60 K, la densit\'e de colonne estim\'ee est alors 2300 $10^{20} \: H \: cm^{-2}$ pour M17 SW, au pic d'intensit\'e.
Si maintenant nous nous int\'eressons aux sources d'incertitudes dans l'estimation continuum, mis \`a part que les barres d'erreur sont larges, nous pouvons mettre l'accent sur le fait que l'opacit\'e \`a 100 $\mic$ que nous avons adopt\'ee pourrait sur-estimer la densit\'e de colonne, particuli\`erement au sein des nuages froids, qui pourraient \^etre le si\`ege d'effets sp\'eciaux tels que la formation de manteaux de glace sur les grains et la coagulation des grains.
Par exemple, ces processus sont pris en compte par le mod\`ele de c{\oe}urs protostellaires d'Ossenkopf \& Henning (1994).
Avec l'opacit\'e \`a 100 $\mic$ de ce mod\`ele, nous d\'erivons des densit\'es de colonne 2.8 fois plus faibles qu'en utilisant l'opacit\'e du mod\`ele de D\'esert \etal (1990).
Cette correction pourrait s'appliquer aux nuages A, B et D.

Nous avons finalement calcul\'e les masses des r\'egions observ\'ees dans ce complexe mol\'eculaire g\'eant.
Pour cela, nous avons int\'egr\'e la densit\'e de colonne d\'eriv\'ee de nos mesures continuum, sur la surface des nuages que nous observons, en supposant une distance de 2200 pc \cite{chini80}.
Nous d\'erivons ainsi une masse de milieu interstellaire total de M17 (sans les nuages \`a l'ouest) de 31000 \msol, dans laquelle M17 SW p\`ese environ 16000 \msol et M17 N 4600 \msol.
Nous d\'erivons \'egalement des estimations des masses des nuages A, B, C et D, que nous pr\'esentons en table \ref{massm17}.
Pour les nuages froids A, B et D, il est vraisemblable que les masses puissent \^etre plus faibles que cela: en utilisant l'opacit\'e \`a 100 $\mic$ d'Ossenkopf \& Henning (1994), nous obtenons des masses d'environ 800 M$_\odot$ pour chacune de ces condensations.
Nous pr\'esentons \'egalement en table \ref{massm17} les masses de Jeans calcul\'ees pour les nuages s\'epar\'es de la r\'egion active M17.
Cette estimation est une indication de la stabilit\'e ou de l'instabilit\'e gravitationnelle de ces objets.
Nous utilisons pour cela l'expression d\'eriv\'ee de l'\'egalit\'e des \'energies gravitationnelle et thermique (\'equation \ref{eqjeans}).

Il y a bien s\^ur de larges incertitudes \`a la fois sur la masse du nuage mesur\'ee et sur les sources non thermiques d'\'energie interne qui peuvent contre-balancer l'effondrement gravitationnel, telles que le champ magn\'etique et la turbulence du gaz.
Malgr\'e ces restrictions, on voit en table \ref{massm17} que les nuages A, B et D pourraient \^etre instables gravitationnellement, particuli\`erement le nuage A qui pourrait \^etre comprim\'e par la pression ext\'erieure due \`a la r\'egion ionis\'ee au nord.

 \subsection{Conclusion sur M17}

Nous avons observ\'e le complexe mol\'eculaire Messier 17 dans le domaine submillim\'etrique.
La large r\'egion observ\'ee (50$'$ x 30$'$, $\approx$ 30 pc x 20 pc) a permis de mettre en \'evidence des condensations de faible intensit\'e, situ\'ees dans les environs imm\'ediats de la r\'egion active, mais qui font appara\^\i tre des temp\'eratures basses et qui pourraient \^etre le si\`ege futur de formation stellaire.
Notre \'etude a montr\'e une variation importante des temp\'eratures et des indices spectraux.
La temp\'erature varie d'environ 10 K \`a 100 K, et l'indice spectral de 1 \`a 2.5.

Nous avons estim\'e les densit\'es de colonne et les masses des r\'egions observ\'ees dans et aux environs de M17, et nous d\'erivons notamment une masse totale du complexe M17 de 31000 \msol.
Bien s\^ur, les barres d'erreur sont larges et difficiles \`a estimer de mani\`ere r\'ealiste, mais il existe n\'eanmoins une tendance claire \`a ce que les nuages froids observ\'es (A, B et D) soient gravitationnellement instables, et donc puissent produire une prochaine g\'en\'eration d'\'etoiles.
Ces r\'esultats pourraient \^etre soutenus par des observations continuum \`a haute sensibilit\'e dans le domaine millim\'etrique, particuli\`erement pour les nuages de faible intensit\'e \'etudi\'es \`a l'ouest de M17, afin de mieux contraindre l'indice spectral de la poussi\`ere froide.
Des observations en infrarouge \`a plus haute r\'esolution angulaire pourraient \'egalement \^etre utiles pour \'etudier la structure interne (\'eventuellement fractionn\'ee, voire fractale) de ces nuages froids.

L'ensemble de cette analyse sur M17 a \'et\'e publi\'ee dans Dupac \etal (2002) et a fait l'objet d'un article d'actes de conf\'erence (Dupac \etal 2001 {\it coll. c}).

\section{ Anticorr\'elation entre la temp\'erature et l'indice spectral des poussi\`eres\label{anticor}}

 \subsection{R\'esultats sur les pics d'intensit\'e}

En plus des variations importantes de temp\'erature et d'indice spectral mises en \'evidence dans ce travail concernant Orion et Messier 17, il appara\^\i t une anticorr\'elation entre les deux param\`etres, ne serait-ce qu'en observant les valeurs dans les tables \ref{paramorion} et \ref{paramm17}.
Les r\'egions froides ($<$ 20 K environ) ont en effet des indices proches de 2 tandis que les r\'egions plus chaudes ont des indices plus faibles.

Nous avons calcul\'e les contours de vraisemblance \`a 68 \% de confiance (1 $\sigma$) des couples (T, $\beta$) d\'eriv\'es par l'ajustement pour chaque r\'egion dans Orion.
Notons que la vraisemblance est proportionnelle \`a $e^{-\chi^{2}\over2}$.
Nous pr\'esentons ces contours de vraisemblance en figure \ref{vraisemborion}.

\begin{figure}
\begin{center}
\caption[Contours de vraisemblance de six r\'egions dans Orion]{Contours de vraisemblance de six r\'egions dans Orion M42.
Les croix marquent les positions des maxima de vraisemblance.}
\label{vraisemborion}
\end{center}
\end{figure}

Nous avons \'egalement calcul\'e le coefficient de corr\'elation entre temp\'erature et indice spectral, pour l'ensemble des pics d'intensit\'e des r\'egions identifi\'ees.
Nous avons plusieurs couples (T, $\beta$) dont nous analysons la corr\'elation par le coefficient:

\begin{equation}
C={\sum{(T_{i}-\bar T) \: (\beta_{i}-\bar \beta)}\over\sqrt{\sum{(T_{i}-\bar T)^{2}}
  \: \sum{(\beta_{i}-\bar \beta)^{2}}}}
\end{equation}

Il s'agit de la d\'efinition standard du coefficient de corr\'elation.
Une corr\'elation positive (une grandeur augmente quand l'autre augmente aussi) est marqu\'ee par une valeur positive entre 0 et 1 de ce coefficient.
Une corr\'elation n\'egative (anticorr\'elation) est marqu\'ee par une valeur n\'egative entre 0 et -1.
Une valeur de ce coefficient proche de -1 marque une tr\`es forte anticorr\'elation entre les deux param\`etres.

Une premi\`ere analyse de la figure \ref{vraisemborion} montre bien l'anticorr\'elation entre temp\'erature et indice spectral.
Le coefficient de corr\'elation des sept couples (T, $\beta$) dans Orion est de -0.92.
Cette corr\'elation est pour partie domin\'ee par la r\'egion chaude d'OMC-1, mais la corr\'elation est toujours forte (-0.83) quand OMC-1 n'est pas inclus dans le calcul.
Pour les six sources de M17 dans la table \ref{paramm17},
le coefficient de corr\'elation vaut -0.60, ce qui est une corr\'elation significative.
Cependant, ces r\'esultats ne sont pas tr\`es significatifs statistiquement (seulement quelques points de donn\'ees).
C'est pourquoi nous avons \'egalement analys\'e la corr\'elation entre temp\'erature et indice spectral sur l'ensemble des pixels des cartes \ref{cartestborion} et \ref{cartestbm17}.

 \subsection{R\'esultats sur les cartes de temp\'erature et d'indice spectral}

Cette analyse nous permet de d\'eriver un coefficient d'anticorr\'elation pour l'ensemble des pixels de chaque carte ayant leur temp\'erature comprise entre 10 K et 80 K, et leur indice spectral compris entre 1 et 2.5.

Pour les cartes d'Orion, ce coefficient vaut -0.56.
69~\% de ces pixels ont des erreurs relatives sur \`a la fois la temp\'erature et l'indice spectral de moins de 20~\%.
Si l'on restreint l'analyse de corr\'elation \`a ces pixels, on trouve un coefficient de -0.68.
Cette augmentation de l'anticorr\'elation avec la restriction sur les erreurs montre la r\'ealit\'e de cet effet.
Nous pr\'esentons en figure \ref{tbsorion} une carte de la r\'epartition des pixels dans les cartes en figure \ref{cartestborion} pour lesquels la temp\'erature est entre 10 et 80 K et l'indice entre 1 et 2.5.
Ceci a \'et\'e r\'ealis\'e en comptant les couples (T, $\beta$) dans des rectangles de taille ($\Delta$T = 3 K, $\Delta \beta$ = 0.1), et en lissant le r\'esultat obtenu.
L'allongement en forme de banane, en figure \ref{tbsorion}, de la r\'epartition des temp\'eratures et indices spectraux, montre clairement l'anticorr\'elation entre les deux param\`etres.
La quasi-totale absence de pixels ayant T $>$ 40 K et $\beta >$ 1.5 est particuli\`erement visible.

\begin{figure}
\begin{center}
\caption[R\'epartition des couples temp\'erature - indice pour Orion]{R\'epartition dans l'espace (T, $\beta$) des couples temp\'erature - indice d\'eriv\'es dans Orion, pour T entre 10 et 80 K et $\beta$ entre 1 et 2.5.
Ceci a \'et\'e fait en comptant les couples (T, $\beta$) dans des rectangles de taille (3 K, 0.1), et en lissant le r\'esultat obtenu.
La couleur rouge correspond \`a 10-12 points par rectangle, le jaune \`a 8-9, le vert \`a 5-7, le bleu clair \`a 4, le bleu fonc\'e \`a 2-3, le violet \`a 1 et le noir \`a 0.}
\label{tbsorion}
\end{center}
\end{figure}

\begin{figure}
\begin{center}
\caption[R\'epartition des couples temp\'erature - indice pour M17]{R\'epartition dans l'espace (T, $\beta$) des couples temp\'erature - indice d\'eriv\'es dans M17, pour T entre 10 et 80 K et $\beta$ entre 1 et 2.5.
Ceci a \'et\'e fait en comptant les couples (T, $\beta$) dans des rectangles de taille (3 K, 0.1), et en lissant le r\'esultat obtenu.
La couleur rouge correspond \`a 20-23 points par rectangle, le jaune \`a 16-19, le vert \`a 11-15, le bleu clair \`a 7-10, le bleu fonc\'e \`a 3-6, le violet \`a 1-2 et le noir \`a 0.}
\label{tbsm17}
\end{center}
\end{figure}

Nous pr\'esentons en figure \ref{tbsm17} une carte dans l'espace (T, $\beta$) montrant la r\'epartition des pixels des cartes de M17 (figure \ref{cartestbm17}).
Cette carte a \'et\'e r\'ealis\'ee de la m\^eme fa\c{c}on que pour Orion.
L'allongement en forme de banane est \'egalement bien visible.
Si l'on consid\`ere tous les pixels des cartes de M17 dont la temp\'erature est comprise entre 10 K et 80 K, et dont l'indice est compris entre 1 et 2.5, le coefficient de corr\'elation d\'eduit vaut -0.56.
Parmi ces 1079 points, 1071 ont des barres d'erreur relatives inf\'erieures \`a 50~\% \`a la fois sur la temp\'erature et sur l'indice spectral, et 801 ont leurs deux barres d'erreur inf\'erieures \`a 20~\% (1 $\sigma$).
Le coefficient de corr\'elation d\'eduit de ces 801 pixels bien ajust\'es est -0.64.
Si l'on consid\`ere uniquement les points de la carte pour lesquels la temp\'erature et l'indice sont d\'eriv\'es avec une pr\'ecision meilleure que 10~\%, nous n'avons plus que 87 points et le coefficient de corr\'elation est fort: -0.74.
Cette augmentation de la corr\'elation avec la restriction sur les erreurs est une preuve de la r\'ealit\'e de cet effet d'anticorr\'elation.

 \subsection{Effet des erreurs sur les param\`etres de l'ajustement}

On peut voir en figure \ref{vraisemborion} que les contours de vraisemblance sont allong\'es, ce qui est principalement d\^u, pour OMC-1, \`a la relative insensibilit\'e aux variations de temp\'erature \'elev\'ee des mesures submillim\'etriques (i.e. dans la queue Rayleigh-Jeans).
Pour les nuages 1, 2, et 3 dans Orion, cet effet est d\^u \`a la relative insensibilit\'e aux variations d'indice des mesures submillim\'etriques de la poussi\`ere froide.
On remarque d'ailleurs que gr\^ace aux donn\'ees millim\'etriques de DIABOLO, le nuage 2 a un indice spectral tr\`es nettement mieux d\'efini que les nuages 1 et 3.
Nous voyons donc qu'\`a cause des incertitudes sus-dites sur les ajustements, une certaine quantit\'e d'anticorr\'elation artificielle peut appara\^\i tre.
Afin de quantifier cet effet, nous avons r\'ealis\'e des simulations en faisant des ajustements r\'ep\'et\'es sur des donn\'ees simul\'ees \`a partir de couples (T, $\beta$) sans corr\'elation intrins\`eque.
Nous simulons des donn\'ees PRONAOS avec des barres d'erreur r\'ealistes, et nous les ajustons de la m\^eme fa\c{c}on que les vraies donn\'ees.
Nous calculons ensuite le coefficient de corr\'elation sur une distribution de points comparable \`a ce que nous observons dans les donn\'ees dans le domaine T-$\beta$ (et sur le m\^eme nombre de points).
En effet, nous observons une anticorr\'elation induite par les erreurs sur l'ajustement, qui se traduit par un coefficient toujours inf\'erieur en valeur absolue \`a -0.4 pour six points de donn\'ees.
Cet effet est donc clairement insuffisant pour expliquer la corr\'elation d\'ecouverte dans les donn\'ees sur les pics d'intensit\'e (-0.92 dans Orion et -0.60 dans M17).

Nous avons \'egalement r\'ealis\'e des ajustements de donn\'ees PRONAOS simul\'ees \`a partir de 801 couples (T, $\beta$) al\'eatoires sans corr\'elation intrins\`eque, dans les m\^emes domaines de temp\'erature et d'indice que ceux d\'eriv\'es des donn\'ees.
Nous avons ajust\'e ces spectres simul\'es de la m\^eme fa\c{c}on que les vraies donn\'ees, et r\'ep\'et\'e ce processus de simulations avec plusieurs distributions T-$\beta$, avec et sans les points \`a 100 $\mic$.
Nous avons obtenu des coefficients de corr\'elation sur les param\`etres ajust\'es, la plupart du temps compris entre 0.0 et -0.2, et jamais au-dessus (en valeur absolue) de -0.3.
C'est pourquoi nous pouvons affirmer que la proc\'edure d'ajustement elle-m\^eme induit un peu d'anticorr\'elation entre la temp\'erature et l'indice spectral d\'eduits.
Il y a un effet de d\'eg\'en\'erescence entre les deux param\`etres, mais cette corr\'elation artificielle est clairement insuffisante pour expliquer le degr\'e d'anticorr\'elation d\'ecouvert sur les vraies donn\'ees (-0.68 dans Orion et -0.64 dans M17).
Nous devons donc conclure \`a une propri\'et\'e physique intrins\`eque des grains observ\'es, o\`u au moins \`a une propri\'et\'e de la poussi\`ere observ\'ee en int\'egrant sur la taille de faisceau et sur la ligne de vis\'ee (colonne densit\'e de milieu interstellaire).

 \subsection{Effet de m\'elange de temp\'eratures}

\subsubsection{Introduction}

En effet, la principale raison d'\^etre sceptique vis-\`a-vis de cet effet d'anticorr\'elation est que nous avons r\'ealis\'e les ajustements en supposant pouvoir r\'eduire l'\'emission de la poussi\`ere observ\'ee \`a une composante dominante ayant une seule temp\'erature (et \'egalement un seul indice spectral).
Or, il n'est pas \'evident {\it a priori} que les spectres observ\'es ne puissent pas \^etre d\'ecrits par deux ou plusieurs composantes de poussi\`eres ayant des temp\'eratures diff\'erentes, le long de la ligne de vis\'ee ou l'une \`a c\^ot\'e de l'autre dans le faisceau.
En effet, il est possible que des composantes de poussi\`eres froides augmentent l'\'emission \`a grande longueur d'onde d'une composante chaude ayant un indice r\'eel de 2.
De ce fait, l'indice apparent de la poussi\`ere, si l'on ne consid\`ere qu'une composante, est r\'eduit.
Cet effet pourrait apparemment expliquer l'anticorr\'elation mesur\'ee, puisque si en r\'ealit\'e toutes les poussi\`eres ont un indice de 2, on peut supposer que dans les zones o\`u l'on observe de la poussi\`ere chaude, il se trouve aussi de la poussi\`ere plus froide qui fait que l'indice mesur\'e diminue.

D'apr\`es les consid\'erations sur OMC-1 et l'ISF en section \ref{vartborion}, il est assez bien exclu qu'un tel effet d'anticorr\'elation soit uniquement d\^u dans ces r\'egions \`a des effets de m\'elange de poussi\`eres \`a diff\'erentes temp\'eratures.
Nous avons r\'ealis\'e des investigations plus g\'en\'erales, en simulant des spectres \`a une composante de poussi\`eres de diverses temp\'eratures assez chaudes et d'indices spectraux entre 1 et 2.
Nous essayons ensuite d'ajuster ces spectres simul\'es avec deux composantes de poussi\`eres, ayant toutes les deux un indice spectral ``standard'' de 2, la composante chaude ayant la m\^eme \'emission \`a 100 $\mic$ et la m\^eme temp\'erature que le spectre \`a une seule composante.
De ce fait, il y a \'egalit\'e entre les deux profondeurs optiques ($C \: \lambda^{-\beta}$) \`a 100 $\mic$, et donc la composante chaude a la m\^eme densit\'e de colonne que celle du spectre \`a une composante (voir le mod\`ele simple de densit\'e de colonne en section \ref{colorion}).
Nous ajustons ainsi les longueurs d'onde courtes du domaine PRONAOS et celles d'IRAS, dans le spectre originel, par la composante chaude de poussi\`eres.
La partie du spectre correspondant aux grandes longueurs d'onde, en revanche, fait appara\^\i tre un manque d'\'emissivit\'e de la composante chaude par rapport \`a ce qui est simul\'e originellement.
Ceci est bien s\^ur d\^u au fait que l'indice spectral standard (\'egal \`a 2) sous-estime l'\'emission \`a grande longueur d'onde.
Nous introduisons donc une composante de poussi\`eres plus froide, avec le m\^eme indice spectral standard, que nous ajoutons \`a la composante chaude pour ajuster le spectre originel.
Nous pouvons alors calculer quelle quantit\'e de mati\`ere froide est n\'ecessaire pour expliquer de cette mani\`ere les indices spectraux faibles que nous observons, en calculant les densit\'es de colonne de gaz d'apr\`es notre mod\`ele.
En effet, avec deux composantes de poussi\`eres ayant le m\^eme indice spectral ($\beta$=2), le rapport des densit\'es de colonne entre les deux composantes est simplement le rapport des param\`etres C (voir \'equation \ref{cnm}).

 \subsubsection{L'effet de m\'elange de temp\'eratures explique-t-il les indices faibles ?}

Il appara\^\i t que les indices faibles (autour de 1) impliqueraient de tr\`es grandes quantit\'es de poussi\`eres froides sur la ligne de vis\'ee (voir table \ref{twocomp}).
Par exemple, afin d'ajuster par cette m\'ethode un spectre ayant T = 50 K, $\beta$ = 1, avec une composante froide de 10 K, il est n\'ecessaire d'invoquer une densit\'e de colonne de gaz froid 100 fois sup\'erieur \`a la densit\'e de colonne de gaz chaud (50 K).
Pour ajuster T = 70 K, $\beta$ = 1, on a besoin de 200 fois plus de gaz froid (10 K) que de gaz chaud.
Nous avons dit que la densit\'e de colonne de gaz chaud dans l'ajustement \`a deux composantes \'etait la m\^eme que celle de l'ajustement \`a une seule composante (chaude).
Comme on peut l'observer en table \ref{massm17}, les densit\'es de colonne estim\'ees semblent avoir le bon ordre de grandeur, compar\'e par exemple aux estimations d'apr\`es le CO.
Il est extr\^emement improbable qu'il y ait de telles quantit\'es de gaz froid (plus de 2 d\'ecades plus massives que le gaz chaud et les estimations d'apr\`es le CO) dans chaque ligne de vis\'ee o\`u nous observons des indices spectraux faibles.
Cependant, en invoquant de la poussi\`ere moins froide (20 K au lieu de 10 K), le r\'esultat pourrait \^etre diff\'erent.
Par exemple, pour ajuster le spectre T = 60 K, $\beta$ = 1, on a besoin de ``seulement'' 20 fois plus de gaz froid que de gaz chaud, mais cette hypoth\`ese est encore extr\^emement peu probable.
Si nous rempla\c{c}ons la composante \`a 20 K par de la poussi\`ere ti\`ede \`a 40 K, alors il n'est plus possible du tout d'ajuster le spectre, il faut vraiment de grandes quantit\'es de poussi\`ere froide pour expliquer l'\'emission aux grandes longueurs d'onde submillim\'etriques.
Ces investigations prouvent clairement que des indices spectraux autour de 1 ne peuvent pas s'expliquer seulement par des m\'elanges de temp\'eratures de poussi\`eres.

\begin{table}
\caption[Ajustement \`a deux composantes de poussi\`eres]{R\'esum\'e des investigations effectu\'ees concernant l'ajustement des spectres par deux composantes de poussi\`eres.
De gauche \`a droite: la temp\'erature du spectre d'origine, l'indice spectral du spectre d'origine, la temp\'erature de la composante froide, qui a un indice spectral $\beta$=2, et le rapport en masse d\'eriv\'e entre les composantes froide et chaude.
}

\begin{center}
\begin{tabular}{llll}
\hline
T (K) & $\beta$ & T$_{froid}$ (K) & Rapport en masse \\
\hline
50 & 1 & 10 & 100 \\
\hline
70 & 1 & 10 & 200\\
\hline
60 & 1 & 20 & 20\\
\hline
60 & 1 & 40 & pas d'ajustement\\
\hline
50 & 1.5 & 10 & 50\\
\hline
50 & 1.5 & 20 & 5\\
\hline
50 & 1.5 & 30 & 3, mauvais ajust.\\
\hline
40 & 1.7 & 10 & 20\\
\hline
\label{twocomp}
\end{tabular}
\end{center}
\end{table}

 \subsubsection{L'effet de m\'elange de temp\'eratures explique-t-il les indices interm\'ediaires ?}

Si nous essayons \`a pr\'esent d'ajuster un spectre ayant T = 50 K et un indice spectral interm\'ediaire de 1.5, nous observons que la quantit\'e de gaz froid \`a 10 K n\'ecessaire est de l'ordre de 50 fois celle de gaz chaud.
Avec une composante froide \`a 20 K au lieu de 10 K, il faut encore une quantit\'e de gaz froid 5 fois plus importante que celle de gaz chaud.
Avec une composante \`a 30 K, il suffit de deux fois plus de gaz froid que de gaz chaud, mais cette distribution n'ajuste pas bien les grandes longueurs d'onde, et m\^eme une densit\'e de colonne trois fois plus importante que l'estimation \`a une composante reste improbable, au vu surtout des estimations CO.
Nous voyons donc qu'y compris pour les indices interm\'ediaires, il est tr\`es difficile d'ajuster le spectre avec seulement des composantes ayant $\beta$ = 2.
Si l'on prend un spectre cr\'e\'e avec un indice encore plus proche de 2, \'egal \`a 1.7, et une temp\'erature de 40 K, il faut encore 20 fois plus de gaz froid \`a 10 K que de gaz chaud pour ajuster le spectre.
Dans cette situation, si l'on invoque une composante \`a 30 K au lieu de 10 K, il est assez ais\'e d'ajuster le spectre, mais ce cas correspond de tr\`es pr\`es \`a l'ajustement avec une composante unique.
Nous montrons donc par ces investigations que m\^eme pour des indices spectraux interm\'ediaires (1.5 - 1.7), il n'est pas vraisemblable qu'ils s'expliquent seulement par des m\'elanges de temp\'eratures avec le m\^eme indice spectral standard.

Un aspect tr\`es spectaculaire de la figure \ref{tbsm17} est que tr\`es peu de pixels apparaissent dans les donn\'ees avec T $>$ 40 K et $\beta >$ 1.5 (le m\^eme effet, encore plus net, peut \^etre not\'e pour Orion en figure \ref{tbsorion}), et qu'aucun pixel ne peut \^etre trouv\'e avec T $<$ 20 K et $\beta <$ 1.6.
Nous avons v\'erifi\'e par les simulations de couples (T, $\beta$) al\'eatoires que ceci n'\'etait pas un art\'efact de l'ajustement.
Nous avons montr\'e auparavant que pour expliquer cet effet d'anticorr\'elation par un m\'elange de temp\'eratures de poussi\`eres, il faudrait faire appel \`a de tr\`es grandes quantit\'es de poussi\`eres froides.
Il est tr\`es improbable que de telles quantit\'es soient pr\'esentes dans chaque lieu de ce complexe mol\'eculaire o\`u nous mesurons des temp\'eratures \'elev\'ees ou moyennes.
Puisque, de fa\c{c}on significative, nous n'observons pas de poussi\`ere chaude associ\'ee \`a des indices \'elev\'es, nous devons conclure \`a une explication plus fondamentale pour cet effet d'anticorr\'elation entre la temp\'erature et l'indice spectral.
De plus, la forme de la r\'epartition des points (T, $\beta$) en figure \ref{tbsm17} est tr\`es difficile \`a expliquer par la seule raison des m\'elanges de temp\'erature.
En effet, ceci impliquerait que plus la composante (chaude) est chaude, plus la composante froide est massive, car l'indice spectral est plus r\'eduit (vers 1) aux hautes temp\'eratures, en plus du fait que pour le m\^eme indice il faut plus de masse froide pour une temp\'erature plus chaude.
Il est inimaginable que ceci soit une r\`egle g\'en\'erale, et cela montre tr\`es clairement que la supposition des m\'elanges de temp\'erature est insuffisante pour justifier des variations d'indice spectral mesur\'ees, non plus que de sa d\'ependance inverse avec la temp\'erature.

 \subsection{Conclusion: un effet d\^u \`a la physique des grains ?}

Nous privil\'egions donc une explication provenant des propri\'et\'es physiques intrins\`eques des grains pour cet effet d'anticorr\'elation.
Des exp\'erimentations en laboratoire (Agladze \etal 1996 et Mennella \etal 1998) ont montr\'e cet effet pour des temp\'eratures descendant jusqu'\`a 10 K, mais il n'avait jamais encore \'et\'e mis en \'evidence sur des observations de milieux astrophysiques avant PRONAOS.

Agladze \etal (1996) ont mesur\'e les spectres d'absorption de grains cristallins et de grains amorphes dans le domaine de longueurs d'onde entre 0.7 et 2.9 mm.
Ils ont montr\'e une anticorr\'elation entre la temp\'erature et l'indice spectral dans l'intervalle de temp\'erature 10-25 K et l'ont attribu\'e \`a des processus quantiques faisant intervenir des effets tunnel \`a deux niveaux.
Leurs mesures ne sont pas suffisantes pour expliquer nos observations dans le domaine submillim\'etrique, car les caract\'eristiques de l'absorption (et donc de l'\'emission) des grains peuvent \^etre assez diff\'erentes dans le domaine millim\'etrique.
Mennella \etal (1998) ont mesur\'e le coefficient d'absorption d'analogues de grains de poussi\`ere cosmique, cristallins et amorphes, entre 20 $\mic$ et 2 mm de longueur d'onde.
Ils ont d\'eduit une anticorr\'elation entre temp\'erature et indice spectral dans le domaine de temp\'erature 24-295 K, et l'ont attribu\'e \`a des processus quantiques de diff\'erences de deux phonons.
Nous observons cet effet jusqu'\`a des temp\'eratures plus basses ($\approx$ 10 K), aussi des donn\'ees sur l'\'emission submillim\'etrique provenant de grains de laboratoire \`a basse temp\'erature pourraient compl\'eter efficacement les donn\'ees d\'ej\`a existantes.

Ceci dit, d'autres causes peuvent faire varier l'indice spectral.
Les variations d'\'emissivit\'e des grains peuvent \^etre dues \`a la composition et la taille des grains.
Les valeurs de $\beta$ invoqu\'ees pour les grains silicat\'es tournent autour de 2, tandis que les carbones amorphes auraient plut\^ot un indice spectral proche de 1.
En revanche, les grains graphitiques auraient $\beta$ = 2 (voir Mennella \etal 1995).
Des indices spectraux tr\`es faibles (0.2 - 1.4) ont \'et\'e observ\'es autour d'\'etoiles jeunes ou \'evolu\'ees (Weintraub \etal 1989, Knapp \etal 1993), et l'on attribue en g\'en\'eral ceci \`a des grains qui grossissent beaucoup dans des enveloppes ou des disques stellaires denses.
Notons aussi que les observations de galaxies proches \cite{dunne01} favorisent $\beta$ = 2, mais que ceci pourrait \^etre d\^u au fait que ces observations moyennent des r\'egions o\`u domine la poussi\`ere froide.
Dunne \& Eales (2001) ajustent des spectres d'\'emission consistant en donn\'ees IRAS \`a 60 $\mic$ et 100 $\mic$, et en donn\'ees SCUBA \`a 450 $\mic$ et 850 $\mic$, en utilisant deux composantes de poussi\`eres ayant $\beta$ = 2.
Ils ajustent donc quatre points avec quatre param\`etres (deux temp\'eratures et deux C), ce qui, assez naturellement, produit un excellent ajustement, mais qui n'est pas tr\`es significatif pour la valeur de l'indice spectral fix\'ee.
Cependant, il semble bien au vu des spectres pr\'esent\'es que leurs observations favorisent $\beta$ proche de 2.
Quoi qu'il en soit, les observations de PRONAOS des r\'egions actives de formation d'\'etoiles massives telles qu'Orion et M17 montrent que des grains de faible indice spectral sont associ\'es aux r\'egions chaudes.

D'autres tendances sont observables en figure \ref{tbsm17}.
Trois concentrations sont visibles: l'une se trouve autour de 25 - 40 K et $\beta$ = 1 - 1.7, une autre autour de 45 - 60 K et $\beta$ = 1 - 1.4.
Ces deux concentrations correspondent \`a la zone chaude dans M17.
La troisi\`eme concentration correspond \`a T = 10 - 18 K avec $\beta$ = 1.6 - 2.5, et traduit essentiellement les nuages froids \`a l'ouest de M17.
Il semble \'egalement que l'anticorr\'elation T-$\beta$ se change en une ligne verticale (temp\'erature constante) quand la temp\'erature devient plus basse que 15 K.
Cependant, cette derni\`ere tendance pourrait \^etre due principalement \`a la relative insensibilit\'e aux variations d'indice spectral de nos mesures submillim\'etriques, \`a ces basses temp\'eratures.
Elle n'est d'ailleurs pas visible dans Orion (figure \ref{tbsorion}).
Des observations millim\'etriques des nuages froids mis en \'evidence dans ce travail seraient tr\`es utiles pour contraindre efficacement l'indice spectral.

\vspace*{1em}

Nous pouvons conclure de cette analyse concernant les variations de temp\'erature et d'indice spectral que l'anticorr\'elation mesur\'ee dans Orion et M17 est tr\`es probablement due majoritairement \`a une propri\'et\'e fondamentale des grains, qui produit une d\'ependance de l'\'emissivit\'e en fonction de la temp\'erature.

\section{ Conclusion: l'apport de PRONAOS pour la connaissance de la structure galactique et la cosmologie}

Nous pouvons d\'egager deux apports principaux du projet PRONAOS, le premier concernant l'\'etude de l'effet Sunyaev-Zeldovich (voir la description de cet effet en section \ref{rfsec}), le second concernant les mesures de temp\'erature et d'indice spectral des poussi\`eres galactiques et leur anticorr\'elation.
Les observations SZ (voir section \ref{rfsec}) sur les amas Abell 2142, 2163 et 478 ont permis de mesurer pour la premi\`ere fois l'effet Sunyaev-Zeldovich positif, c'est-\`a-dire l'incr\'ement de flux du rayonnement fossile dans la direction de l'amas dans le domaine submillim\'etrique.
Cette mesure effectu\'ee avec diff\'erents instruments (PRONAOS, SuZIE, DIABOLO et ISO-PHOT) sur l'amas Abell 2163 \cite{lamarre98} a permis de d\'ecrire pr\'ecis\'ement le spectre de l'infrarouge lointain au millim\'etrique dans la direction d'un amas de galaxies.
Elle a montr\'e l'effet Sunyaev-Zeldovich positif et n\'egatif, ainsi que la contribution de la poussi\`ere galactique \`a l\'emission submillim\'etrique.
Cette mesure de l'effet SZ positif est sans aucun doute l'un des deux r\'esultats majeurs de l'exp\'erience PRONAOS.

La plus grande partie de la science r\'ealis\'ee gr\^ace aux observations de PRONAOS concerne la poussi\`ere, dans des r\'egions tr\`es diverses de la Galaxie.
La mesure simultan\'ee de la temp\'erature et de l'indice spectral est une originalit\'e d\'ecisive de ce projet, puisqu'elle permet de mettre en \'evidence des variations de temp\'erature et d'indice importantes.
La d\'ecouverte de l'anticorr\'elation qui appara\^\i t entre ces deux param\`etres est une contribution observationnelle majeure \`a la connaissance des grains et de leur r\^ole dans le milieu interstellaire, mais cet effet reste \`a \^etre bien compris du point de vue des m\'ecanismes physiques qui en sont la cause.
Cette contribution sera \'egalement tr\`es utile \`a l'am\'elioration de la pr\'ecision sur les densit\'es de colonne et les masses d\'eriv\'ees des mesures continuum.
Les premi\`eres publications de cet effet sur Orion et M17 (Dupac \etal 2001, Dupac \etal 2002) devraient \^etre suivies par d'autres concernant des objets aussi divers que $\rho$ Ophiuci (Ristorcelli {\it et al.}, \enprep), Cygnus B (M\'eny {\it et al.}, \enprep), etc.

\section{ Projets futurs}

 \subsection{Introduction}

La meilleure connaissance du spectre d'\'emission des poussi\`eres, et \'egalement de la polarisation des grains, sont des \'el\'ements qui permettraient de mieux appr\'ehender la physico-chimie du milieu interstellaire, la structure \`a toutes les \'echelles de la Galaxie, et \'egalement la physique des autres galaxies.
Aussi est-ce un domaine qui suscite de nombreux projets d'observation, notamment par de futurs instruments spatiaux (Herschel, Planck), mais \'egalement par des projets ballon tels qu'ELISA.
Nous d\'ecrivons \'egalement tr\`es succintement le projet SNOOPY, qui n'est pas (encore ?) \`a proprement parler un projet mais plut\^ot un exercice d'\'ecole d'\'et\'e approfondi.
Bien entendu, ce qui suit ne repr\'esente pas une liste exhaustive des projets futurs dans ce domaine.

 \subsection{ELISA}

ELISA ({\it Experiment for Large Infrared Survey Astronomy}) est un projet en cours d'\'evaluation par le CNES, dont le PI ({\it prime investigator}) est Jean-Philippe Bernard au CESR.
Il s'agit d'un projet ballon de cartographie compl\`ete du plan galactique dans quatre bandes larges submillim\'etriques.
La nacelle comporterait un miroir primaire de 1 m de diam\`etre, donc deux fois plus petit que PRONAOS, ce qui permet d'all\'eger consid\'erablement le poids (500 kg au lieu de 3 t) et les co\^uts, et d'en faire un projet plus flexible que PRONAOS.
Les longueurs d'onde d'observation pr\'evues sont 170, 240, 400 et 650 $\mic$, ce qui est tr\`es semblable mais sur un domaine un peu plus large que PRONAOS.
Ces longueurs d'onde couvrent bien le domaine submillim\'etrique, mais elles ne permettent toujours pas d'intercalibrer avec les donn\'ees IRAS.
On peut aussi regretter que le nombre de mesures ind\'ependantes dans le domaine submillim\'etrique ne soit pas plus important, de sorte \`a mieux contraindre le spectre d'\'emission des poussi\`eres, en particulier vis-\`a-vis des probl\`emes de m\'elange de temp\'eratures.
Cependant, le fait de pouvoir cartographier le plan galactique en d\'erivant des temp\'eratures et indices spectraux aussi bien que PRONAOS est extr\^emement int\'eressant.
La r\'esolution angulaire du projet est de 3.5$'$, ce qui est exactement la r\'esolution de la bande de longueur d'onde la plus grande de PRONAOS.
Gr\^ace \`a l'utilisation de matrices de bolom\`etres r\'ealis\'ees par le CEA-LETI pour Herschel-PACS, le gain en sensibilit\'e est tr\`es important par rapport \`a PRONAOS.
En effet, ELISA utiliserait 1024 bolom\`etres par bande spectrale au lieu d'un, et chaque d\'etecteur est lui m\^eme dix fois plus sensible.
Ceci permet d'envisager une cartographie compl\`ete du plan galactique en deux vols.

\`A part cet objectif premier, ELISA devrait \^etre capable de r\'ealiser un catalogue de sources submillim\'etriques, d'\'etudier de jeunes \'etoiles enfouies, de mesurer la polarisation des grains, d'\'etudier la poussi\`ere dans les galaxies proches, de mesurer l'effet SZ positif sur quelques amas de galaxies, et de caract\'eriser le fond diffus infrarouge.

Pour que les donn\'ees arrivent au bon moment par rapport \`a la pr\'eparation des missions Planck et Herschel, il faudrait qu'ELISA r\'ealise ses vols entre 2004 et 2006, dont au moins un dans chaque h\'emisph\`ere, afin de couvrir tout le plan galactique.


 \subsection{SNOOPY\label{snoopy}}

SNOOPY ({\it Submillimeter Observatory Of PolarimetrY}) est un projet con\c{c}u lors de l'\'ecole d'\'et\'e d'Alpbach, Autriche, en juillet 2000, par un groupe de treize jeunes chercheurs en formation (voir figure \ref{figsnoopy}) encadr\'es par Jean-Michel Lamarre (IAS, Orsay).
Cet exercice d'\'ecole a \'et\'e pr\'esent\'e \`a Roger Bonnet, directeur scientifique de l'ESA (Agence Spatiale Europ\'eenne), et a suscit\'e un grand int\'er\^et dans la communaut\'e astronomique, aussi se pourrait-il qu'il devienne plus qu'un simple exercice.
La pr\'esentation de ce projet peut \^etre consult\'ee \`a l'adresse suivante: {\it http:// www.lal.in2p3.fr/recherche/planck/ Sophie/Snoopy}.

Il s'agit d'un projet de mission satellite bon march\'e, en orbite basse (Low Earth Orbit ou LEO), utilisant la technologie d'Herschel en ce qui concerne le t\'elescope (3.5 m de diam\`etre), ayant pour but de r\'ealiser une cartographie compl\`ete du ciel submillim\'etrique, afin d'observer la polarisation de la poussi\`ere galactique.
La polarisation est en effet un param\`etre clef pour comprendre la structure de la Galaxie, et en particulier le r\^ole des champs magn\'etiques.
Un objectif secondaire consiste \`a effectuer un large relev\'e d'amas de galaxies en effet Sunyaev-Zeldovich.

L'observation se ferait dans cinq bandes spectrales larges dans le domaine 100 - 700 $\mic$, plus une bande \`a 2 mm pour l'observation de l'effet Sunyaev-Zeldovich n\'egatif.
Il est pr\'evu tout d'abord de r\'ealiser quatre fois de suite (afin d'augmenter le rapport signal sur bruit) la cartographie compl\`ete du ciel (2 ans), puis de r\'ealiser des observations point\'ees en vue d'augmenter le temps d'int\'egration sur des sources galactiques et extragalactiques particuli\`erement int\'eressantes (1 an).
Les d\'etecteurs seraient des matrices de bolom\`etres du CEA, polaris\'es, refroidis \`a 0.3 K.
Le cryostat est le m\^eme que celui d'Herschel, ainsi que les optiques.
La r\'esolution angulaire est de 50$''$ pour les cinq bandes submillim\'etriques (100, 200, 350, 500 et 700 $\mic$) et de 2.5$'$ pour la bande \`a 2 mm.
Une mission telle que SNOOPY pourrait \^etre d\'evelopp\'ee rapidement et \^etre lanc\'ee peu de temps apr\`es Herschel/Planck.
L'int\'er\^et scientifique est consid\'erable, \`a la fois sur la connaissance de la Galaxie (polarisation, champ magn\'etique, temp\'eratures, masses du milieu interstellaire, objets protostellaires...) et du monde extragalactique (amas de galaxies, galaxies infrarouge, etc).

\begin{figure}
\begin{center}
\caption[L'\'equipe SNOOPY]{L'\'equipe SNOOPY \`a l'\'ecole d'\'et\'e d'Alpbach, Autriche, juillet 2000.}
\label{figsnoopy}
\end{center}
\end{figure}

 \subsection{Herschel}

Herschel ({\it http://www.sci.esa.int/home/herschel}) est l'une des quatre pierres angulaires de l'Agence Spatiale Europ\'eenne (ESA) \`a l'horizon 2000.
FIRST, {\it Far InfraRed and Submillimeter Telescope}, est l'ancien nom de Herschel.
Il s'agit d'un observatoire spatial d\'edi\'e \`a la connaissance du ciel infrarouge et submillim\'etrique, pour toutes sortes d'objets galactiques et extragalactiques.
Le satellite fait 7 m\`etres de long pour plus de 4 m\`etres de large, et p\`ese un peu plus de 3 tonnes.
Il doit \^etre lanc\'e en 2007 conjointement avec le satellite Planck, par une fus\'ee Ariane V, pour \^etre amen\'e au point de Lagrange L2.

\begin{figure}
\begin{center}
\caption[Le satellite Herschel]{Le satellite Herschel.}
\label{figherschel}
\end{center}
\end{figure}

Le t\'elescope Ritchey-Chr\'etien a un miroir primaire de 3.5 m de diam\`etre.
Trois instruments sont install\'es sur le satellite Herschel: HIFI, PACS et SPIRE.
HIFI ({\it Heterodyne Instrument for the Far-Infrared}, voir de Graauw \& Helmich 2001) est un spectrom\`etre h\'et\'erodyne \`a haute r\'esolution spectrale, qui permettra de mieux conna\^\i tre la physique et la cin\'ematique des r\'egions de formation d'\'etoiles, \`a travers leurs raies de refroidissement mol\'eculaire, et ce dans la Voie Lact\'ee et les galaxies proches.
PACS ({\it Photoconductor Array Camera and Spectrometer}, voir Poglitsch \etal 2001) est une cam\'era bi-bande de photoconducteurs, qui a pour objectifs de mieux conna\^\i tre la formation des galaxies dans l'Univers, ainsi que la formation stellaire dans la Galaxie et l'Histoire du syst\`eme solaire.
SPIRE ({\it Spectral and Photometric Imaging REceiver}, voir Griffin \etal 2001) est un instrument spectro-photom\'etrique submillim\'etrique, utilisant des bolom\`etres, qui a pour but d'\'etudier la formation des galaxies et la formation stellaire.

 \subsection{Planck\label{galplanck}}

Planck ({\it http://www.sci.esa.int/home/planck} ou Tauber 2000) est une mission ESA de taille moyenne qui doit \^etre lanc\'ee en 2007 avec Herschel.
Plac\'e au point de Lagrange L2, il doit effectuer une cartographie compl\`ete du ciel millim\'etrique, avec pour objectif majeur l'\'etude des fluctuations du rayonnement fossile (voir section \ref{rfsec}), et la contrainte tr\`es efficace des param\`etres cosmologiques de l'Univers.
Nous d\'ecrivons l'aspect cosmologique de ce projet en section \ref{rfplanck}.

\begin{figure}
\begin{center}
\caption[Le satellite Planck]{Le satellite Planck.}
\label{figplanck}
\end{center}
\end{figure}

En ce qui concerne l'aspect galactique de la science de Planck, il faut souligner qu'il s'agit de la premi\`ere cartographie compl\`ete du ciel dans le domaine submillim\'etrique - millim\'etrique depuis COBE (1992), et que la r\'esolution angulaire est am\'elior\'ee d'un facteur 80 (5$'$ environ au lieu de 7 degr\'es).
De ce fait, la connaissance spectrale de l'\'emission continuum de la poussi\`ere galactique serait compl\`ete en prenant en compte les donn\'ees IRAS (10 - 100 $\mic$), ELISA ou SNOOPY (100 $\mic$ - 700 $\mic$ environ), et Planck (350 $\mic$ - 1 cm).
L'\'etude des propri\'et\'es physico-chimiques des grains, de la structure de la Galaxie, des phases du milieu interstellaire, des premi\`eres \'etapes de la formation des \'etoiles, sera donc bien contrainte gr\^ace aux donn\'ees Planck.
En particulier, un grand nombre de c{\oe}urs froids devraient \^etre d\'etect\'es par Planck.
De plus, l'\'etude des \'emissions libre-libre et synchrotron devrait permettre de cartographier la distribution de rayons cosmiques et le champ magn\'etique galactique.

Le satellite est compos\'e d'un t\'elescope gr\'egorien de diam\`etre efficace 1.5~m, et de deux instruments focaux, HFI et LFI.
Le t\'elescope est refroidi passivement par son environnement spatial \'eloign\'e de la Terre (L2), jusqu'\`a environ 60 K.
HFI ({\it High Frequency Instrument}) poss\`ede six bandes spectrales bolom\'etriques centr\'ees sur 350 $\mic$, 550 $\mic$, 850 $\mic$, 1.38 mm, 2.1 mm et 3 mm.
Les bolom\`etres de cet instrument sont refroidis \`a 0.1 K gr\^ace \`a une cha\^\i ne de refroidissement \`a quatre \'etages.
LFI ({\it Low Frequency Instrument}) observe gr\^ace \`a 56 d\'etecteurs radiom\'etriques r\'epartis dans quatre bandes millim\'etriques (3, 4.3, 6.8 et 10 mm).
La r\'esolution angulaire de Planck varie entre environ 5$'$ et 30$'$ en fonction de la longueur d'onde.
Certains bolom\`etres de HFI sont polaris\'es, ainsi que les d\'etecteurs de LFI, ce qui permettra d'\'etudier la polarisation du CMB et celle des \'emissions galactiques.

\chapter{Le rayonnement fossile\label{rf}}
{\flushright\it ... To see Light without knowing Darkness. It cannot be.\\
... Voir la Lumi\`ere sans conna\^\i tre les T\'en\`ebres. Ce ne peut \^etre. \\}
{\flushright Frank Herbert, {\it Dune}\\}
\vspace*{1em}

\section{ Les fluctuations du rayonnement fossile\label{rfsec}}

Comme expliqu\'e en section \ref{rfint}, le rayonnement fossile ou CMB ({\it Cosmic Microwave Background}) est une relique des premiers \^ages de l'Univers.
Le CMB est tr\`es isotrope: on l'observe dans toutes les directions
avec la m\^eme intensit\'e et les m\^emes caract\'eristiques spectrales, \`a
ceci pr\`es que le mouvement local de l'observateur (la Terre) par rapport au
fluide cosmique introduit un dip\^ole (le dip\^ole cosmologique), et que de
tr\`es faibles fluctuations (10$^{-4}$) sont observables.
Ces fluctuations de temp\'erature sont d'origines diverses, et sont particuli\`erement int\'eressantes pour la cosmologie.

\subsection{Les fluctuations primordiales}

Les anisotropies primordiales du CMB sont \`a l'origine des grandes structures observables dans l'Univers, telles que galaxies, amas de galaxies, etc.
Il suffit en effet de tr\`es faibles fluctuations initiales pour engendrer par effondrement gravitationnel des structures aujourd'hui tr\`es inhomog\`enes.

Dans le cadre de la th\'eorie de l'inflation (voir section \ref{bigbang}), on pense que les anisotropies primordiales de densit\'e sont initi\'ees par de tr\`es faibles fluctuations al\'eatoires quantiques.
Les ph\'enom\`enes dominants \`a l'origine de leur amplification sont les oscillations acoustiques dans le fluide photons - baryons.
Il s'agit de la comp\'etition entre l'\'energie gravitationnelle qui tend \`a comprimer le fluide cosmique, et la pression de radiation des photons qui tend \`a le dilater.
Il y a donc des fluctuations de densit\'e dans le fluide cosmique, qui se traduisent par des fluctuations de temp\'erature sur la surface de derni\`ere diffusion, car:

- d'une part, les r\'egions les plus denses sont aussi les plus chaudes,

- de plus, les photons qui sortent d'un puits de potentiel gravitationnel lors de la derni\`ere diffusion prennent un d\'ecalage vers le rouge gravitationnel,

- enfin, l'effet Doppler d\^u aux mouvements locaux induit un d\'ecalage vers le rouge/bleu suppl\'ementaire.

Ces trois effets sont exprim\'es par l'\'equation suivante:

\begin{equation}
{\Delta T \over T} ({\bf \Hat r}) = \phi ({\bf r}) \: - \: {\bf \Hat r} . {\bf v} ({\bf r}) \: + \: {\delta ({\bf r}) \over 3}
\end{equation}

o\`u la norme du vecteur {\bf r} est la distance comobile \`a la surface de derni\`ere diffusion, T est la temp\'erature, $\phi$ le potentiel gravitationnel, {\bf v} la vitesse particuli\`ere et $\delta$ la surdensit\'e.

Les \'echelles angulaires auxquelles apparaissent les anisotropies de temp\'erature d\'ependent de la taille de l'horizon au moment de la derni\`ere diffusion.
Ainsi, il appara\^\i t des ``pics Doppler'' dans le spectre de puissance des anisotropies.
Si les conditions initiales sont adiabatiques, les positions des surdensit\'es correspondent \`a celles des puits de potentiel gravitationnels.
Dans ce cas, l'effet de densit\'e compense partiellement l'effet de gravit\'e, et la diff\'erence, aux grandes \'echelles, s'exprime comme le tiers de l'effet gravitationnel.
On parle alors d'effet Sachs-Wolfe.

\`A ces trois causes d'anisotropies primordiales (densit\'e, gravit\'e, Doppler) s'ajoute le fait que la surface de derni\`ere diffusion n'est pas infiniment fine, et donc que les anisotropies dans le rayonnement fossile sont en fait une moyenne pond\'er\'ee le long de l'\'epaisseur de la surface de derni\`ere diffusion.
Les fluctuations dont l'\'echelle est inf\'erieure \`a l'\'epaisseur de cette surface sont donc gomm\'ees par ce moyennage.
Cet effet est appel\'e l'amortissement ({\it damping}).

Plus de d\'etails concernant la physique des fluctuations primordiales peuvent \^etre trouv\'es notamment dans Seljak (1994), Jorgensen \etal (1995), Hu \& Sugiyama (1995) et Hu \etal (1997).

Les causes d'anisotropies que nous avons expos\'ees jusqu'\`a maintenant appartiennent au cadre de la th\'eorie de l'inflation (voir section \ref{bigbang}), ont une statistique gaussienne et sont aujourd'hui les plus favoris\'ees, essentiellement parce que des exp\'eriences r\'ecentes ont mis en \'evidence ces pics dans le spectre du CMB.
Cependant, il faut aussi mentionner la th\'eorie des d\'efauts topologiques, qui pr\'evoit l'existence de cordes et de textures cosmiques, qui se retrouveraient dans la non-gaussianit\'e \'eventuelle du CMB.
Cette th\'eorie est aujourd'hui mise \`a l'\'ecart en ce qui concerne les fluctuations dominantes du CMB.

\subsection{Les fluctuations secondaires}

Le rayonnement fossile parcourt environ 15 milliards d'ann\'ees-lumi\`ere avant de nous parvenir, aussi rencontre-t-il en chemin de nombreuses sources de perturbation ou de contamination, qui sont un probl\`eme r\'eel pour qui s'int\'eresse \`a la signature fossile du Big Bang qu'est le CMB, mais qui n'en sont pas moins des sources d'information pr\'ecieuses et essentielles \`a la connaissance du milieu intergalactique, des amas de galaxies, de notre galaxie, etc.

Les fluctuations secondaires sont des fluctuations de temp\'erature du CMB lui-m\^eme (et non des contaminations d'avant-plans qui viennent se superposer \`a l'observation), mais dues \`a des processus physiques qui ont lieu apr\`es la derni\`ere diffusion.
Elles se classent en trois cat\'egories: les fluctuations secondaires dues \`a la gravit\'e, l'effet Sunyaev-Zeldovich et la r\'e-ionisation globale.
Les effets gravitationnels sont d'une part l'effet Sachs-Wolfe int\'egr\'e (ISW), et d'autre part le {\it lensing} (``lentillage'').

L'ISW est d\^u aux variations dans le temps du potentiel gravitationnel, qui impriment des anisotropies dans le CMB.
En effet, quand un photon traverse un puits de potentiel, son \'energie variera si pendant le temps de la travers\'ee, le dit potentiel a vari\'e.
L'ISW se traduit par trois effets:

- peu de temps apr\`es la combinaison des atomes, la contribution des photons \`a la densit\'e d'\'energie de l'Univers n'est pas encore n\'egligeable.
De ce fait, le potentiel gravitationnel diminue quelque peu, causant l'ISW primaire.

- s'il existe une constante cosmologique, ou que l'Univers n'est pas plat, le potentiel gravitationnel varie dans le temps (Univers domin\'e par le vide, ou domin\'e par la courbure).
Cet effet induit l'ISW tardif, aux redshifts faibles.

- lorsque des structures non lin\'eaires telles que les galaxies apparaissent, il y a des effets non lin\'eaires induits par rapport \`a la th\'eorie des perturbations lin\'eaires.
C'est l'effet Rees-Sciama.

L'autre effet gravitationnel marquant est le {\it lensing}, effet de d\'eviation de la trajectoire des photons du CMB d\^u aux fluctuations de potentiel gravitationnel.

La r\'e-ionisation des baryons dans l'Univers, due aux photons ultraviolets issus de la formation stellaire ou au chauffage li\'e \`a un effondrement gravitationnel, produit des \'electrons libres capables de diffuser les photons du CMB.
En particulier, la diffusion Compton inverse des photons du CMB par le gaz interne (chaud: 10$^6$ \`a 10$^8$ K) des amas de galaxies produit l'effet Sunyaev-Zeldovich \cite{sunyaev70}, qui consiste en une distorsion du spectre du CMB dans la direction des amas.
Cet effet SZ thermique se traduit par un gain d'\'energie des photons du CMB, c'est-\`a-dire un d\'ecr\'ement d'intensit\'e dans le domaine millim\'etrique, et un incr\'ement dans le submillim\'etrique.
Il existe aussi un effet Doppler d\^u \`a la vitesse particuli\`ere des amas ionis\'es diffusants, qui se traduit par une alt\'eration de la temp\'erature du CMB dans la direction de l'amas: c'est l'effet Sunyaev-Zeldovich cin\'etique.
Plus de pr\'ecisions concernant l'effet Sunyaev-Zeldovich peuvent \^etre trouv\'es dans Pointecouteau (1999).

L'effet SZ traduit les effets de la r\'e-ionisation locale, mais il existe aussi des effets tels que la suppression de puissance aux petites \'echelles qui sont d\^us \`a la r\'e-ionisation globale des baryons dans l'Univers.

Plus de renseignements concernant la physique des anisotropies du CMB peuvent \^etre trouv\'es par exemple dans Tegmark (1995) et r\'ef\'erences attenant.

\subsection{Les avant-plans}

En plus des fluctuations du CMB observables sur le ciel millim\'etrique, d'autres composantes se superposent en tant qu'avant-plans sur la ligne de vis\'ee.
Il s'agit des contributions extragalactiques (sources ponctuelles radio ou infrarouge), galactiques (poussi\`eres, libre-libre, synchrotron), locales (Syst\`eme Solaire), et \'eventuellement terrestres (atmosph\`ere).

Les sources extragalactiques de contamination du CMB ne sont pas les plus ennuyeuses, car elles sont ponctuelles.
Si l'on s'en tient \`a une m\'ethode sauvage de traitement de ces sources, il suffit d'\'eliminer les pixels des cartes les contenant.
On peut aussi utiliser l'information spectrale de ces sources d'apr\`es des catalogues de sources ponctuelles, pour les retirer des cartes.

La poussi\`ere galactique est l'avant-plan le plus important et donc le plus ennuyeux pour les observations CMB, \`a grande \'echelle particuli\`erement.
En effet, la zone autour du plan galactique est en pratique inutilisable pour l'observation du rayonnement fossile, quelle que soit la longueur d'onde, et m\^eme en dehors de cette zone, la poussi\`ere contamine quelque peu les mesures millim\'etriques du CMB.
La connaissance de l'\'emissivit\'e de la poussi\`ere dans le millim\'etrique (i.e. de l'indice spectral $\beta$) permet de s\'eparer la poussi\`ere du CMB, mais cette connaissance est tr\`es imparfaite.

L'\'emission {\it bremsstrahlung} (rayonnement de freinage) ou libre-libre des \'electrons libres galactiques est un effet de diffusion des \'electrons par les ions (r\'egions ionis\'ees quand la temp\'erature est sup\'erieure \`a environ 10000 K).
Le meilleur traceur de l'\'emission libre-libre semble \^etre la raie H$_\alpha$ (Lyman $\alpha$ de l'hydrog\`ene atomique).
L'\'emission synchrotron est un effet de l'acc\'el\'eration d'\'electrons relativistes dans un champ magn\'etique.

Les contaminants provenant du syst\`eme solaire concernent la poussi\`ere zodiacale, c'est-\`a-dire la poussi\`ere situ\'ee dans le plan \'ecliptique.
Les contaminations atmosph\'eriques concernent les observations au sol, et, dans une moindre mesure, les observations en ballon stratosph\'erique.
Les missions spatiales ne sont pas concern\'ees par ce probl\`eme.

R\'ep\'etons enfin que ces ``contaminants'' du CMB sont d'un grand int\'er\^et scientifique, et qu'en particulier la connaissance de la Galaxie peut largement b\'en\'eficier des observations d'exp\'eriences CMB.

\section{ Observations des fluctuations du CMB\label{rfobs}}

La premi\`ere observation des fluctuations du CMB a \'et\'e r\'ealis\'ee par la mission COBE \cite{boggess92}.
Nous d\'ecrivons cette exp\'erience en section \ref{obspousgal}.
Celle-ci a mesur\'e pr\'ecis\'ement la temp\'erature du CMB (2.728 $\pm$ 0.004 K) et mis en \'evidence les fluctuations \`a grande \'echelle de ce rayonnement (r\'esolution angulaire de 7 degr\'es).
Nous pr\'esentons la carte compl\`ete du ciel r\'ealis\'ee par COBE en figure \ref{cobemap}.

\begin{figure}
\begin{center}
\caption[Carte COBE]{Carte COBE de tout le ciel des fluctuations du CMB et de la Galaxie.}
\label{cobemap}
\end{center}
\end{figure}

Les fluctuations du rayonnement fossile se caract\'erisent par leur spectre de puissance.
Ce spectre sur la sph\`ere c\'eleste s'exprime en harmoniques sph\'eriques.
Les variations de temp\'erature s'expriment alors par:

\begin{equation}
{\Delta T \over T} = \sum_l \sum_{m=-l}^la_{lm}Y_{lm}
\end{equation}

Les $a_{lm}$ constituent l'ensemble discret des coefficients de la d\'ecomposition en harmoniques sph\'eriques.
D'apr\`es la th\'eorie de l'inflation, ces fluctuations forment un champ al\'eatoire gaussien.
Dans ce cas, toute l'information sur ces anisotropies est contenue dans les $C_l$:

\begin{equation}
C_l = <|a_{lm}|^2>
\end{equation}

Le spectre de puissance en $C_l$ est donc un outil naturel - mais pas le seul - pour caract\'eriser les fluctuations du CMB et comparer les observations aux mod\`eles cosmologiques.

De nombreuses exp\'eriences d\'edi\'ees au rayonnement fossile ont eu lieu depuis COBE, et d'autres sont en projet.
En voici une liste assez exhaustive:

- exp\'eriences au sol: Tenerife, South Pole, Saskatoon, Python, IAB, White Dish, CAT, OVRO, ATCA, SuZIE, Viper, COBRA, Jodrell Bank, Brown/Wisc Polarization, MAT, DASI, VSA ({\it Very Small Array}), CBI, POLAR, Polatron.
Ces exp\'eriences au sol observent dans des fr\'equences d'environ 10 \`a 100 GHz, soit environ 3 mm - 3 cm de longueur d'onde.
Bien que l'\'emission du CMB soit plus faible dans ce domaine que dans le domaine 1 mm - 3 mm, c'est aux grandes longueurs d'onde qu'est accessible la mesure du CMB pour les exp\'eriences au sol, \`a cause de l'absorption atmosph\'erique.

- exp\'eriences ballon: Archeops, FIRS, ARGO, MAX, MSAM, BAM, QMAP, BOOMERANG, MAXIMA, Top Hat, ACE, BEAST.
Les longueurs d'onde d'observation des exp\'eriences CMB ballon sont g\'en\'eralement comprises entre 0.5 et 3 mm.

- satellites: COBE (1992), MAP et Planck (lancement pr\'evu en 2007).
Nous avons pr\'esent\'e le projet Planck en section \ref{galplanck}.
Nous pr\'esentons l'aspect scientifique cosmologique de Planck en section \ref{rfplanck}.
MAP ({\it Microwave Anisotropy Probe}, {\it http://map.gsfc.nasa.gov} ou Wright 1999) est un satellite lanc\'e en juin 2001 et arriv\'e sans encombre \`a L2 en octobre 2001.
Il a pour but de r\'ealiser la cartographie compl\`ete du ciel avec une r\'esolution angulaire d'environ 20$'$.

- station spatiale: Submillimetron (projet russe de t\'elescope submillim\'etrique sur l'ISS - {\it International Space Station}).

Les r\'ef\'erences de ces exp\'eriences CMB peuvent \^etre trouv\'ees sur le site de Martin White: {\it http://cfa-www.harvard.edu/.mwhite/cmbexptlist.html}.

\begin{figure}
\begin{center}
\caption[Spectre observationnel du CMB avant BOOMERANG]{Spectre de puissance observationnel du CMB avant BOOMERANG.
Figure tir\'ee du site web de Max Tegmark: {\it http://www.hep.upenn.edu/.max}.}
\label{cl1}
\end{center}
\end{figure}

Nous pr\'esentons en figure \ref{cl1} l'\'etat des contraintes observationnelles avant les r\'esultats de BOOMERANG.
Les exp\'eriences ballon BOOMERANG ({\it Balloon Observations Of Millimetric Extragalactic Radiation ANd Geophysics}, voir {\it http://www.physics.ucsb.edu/.boomerang} et de Bernardis \etal 2000) et MAXIMA ({\it Millimeter Anisotropy eXperiment IMaging Array}, voir {\it http://cosmology.berkeley.edu/group/cmb} et Hanany \etal 2000) ont apport\'e une contribution majeure \`a la cosmologie en mesurant le spectre de puissance du CMB avec une bonne pr\'ecision.
Nous pr\'esentons en figure \ref{boommap} la carte des fluctuations du CMB produite par BOOMERANG (voir Netterfield \etal 2002), et en figure \ref{cl2} le spectre observationnel des fluctuations du CMB produit par BOOMERANG (voir Netterfield \etal 2002 et de Bernardis \etal 2002), MAXIMA \cite{lee01}, DASI \cite{halverson02}, CBI \cite{padin01} et COBE/DMR \cite{smoot92}.
Ce graphique traduit l'\'etat de la connaissance des fluctuations du CMB \`a l'heure o\`u nous \'ecrivons ces lignes.

\begin{figure}
\begin{center}
\caption[Carte BOOMERANG des fluctuations du CMB]{Carte BOOMERANG des fluctuations du CMB mesur\'ees \`a la fr\'equence de 150 GHz, tir\'ee de Netterfield \etal (2002).
La taille de la carte est de 110 X 35 degr\'es.}
\label{boommap}
\end{center}
\end{figure}

\begin{figure}
\begin{center}
\caption[Spectre observationnel du CMB aujourd'hui]{Spectre observationnel du CMB par BOOMERANG, MAXIMA, Dasi, CBI et COBE/DMR tir\'e du site web du groupe de cosmologie de UC Berkeley: {\it http:// cosmology.berkeley.edu/ group/cmb}.}
\label{cl2}
\end{center}
\end{figure}

Les exp\'eriences BOOMERANG et MAXIMA ont permis de mesurer pr\'ecis\'ement la position et la hauteur du premier pic Doppler, de mettre en \'evidence l'existence des deux pics suivants, et ainsi de contraindre certains param\`etres cosmologiques tels que le param\`etre de densit\'e $\Omega$, d\'eduit proche de 1.
Netterfield \etal (2002) d\'erivent des donn\'ees BOOMERANG les valeurs suivantes des param\`etres cosmologiques, sous l'{\it a priori} faible ({\it weak prior}) supposant que H$_0$ est compris entre 45 et 90 km/s/Mpc et que l'\^age de l'Univers est sup\'erieur \`a 10 milliards d'ann\'ees.
Les barres d'erreur sont donn\'ees pour l'intervalle de confiance \`a 1 $\sigma$.

- $\Omega$ = 1.03 $\pm$ 0.06

Ceci favorise l'hypoth\`ese d'un Univers plat, et donc la th\'eorie de l'inflation.

- n = 0.93 $^{+0.10} _{-0.08}$

- $\Omega_{CDM} \: h^2$ = 0.12 $\pm$ 0.05

- $\Omega_{bar} \: h^2$ = 0.021 $^{+0.004} _{-0.003}$

o\`u h est le centi\`eme de H$_0$.

- L'\^age de l'Univers vaudrait 15.4 $\pm$ 2.1 milliards d'ann\'ees.

En combinant les r\'esultats des exp\'eriences CMB avec ceux des exp\'eriences portant sur les supernov{\ae}, les pr\'ecisions sur les param\`etres cosmologiques sont encore am\'elior\'ees.
La prochaine g\'en\'eration de r\'esultats majeurs pourrait \^etre apport\'ee par Archeops et MAP, en attendant Planck.

\section{ L'exp\'erience Archeops\label{archeops}}

Dans le contexte d\'ecrit en section pr\'ec\'edente, nous d\'ecrivons l'exp\'erience Archeops, \`a laquelle nous avons 
activement particip\'e en ce qui concerne les campagnes de lancement, le traitement et l'analyse des donn\'ees.

\subsection{Pourquoi Archeops ?}

BOOMERANG et MAXIMA, comme beaucoup d'autres exp\'eriences CMB, observent les petites \'echelles angulaires des anisotropies du CMB, afin de mesurer la position et la hauteur des pics.
Cependant, les grandes \'echelles angulaires (i.e. les bas l dans le spectre de puissance) ne sont pas tr\`es bien connues.
Entre la mesure des grandes \'echelles de COBE et le premier pic mesur\'e par BOOMERANG et MAXIMA, il existe une gamme de multip\^oles int\'eressants (l = 10 - 100) pour d\'eterminer la hauteur du plateau avant les pics.
L'originalit\'e d'Archeops est que cette exp\'erience permet de mesurer \`a la fois les bas l (10 - 100) et les hauts l (100 - 1000).
La possibilit\'e d'intercalibrer photom\'etriquement les exp\'eriences ballon avec COBE (aux grandes \'echelles angulaires) est aussi un avantage d'Archeops.
En effet, la couverture spatiale d'Archeops (environ 30 \% du ciel total) est bien sup\'erieure \`a celles de BOOMERANG ou MAXIMA.
En contrepartie, le temps d'int\'egration par pixel est moins grand.
L'objectif principal d'Archeops est donc la cartographie d'une grande partie du ciel, afin de contraindre le spectre de puissance du CMB sur un large domaine de multip\^oles, et donc de contraindre les param\`etres cosmologiques.
Gr\^ace \`a sa large couverture sur le ciel, Archeops peut \'egalement observer la Galaxie en temps qu'objectif secondaire.

\subsection{Qu'est-ce qu'Archeops ?}

Archeops est un instrument embarqu\'e en ballon stratosph\'erique.
La collaboration Archeops regroupe une soixantaine de personnes en France, en Italie, au Royaume-Uni et aux \'Etats-Unis.
Le PI d'Archeops est Alain Beno\^\i t au CRTBT ({\it Centre de Recherche sur les Tr\`es Basses Temp\'eratures}) \`a Grenoble.
L'instrument est constitu\'e d'un t\'elescope de 1.5~m de diam\`etre, et de d\'etecteurs bolom\'etriques.
La technologie d\'evelopp\'ee pour le projet Planck est largement utilis\'ee pour Archeops.

\begin{figure}
\begin{center}
\caption[La nacelle Archeops]{La nacelle Archeops.}
\label{figarcheops}
\end{center}
\end{figure}

Le t\'elescope d'Archeops, d\'evelopp\'e par l'Universit\'e de Minnesota (\'Etats-Unis), est de type gr\'egorien, hors axe, et est constitu\'e de deux miroirs, le primaire (1.5~m) parabolique et le secondaire elliptique.
Le syst\`eme gr\'egorien hors axe permet une meilleure efficacit\'e d'ouverture et une r\'eponse plus faible aux lobes lointains que le syst\`eme \'equivalent align\'e.

Les d\'etecteurs sont des bolom\`etres toile d'araign\'ee d\'evelopp\'es par CalTech et JPL (Pasadena, \'Etats-Unis) pour Planck - HFI.
Archeops poss\`ede des bolom\`etres qui observent dans quatre canaux de fr\'equences identiques \`a ceux de Planck: 143 GHz, 217 GHz, 353 GHz (bolom\`etres polaris\'es) et 545 GHz, ce qui correspond \`a des longueurs d'onde de 2.1 mm, 1.4 mm, 850 $\mic$ et 550 $\mic$.
Ces d\'etecteurs ont une r\'eponse rapide, ce qui est n\'ecessaire pour la strat\'egie d'observation d'Archeops, et une grande sensibilit\'e.
Un autre avantage des bolom\`etres toile d'araign\'ee est qu'ils ont une faible section efficace d'interaction avec les rayons cosmiques, ce qui permet de minimiser le nombre de {\it glitches} (pics d\^us \`a des impacts de rayons cosmiques sur un bolom\`etre) dans les donn\'ees temporelles.
Le refroidissement du plan focal (jusqu'\`a moins de 0.1 K) est assur\'e par un cryostat d\'evelopp\'e au CRTBT \`a Grenoble.
Le syst\`eme utilise un cycle de dilution de l'h\'elium d\'evelopp\'e par Beno\^\i t \& Pujol (1994).
L'\'etage de dilution est plac\'e dans une bo\^\i te \`a basse temp\'erature, situ\'ee sur un r\'eservoir d'h\'elium liquide \`a 4.2 K.
La partie sup\'erieure de cette bo\^\i te contient les cornets de r\'eception du rayonnement.
Les bolom\`etres sont plac\'es sur l'\'etage \`a 0.1 K, maintenus par des fils de kevlar.
Les optiques froides sont d\'evelopp\'ees par le Queen Mary College \`a Londres.
Les \'electroniques de lecture des bolom\`etres et thermom\`etres utilisent un syst\`eme original de modulation \'electrique d\'evelopp\'e par le CESR et le CRTBT.
Il permet d'obtenir une bonne stabilit\'e de la cha\^\i ne de lecture y compris aux basses fr\'equences (0.01 Hz).

Un senseur stellaire, petit t\'elescope optique, permet de rep\'erer les \'etoiles afin de conna\^\i tre \`a tout moment le pointage pr\'ecis du t\'elescope principal.
La nacelle et le senseur stellaire ont \'et\'e r\'ealis\'es par les \'equipes italiennes des universit\'es de Rome et Florence.
Les bords lat\'eraux et inf\'erieur de la nacelle sont couverts (bafflage), afin d'emp\^echer les rayonnements terrestre, solaire, etc, de p\'en\'etrer dans le photom\`etre.

L'\'etalonnage photom\'etrique de l'instrument est r\'ealis\'e au sol gr\^ace \`a un corps noir froid de temp\'erature variable, install\'e devant la fen\^etre d'entr\'ee du photom\`etre.
L'\'etalonnage optique est r\'ealis\'e gr\^ace \`a une table de pointage construite \`a l'ISN (Grenoble).
Celle-ci permet de caract\'eriser le lobe principal du syst\`eme optique.
Le syst\`eme de pointage balaie une zone dans laquelle se trouve une source thermique, plac\'ee habituellement sur une colline \`a 1 km de distance.

Le traitement et l'analyse des donn\'ees est r\'ealis\'e par les \'equipes fran\c{c}aises \`a Grenoble (CRTBT, ISN, LAOG), Paris (CdF, CEA, IAP, IAS, LAL) et Toulouse (CESR, LAT).

La description pr\'ecise de l'instrument Archeops a fait l'objet d'une publication \cite{benoit02}, ainsi que de nombreuses pr\'esentations dans des conf\'erences (liste en sections \ref{collpaut} et \ref{collcoaut}).

\subsection{Comment observe Archeops ?\label{commarch}}

La strat\'egie d'observation d'Archeops consiste \`a r\'ealiser un balayage du ciel en cercles larges.
Pour cela, la nacelle tourne sous la cha\^\i ne de vol du ballon gr\^ace \`a un pivot.
La fr\'equence de rotation est de 2 ou 3 tours par minute suivant les vols, et l'angle par rapport \`a l'horizontale est de 41 degr\'es.
Les cercles sur le ciel se d\'ecalent progressivement gr\^ace \`a la rotation terrestre, ce qui permet de balayer une grande zone de ciel.
L'acquisition des donn\'ees se fait avec une fr\'equence d'\'echantillonnage d'environ 150 Hz.

Afin de minimiser les perturbations li\'ees \`a la lumi\`ere du soleil, Archeops est lanc\'e depuis une zone polaire, en hiver, pour b\'en\'eficier de la nuit polaire et donc du maximum de temps de nuit noire.
C'est pourquoi la base de lancement pour les vols scientifiques d'Archeops est Esrange, pr\`es de Kiruna, en Laponie su\'edoise.
Le temps de vol, et donc le temps d'int\'egration, est essentiel pour cette exp\'erience, car il est n\'ecessaire de r\'eduire le plus possible le bruit instrumental, afin de diminuer les erreurs sur la temp\'erature du CMB dans chaque pixel, et donc sur le spectre de puissance.
Pour pouvoir r\'ealiser des vols de longue dur\'ee (jusqu'\`a 24 h), il faut que les vents stratosph\'eriques poussent le ballon dans la bonne direction (vers la Russie), mais pas trop vite.
Or, les conditions atmosph\'eriques dans la stratosph\`ere sont tr\`es capricieuses, ainsi que les conditions au sol, qui lorsqu'elles sont mauvaises (vent, neige), emp\^echent le lancement.
Ceci, ajout\'e aux contraintes li\'ees aux astres (la lumi\`ere de la lune ou du soleil perturbent les mesures) et aux sources d'\'echec techniques, fait que les campagnes de lancement d'Archeops ont \'et\'e difficiles mais toujours fructueuses, au bout du suspense (voir table \ref{volsarch}).

\begin{table}
\begin{center}
\caption[Vols r\'ealis\'es par l'exp\'erience Archeops]{Vols r\'ealis\'es par l'exp\'erience Archeops.
Date du lancement, nom du vol, lieu du lancement, lieu d'atterrissage, temps de donn\'ees (heures), nombre de bolom\`etres, qualit\'e des donn\'ees.}
\begin{tabular}{lllllll}
\hline
Date & Vol & Lancement & Atterrissage & Temps (h) & Bolom\`etres & Qualit\'e \\
\hline
17/07/99 & T & Trapani & Badajoz & 4 & 6 & Moyenne \\
\hline
12/01/01 & & Esrange & Finlande & 0 & 24 & Pas de donn\'ees \\
29/01/01 & KS1 & Esrange & Syktyvkar & 7.5 & 24 & Bonne \\
\hline
17/01/02 & KS2 & Esrange & Finlande & 1.5 & 24 & Bonne \\
07/02/02 & KS3 & Esrange & Noril'sk & 19 & 24 & Excellente \\
\hline
\end{tabular}
\end{center}
\label{volsarch}
\end{table}

Le premier vol a \'et\'e r\'ealis\'e depuis Trapani (Sicile), le 17 juillet 1999.
La nacelle a atterri dans le sud de l'Espagne.
Ce vol technologique avec seulement 6 bolom\`etres a fourni 4 heures de donn\'ees, exploitables essentiellement pour la science galactique, mais aussi quelque peu pour le CMB, malgr\'e le haut niveau de bruit.

Un vol test sur une nacelle du CNES (sans le t\'elescope Archeops) a \'et\'e r\'ealis\'e \`a Kiruna le 3 mars 2000, afin de valider la t\'el\'em\'etrie.
Un autre vol, le 4 avril 2000, a permis de tester le senseur stellaire.

L'hiver 2000-2001 a vu les deux premi\`eres campagnes scientifiques \`a Kiruna.
J'ai moi-m\^eme particip\'e \`a ces deux campagnes.
La premi\`ere, en novembre et d\'ecembre 2000, n'a pas permis de lancer Archeops, \`a cause de mauvaises conditions m\'et\'eorologiques dans la stratosph\`ere: le vortex polaire mal plac\'e aurait pouss\'e Archeops vers la banquise.
La deuxi\`eme campagne (janvier-f\'evrier) a permis de lancer Archeops le 12 janvier, mais le vol a avort\'e \`a cause d'un probl\`eme technique sur le cryostat.
Un deuxi\`eme lancement a eu lieu le 29 janvier, auquel j'ai particip\'e.
Il a permis d'obtenir 7h30 de donn\'ees scientifiques de bonne qualit\'e, malgr\'e de nombreux bruits instrumentaux.
Archeops a vol\'e plus bas que pr\'evu (31.5 km d'altitude), car les trajectoires pr\'evues \`a plus haute altitude risquaient de pousser le ballon vers l'Oc\'ean Arctique.
L'atterrissage a eu lieu en Russie, l\'eg\`erement \`a l'ouest des Monts Oural, pr\`es de Syktyvkar.

Apr\`es un vol test t\'el\'em\'etrique le 29 ao\^ut 2001, trois campagnes scientifiques ont pris place durant l'hiver 2001-2002.
Comme l'ann\'ee pr\'ec\'edente, la premi\`ere n'a pas permis de lancer Archeops, pour les m\^emes raisons m\'et\'eorologiques.
La campagne de janvier a permis de lancer Archeops dans de tr\`es bonnes conditions, le 17 janvier.
Apr\`es 4 heures de plafond \`a 34.5 km, un probl\`eme technique au niveau du ballon a condamn\'e le vol.
Archeops a atterri en Finlande, apr\`es avoir obtenu 4 heures de donn\'ees de bonne qualit\'e.
Une troisi\`eme campagne a alors \'et\'e organis\'ee en f\'evrier, qui a enfin permis un vol de longue dur\'ee, le 7 f\'evrier.
Archeops a atterri au-del\`a de l'Oural, \`a l'extr\'emit\'e orientale de la plaine de Sib\'erie Occidentale, pr\`es de Noril'sk (voir la trajectoire en figure \ref{trajarch}), apr\`es avoir obtenu 19 h de donn\'ees scientifiques de tr\`es bonne qualit\'e.

\begin{figure}
\begin{center}
\caption[Trajectoire suivie par Archeops le 7 f\'ev. 02]{Trajectoire suivie par Archeops le 7 f\'evrier 2002.
Le lancement a eu lieu \`a Esrange, pr\`es de Kiruna, en Laponie su\'edoise, et l'atterrissage pr\`es de Noril'sk, en Sib\'erie Occidentale.}
\label{trajarch}
\end{center}
\end{figure}

\begin{figure}
\begin{center}
\caption[L'\'equipe scientifique Archeops \`a Kiruna]{L'\'equipe scientifique Archeops \`a Kiruna lors du vol du 29 janvier 2001.
Derri\`ere appara\^\i t Archeops sur le pas de lancement.}
\label{eqarch}
\end{center}
\end{figure}

Nous d\'ecrivons notre apport au traitement des donn\'ees d'Archeops en sections \ref{traitarch} et \ref{analarch}.

\section{ L'exp\'erience Planck et le rayonnement fossile\label{rfplanck}}

Nous avons d\'ecrit l'exp\'erience Planck ({\it http://www.sci.esa.int/home/planck} ou Tauber 2000) et ses implications pour l'\'etude de la poussi\`ere galactique en section \ref{galplanck}.
L'objectif principal de ce projet est la mesure pr\'ecise des anisotropies du CMB, afin de contraindre de fa\c{c}on efficace les param\`etres cosmologiques.
Nous pr\'esentons en figure \ref{plancksimucl} une simulation du spectre de puissance du CMB mesurable par Planck, ainsi que le spectre simul\'e de la polarisation du CMB.
Il est clair qu'apr\`es Planck, le spectre de puissance devrait \^etre extr\^ement bien d\'etermin\'e.

\begin{figure}
\begin{center}
\caption[Simulation du spectre du CMB visible par Planck]{Simulation du spectre de puissance du CMB tel que mesurable par Planck.
Tir\'e du site web de Wayne Hu: {\it http://background.uchicago.edu/.whu}.}
\label{plancksimucl}
\end{center}
\end{figure}

Ceci signifie que les param\`etres cosmologiques (voir \ref{params}) devraient \^etre d\'etermin\'es avec une grande pr\'ecision par Planck.
En adoptant certaines suppositions g\'en\'erales, $\Omega$, $H_0$, $\Lambda$, $\Omega_{bar}$, $\Omega_{CDM}$, n (l'indice spectral scalaire des fluctuations), etc, devraient \^etre d\'etermin\'es \`a quelques pourcent pr\`es (voir Bond \etal 1997, Zaldarriaga \etal 1997, Efstathiou \& Bond 1999).
Ceci serait crucial pour le test des th\'eories d'inflation.
En plus de la d\'etermination pr\'ecise des param\`etres, les mesures de Planck devraient permettre de mieux comprendre les conditions initiales de la formation des structures, de contraindre la physique des particules \`a haute \'energie (en particulier de mieux comprendre l'origine des fluctuations initiales), de mieux comprendre la nature de la mati\`ere noire et de d\'eterminer sa quantit\'e, etc.
La mesure pr\'ecise de la polarisation des anisotropies du CMB devrait permettre de discriminer entre modes scalaires et tensoriels des fluctuations, et donc de contraindre les th\'eories cosmologiques (voir par exemple Kamionkowski \etal 1997, Zaldarriaga \etal 1997).

Les mesures de Planck devraient \'egalement permettre de mesurer l'effet Sunyaev-Zeldovich sur plus de 10000 amas de galaxies, et ainsi de mieux comprendre leur structure et \'evolution, et de contraindre la constante de Hubble.
Planck devrait aussi cataloguer des milliers de sources extragalactiques telles que galaxies infrarouge et radio, noyaux actifs de galaxies, quasars et autres blazars, etc.
Planck devrait \'egalement \^etre capable de mesurer les fluctuations du fond diffus infrarouge, qui est form\'e par l'ensemble des galaxies lointaines, ainsi que de contraindre l'\'evolution des comptages de galaxies.

\section{ Introduction sur la construction de cartes pour les exp\'eriences CMB\label{intmapmakcmb}}

Nous nous sommes particuli\`erement int\'eress\'es \`a la construction de cartes pour les exp\'eriences qui observent le rayonnement fossile.
Les m\'ethodes que nous avons d\'evelopp\'ees et test\'ees sont fond\'ees sur la th\'eorie de l'inversion lin\'eaire d\'ecrite en chapitre \ref{chapconst}.

Le rayonnement fossile est \`a l'heure actuelle le sujet d'une \'etude intensive, gr\^ace \`a l'am\'elioration des performances des instruments et des d\'etecteurs.
Depuis le satellite COBE \cite{smoot92}, qui r\'ealisa la premi\`ere observation des anisotropies du CMB, la taille des trains de donn\'ees ordonn\'ees dans le temps (TOI ci-apr\`es, pour {\it time-ordered information}) a grandement augment\'e.
Le traitement et l'analyse des donn\'ees est donc toujours un d\'efi pour les TOI larges, d\'ej\`a en train d'\^etre trait\'ees ou \`a venir.
En particulier, les exp\'eriences satellite MAP (voir par exemple Wright 1999) et Planck (voir par exemple Tauber 2000) vont produire des TOI immenses, compos\'ees de centaines de millions de points ou plus.
M\^eme les exp\'eriences ballon telles que BOOMERANG \cite{debernardis00}, MAXIMA \cite{hanany00}, Archeops \cite{benoit02}, etc, produisent de larges TOI pour de nombreux canaux de d\'etection.

Le traitement des donn\'ees CMB consiste \`a travailler les TOI \`a une dimension, afin de r\'eduire le bruit (d\'eglitchage, r\'eduction du bruit par d\'ecorr\'elation avec les thermom\`etres ou entre bolom\`etres, etc, voir section \ref{traitarch}).
L'analyse des donn\'ees CMB est habituellement r\'ealis\'ee en quatre \'etapes, la premi\`ere \'etant la construction de cartes, la seconde la s\'eparation de composantes (CMB, Galaxie, etc), la troisi\`eme l'estimation du spectre de puissance en $C_l$ d'apr\`es la carte et la matrice de covariance du bruit dans la carte, et la derni\`ere l'estimation des param\`etres cosmologiques par comparaison des spectres en $C_l$ tir\'es des mod\`eles th\'eoriques avec le spectre observationnel.
Bien entendu, apr\`es la deuxi\`eme \'etape, d'autres analyses scientifiques peuvent \^etre effectu\'ees: science galactique, sources ponctuelles, etc.
Chaque \'etape est importante, mais il est clair que la construction de cartes est particuli\`erement cruciale dans ce processus d'analyse, car il s'agit d'une part de r\'ealiser des produits finis (carte du ciel par longueur d'onde), et \'egalement de pr\'eserver toute l'information cosmologique, tout en r\'eduisant et en caract\'erisant correctement le bruit, ce qui est n\'ecessaire pour l'estimation du spectre de puissance du CMB.

Nous nous sommes concentr\'es sur l'\'etape de la construction de cartes.
Dans le contexte des exp\'eriences CMB, la grande taille des TOI fait qu'il n'est pas possible d'appliquer les m\'ethodes matricielles optimales d\'ecrites au chapitre \ref{chapconst} de fa\c{c}on brutale.
M\^eme pour des TOI aussi courtes que celles de PRONAOS (quelques milliers de points de donn\'ees par objet), le calcul matriciel est un peu lourd (15 mn sur un processeur Pentium III \`a 500 MHz), bien que l'on ne calcule pas d'inversion matricielle.
Aussi, il est parfaitement clair que pour des dizaines de millions de points de donn\'ees comme pour Archeops (voir section \ref{traitarch}, il est tout \`a fait inenvisageable d'inverser, de multiplier ou m\^eme de stocker des matrices de $10^{14}$ \'el\'ements (environ un million de giga-octets de m\'emoire).
Nous nous sommes donc attach\'es \`a d\'evelopper des m\'ethodes alternatives, mais tout aussi optimales.
Nous avons \'egalement explor\'e comment la strat\'egie d'observation influence le processus de construction de cartes et la quantit\'e finale de bruit dans les cartes.

Il faut mentionner que des m\'ethodes de construction de cartes non optimales, d'apr\`es les crit\`eres indiqu\'es au chapitre  \ref{chapconst}, mais efficaces, ont \'et\'e d\'evelopp\'ees.
Il est possible par exemple d'utiliser directement les propri\'et\'es de croisement des balayages du faisceau sur le ciel afin de d\'estrier les cartes du bruit corr\'el\'e qui s'y trouve (voir section \ref{moypix}).
Voir \`a ce sujet Delabrouille (1998), Giard \etal (1999), Dupac \& Giard (2001, {\it coll. b}).
Nous nous sommes plut\^ot concentr\'es sur les m\'ethodes optimales, appliqu\'ees aux larges TOI, en d\'eveloppant des algorithmes pour contourner les probl\`emes li\'es \`a la taille des donn\'ees.
Les m\'ethodes lin\'eaires optimales, d\'ecrites en section \ref{invlin}, n\'ecessitent d'utiliser d'une fa\c{c}on ou d'une autre les propri\'et\'es statistiques du bruit (matrice de covariance dans les TOI: {\bf N}) et du signal pour les m\'ethodes de Wiener (matrice de covariance du ciel dans les pixels: {\bf S}).
Si l'on ne peut pas \'ecrire les matrices, il faut trouver un moyen autre de traiter le probl\`eme de la construction de cartes, en utilisant uniquement des calculs sur des vecteurs.
Nous appelons ``m\'ethode vecteurs-seuls'' toute m\'ethode qui ne n\'ecessite \`a aucun moment de calculer ou m\^eme d'\'ecrire les matrices {\bf A}, {\bf N}, {\bf S}, etc.
Dans ce contexte, la mani\`ere de cr\'eer les TOI, c'est-\`a-dire la strat\'egie d'observation, peut \^etre importante pour la construction de cartes et la qualit\'e des produits finis.
Nous pr\'esentons en section \ref{simustrat} diverses strat\'egies d'observation par des exp\'eriences CMB ballon et satellite.
Ces exemples de strat\'egies d'observation ont pour but d'\^etre assez repr\'esentatifs des diff\'erentes priorit\'es sur lesquelles on peut se focaliser: redondance des observations par pixel, couverture sur le ciel, cartographie de tout le ciel, balayages crois\'es, etc.
Elles ont \'egalement pour but de montrer comment la strat\'egie d'observation choisie pour une exp\'erience sur le CMB influence l'efficacit\'e du processus de construction de cartes.
En section \ref{simudon}, nous montrons comment nous simulons les donn\'ees temporelles (TOI) de ces exp\'eriences.
Nous pr\'esentons nos m\'ethodes de construction de cartes vecteurs-seuls en section \ref{mapmakk}, et discutons leurs conditions d'application.
En section \ref{applimak}, nous appliquons ces m\'ethodes aux donn\'ees simul\'ees, et comparons les r\'esultats pour les diff\'erentes m\'ethodes et les diff\'erentes strat\'egies d'observation.

\section{ Simulations de donn\'ees d'exp\'eriences CMB\label{simudo}}

\subsection{Strat\'egies d'observation\label{simustrat}}

L'observation du rayonnement fossile \`a grande \'echelle angulaire est possible depuis l'espace (satellites) ou depuis la stratosph\`ere (ballons).
Nous simulons des strat\'egies d'observation d'apr\`es quelques n\'ecessit\'es techniques simples et habituelles.

\subsubsection{Exp\'eriences en ballon stratosph\'erique}

Nous simulons des exp\'eriences ballon qui balaient le ciel en faisant des cercles \`a hauteur constante, et avec une vitesse de rotation \`a peu pr\`es constante, comme font Archeops (voir section \ref{commarch}) et Top Hat ({\it http://topweb.gsfc.nasa.gov}).
Cette technique simple permet d'observer une large zone du ciel, gr\^ace \`a la rotation de la Terre, et, \'eventuellement, au mouvement du ballon \`a la surface de la Terre.
Dans ce contexte, il est clair que le lieu de lancement du ballon et la dur\'ee du vol auront une influence cruciale sur la couverture du ciel r\'ealis\'ee par l'exp\'erience et la redondance des observations dans les pixels de la carte.
Les exp\'eriences qui utilisent cette strat\'egie de balayage \`a \'el\'evation constante doivent trouver un \'equilibre entre la couverture et la redondance.
Le maximum de redondance est obtenu en prenant de grands angles d'\'el\'evation et des sites de lancement \`a haute latitude.
Inversement, le meilleur rapport couverture sur temps de vol est obtenu pour des sites de lancement pr\`es de l'\'equateur et avec des angles d'\'el\'evation faibles.
Les facteurs qui contr\^olent les caract\'eristiques de l'observation sont le lieu et la date de lancement, la trajectoire du ballon au-dessus de la surface terrestre, la dur\'ee du vol, l'angle d'\'el\'evation du balayage, la fr\'equence d'acquisition des donn\'ees, et la vitesse de rotation de la nacelle (particuli\`erement si elle est constante ou pas).

Nous pr\'esentons en figure \ref{couvballons} la couverture du ciel obtenue pour trois exp\'eriences ballon simul\'ees r\'ealisant un vol de 24 heures depuis Kiruna (Laponie su\'edoise) vers la Russie, utilisant une strat\'egie d'observation \`a hauteur constante sur le ciel, dont l'angle d'\'el\'evation varie en fonction de l'exp\'erience: 10$^o$, 45$^o$ et 80$^o$.
La fr\'equence d'\'echantillonnage est de 100 Hz et la vitesse de rotation de 2 tours par minute.
Comme on le voit en figure \ref{couvballons}, la couverture sur le ciel et la redondance par pixel varient consid\'erablement.
L'exp\'erience ayant un angle d'\'el\'evation de 10$^o$ obtient une couverture de 43 \% du ciel total, celle ayant un angle de 45$^o$ obtient une couverture de 31 \%, et celle ayant un angle de 80$^o$ obtient une couverture de seulement 8 \%.
Inversement, le poids moyen par pixel de 13.5$'$ de c\^ot\'e que l'on observe pour l'exp\'erience \`a l'angle de 10$^o$ est seulement de 26, alors qu'il est de 36 pour un angle de 45$^o$ et de 141 pour un angle de 80$^o$.
En fonction des objectifs recherch\'es en terme de sensibilit\'e par pixel et de couverture sur le ciel (domaine de multip\^oles observable), la conception d'une exp\'erience CMB ballon peut donc jouer sur ce param\`etre qu'est l'angle d'\'el\'evation.

\begin{figure}
\begin{center}
\caption[Cartes de poids d'exp\'eriences ballon polaires]{Cartes de poids (nombre d'observations par pixel de 13.5$'$ de c\^ot\'e, {\it i.e.} HEALPIX 256) obtenues avec des vols ballon simul\'es de 24 heures partant de Kiruna, en faisant varier l'angle d'\'el\'evation du faisceau tournant: 10$^o$ (haut), 45$^o$ (centre), 80$^o$ (bas).
Projections Mollweide galactiques centr\'ees sur l'anticentre galactique.
}
\label{couvballons}
\end{center}
\end{figure}

Nous pr\'esentons \'egalement en figure \ref{couveq} la couverture du ciel obtenue pour trois exp\'eriences ballon simul\'ees r\'ealisant un vol de 12 heures \`a l'\'equateur.
Elles utilisent une strat\'egie d'observation \`a hauteur constante sur le ciel, dont l'angle d'\'el\'evation varie en fonction de l'exp\'erience: 10$^o$, 45$^o$ et 80$^o$.
La fr\'equence d'\'echantillonnage est de 100 Hz et la vitesse de rotation de 2 tours par minute.
Comme pour les vols polaires, la couverture sur le ciel et le poids par pixel varient consid\'erablement en fonction de l'angle d'\'el\'evation choisi.
Il est notamment possible, si l'on peut observer sans dommages 10 degr\'es au-dessus de l'horizon (la sensibilit\'e peut \^etre r\'eduite par l'atmosph\`ere), de r\'ealiser une cartographie quasi-compl\`ete du ciel (91 \%) en 12 heures de vol, avec un poids moyen de 6 observations par pixel de 13.5$'$ de c\^ot\'e.
Avec un angle d'\'el\'evation de 45 degr\'es, on observe 50 \% du ciel avec un poids moyen de 11 observations par pixel.
Avec un angle d'\'el\'evation de 80 degr\'es, on n'observe que 10 \% du ciel, mais avec un poids moyen de 57 observations par pixel.

\begin{figure}
\begin{center}
\caption[Cartes de poids d'exp\'eriences ballon \'equatoriales]{Cartes de poids (nombre d'observations par pixel de 13.5$'$ de c\^ot\'e, {\it i.e.} HEALPIX 256) obtenues avec des vols \'equatoriaux simul\'es de 12 heures, en faisant varier l'angle d'\'el\'evation du faisceau tournant: 10$^o$ (haut), 45$^o$ (centre), 80$^o$ (bas).
Projections Mollweide galactiques centr\'ees sur le centre galactique.
}
\label{couveq}
\end{center}
\end{figure}

Nous avons simul\'e en particulier deux trains de donn\'ees d'exp\'eriences ballon, avec une \'el\'evation constante de balayage de 35 degr\'es au-dessus de l'horizon, une fr\'equence d'\'echantillonnage de 100 Hz et une vitesse de rotation de 2 tours par minute.
La premi\`ere est con\c{c}ue pour maximiser la redondance par pixel, tout en couvrant une large partie du ciel.
Il s'agit d'un vol polaire de 24 heures depuis Kiruna, Laponie su\'edoise (voir \`a ce propos la description des vols d'Archeops en section \ref{commarch}).
La nuit polaire permet en effet d'avoir 24 h ou plus de nuit, et ainsi de faire voler l'exp\'erience en \'evitant les perturbations li\'ees \`a la lumi\`ere du soleil.
La couverture du ciel correspondant \`a ce vol polaire est de 35 \%.
Le deuxi\`eme train de donn\'ees simul\'e est celui d'un vol \'equatorial de 12 h.
La couverture angulaire pour ce vol est de 61 \% du ciel total, et serait de 82 \% avec un vol de 24 h, ou deux vols de 12 h, le deuxi\`eme ayant lieu six mois apr\`es le premier afin d'\'eviter la lumi\`ere du jour.
Le vol polaire que nous avons pr\'esent\'e est particuli\`erement bon pour la redondance, tandis que ce vol \'equatorial a une couverture nettement meilleure mais une redondance bien plus faible.
La comparaison des m\'ethodes de construction de cartes sur ces deux vols pourra donc donner des r\'esultats int\'eressants sur la fa\c{c}on dont, en pratique, la strat\'egie d'observation contr\^ole la qualit\'e finale des cartes.

\subsubsection{Exp\'eriences satellite}

Nous simulons \'egalement des exp\'eriences satellite, qui balaient le ciel en faisant des cercles orthogonaux au plan \'ecliptique.
L'angle constant du faisceau par rapport \`a l'axe Soleil-Terre d\'etermine la couverture et la redondance, comme la hauteur pour les exp\'eriences ballon pr\'ec\'edemment d\'ecrites.
Nous pr\'esentons en figure \ref{couvsat} les cartes de poids par pixel du ciel obtenues pour quatre exp\'eriences satellite balayant le ciel par cercles orthogonaux au plan \'ecliptique, l'une ayant un angle de 20$^o$, une autre de 45$^o$, une autre de 85$^o$, et une autre de 90$^o$, c'est-\`a-dire qu'elle r\'ealise des grands cercles le long des m\'eridiens \'ecliptiques.
L'exp\'erience Planck correspondra \`a la strat\'egie \`a 85$^o$ s'il n'est pas r\'ealis\'e de balancements en plus des cercles sur le ciel.
\`A nombre \'egal de points de donn\'ees, nous observons que bien \'evidemment le poids moyen par pixel observ\'e est plus important (123 par pixel de 27$'$ de c\^ot\'e) pour la strat\'egie ayant un angle de 20$^o$.
Le poids moyen est de 60 pour 45$^o$, de 43 pour 85$^o$ et 90$^o$.
La couverture sur le ciel est, inversement, de 35 \% pour un angle de 20$^o$, de 71 \% pour un angle de 45$^o$, de 99.7 \% pour un angle de 85$^o$, et bien s\^ur la couverture compl\`ete est atteinte pour un angle de 90$^o$ (cercles dans les m\'eridiens \'ecliptiques).

\begin{figure}
\begin{center}
\epsfxsize 10.0 true cm
\leavevmode
\caption[Cartes de poids d'exp\'eriences satellite]{Cartes de poids (nombre d'observations par pixel de 27$'$ de c\^ot\'e, {\it i.e.} HEALPIX 128) obtenues avec des missions satellite simul\'ees, en faisant varier l'angle d'inclinaison du faisceau par rapport \`a l'axe Soleil-Terre: 20$^o$ (bas), 45$^o$, 85$^o$ et 90$^o$ (haut).
Projections Mollweide galactiques centr\'ees sur le centre galactique.
}
\label{couvsat}
\end{center}
\end{figure}

En particulier, nous avons simul\'e un train de donn\'ees d'une exp\'erience satellite observant par grands cercles (angle d'inclinaison du faisceau de 90 degr\'es) sur le ciel le long des m\'eridiens \'ecliptiques.
Pour un satellite en orbite basse (LEO pour {\it Low Earth Orbit}), cette strat\'egie d'observation permet de pointer le t\'elescope en permanence vers le z\'enith, \'evitant ainsi la contamination terrestre.
Les grands cercles que d\'ecrit ce satellite sont orthogonaux \`a l'axe Terre-Soleil, ce qui permet de s'affranchir de la plus grande partie de la lumi\`ere solaire - et de la lumi\`ere terrestre si le satellite est \`a L2.
En r\'ealit\'e, sur une orbite basse, la trajectoire exacte suivie par le satellite ne serait pas aussi simple, \`a cause des perturbations gravitationnelles qui l'affecteraient, mais notre train de donn\'ees simul\'ees est proche de ce que cela serait.
Cette strat\'egie d'observation fonctionne aussi pour un satellite \`a L2, comme Planck (voir section \ref{galplanck}).
Dans ce cas, la strat\'egie d'observation est simplement un grand cercle constamment dans le m\'eridien orthogonal \`a l'axe Terre-Soleil, qui se d\'ecale lentement avec la r\'evolution de la Terre.
La cartographie compl\`ete du ciel est ainsi termin\'ee en six mois.
Nous avons simul\'e notre train de donn\'ees satellite en ajoutant un ingr\'edient qui complique cette strat\'egie d'observation.
En effet, nous avons ajout\'e une d\'ependance en ascension droite qui peut se comprendre comme le d\'ecalage d'une orbite LEO autour de la Terre.
Nous avons simul\'e une cartographie rapide de la totalit\'e du ciel, faite gr\^ace \`a cette strat\'egie et contenant huit million de points de donn\'ees.
Bien que cette quantit\'e de donn\'ees ne correspondrait bien s\^ur pas au produit d'une mission satellite compl\`ete, ceci nous permet de comparer la strat\'egie de balayage et les propri\'et\'es finales des cartes avec celles des deux exp\'eriences ballon.
On peut consid\'erer que ce train de donn\'ees ``satellite'' correspond \`a des donn\'ees sous-\'echantillonn\'ees.
Les caract\'eristiques principales des trois observations simul\'ees sont r\'esum\'ees en table \ref{tabstrat}.

\begin{table}
\begin{center}
\caption[Caract\'eristiques des trois trains de donn\'ees simul\'es]{\label{tabstrat}
Caract\'eristiques des trois trains de donn\'ees simul\'es.
}
\begin{tabular}{llll}
\hline
 & Polaire & \'Equatorial & Satellite\\
\hline
\hline
Temps d'obs. & 24 h & 12 h & {\small (sous-\'echant.)}\\
\hline
Couverture & 35 \%  & 61 \% & 100 \%\\
\hline
Nb de points & 8 & 4 & 8\\
(millions) & & & \\
\hline
Redondance moy. & 30.5 & 8.7 & 11.8\\
par pixel de 13.5$'$ & & & \\
\hline
\end{tabular}
\end{center}
\end{table}

\subsection{Simulation des donn\'ees\label{simudon}}

Nous avons simul\'e le ciel millim\'etrique (2 mm de longueur d'onde) avec trois composantes: les fluctuations du CMB, le dip\^ole, et la Galaxie (\'emission thermique des poussi\`eres).
L'Univers que nous avons simul\'e est proche de celui qui est favoris\'e par les observations actuelles telles que celles de BOOMERANG \cite{debernardis00} et MAXIMA \cite{hanany00}.
Il est domin\'e par la constante cosmologique, avec $\Omega_\Lambda$ = 0.7, $\Omega_{CDM}$ = 0.25, $\Omega_{bar}$ = 0.05, $H_0$ = 50, et un indice spectral scalaire des fluctuations \'egal \`a 1.
Les valeurs pr\'ecises des param\`etres cosmologiques utilis\'es ne sont pas importantes pour les m\'ethodes de construction de cartes, mais en revanche, la gaussianit\'e des fluctuations du CMB est, {\it a priori}, importante pour les m\'ethodes de Wiener que nous voulons tester, car le filtre de Wiener est optimal quand le signal comme le bruit sont des champs al\'eatoires gaussiens (voir section \ref{introwie}).
Le spectre de puissance harmonique du CMB est simul\'e gr\^ace au logiciel CMBFAST de Seljak \& Zaldarriaga (1996).
Ce spectre ne tient pas compte de la variance cosmique, c'est-\`a-dire des fluctuations al\'eatoires qui existent dans un Univers; en fait le spectre th\'eorique ainsi simul\'e est une moyenne statistique sur les r\'ealisations d'univers, \'etant donn\'es les param\`etres cosmologiques.
Le champ al\'eatoire gaussien des fluctuations du CMB (carte repr\'esentant une r\'ealisation d'univers) est alors simul\'e gr\^ace \`a l'outil SYNFAST du package HEALPIX ({\it http://www.eso.org/science/healpix} ou G\'orski \etal 1999).
Le dip\^ole est ajout\'e gr\^ace \`a sa d\'etermination par COBE-DMR \cite{lineweaver96}.
La Galaxie \`a 2 mm de longueur d'onde est extrapol\'ee \`a partir de l'ensemble de donn\'ees composite sur tout le ciel 100 $\mic$ IRAS - COBE-DIRBE \cite{schlegel98}.
Nous pr\'esentons en figure \ref{simusallsky} une carte des fluctuations du rayonnement fossile sur tout le ciel, ainsi qu'une autre carte de tout le ciel, contenant les fluctuations du CMB, le dip\^ole, et l'\'emission des poussi\`eres galactiques, \`a 2 mm de longueur d'onde.

\begin{figure}
\begin{center}
\caption[Ciel simul\'e]{Haut: carte simul\'ee de tout le ciel contenant les fluctuations du CMB avec $\Omega_\Lambda$ = 0.7, $\Omega_{CDM}$ = 0.25, $\Omega_{bar}$ = 0.05, $H_0$ = 50, et un indice spectral scalaire des fluctuations \'egal \`a 1.
Bas: carte simul\'ee de tout le ciel contenant le CMB et la Galaxie \`a 2 mm.
Cartes en unit\'e de temp\'erature du CMB (${\mu}K_{CMB}$), projections Mollweide galactiques centr\'ees sur le centre galactique.}
\label{simusallsky}
\end{center}
\end{figure}

Nous avons simul\'e des trains de donn\'ees bruit\'es \`a partir de ces ciels simul\'es.
Cependant, il faut bien se rendre compte que les vraies donn\'ees provenant directement d'un vrai d\'etecteur n'ont pas forc\'ement la m\^eme allure.
Habituellement, il y a des {\it glitches}, des effets syst\'ematiques et du bruit qui peuvent \^etre soustraits avant d'appliquer la construction de cartes optimale.
Les {\it glitches} sont des rayons cosmiques \'energ\'etiques qui atteignent un bolom\`etre, produisant ainsi un pic fin et \'elev\'e dans le train de donn\'ees.
En g\'en\'eral, ils peuvent \^etre rep\'er\'es et enlev\'es des donn\'ees.
Les effets syst\'ematiques sont des perturbations qui agissent de mani\`ere non al\'eatoire sur les donn\'ees, contrairement aux bruits.
Pour adopter une d\'efinition plus rigoureuse, nous pouvons dire que l'on d\'efinit un bruit comme toute perturbation stationnaire, corr\'el\'ee ou pas avec elle-m\^eme, et comme syst\'ematique toute perturbation non stationnaire.
Une perturbation est stationnaire si et seulement si son spectre de puissance ne varie pas au cours du temps, autrement dit, si ses propri\'et\'es d'autocorr\'elation sont constantes.
Certains bruits et syst\'ematiques peuvent \'eventuellement \^etre enlev\'es des donn\'ees temporelles en d\'ecorr\'elant avec des signaux provenant des thermom\`etres ou d'autres bolom\`etres dans le plan focal.
Nous ne nous sommes pas essentiellement int\'eress\'es \`a ces probl\`emes, mais il est important de se rappeler que la construction de cartes que nous pr\'esentons en section \ref{mapmakk} n'est pas connue pour bien fonctionner \`a tout coup sur des donn\'ees brutes.
En ce qui concerne nos simulations pr\'esentes, nous souhaitons tester nos m\'ethodes de construction de cartes sur du bruit blanc (non autocorr\'el\'e) et du bruit 1/f (autocorr\'el\'e).
Ce bruit que nous introduisons dans les trains de donn\'ees simul\'es peut \^etre caract\'eris\'e par son spectre de puissance statistique, qui suit une loi ``1/f'':

\begin{equation}
S_n = l_{inf} \: [1 + (f_c/f)^n]
\label{unsurf}
\end{equation}

o\`u $S_n$ est le spectre de puissance du bruit, $l_{inf}$ le niveau de bruit \`a fr\'equence infinie, c'est-\`a-dire le niveau de bruit blanc, $f_c$ la fr\'equence de coupure du bruit corr\'el\'e et n l'indice spectral.
Ces sources de bruit g\'en\'er\'e de fa\c{c}on stationnaire sont, habituellement, ce qu'il reste dans un train de donn\'ees nettoy\'e, apr\`es avoir \'elimin\'e les autres sources possibles de bruits et de syst\'ematiques sans modifier le signal.
Ce bruit 1/f est tout \`a fait caract\'eristique des mesures bolom\'etriques, et m\^eme si, aujourd'hui, les progr\`es technologiques tendent \`a diminuer la d\'erive 1/f, celle-ci reste une r\`egle tr\`es habituelle dans les donn\'ees.
Afin de comparer utilement nos diff\'erentes strat\'egies d'observation, nous appliquons nos m\'ethodes de construction de cartes \`a des trains de donn\'ees simul\'es ayant le m\^eme niveau de bruit et les m\^emes propri\'et\'es statistiques, i.e. n=1, $f_c$=0.1 Hz et $l_{inf}$ = 100 $\ukc$ rms.
(rms: {\it root mean square} ou \'ecart quadratique moyen.)
Cet \'ecart quadratique moyen de bruit blanc est approximativement le niveau attendu pour les bolom\`etres de Planck \cite{tauber00} avec un taux d'\'echantillonnage de 100 Hz.
Bien entendu, cette valeur relativement faible ne repr\'esente que le bruit non corr\'el\'e (bruit blanc), alors que la quantit\'e totale de bruit introduit dans la simulation des trains de donn\'ees est bien sup\'erieure.

\section{ M\'ethodes de construction de cartes pour les exp\'eriences CMB\label{mapmakk}}

\subsection{Pix\'elisation}

Pour construire des cartes de tout le ciel, ou d'une grande partie du ciel, il nous est tout d'abord n\'ecessaire de d\'efinir une pix\'elisation de la sph\`ere c\'eleste.
Nous utilisons pour cela le sch\'ema HEALPIX RING, dont la r\'esolution angulaire est d\'efinie par le param\`etre $N_{side}$.
La taille angulaire du c\^ot\'e du pixel diminue proportionnellement quand $N_{side}$ augmente.
Nous pr\'esentons nos cartes, simul\'ees et reconstruites, avec $N_{side}$ = 256, ce qui d\'ecoupe la sph\`ere en 786432 pixels.
Ce nombre de pixels, tous de m\^eme surface et approximativement de m\^eme forme, correspond \`a une surface de 13.5 x 13.5 minutes d'arc carr\'ees par pixel.
Ceci correspond bien aux strat\'egies de balayage que nous avons utilis\'ees.
En effet, utiliser $N_{side}$ = 512 pixels laisserait beaucoup de trous dans les cartes, \`a cause du manque d'\'echantillons de donn\'ees qui se produirait pour une r\'esolution angulaire si pr\'ecise, et de ce fait nuirait \`a leur visibilit\'e et \`a leur analyse.
Afin de pix\'eliser correctement les cartes d'une exp\'erience donn\'ee, et afin de r\'ealiser l'analyse qui suit la construction de cartes, il est n\'ecessaire d'utiliser des pixels 2 \`a 3 fois plus petits (il n'y a pas de crit\`ere de Nyquist parfaitement d\'efini sur la sph\`ere) que le diam\`etre du faisceau.
Les cartes de beaucoup d'exp\'eriences CMB modernes seraient bien pix\'elis\'ees avec des pixels l\'eg\`erement plus petits que ceux que l'on utilise, mais cependant, il faudrait stocker des cartes plus lourdes et le temps de calcul serait plus long (car, comme nous le verrons, la convergence des m\'ethodes it\'eratives serait plus lente), sans apporter de r\'esultat significatif suppl\'ementaire \`a ces investigations.
Ici, nous ne traitons pas le probl\`eme du faisceau exp\'erimental, et nous ne chercherons pas \`a d\'econvoluer l'effet du faisceau sur les donn\'ees, contrairement \`a ce que nous avions fait pour PRONAOS (voir section \ref{wieglob}).

\subsection{Moyennage par pixel\label{moypix}}

Projeter un train de donn\'ees sur une carte n'est pas seulement une simple transformation d'un domaine \`a un autre (1-D vers 2-D), mais surtout la mani\`ere simple d'estimer la vraie carte du ciel, en moyennant les \'echantillons d'un m\^eme pixel sur le ciel.
Le bruit est ainsi r\'eduit d'un facteur racine carr\'ee du poids, o\`u le poids est le nombre d'\'echantillons de donn\'ees tombant dans un pixel.
De ce fait, pour accro\^\i tre le rapport signal sur bruit dans les pixels, il suffit d'augmenter le poids par pixel, c'est-\`a-dire qu'il faut augmenter la taille des pixels.
Cependant, bien entendu, de ce fait on perd de la r\'esolution angulaire dans la carte, et surtout on limite le multip\^ole l maximum que l'on peut d\'eduire.
Pour une pix\'elisation de type HEALPIX on a:

\begin{equation}
l_{max} = 2 \: N_{side}
\end{equation}

Chaque \'el\'ement de la matrice de convolution {\bf A} traduit la mani\`ere dont un pixel du ciel contribue \`a un point de mesure, et, inversement, chaque \'el\'ement de la matrice {\bf A}$^t$ traduit comment un point de donn\'ees contribue \`a un pixel de la carte reconstruite.
Si l'on consid\`ere des tailles de pixel plus grandes que le faisceau, ou que l'on ne tient pas compte de l'effet du faisceau, alors la matrice {\bf A}$^t$ est simplement l'addition des points de donn\'ees dans les pixels, ce qui veut dire que cette matrice est constitu\'ee uniquement de 0s et de 1s.
Ceci est beaucoup plus facile \`a manipuler que la matrice r\'eponse (convolution) compl\`ete, puisque dans ce cas, la construction de carte simple d'une TOI (voir \'equation \ref{pixelavg}) est un moyennage par pixel.
Dans l'\'equation \ref{pixelavg}, {\bf A$^t$A} est alors un simple compteur d'observations par pixel de la carte.
Dans ce cas, {\bf A} est simplement la d\'eprojection d'une carte (2-D) sur un train de donn\'ees (1-D).
Nous n'avons pas fait d'investigations concernant la d\'econvolution du faisceau pour les larges TOI, et donc nous consid\'erons des matrices {\bf A} et {\bf A}$^t$ compos\'ees uniquement de 0s et de 1s.
Nous pr\'esentons en figure \ref{figpoids} les cartes de poids pour les trois strat\'egies d'observation que nous avons d\'ecrites.
Il est clair que le vol polaire de 24 heures a la meilleure redondance.
Nous verrons si ces diff\'erences dans la strat\'egie d'exp\'erimentation et d'observation produisent des diff\'erences importantes dans les cartes reconstruites.

\begin{figure}
\begin{center}
\caption[Cartes de poids]{Cartes de poids (nombre d'observations par pixel de 13.5$'$ de c\^ot\'e): en haut, le vol ballon polaire, au centre, le vol ballon \'equatorial, en bas, la mission satellite, en unit\'e de nombre de points de donn\'ees par pixel de la carte.
Projections Mollweide galactiques centr\'ees sur le centre galactique.}
\label{figpoids}
\end{center}
\end{figure}

Nous pr\'esentons en figure \ref{figsimusdon} les trois jeux de donn\'ees simul\'es projet\'es sur une carte en moyennant les donn\'ees dans les pixels.
Les cartes contiennent les fluctuations du CMB, le dip\^ole et la Galaxie, ainsi que beaucoup de bruit corr\'el\'e qui se projette sur les cartes par les stries et draperies bien visibles.
Nous pr\'esentons \'egalement en figure \ref{figsimuscmb} les cartes construites en moyennant par pixel uniquement les fluctuations du CMB et le bruit, sans le dip\^ole et la Galaxie.
En effet, il est important de bien se rendre compte que le bruit corr\'el\'e introduit noie compl\`etement les fluctuations du CMB.
En ce qui concerne cette m\'ethode simple de moyennage par pixel, il est donc clair que la construction des cartes n'est pas satisfaisante.
Les strat\'egies d'observation pr\'esent\'ees, qui privil\'egient plut\^ot la couverture sur le ciel que la redondance par pixel, ne peuvent donc pas se passer de m\'ethodes de construction de cartes avanc\'ees pour r\'eduire le bruit corr\'el\'e.

\begin{figure}
\begin{center}
\caption[Cartes avec le bruit]{Cartes r\'ealis\'ees par moyennage par pixel des trois trains de donn\'ees simul\'es: en haut, le vol ballon polaire de 24 h, au centre, le vol ballon \'equatorial de 12 h, en bas, la mission satellite.
Les cartes contiennent les fluctuations du CMB, le dip\^ole et la Galaxie, ainsi que beaucoup de bruit corr\'el\'e qui se projette sur les cartes par les stries et draperies bien visibles.
Cartes en unit\'e de temp\'erature du CMB (${\mu}K_{CMB}$), projections Mollweide galactiques centr\'ees sur le centre galactique.}
\label{figsimusdon}
\end{center}
\end{figure}

\begin{figure}
\begin{center}
\caption[Cartes du CMB avec le bruit]{Cartes r\'ealis\'ees par moyennage par pixel des trois trains de donn\'ees simul\'es, avec uniquement les fluctuations du CMB et le bruit: en haut, le vol ballon polaire de 24 h, au centre, le vol ballon \'equatorial de 12 h, en bas, la mission satellite.
Le bruit corr\'el\'e se projette sur les cartes par les stries et draperies bien visibles.
Cartes en unit\'e de temp\'erature du CMB (${\mu}K_{CMB}$), projections Mollweide galactiques centr\'ees sur le centre galactique.}
\label{figsimuscmb}
\end{center}
\end{figure}

\subsection{Blanchissage et moyennage par pixel\label{blanch}}

Puisque la quantit\'e de bruit corr\'el\'e n'est pas diminu\'ee efficacement par le processus de moyennage par pixel, il est n\'ecessaire d'appliquer une m\'ethode de construction de cartes optimale (voir section \ref{wiedier} et suivantes), ou de chercher \`a r\'eduire la quantit\'e de bruit corr\'el\'e dans les donn\'ees avant de construire la carte.
Comme le bruit 1/f se trouve principalement, par d\'efinition, aux basses fr\'equences, il est envisageable de filtrer d'une fa\c{c}on ou d'une autre les basses fr\'equences, afin de laver les donn\'ees de leur bruit corr\'el\'e.
Cependant, puisque le bruit 1/f se trouve en quantit\'e importante jusqu'\`a des fr\'equences relativement \'elev\'ees, l'application sauvage d'un filtre de ``blanchissage'' (filtre dont le but est d'\'eliminer le bruit corr\'el\'e, de sorte que le bruit dans les TOI apr\`es filtrage soit blanc) suivie d'un moyennage par pixel d\'etruit efficacement le signal en m\^eme temps que le bruit corr\'el\'e, et ce dans une bande tr\`es large de fr\'equences, et donc de multip\^oles.
Nous pr\'esentons en figure \ref{sauvage} l'application d'un filtre de blanchissage, construit comme l'inverse de l'ajustement 1/f du spectre de puissance du bruit, aux donn\'ees du vol ballon polaire.
Ce r\'esultat, \`a comparer avec la carte des fluctuations du CMB en figure \ref{simusallsky}, montre tr\`es clairement l'effet d\'evastateur de ce filtrage non optimal.

\begin{figure}
\begin{center}
\epsfxsize 10.0 true cm
\leavevmode
\caption[Filtrage de blanchissage]{Carte des fluctuations du CMB construite gr\^ace \`a un filtre de ``blanchissage'', sur les donn\'ees simul\'ees du vol ballon polaire.
Le signal est largement d\'etruit par ce filtre non optimal.
Carte en unit\'e de temp\'erature du CMB (${\mu}K_{CMB}$), projection Mollweide galactique centr\'ee sur le centre galactique.}
\label{sauvage}
\end{center}
\end{figure}

Il semble donc indispensable de se servir plus subtilement des propri\'et\'es de corr\'elation du bruit pour construire la carte.

\subsection{Propri\'et\'es de corr\'elation\label{propcorr}}

Les m\'ethodes optimales de construction de cartes utilisent les matrices de covariance du bruit dans la TOI ({\bf N}) et \'eventuellement du ciel dans la carte ({\bf S}).
Elles ne peuvent pas \^etre invers\'ees ou m\^eme stock\'ees en m\'emoire pour des donn\'ees aussi massives.

Cependant, si le bruit est stationnaire dans le domaine temporel, comme c'est le cas pour le bruit 1/f et le bruit blanc (les bruits bolom\'etriques habituels), la matrice de corr\'elation du bruit dans le domaine temporel ({\bf N}) est circulante.
En clair, ceci signifie que si les propri\'et\'es statistiques du bruit, c'est-\`a-dire le spectre de puissance, ne changent pas au cours du temps (stationnarit\'e), alors la matrice {\bf N} s'\'ecrit ligne par ligne en d\'ecalant les \'el\'ements de la gauche vers la droite \`a chaque changement de ligne (circularit\'e).
Puisque {\bf N} est circulante, il en va de m\^eme de {\bf N}$^{-1}$.
En cons\'equence, multiplier un vecteur par une telle matrice est \'equivalent \`a appliquer une op\'eration de convolution, ce qui signifie r\'ealiser un filtrage dans l'espace de Fourier.
En d'autres termes, une matrice circulante dans l'espace r\'eel est diagonale dans l'espace de Fourier.

La matrice de corr\'elation du bruit doit \^etre une moyenne statistique sur les r\'ealisations, ce qui veut dire qu'un ajustement doit \^etre r\'ealis\'e sur le vecteur de bruit dans l'espace de Fourier, afin d'\'eliminer les fluctuations gaussiennes.
Le vecteur de bruit dans l'espace de Fourier peut \^etre estim\'e \`a partir des donn\'ees, puisque g\'en\'eralement la queue 1/f est bien visible dans le spectre de puissance.
Nous ajustons ce spectre du bruit avec la loi en 1/f$^n$ d\'ecrite par l'\'equation \ref{unsurf}.

En ce qui concerne la matrice de covariance du signal, le probl\`eme est diff\'erent.
Pour les fluctuations du CMB uniquement, cette matrice est stationnaire dans le domaine de la carte, \`a condition que les fluctuations du CMB forment un champ al\'eatoire gaussien.
De ce fait, la matrice {\bf S} (covariance du signal dans la carte) est circulante pour les cartes des fluctuations du CMB uniquement.
Cependant, ceci ne signifie pas que la matrice de covariance du signal dans le domaine temporel est elle aussi circulante.
Le signal est stationnaire dans les donn\'ees temporelles si en plus, l'\'echantillonnage des donn\'ees sur le ciel est constant, ce qui signifie en pratique que d'une part la vitesse de balayage sur le ciel doit \^etre constante, et que d'autre part la fr\'equence d'\'echantillonnage des donn\'ees doit \'egalement \^etre constante.
Il n'y a pas de raison particuli\`ere pour que la fr\'equence d'\'echantillonnage ne soit pas constante.
La vitesse de balayage sur le ciel est constante si l'angle de balayage est constant, et que la vitesse de rotation est \'egalement constante, ce qui est le cas pour nos exp\'eriences simul\'ees.
Si la vitesse de rotation n'est pas constante, la stationnarit\'e du signal CMB est perdue dans le domaine temporel.
Cependant, le seuil de tol\'erance \`a cette condition doit \^etre \'etudi\'e.

Si l'on consid\`ere que les TOI de signal et de bruit sont stationnaires, alors multiplier par {\bf ASA}$^t$, matrice de covariance du signal dans les TOI, ou {\bf N}$^{-1}$, inverse de la matrice de covariance du bruit dans la TOI, est une convolution, que l'on r\'ealisera comme une multiplication dans l'espace de Fourier.

NB: {\bf A S A}$^t$ = ${\bf \Sigma}$, la matrice de covariance du signal dans les TOI, car:

\begin{equation}
{\bf \Sigma} = <({\bf A \: x}) \: ({\bf A \: x})^t> = < {\bf A \: x \: x}^t \: {\bf A}^t> = {\bf A} \: <{\bf x \: x}^t> \: {\bf A}^t = {\bf A \: S \: A}^t
\end{equation}

\subsection{M\'ethode de Wiener directe\label{wiedier}}

Comme mentionn\'e en section \ref{introwie}, la m\'ethode de Wiener fait l'hypoth\`ese de l'ind\'ependance du signal et du bruit.
Pour des donn\'ees concernant le rayonnement fossile, cette hypoth\`ese est tout \`a fait valable, mais comme nous l'avons vu pour PRONAOS (section \ref{wieglob}), la r\'eponse des bolom\`etres aux forts contrastes de flux peut \'eventuellement nuire \`a cette supposition dans le cas de donn\'ees int\'egrant des sources galactiques tr\`es intenses.

La matrice de Wiener 1 (\'equation \ref{wie1}) se r\'eduit au filtrage de Wiener \cite{wiener49} dans le domaine temporel, si quelques conditions sont v\'erifi\'ees.
En effet l'\'equation \ref{wie1} s'\'ecrit aussi:

\begin{equation}
{\bf W} = [{\bf A}^t{\bf A}]^{-1} {\bf A}^t {\bf ASA}^{t} [{\bf ASA}^{t} + {\bf N}]^{-1}
\end{equation}

o\`u {\bf ASA}$^t$ est la matrice de covariance du signal dans le train de donn\'ees temporelles, et {\bf N} est la matrice de covariance du bruit dans le train de donn\'ees.
De ce fait, si le bruit et le signal sont tous les deux stationnaires dans le domaine temporel, alors la m\'ethode de Wiener 1 consiste simplement \`a filtrer les donn\'ees avant de coadditionner dans les pixels.
Le filtre optimal de Wiener est alors $\sigma^2\over{\sigma^2+\nu^2}$ dans le domaine de Fourier, o\`u $\sigma$ et $\nu$ d\'esignent respectivement les transform\'ees de Fourier du signal et du bruit, ajust\'ees pour \'eliminer les fluctuations al\'eatoires gaussiennes.
Il faut bien noter que deux conditions sont n\'ecessaires pour que le signal soit stationnaire dans le train de donn\'ees: le signal doit \^etre stationnaire dans la carte et l'observation doit \^etre stationnaire par rapport \`a la carte, ce qui signifie que la vitesse de balayage et la fr\'equence d'\'echantillonnage doivent \^etre constantes.
L'ajustement du spectre de bruit dans l'espace de Fourier est r\'ealis\'e comme d\'ecrit en section \ref{propcorr}.
Le spectre de puissance du signal dans les TOI est domin\'e par les pics de rotation, et donc approximer la moyenne sur les r\'ealisations par le spectre lui-m\^eme n'est pas une mauvaise approximation.
Ceci peut \^etre fait facilement m\^eme pour des vraies donn\'ees en simulant l'observation du ciel sur la m\^eme zone (sans bruit).
Nous pr\'esentons en figure \ref{specnoisesky} le spectre de puissance du bruit simul\'e dans le train de donn\'ees, ainsi que celui du signal dans les donn\'ees.

\begin{figure}
\begin{center}
\epsfxsize 10.0 true cm
\leavevmode
\caption[Spectres du bruit et du signal]{Spectres de puissance du bruit dans la TOI du vol ballon polaire (noir) et du signal (gris).
Les pics de rotation sont bien visibles sur le spectre de puissance du signal.}
\label{specnoisesky}
\end{center}
\end{figure}

La m\'ethode directe de Wiener est optimale quand elle est appliqu\'ee \`a des signaux al\'eatoires gaussiens stationnaires, mais nous la testons \'egalement sur des ciels non gaussiens et non stationnaires, le premier \'etant non stationnaire y compris dans la carte, du fait de la pr\'esence du signal galactique, le second \'etant non stationnaire dans le train de donn\'ees, bien que stationnaire sur la carte, \`a cause de larges variations de la vitesse de rotation de la nacelle.
Nous pr\'esentons les r\'esultats de cette \'etude en section \ref{etupol}.

\subsection{M\'ethode COBE it\'erative}

L'\'equation \ref{cob} correspondant \`a la m\'ethode dite ``COBE'' ne peut pas \^etre appliqu\'ee directement avec des algorithmes vecteurs-seuls, contrairement \`a la m\'ethode Wiener 1.
En effet, il y a des inversions matricielles n\'ecessaires \`a l'application directe de cette m\'ethode COBE.
Il faut donc trouver une m\'ethode qui permette de r\'esoudre le m\^eme probl\`eme (d\'ecrit par la m\^eme \'equation) de fa\c{c}on indirecte.
L'astuce consiste \`a r\'esoudre:

\begin{equation}
[{\bf A}^{t} {\bf N}^{-1} {\bf A}] \: {\bf \Tilde x} = {\bf A}^{t} {\bf N}^{-1} {\bf y}
\end{equation}

Cette forme permet d'\'eviter la lourde inversion matricielle, mais elle n\'ecessite un sch\'ema it\'eratif pour r\'esoudre ce syst\`eme lin\'eaire.
Le sch\'ema it\'eratif g\'en\'eral correspondant \`a cette \'equation est:

\begin{equation}
\alpha \: {\bf \Tilde x}_{n+1} = \alpha \: {\bf \Tilde x}_n + {\bf A}^{t} {\bf N}^{-1} {\bf y} - [{\bf A}^{t} {\bf N}^{-1} {\bf A}] \: {\bf \Tilde x}_n
\label{iter1}
\end{equation}

o\`u $\alpha$ est un op\'erateur lin\'eaire quelconque agissant sur un vecteur, c'est-\`a-dire une matrice carr\'ee quelconque.
Nous avons test\'e cet algorithme sur des simulations et les donn\'ees Archeops (voir section \ref{mapmakark}), avec $\alpha$ \'etant un scalaire.
En testant la m\'ethode avec diff\'erents $\alpha$, nous concluons que la matrice identit\'e est le meilleur it\'erateur, au vu de la convergence (ou pas) et de sa rapidit\'e.

Un autre sch\'ema peut \^etre d\'evelopp\'e, en faisant converger la carte de bruit plut\^ot que la carte du ciel reconstruit, comme d\'ecrit par Prunet (2001).
Ceci peut \^etre meilleur pour la stabilit\'e de la m\'ethode it\'erative, car le signal peut \^etre plus ennuyeux que le bruit instrumental: des points intenses galactiques, par exemple, peuvent mal r\'eagir au filtrage et cr\'eer des stries sur les cartes.
Le sch\'ema d'it\'eration sur le bruit fonctionne gr\^ace \`a l'astuce consistant \`a faire le changement de variable de ${\bf \Tilde x}$ \`a ${\bf \Hat x}$:

\begin{equation}
{\bf \Hat x} = [{\bf A}^t{\bf A}]^{-1} {\bf A}^t {\bf y} - {\bf \Tilde x}
\label{chgvar}
\end{equation}

On peut montrer imm\'ediatement que cette expression correspond \`a la carte de bruit \`a laquelle on ajoute l'erreur de reconstruction.
L'\'equation \ref{chgvar} conduit \`a r\'e\'ecrire l'\'equation \ref{iter1}:

\begin{equation}
\alpha \: {\bf \Hat x}_{n+1} = \alpha \: {\bf \Hat x}_n + {\bf A}^{t} {\bf N}^{-1} {\bf z} - [{\bf A}^{t} {\bf N}^{-1} {\bf A}] \: {\bf \Hat x}_n
\label{iter2}
\end{equation}

o\`u l'on a:

\begin{equation}
{\bf z} = {\bf A}[{\bf A}^t{\bf A}]^{-1}{\bf A}^t {\bf y} - {\bf y}
\end{equation}

{\bf z} est donc la diff\'erence entre la TOI cartographi\'ee par moyennage par pixel et d\'eprojet\'ee, et la TOI.
{\bf z} est donc un train de donn\'ees repr\'esentant l'oppos\'e du bruit \'eliminable par moyennage par pixel.
Nous voyons donc bien que l'\'equation \ref{iter2} consiste \`a it\'erer sur le bruit plut\^ot que sur le signal.
Si cet algorithme converge, alors la solution donn\'ee par la convergence est exactement la solution optimale du probl\`eme de construction de cartes.
La preuve en est que si cela converge, alors on a ${\bf \Hat x}_{n+1} = {\bf \Hat x}_n$ quand n tend vers l'infini, et donc l'\'equation \ref{iter2} se r\'eduit \`a l'\'equation COBE (\ref{cob}) sur le bruit.
Bien entendu, cette unique solution du probl\`eme de construction de cartes est quelque peu diff\'erente de la carte du vrai ciel, la diff\'erence \'etant caract\'eris\'ee par la matrice de covariance du bruit d\'ecrite en section \ref{matcovb}.
Le cas it\'eratif $\alpha$ = 1 est en fait l'it\'erateur le plus simple que l'on puisse imaginer, mais il fonctionne tr\`es bien sur les simulations et les vraies donn\'ees que nous avons trait\'ees.

Il faut noter qu'une m\'ethode similaire \`a la n\^otre a \'et\'e d\'evelopp\'ee simultan\'ement au sein de la collaboration Archeops, par Dor\'e \etal (2001).
Bien que leur m\'ethode soit d\'eriv\'ee diff\'eremment, par une approximation de l'it\'erateur de Jacobi, et formul\'ee diff\'eremment, il s'agit d'apr\`es nos investigations ult\'erieures du m\^eme it\'erateur que le n\^otre lorsque l'on utilise $\alpha$ = 1.

Nous pr\'esentons en section \ref{applimak} les cartes construites par la m\'ethode COBE it\'erative sur le bruit.

\subsection{Matrice de covariance du bruit dans la carte\label{matcovb}}

La matrice de covariance du bruit r\'esiduel dans la carte reconstruite, dans le cas de la m\'ethode COBE, est d\'ecrite par l'\'equation suivante:

\begin{equation}
{\bf B} = [{\bf A}^t{\bf N}^{-1}{\bf A}]^{-1}
\end{equation}

Elle correspond \`a la quantit\'e de bruit restant dans la carte finale, bruit qui poss\`ede \'eventuellement des corr\'elations r\'esiduelles.
Calculer cette matrice par la force brute n'est bien entendu pas possible non plus, pour les larges TOI des exp\'eriences CMB modernes.
Une fa\c{c}on de la calculer exactement consiste \`a appliquer le filtrage optimal d\'ecrit en section \ref{propcorr} pour effectuer l'op\'eration ${\bf N}^{-1}{\bf A}$ colonne par colonne de {\bf A}.
Le nombre de colonnes de la matrice r\'eponse de l'instrument {\bf A} est le nombre de pixels dans la carte \`a reconstruire.
Si ce nombre de pixels est suffisamment faible, il est donc possible d'effectuer cette op\'eration, d'appliquer {\bf A}$^t$ c'est-\`a-dire d'additionner dans les pixels chaque colonne filtr\'ee de {\bf A}.
Il faut donc r\'ealiser n filtrages et n cartographies, n \'etant le nombre de pixels, puis inverser la matrice n x n ainsi cr\'e\'ee.
Pour des cartes de quelques centaines ou quelques milliers de pixels, cette op\'eration est r\'ealisable quelle que soit la taille des colonnes, c'est-\`a-dire des TOI.
En revanche, des cartes \`a pleine r\'esolution d'Archeops ou Planck ne peuvent clairement pas donner lieu au calcul exact de la matrice de covariance du bruit.
Celle-ci peut alors \^etre estim\'ee par bouts \`a petite \'echelle et exactement \`a grande \'echelle.
En effet, pour des cartes r\'ealis\'ees avec de larges pixels, il est possible de calculer exactement cette matrice de covariance du bruit.
Il est aussi possible de la calculer pour des morceaux de cartes r\'ealis\'ees \`a pleine r\'esolution.

\subsection{M\'ethode de Wiener it\'erative\label{wieit}}

Bien qu'identique \`a la matrice Wiener 1 (\'equation \ref{wie1}), la matrice Wiener 2 (\'equation \ref{wie2}) est d'un point de vue algorithmique plus proche de la matrice COBE (\'equation \ref{cob}).
De ce fait, le sch\'ema it\'eratif que nous retenons pour mettre en {\oe}uvre une m\'ethode de Wiener it\'erative est proche de celui de la m\'ethode COBE it\'erative.
Voici ce nouveau sch\'ema it\'eratif:

\begin{equation}
\alpha \: {\bf \Hat x}_{n+1} = \alpha \: {\bf \Hat x}_n + {\bf u} - [{\bf S}^{-1} + {\bf A}^{t} {\bf N}^{-1} {\bf A}] {\bf \Hat x}_n
\end{equation}

o\`u l'on a:

\begin{equation}
{\bf u} = {\bf A}^t {\bf N}^{-1}[{\bf A}[{\bf A}^t{\bf A}]^{-1} {\bf A}^t {\bf y} - {\bf y}] + {\bf S}^{-1}[{\bf A}^t{\bf A}]^{-1} {\bf A}^t {\bf y}
\end{equation}

Comme nous l'avons montr\'e pour la m\'ethode COBE it\'erative, la limite de convergence est l'exacte solution du probl\`eme de construction de cartes de Wiener.
Ce sch\'ema it\'eratif demande de manipuler \`a la fois la matrice {\bf N}, covariance du bruit dans le train de donn\'ees, que nous traitons comme un filtre dans l'espace de Fourier comme d'habitude, mais aussi la matrice {\bf S} de covariance du signal dans la carte.
La manipuler en tant que matrice n'est pas possible pour la pix\'elisation \`a petite \'echelle que nous utilisons, et qui est n\'ecessaire pour les exp\'eriences CMB modernes.
Cependant, le sch\'ema HEALPIX RING est stationnaire par rapport \`a la sph\`ere c\'eleste, car il la pix\'elise en faisant un anneau autour de la sph\`ere du p\^ole nord au p\^ole sud, avec des pixels de surfaces \'egales.
Le signal du CMB \'etant stationnaire sur la sph\`ere, il l'est \'egalement dans la pix\'elisation HEALPIX RING.
Ainsi, filtrer un vecteur HEALPIX ({\it i.e.} une carte) dans l'espace de Fourier par un filtre cr\'e\'e \`a partir de l'autocorr\'elation du signal CMB est optimal.
Nous traitons donc le probl\`eme de la multiplication par {\bf S}$^{-1}$ comme un filtrage dans le sch\'ema HEALPIX RING.
Cependant, il ne faut pas qu'il y ait de grands trous dans la carte des donn\'ees, car ceci nuirait \`a la stationnarit\'e du ciel dans la sch\'ema HEALPIX.
Cela signifie que pour des cartes telles que celles des vols ballons simul\'es, ou les cartes d'Archeops, le filtrage dans le sch\'ema HEALPIX RING doit \^etre effectu\'e par bouts, s\'epar\'ement pour chaque partie du ciel observ\'ee qui est continue dans le sch\'ema de pix\'elisation.
Une autre m\'ethode consiste \`a d\'efinir une autre pix\'elisation, adapt\'ee \`a une observation donn\'ee.
Cependant, notre exp\'erience satellite de cartographie compl\`ete du ciel est particuli\`erement bien adapt\'ee pour tester cette m\'ethode de construction de cartes, puisque les trous dans la carte \`a N$_{side}$ = 256 sont tr\`es peu nombreux.
Nous pouvons donc filtrer la carte en entier.

Cette m\'ethode de Wiener it\'erative est une originalit\'e de notre travail, par rapport \`a la m\'ethode COBE it\'erative qui a \'et\'e d\'evelopp\'ee simultan\'ement par une autre \'equipe.
Nous pr\'esentons les r\'esultats de l'application de cette m\'ethode de Wiener en section \ref{appliwie}.

\subsection{G\'en\'eralisation des m\'ethodes it\'eratives\label{genmet}}

Le sch\'ema it\'eratif sur le bruit peut \^etre g\'en\'eralis\'e de sorte \`a \'equilibrer l'influence de la connaissance {\it a priori} de la corr\'elation du signal dans le processus de construction de cartes.
Ceci est r\'ealis\'e par la m\'ethode de construction de cartes d\'ecrite en section \ref{metadv} par l'\'equation \ref{sask}.
Il s'agit de la m\'ethode Saskatoon (Tegmark \etal 1997, Tegmark 1997) qui permet de choisir le rapport signal sur bruit dans la carte reconstruite.
Nous construisons la m\'ethode it\'erative correspondante en appliquant le m\^eme sch\'ema sur le bruit que pour la m\'ethode de Wiener it\'erative, \`a ceci pr\`es que le facteur $\eta$ (scalaire) s'applique sur la matrice {\bf S}$^{-1}$.
Nous pr\'esentons \'egalement les r\'esultats de cette m\'ethode en section \ref{appliwie}.

\section{ Application des m\'ethodes de construction de cartes aux donn\'ees simul\'ees\label{applimak}}

\subsection{\'Etude des m\'ethodes sur l'exp\'erience ballon polaire\label{etupol}}

\subsubsection{\'Etude de la m\'ethode COBE it\'erative}

Nous avons r\'ealis\'e 1000 it\'erations avec la m\'ethode COBE pour les donn\'ees simul\'ees du vol ballon polaire contenant les fluctuations du CMB, le dip\^ole, la Galaxie et le bruit instrumental (blanc et 1/f).
Nous pr\'esentons en figure \ref{plotconv} (haut) l'\'evolution du rms de la carte de diff\'erence entre le ciel reconstruit \`a l'it\'eration i (abscisse) et le vrai ciel simul\'e.
Il s'agit de l'\'evolution du bruit r\'esiduel global dans la carte.
Comme on le voit, la m\'ethode converge vers une valeur faible bien d\'efinie de ce rms de bruit.
Le niveau de bruit blanc moyenn\'e dans les pixels appara\^\i t comme un trait horizontal \`a 21.1 $\ukc$ en rms.
En figure \ref{plotconv} (bas), nous pr\'esentons le graphe d'\'evolution du pire pixel de la carte, c'est-\`a-dire que nous consid\'erons \`a chaque it\'eration le pixel le moins bien reconstruit de la carte, i.e. celui qui a le bruit r\'esiduel le plus grand en valeur absolue.

\begin{figure}
\begin{center}
\epsfxsize 10.0 true cm
\leavevmode
\caption[Convergence des it\'erations]{Haut: \'evolution du rms de la carte de diff\'erence entre le ciel reconstruit \`a l'it\'eration i (abscisse) et le vrai ciel simul\'e.
Le niveau de bruit blanc correspond au trait horizontal \`a 21.1 $\ukc$.
Bas: \'evolution du maximum de la carte de bruit r\'esiduel.}
\label{plotconv}
\end{center}
\end{figure}

\begin{table}
\begin{center}
\caption[Bruit r\'esiduel]{
Cette table pr\'esente les quantit\'es de bruit dans les cartes provenant de donn\'ees contenant les fluctuations du CMB et le bruit instrumental.
La premi\`ere ligne pr\'esente le rms du bruit dans les cartes coadditionn\'ees, la seconde le rms du bruit blanc (non corr\'el\'e), la troisi\`eme le rms du bruit r\'esiduel apr\`es convergence de la m\'ethode COBE it\'erative, la quatri\`eme celui de la m\'ethode de Wiener it\'erative.
La derni\`re ligne montre le rms du bruit r\'esiduel pour la m\'ethode de Wiener directe.
Les unit\'es sont des $\ukc$.
}
\begin{tabular}{lllll}

\hline
 & Polaire & Polaire 24 h &
 \'Equatorial & Satellite \\
 & 24 h & vitesse non cste & 12 h & \\
\hline
\hline

{\small Bruit total rms} & 349.7 & 352.5 & 498.6 & 378.4\\
{\small dans la carte} & & & & \\
\hline

{\small Bruit blanc rms} & 21.06 & 21.37 & 40.84 & 38.14\\
{\small dans la carte} & & & & \\
\hline

{\small R\'esidu rms} & 21.61 & - & 42.13 & 38.93\\
{\small COBE it\'er.} & & & & \\
\hline

{\small R\'esidu rms} & - & - & - & 38.93\\
{\small Wiener it\'er.} & & & & \\
\hline

{\small R\'esidu rms} & 22.82 & 33.82 & 32.34 & 41.40\\
{\small Wiener direct} & & & & \\
\hline

\end{tabular}
\end{center}
\label{rmsbru}
\end{table}

La convergence est atteinte apr\`es environ 50 it\'erations.
Le bruit r\'esiduel atteint est de 21.6 $\ukc$ rms.
Nous pr\'esentons les r\'esultats en termes de rms de bruit r\'esiduel en table \ref{rmsbru}.
Nous v\'erifions que nous avons bien atteint la convergence en observant l'\'evolution du rms global du bruit r\'esiduel, mais \'egalement l'\'evolution individuelle par pixel, l'aspect de la carte et le spectre de puissance en C$_l$.

Nous devons comparer ce r\'esultat sur le bruit r\'esiduel \`a la quantit\'e de bruit blanc moyenn\'e dans la carte.
On peut calculer th\'eoriquement le rms du bruit blanc dans la carte par l'expression:

\begin{equation}
n_{rms}^2 . \Sigma_i w_i^{-1}/N_{pix}
\end{equation}

o\`u $n_{rms}$ est le niveau rms de bruit blanc dans le train de donn\'ees, $w_i$ le poids dans le pixel i et $N_{pix}$ le nombre total de pixels observ\'es.
La somme se fait bien s\^ur sur les pixels observ\'es.
Afin de d\'eduire le rms du bruit blanc dans la carte, on peut \'egalement simuler un train de donn\'ees de bruit blanc et faire une carte.
Le niveau rms de bruit blanc dans la carte du vol polaire est de 21.1 $\ukc$, ce qui est tr\`es proche de la quantit\'e de bruit r\'esiduel.
Ceci montre que la reconstruction effectu\'ee est tr\`es performante, puisque la quasi-totalit\'e du bruit corr\'el\'e pr\'esent dans le train de donn\'ees est \'elimin\'ee dans la carte optimale.
La carte reconstruite ne fait pas appara\^\i tre de diff\'erences \`a l'{\oe}il par rapport \`a la carte du ciel simul\'ee (fig. \ref{simusallsky}).
Nous pr\'esentons en figure \ref{bruitres} la carte de bruit r\'esiduel.
Les structures de bruit visibles dans cette carte sont produites par la granularit\'e du bruit blanc, dont l'amplitude d\'epend du poids par pixel.
Aucune strie, signature d'\'eventuels bruits corr\'el\'es, n'est visible dans cette carte.
Ceci confirme que la m\'ethode it\'erative \'elimine presque compl\`etement le bruit corr\'el\'e.
De plus, le graphe en bas de la figure \ref{plotconv} montre qu'il n'y a pas de pixels exclus de la convergence g\'en\'erale.

\begin{figure}
\begin{center}
\epsfxsize 10.0 true cm
\leavevmode
\caption[Carte de bruit r\'esiduel du vol polaire]{Carte de bruit r\'esiduel obtenue par la m\'ethode it\'erative COBE appliqu\'ee sur les donn\'ees simul\'ees du vol polaire.
La quasi-totalit\'e de ce r\'esidu est clairement le bruit blanc moyenn\'e.
Projection Mollweide galactique centr\'ee sur le centre galactique.}
\label{bruitres}
\end{center}
\end{figure}

\begin{figure}
\begin{center}
\epsfxsize 10.0 true cm
\leavevmode
\caption[Spectres en C$_l$ du bruit]{Haut: spectre de puissance en C$_l$ de la carte de bruit r\'esiduel du vol polaire construite par moyennage par pixel.
Bas: spectre de puissance en C$_l$ de la carte de bruit r\'esiduel apr\`es convergence, et de la carte de bruit blanc (courbe la plus basse, en gris).
Chaque spectre est corrig\'e de l'effet de la fraction de ciel couverte.
Les \'echelles en ordonn\'ee des deux graphes montrent tr\`es bien \`a quel point la m\'ethode it\'erative est sup\'erieure au moyennage par pixel.
Le spectre de bruit est tr\`es proche de celui d'un bruit blanc (c'est-\`a-dire plat en C$_l$) au-dessus de l $\approx$ 50, mais montre une corr\'elation r\'esiduelle (extr\^emement faible) aux plus grandes \'echelles.
}
\label{clsim}
\end{center}
\end{figure}

En figure \ref{clsim}, nous pr\'esentons le spectre de puissance en C$_l$ de la carte construite par moyennage par pixel, de la carte de bruit r\'esiduel apr\`es convergence de la m\'ethode it\'erative, et de la carte de bruit blanc.
Tous les spectres sont corrig\'es de l'effet de la fraction de ciel couverte.
Les \'echelles en ordonn\'ee des deux graphes montrent tr\`es bien \`a quel point la m\'ethode it\'erative est sup\'erieure au moyennage par pixel.
Le spectre de bruit est tr\`es proche de celui d'un bruit blanc (c'est-\`a-dire plat en C$_l$) au-dessus de l $\approx$ 50, mais montre une corr\'elation r\'esiduelle faible aux plus grandes \'echelles.
Puisque la convergence a \'et\'e atteinte, la carte obtenue est la solution du probl\`eme de construction de cartes, pourtant nous observons qu'une petite quantit\'e de bruit corr\'el\'e se tapit toujours aux grandes \'echelles.
Ceci s'explique par le fait que ces strat\'egies d'observation ne contraignent pas parfaitement les grandes \'echelles, car il y a relativement peu de balayages crois\'es.
Ces grandes \'echelles \'etant au d\'epart les plus bruit\'ees, par d\'efinition du bruit 1/f, il appara\^\i t naturel qu'une partie du bruit corr\'el\'e soit toujours pr\'esent apr\`es convergence.
Une m\'ethode telle que MASTER \cite{hivon02} peut \^etre capable de prendre en compte cet exc\`es de bruit \`a grande \'echelle dans le calcul du spectre de puissance angulaire.
Cependant, ce bruit r\'esiduel corr\'el\'e est extr\^emement faible, m\^eme aux plus grandes \'echelles, et ne pourrait avoir \`a \^etre pris en compte que pour des mesures extr\^emement pr\'ecises des bas l, avec une pr\'ecision de mieux que 0.05 $\uk^2$ sur les C$_l$ aux grandes \'echelles (pr\'ecision utile uniquement si l'on s'int\'eresse \`a mesurer extr\^emement pr\'ecis\'ement la variance cosmique !)

NB: la variance cosmique, signature de la r\'ealisation d'Univers dans laquelle nous vivons, se traduit par des fluctuations du spectre de puissance du CMB autour de la valeur moyenne th\'eorique.

Nous avons \'egalement test\'e la m\'ethode COBE it\'erative sur des trains de donn\'ees contenant uniquement les fluctuations du CMB et le bruit instrumental.
Pour obtenir une TOI de vraies donn\'ees sans la Galaxie, il est possible, par exemple, de d\'ecorr\'eler d'avec un bolom\`etre \`a plus courte longueur d'onde.
En effet, beaucoup d'exp\'eriences CMB ont des canaux \`a haute fr\'equence (par exemple, les bandes de Planck et Archeops \`a 353 et 545 GHz) dans lesquels les fluctuations du CMB sont n\'egligeables.
Ces canaux sont utilis\'es comme traceurs de l'\'emission de la poussi\`ere galactique.
La fa\c{c}on de d\'ecorr\'eler les canaux CMB de l'\'emission de la poussi\`ere est habituellement de le faire apr\`es le processus de construction de cartes, dans le processus de s\'eparation de composantes.
Cependant, il est \'egalement possible de d\'ecorr\'eler l'\'emission galactique dans le train de donn\'ees, bien que cela puisse causer de la propagation de bruit dans la TOI du canal CMB.
La fa\c{c}on correcte de le faire nous semble d'obtenir tout d'abord une carte galactique aussi propre que possible (d'apr\`es des bolom\`etres \`a haute fr\'equence, ou m\^eme d'apr\`es des cartes galactiques provenant d'autres observations telles qu'IRAS, etc), puis de r\'e-\'echantillonner cette carte avec la strat\'egie d'observation correspondant au train de donn\'ees que l'on veut d\'ecorr\'eler.
Cependant, nous devons aussi mentionner que beaucoup d'exp\'eriences CMB n'observent pas du tout la Galaxie, et donc obtiennent directement des trains de donn\'ees contenant presque uniquement le CMB.

Le bruit r\'esiduel obtenu dans la carte reconstruite des fluctuations du CMB uniquement est le m\^eme que celui que nous obtenons pour les donn\'ees compl\`etes (avec la Galaxie et le dip\^ole).
Ceci montre que la m\'ethode COBE it\'erative est aussi efficace avec des signaux tr\`es diff\'erents les uns des autres, comme nous pouvions l'attendre d'apr\`es la th\'eorie, puisque les propri\'et\'es de corr\'elation du signal n'interviennent pas dans la m\'ethode COBE (\'equation \ref{cob}).

Cet algorithme it\'eratif est tr\`es efficace mais un peu lent: le temps CPU n\'ecessaire sur un processeur Pentium III \`a 500 MHz est d'environ une heure pour 25 it\'erations, pour les huit millions de points de donn\'ees trait\'es.
Cependant, nous verrons un peu plus loin que le nombre d'it\'erations n\'ecessaire pour atteindre la convergence peut \^etre consid\'erablement r\'eduit si l'on utilise une premi\`ere estimation non nulle de la carte du ciel.
Il est aussi possible d'acc\'el\'erer ce genre de m\'ethodes it\'eratives par des algorithmes multi-\'echelles \cite{dore01}, qui utilisent le fait que les cartes de r\'esolution plus grossi\`ere ont besoin de moins d'it\'erations pour converger.

\subsubsection{\'Etude de la m\'ethode de Wiener directe}

Nous avons appliqu\'e la m\'ethode directe Wiener 1 aux m\^emes trains de donn\'ees.
Cette m\'ethode est particuli\`erement bien adapt\'ee \`a des signaux uniquement CMB, puisque la stationnarit\'e est alors parfaitement v\'erifi\'ee, et plus fondamentalement car la m\'ethode de Wiener est optimale quand \`a la fois le signal et le bruit sont des champs al\'eatoires gaussiens.
La m\'ethode directe de Wiener fonctionne de fa\c{c}on tr\`es satisfaisante sur les donn\'ees contenant uniquement les fluctuations du CMB (et bien s\^ur le bruit): le rms de la carte de diff\'erence entre la carte reconstruite ainsi et le vrai ciel simul\'e est de 22.8 $\ukc$, ce qui est proche du niveau de bruit blanc.
Nous pr\'esentons en figure \ref{wiedipol} la carte de bruit r\'esiduel obtenue en utilisant cette m\'ethode de Wiener directe.
Ce r\'esultat est \`a comparer avec l'application d'un filtre de blanchissage en section \ref{blanch} (figure \ref{sauvage}).
La diff\'erence montre tr\`es clairement que le choix du filtre est particuli\`erement crucial pour ces m\'ethodes directes.
Le filtre optimal de Wiener est clairement le mieux adapt\'e.

\begin{figure}
\begin{center}
\epsfxsize 10.0 true cm
\leavevmode
\caption[Carte de bruit r\'esiduel Wiener directe]{Carte de bruit r\'esiduel obtenue par la m\'ethode de Wiener directe appliqu\'ee sur les donn\'ees CMB simul\'ees du vol polaire.
Il y a un peu de striage r\'esiduel, contrairement \`a ce qui est obtenu apr\`es it\'erations.
Projection Mollweide galactique centr\'ee sur le centre galactique.}
\label{wiedipol}
\end{center}
\end{figure}

N\'eanmoins, un peu de striage r\'esiduel est observable sur la carte de bruit en figure \ref{wiedipol}.
Ceci montre que le filtrage de Wiener traite les donn\'ees d'une mani\`ere tr\`es diff\'erente de la m\'ethode it\'erative.
Celle-ci enl\`eve le bruit corr\'el\'e graduellement, pour obtenir une carte finale parfaitement destri\'ee, alors que le filtre direct de Wiener traite les donn\'ees de sorte \`a construire proprement la carte du ciel.
Dans cette m\'ethode, c'est le signal qui est restaur\'e, non le bruit corr\'el\'e qui est enlev\'e.
Ceci peut expliquer pourquoi du bruit corr\'el\'e se trouve toujours dans la carte finale, alors que la quantit\'e totale de bruit dans la carte construite est tr\`es faible.
En fait, l' important n'est pas seulement la quantit\'e de bruit restant dans la carte finale, mais surtout sa ``blancheur'', c'est-\`a-dire son caract\`ere non autocorr\'el\'e.
Si le bruit r\'esiduel est corr\'el\'e de fa\c{c}on significative, alors la matrice du bruit dans les pixels de la carte reconstruite n'est pas diagonale, ce qui peut poser probl\`eme pour l'estimation du spectre en C$_l$.
Le spectre en C$_l$ de la carte reconstruite par la m\'ethode de Wiener directe montre que de la puissance est supprim\'ee \`a toutes les \'echelles, alors que ce n'est pas le cas pour la m\'ethode COBE it\'erative, pour laquelle le spectre de puissance ne montre pas de diff\'erence significative par rapport au spectre du vrai ciel (voir figure \ref{clcar}).

\begin{figure}
\begin{center}
\epsfxsize 10.0 true cm
\leavevmode
\caption[Spectres en C$_l$ des cartes]{Spectres de puissance en C$_l$ de la carte du CMB (noir), de celle reconstruite par la m\'ethode COBE (courbe grise quasiment confondue avec la premi\`ere) et de celle construite par la m\'ethode de Wiener directe (grise, en bas).
}
\label{clcar}
\end{center}
\end{figure}

Afin d'acc\'el\'erer la convergence des m\'ethodes it\'eratives, nous pouvons utiliser une premi\`ere estimation non nulle de la carte, la meilleure \'etant la carte produite par la m\'ethode directe de Wiener.
Il est important de bien comprendre que quelle que soit la premi\`ere estimation, ce qui d\'efinit la solution exacte est uniquement l'\'equation de construction de cartes choisie.
Choisir une carte produite par une m\'ethode de Wiener comme premier estimateur d'une m\'ethode COBE acc\'el\`ere la convergence mais n'influence en aucune mani\`ere la carte it\'er\'ee finale.

Nous souhaitons enfin tester comment le manque de stationnarit\'e du signal peut nuire au filtrage direct de Wiener.
Nous testons donc la m\'ethode sur un train de donn\'ees contenant le CMB et la Galaxie, plus le bruit instrumental.
Nous la testons \'egalement sur une TOI de CMB avec bruit, souffrant d'une vitesse de balayage extr\^emement non constante, le faisceau faisant des \'ecarts sinuso\"\i daux de 30 degr\'es d'amplitude 17 fois par minute.
La carte Wiener directe du train de donn\'ees incluant la Galaxie est assez mal reconstruite.
En effet, des stries sont toujours visibles et le bruit r\'esiduel atteint 257 $\ukc$ en rms.
Ceci n'est pas surprenant dans la mesure o\`u le signal galactique tr\`es intense nuit tr\`es clairement \`a la stationnarit\'e du signal.
Le filtrage direct de Wiener n'est donc pas adapt\'e \`a des trains de donn\'ees contenant du signal galactique de fa\c{c}on importante.
C'est pourquoi il devrait \^etre appliqu\'e \`a des trains de donn\'ees dont les avant-plans ont \'et\'e enlev\'es auparavant, ou \`a des observations r\'ealis\'ees loin du plan galactique.
En revanche, la carte Wiener directe est bien reconstruite pour les donn\'ees CMB ayant une vitesse de rotation extr\^emement non stationnaire.
Le bruit r\'esiduel est de 33.8 $\ukc$ en rms, ce qui n'est pas mauvais compar\'e aux 21.4 $\ukc$ rms de bruit blanc dans la carte, mais qui est bien entendu moins bon que les 22.8 $\ukc$ rms de bruit r\'esiduel obtenu avec le m\^eme traitement sur des donn\'ees b\'en\'eficiant d'une vitesse de rotation constante.
Cependant, le bon r\'esultat que nous obtenons avec ces donn\'ees \'echantillonn\'ees d'une mani\`ere catastrophique montre la robustesse de la m\'ethode de Wiener directe.

\subsection{\'Etude compar\'ee des strat\'egies d'observation}

Nous avons appliqu\'e la m\'ethode COBE it\'erative avec la carte Wiener directe comme premier estimateur, aux donn\'ees du vol ballon \'equatorial et du satellite, avec les fluctuations du CMB et le bruit instrumental.
Encore une fois, cette m\'ethode peut \'egalement \^etre appliqu\'ee aux signaux incluant la Galaxie.
La seule diff\'erence consiste dans le fait que puisque l'estimateur direct de Wiener n'est pas aussi bon que pour les donn\'ees CMB + bruit, le nombre d'it\'erations n\'ecessaires pour atteindre la convergence et donc une carte parfaitement destri\'ee est plus important.
Les cartes produites par cette m\'ethode it\'erative sur les donn\'ees du vol \'equatorial et du satellite sont \'egalement tr\`es satisfaisantes.
Pour les donn\'ees simul\'ees du vol \'equatorial, la carte de bruit r\'esiduel a un rms de 42.1 $\ukc$, avec 40.8 $\ukc$ rms de bruit blanc.
Pour les donn\'ees du satellite, la carte de bruit r\'esiduel a un rms de 38.9 $\ukc$, avec 38.1 $\ukc$ rms de bruit blanc.
Les r\'esidus de la m\'ethode COBE it\'erative sont proches de la quantit\'e de bruit blanc, et les rapports r\'esidu sur bruit blanc sont quasiment constants.
Bien s\^ur, le bruit r\'esiduel est plus important pour les exp\'eriences de plus faible redondance, mais l'efficacit\'e de la m\'ethode it\'erative n'est pas diminu\'ee, ni la blancheur du bruit final dans les cartes.
En particulier, ceci montre que la pr\'esence de nombreux balayages crois\'es n'est pas cruciale pour le processus de construction de cartes utilis\'e. 

Il est int\'eressant de noter que dans le cas du vol \'equatorial, le filtre direct de Wiener produit un bruit r\'esiduel plus faible que la m\'ethode it\'erative, et m\^eme plus faible que le bruit blanc moyenn\'e dans les pixels.
Cependant, la carte de bruit r\'esiduel que nous pr\'esentons en figure \ref{wiedieq} fait appara\^\i tre du striage r\'esiduel.
Ceci peut s'expliquer par les propri\'et\'es du filtrage de Wiener, qui cherche \`a reconstruire la carte en minimisant l'erreur globale de reconstruction, mais sans que les propri\'et\'es de corr\'elation du bruit soient prises en compte de mani\`ere aussi pr\'ecise que pour la m\'ethode it\'erative.
Il reste donc des stries, m\^eme si la solution trouv\'ee est optimale pour la r\'eduction du bruit global.

\begin{figure}
\begin{center}
\epsfxsize 10.0 true cm
\leavevmode
\caption[Carte de bruit r\'esiduel Wiener directe]{Carte de bruit r\'esiduel obtenue par la m\'ethode de Wiener directe appliqu\'ee sur les donn\'ees CMB simul\'ees du vol \'equatorial.
Il y a un peu de striage r\'esiduel, contrairement \`a ce qui est obtenu apr\`es it\'erations, et m\^eme si la quantit\'e de bruit r\'esiduel total est extr\^emement faible.
Projection Mollweide galactique centr\'ee sur le centre galactique.}
\label{wiedieq}
\end{center}
\end{figure}

\subsection{\'Etude des m\'ethodes it\'eratives Wiener et g\'en\'eralis\'ee\label{appliwie}}

Comme expliqu\'e en section \ref{wieit}, notre cartographie satellite de tout le ciel est particuli\`erement int\'eressante pour tester les m\'ethodes it\'eratives de Wiener ou associ\'ees.
La m\'ethode fonctionne tr\`es bien, m\^eme sans premier estimateur du ciel, mais est nettement plus rapide en utilisant une carte Wiener 1 directe comme premier estimateur.
Nous pr\'esentons en figure \ref{wieitres} la carte de bruit r\'esiduel produite apr\`es avoir atteint la convergence.
Cette carte est tr\`es proche de celle de la m\'ethode COBE it\'erative.
En effet, il n'y a pas de diff\'erence visible \`a l'{\oe}il sur les cartes, et les rms globaux du bruit r\'esiduel sont tr\`es proches: 38.933 $\ukc$ pour la m\'ethode COBE it\'erative, 38.931 $\ukc$ pour la m\'ethode de Wiener it\'erative.
Encore une fois, nous observons que le bruit corr\'el\'e est presque totalement \'elimin\'e par les it\'erations.
La m\'ethode de Wiener est connue pour minimiser l'erreur de reconstruction: nous le v\'erifions, mais la diff\'erence par rapport \`a la m\'ethode COBE est tr\`es faible.
De plus, les spectres en C$_l$ du vrai ciel et des deux cartes reconstruites par les m\'ethodes it\'eratives COBE et Wiener ne montrent pas de diff\'erence significative.

\begin{figure}
\begin{center}
\epsfxsize 10.0 true cm
\leavevmode
\caption[Carte de bruit r\'esiduel Wiener it\'eratif]{Carte de bruit r\'esiduel obtenue par la m\'ethode de Wiener it\'erative appliqu\'ee sur les donn\'ees CMB simul\'ees du satellite.
La granularit\'e du bruit blanc est clairement visible et domine de tr\`es loin dans le r\'esidu.
Aucune strie r\'esiduelle n'est visible sur cette carte de bruit r\'esiduel.
Projection Mollweide galactique centr\'ee sur le centre galactique.}
\label{wieitres}
\end{center}
\end{figure}

Nous appliquons au m\^eme train de donn\'ees la m\'ethode de Wiener it\'erative g\'en\'eralis\'ee (Saskatoon), que nous avons d\'ecrite en sections \ref{metadv} et \ref{genmet}.
Pour $\eta$ = 0.5 par exemple, nous obtenons une carte de bruit r\'esiduel ayant un rms de 38.932 $\ukc$, ce \`a quoi nous pouvions nous attendre au vu des valeurs de ce r\'esidu pour les m\'ethodes COBE (38.933 $\ukc$) et Wiener (38.931 $\ukc$) it\'eratives.
La m\'ethode devient instable pour $\eta > 1$.

Nous voyons donc qu'une diff\'erence nettement plus grande existe, en termes de r\'esultats sur les cartes, entre les m\'ethodes it\'eratives et les m\'ethodes directes, qu'entre les m\'ethodes COBE et Wiener it\'eratives.
En effet, les m\'ethodes it\'eratives sont tr\`es pr\'ecises, \`a la fois par rapport \`a l'aspect de la carte et au spectre de puissance, et ne semblent pas autoriser de diff\'erences significatives entre les reconstructions COBE et Wiener, avec ce rapport signal sur bruit blanc plut\^ot \'elev\'e.
Pourtant, nous pouvons \^etre confiants quant \`a la r\'ealit\'e de la m\'ethode de Wiener it\'erative (compar\'ee \`a la m\'ethode COBE), car le facteur $\eta$ est tr\`es sensible dans le sch\'ema it\'eratif (quand $\eta > 1$, la m\'ethode diverge).

\subsection{Construction de cartes en pr\'esence de syst\'ematiques\label{mapmaksyst}}

Il est important de se rendre compte que ces donn\'ees simul\'ees ont \'et\'e r\'ealis\'ees avec des bruits stationnaires, ce qui signifie que le train de donn\'ees ne contient pas de syst\'ematiques.
N\'eanmoins, nous souhaitons savoir quel est l'efficacit\'e de la m\'ethode COBE it\'erative quand elle est confront\'ee \`a un train de donn\'ees contenant du bruit non stationnaire.
Pour cela, nous simulons un train de bruit en 1/f$^n$ dont les caract\'eristiques spectrales: l'indice de puissance n et la fr\'equence de coupure f$_c$, changent au cours du temps de mani\`ere al\'eatoire.
Un r\'esum\'e des tests r\'ealis\'es appara\^\i t en table \ref{nstattab}.

\begin{table}
\begin{center}
\caption[Bruits non stationnaires]{Caract\'eristiques des trains de donn\'ees simul\'ees avec des bruits non stationnaires et les fluctuations du CMB, sur la strat\'egie d'observation du vol ballon \'equatorial.
l$_{inf}$ est le niveau de bruit \`a fr\'equence infinie ($\ukc$), f$_c$ la fr\'equence de coupure variable (Hz) et n l'indice spectral variable (voir l'\'equation \ref{unsurf}).
Les variations al\'eatoires sont uniformes dans les intervalles indiqu\'es dans la table.
N$_p$ est le nombre de points de donn\'ees cons\'ecutifs pour lesquels le bruit est stationnaire.
Le r\'esultat de la m\'ethode COBE it\'erative est indiqu\'e en derni\`ere colonne en termes d'erreur quadratique moyenne de reconstruction sur la carte ($\ukc$ rms).
}
\begin{tabular}{llllll}
\hline
 & l$_{inf}$ & f$_c$ & n & N$_p$ & rms (m\'ethode COBE it.) \\
\hline
Bruit A & 100 & 0.1-0.2 & 1-1.2 & 16384 & 42.4 \\
\hline
Bruit B & 100 & 0.1-0.2 & 1-2 & 4096 & 41.9 \\
\hline
Bruit C & 500 & 0.1-0.2 & 1-2 & 4096 & 209.5 \\
\hline
\end{tabular}
\label{nstattab}
\end{center}
\end{table}

Cette table montre que la m\'ethode it\'erative n'est pas perturb\'ee par la non-stationnarit\'e du bruit dans ces trains de donn\'ees.
En effet, les r\'esidus de bruit apr\`es convergence correspondent \`a celui du bruit stationnaire (42.13 $\ukc$ rms, voir table \ref{rmsbru}) pour les trains de donn\'ees ayant le m\^eme niveau de bruit blanc de 100 $\ukc$ (A et B).
Ils sont l\'eg\`erement sup\'erieurs au niveau de bruit blanc dans la carte (40.84 $\ukc$ rms).
En ce qui concerne le train de donn\'ees C, on remarque que son niveau de bruit plus \'elev\'e n'influe pas sur les performances de la m\'ethode puisque le r\'esidu est le m\^eme que celui du train de donn\'ees B, au facteur 5 pr\`es qui caract\'erise la diff\'erence des deux niveaux de bruit \`a fr\'equence infinie.
Les bonnes performances de la m\'ethode sur les bruits B et C, pourtant souffrant clairement de plus de non stationnarit\'e que le bruit A, montrent la robustesse de cette m\'ethode it\'erative.

Pour de tels bruits, non stationnaires sur une longue p\'eriode mais stationnaires par morceaux, il est \'egalement possible de filtrer bout par bout \`a chaque \'etape de la reconstruction n\'ecessitant un filtrage du train de donn\'ees.
On voit donc que la pr\'esence de non-stationnarit\'e raisonnable dans les trains de donn\'ees n'est pas un probl\`eme pour la m\'ethode it\'erative.

Cependant, s'il se trouve dans les donn\'ees des syst\'ematiques p\'eriodiques ayant la p\'eriode de rotation du faisceau par exemple, nous observons en figure \ref{camelres} que les m\'ethodes it\'eratives ne fonctionnent pas.
En effet, une syst\'ematique de p\'eriode \'egale \`a celle de la rotation se projette sur le ciel \`a peu pr\`es comme un signal, sauf que la lente rotation de la Terre (dans le cas d'un ballon) fait d\'ecaler ce ``signal''.
Les it\'erations ne permettent pas d'\'eliminer ce genre de syst\'ematiques clairement non stationnaires, comme le montre la figure \ref{camelres}.
La solution pour ce genre de syst\'ematiques est donc de les \'eliminer dans le train de donn\'ees (voir section \ref{traitt} pour les donn\'ees Archeops).

\begin{figure}
\begin{center}
\epsfxsize 10.0 true cm
\leavevmode
\caption[Carte du ciel reconstruite avec syst\'ematique]{Carte du ciel reconstruite par la m\'ethode COBE it\'erative sur des donn\'ees du vol ballon \'equatorial contenant les anisotropies du CMB, du bruit blanc, du bruit 1/f stationnaire et une syst\'ematique p\'eriodique avec la rotation de la nacelle.
Comme on le voit, les it\'erations ne permettent pas d'\'eliminer ce genre de syst\'ematiques clairement non stationnaires.
Les fluctuations du CMB sont totalement noy\'ees sous l'effet dip\^olaire de la syst\'ematique.
Projection Mollweide galactique centr\'ee sur le centre galactique.}
\label{camelres}
\end{center}
\end{figure}

\subsection{Conclusion}

Dans les sections pr\'ec\'edentes, nous avons pr\'esent\'e les d\'eveloppements et les tests de m\'ethodes optimales de construction de cartes pour les trains de donn\'ees de grande taille d'exp\'eriences CMB.
La mani\`ere directe de construire une carte du CMB \`a partir d'un train de donn\'ees bruit\'e est le filtrage de Wiener, qui est robuste par rapport aux effets de non-stationnarit\'e d\^us \`a la strat\'egie d'observation, mais qui n\'ecessite de traiter un signal stationnaire.
Ceci demande d'enlever autant que possible le signal galactique avant d'appliquer cette m\'ethode, ou de l'appliquer sur des observations hors du plan galactique.
Les m\'ethodes it\'eratives COBE, Wiener, et g\'en\'eralis\'ee, sont tr\`es efficaces pour \'eliminer le bruit corr\'el\'e, et permettent d'obtenir une tr\`es bonne pr\'ecision sur la carte reconstruite.
Le bruit r\'esiduel est en effet presque uniquement le bruit blanc moyenn\'e.
Bien que les r\'esultats des m\'ethodes it\'eratives COBE, Wiener et Saskatoon soient tr\`es proches, nous v\'erifions que la reconstruction de Wiener r\'eduit le plus l'erreur globale de reconstruction.
La diff\'erence devrait \^etre plus importante si les pixels \'etaient plus bruit\'es, et plus in\'egalement bruit\'es.

Les diff\'erentes strat\'egies d'observation que nous avons simul\'ees permettent une construction de cartes aussi efficace, m\^eme si les balayages crois\'es sont peu nombreux.
De plus, nous avons pu v\'erifier que ces m\'ethodes fonctionnent \'egalement avec diverses quantit\'es de bruit.
En g\'en\'eral, plus le bruit est important, plus la convergence est rapide.
Cependant, en pr\'esence de syst\'ematiques p\'eriodiques, les m\'ethodes it\'eratives ne fonctionnent pas correctement.
Malgr\'e cette restriction, ces m\'ethodes ont un domaine d'application tr\`es large et une grande facilit\'e d'utilisation.
Ce genre de m\'ethodes vecteurs-seuls nous para\^\i t in\'evitable pour construire des cartes optimales des exp\'eriences CMB d'aujourd'hui et du futur.
La r\'eduction de l'information dans les donn\'ees CMB est un travail lourd, depuis les giga-octets de donn\'ees ordonn\'ees dans le temps jusqu'\`a, essentiellement, 12 nombres cosmologiques avec leurs barres d'erreur.
Puisque les moyens informatiques sont limit\'es et insuffisants pour des approches utilisant la force brute, et que ce sera toujours le cas pour la r\'eduction des donn\'ees Planck, c'est un d\'efi int\'eressant de traiter chaque \'etape de l'analyse sans perdre d'information.
Utiliser les propri\'et\'es de stationnarit\'e d'un signal ou du bruit dans un domaine donn\'e (sph\`ere c\'eleste, carte, train de donn\'ees, etc) pour transformer un probl\`eme d'inversion matricielle en une solution vecteurs-seuls, pourrait \'egalement \^etre d\'evelopp\'e pour des \'etapes ult\'erieures de l'analyse des donn\'ees CMB, telles que la s\'eparation de composantes.

Pour conclure, mentionnons qu'une grande partie de cette \'etude a fait l'objet d'une publication \cite{dupac02} et d'actes de conf\'erences (Dupac \& Giard 2001 {\it coll. b}, Dupac 2002 {\it coll. d}).

\section{ Traitement 1-D et construction de cartes d'Archeops\label{traitarch}}

Nous avons particip\'e au traitement et \`a l'analyse des donn\'ees Archeops (voir section \ref{archeops} et Beno\^\i t \etal 2002) du vol de Trapani et de celui de Kiruna en 2001 (KS1, voir table \ref{volsarch}).
Pour cause de r\'edaction de th\`ese, nous n'avons pas encore pu participer activement au traitement des donn\'ees des deux vols de 2002.

Notre travail, en particulier sur les m\'ethodes de construction de cartes, s'est fait en interaction permanente avec les autres membres de la collaboration Archeops, \`a travers des r\'eunions mensuelles o\`u nous avons pu exposer nos avanc\'ees.
Ce travail de construction de cartes s'est ins\'er\'e dans la cha\^\i ne d'analyse des donn\'ees Archeops.
Nous avons ainsi pu b\'en\'eficier du travail fait en amont sur le traitement du signal temporel, notamment par P. Filliatre \cite{filliatre02}, pour produire des cartes, qui ont ensuite pu \^etre utilis\'ees dans la collaboration pour leur analyse.
D'autres personnes dans la collaboration ont travaill\'e sur l'\'etape de la construction de cartes (O. Dor\'e \etal, D. Yvon, P. Filliatre \etal, J.-C. Hamilton...), ce qui a permis de confronter les m\'ethodes et d'am\'eliorer les r\'esultats.

Nous avons \'egalement apport\'e une contribution au traitement des donn\'ees temporelles et \`a l'analyse des cartes d'Archeops.

 \subsection{Traitement du signal temporel\label{traitt}}

Le traitement du signal \`a une dimension repr\'esente la premi\`ere \'etape du traitement des donn\'ees.
Dans le cas d'Archeops, et en particulier pour les deux premiers vols, elle est v\'eritablement une \'etape cruciale, au vu de la quantit\'e de bruits dans les donn\'ees et surtout de leurs corr\'elations.
Les syst\'ematiques (bruits non stationnaires) \'etant difficiles \`a traiter par les m\'ethodes de construction de cartes optimales, il est n\'ecessaire d'en enlever le plus possible directement dans le train de donn\'ees.
Parmi les m\'ethodes utilisables pour \'eliminer des syst\'ematiques dans le train de donn\'ees, l'on peut citer les m\'ethodes de d\'ecorr\'elation d'avec les thermom\`etres.
Ces thermom\`etres sont plac\'es sur le plan focal \`a proximit\'e des bolom\`etres, et ont donc beaucoup de syst\'ematiques et de bruits communs.
D\'ecorr\'eler le signal bolom\'etrique du signal (bruit) d'un thermom\`etre consiste \`a appliquer:

\begin{equation}
y_{corr} = y - k \: y_{ther}
\end{equation}

o\`u y est le train de donn\'ees plus ou moins brutes, $y_{ther}$ le train de donn\'ees thermom\'etriques, $y_{corr}$ les donn\'ees bolom\'etriques corrig\'ees, et k le coefficient de corr\'elation, scalaire.
Nous n'avons pass\'e que peu de temps sur ce traitement, mais nous pouvons dire que pour des bruits \`a basse fr\'equence (d\'erives), la pr\'esence d'un bruit commun \`a deux trains de donn\'ees est souvent bien visible, et que le coefficient k peut \^etre trouv\'e simplement en comparant les deux courbes.

Signalons qu'un certain nombre de personnes dans la collaboration Archeops se sont pench\'ees s\'erieusement sur le sujet du traitement du signal \`a une dimension.
Notamment, la pr\'esence d'une syst\'ematique p\'eriodique dans les donn\'ees des vols Trapani et KS1, de fr\'equence \'egale \`a la fr\'equence de rotation de la nacelle, a rendu particuli\`erement d\'elicate cette \'etape.
Ce ``chameau'', ainsi nomm\'e \`a cause de sa forme \`a deux bosses dans les donn\'ees du vol \`a Trapani, est une syst\'ematique due \`a des causes longtemps rest\'ees myst\'erieuses, probablement li\'ees au moteur du pivot.
Comme expliqu\'e en section \ref{mapmaksyst}, et comme nous le verrons en section \ref{mapmakark}, ce genre de choses est particuli\`erement d\'esagr\'eable pour les m\'ethodes de construction de cartes it\'eratives.
Le chameau a donc \'et\'e l'objet d'une dissection intense, afin de pouvoir le r\'eduire consid\'erablement.
L'autre solution consiste \`a couper sauvagement les basses fr\'equences, ce qui \'elimine la plupart des bruits corr\'el\'es, mais qui \'elimine aussi l'information cosmologique \`a bas l, justement l'un des buts principaux d'Archeops.

Notons que la plupart des trains de donn\'ees que nous avons utilis\'es pour la construction de cartes avaient \'et\'e pr\'ec\'edemment trait\'es \`a une dimension par P. Filliatre (ISN, Grenoble, voir Filliatre 2002).

 \subsection{Construction de cartes\label{mapmakark}}

Au d\'epart, nous avons d\'evelopp\'e nos m\'ethodes it\'eratives sp\'ecifiquement pour les donn\'ees d'Archeops.
Nous avons utilis\'e deux approches compl\'ementaires: l'une utilisant une m\'ethode simple et directe pour cartographier les donn\'ees de fa\c{c}on rapide mais n\'eanmoins efficace, l'autre utilisant une m\'ethode optimale mais qui se heurte \`a beaucoup de difficult\'es du fait de la pr\'esence de syst\'ematiques dans les donn\'ees.
Notons qu'\`a l'heure o\`u nous \'ecrivons ces lignes, la collaboration Archeops n'a pas encore officialis\'e ni publi\'e de cartes d'aucun vol.
Les cartes que nous pr\'esentons ici sont donc issues de notre travail personnel, et ne repr\'esentent bien \'evidemment pas la collaboration Archeops de fa\c{c}on g\'en\'erale.

 \subsubsection{Donn\'ees Trapani}

Le vol technologique de Trapani a fourni des donn\'ees tr\`es bruit\'ees mais d'une qualit\'e largement suffisante pour travailler sur l'\'emission galactique.
En revanche, en ce qui concerne les fluctuations du CMB, ce vol est tr\`es limit\'e.
Des six bolom\`etres pr\'esents pendant ce vol, le bolom\`etre 13 (num\'ero du canal \'electronique) \`a 143 GHz est le plus performant.
Nous avons fait des cartes de chacun des bolom\`etres, tout en nous concentrant sur la construction de cartes du canal 13.
La pr\'esence de beaucoup de bruits et syst\'ematiques (non stationnaires) pousse \`a couper les basses fr\'equences, ce qui tue le principal int\'er\^et d'application de m\'ethodes optimales de construction de cartes par la suite, \`a savoir l'\'elimination du bruit en 1/f.
Nous avons trait\'e les donn\'ees en appliquant un filtre passe-haut ad\'equat, r\'ealis\'e \`a partir du spectre de puissance du signal bolom\'etrique, plus ou moins pr\'e-trait\'e en fonction des progr\`es faits par ailleurs dans la collaboration Archeops pour le traitement du signal \`a une dimension.
Nous pr\'esentons en figure \ref{bol13map} une carte de l'observation du bolom\`etre 13 (non officielle), que nous avons r\'ealis\'ee en moyennant dans les pixels les donn\'ees filtr\'ees.
La carte utilise des pixels de 54' de c\^ot\'e (HEALPIX 64), afin de r\'eduire le bruit.
Nous avons utilis\'e l'\'etalonnage sur le dip\^ole de J.-P. Bernard, f\'evrier 2000: 22360 $\ukc$/$\mu$V.

\begin{figure}
\begin{center}
\epsfxsize 10.0 true cm
\leavevmode
\caption[Carte Archeops Trapani]{Carte personnelle non officielle du ciel \`a 143 GHz vu par le bolom\`etre 13 d'Archeops lors du vol de Trapani.
La carte utilise des pixels de 54' de c\^ot\'e (HEALPIX 64), afin de r\'eduire le bruit.
Nous avons utilis\'e l'\'etalonnage sur le dip\^ole de J.-P. Bernard, f\'evrier 2000: 22360 $\ukc$/$\mu$V.
Projection Mollweide galactique centr\'ee sur le centre galactique.}
\label{bol13map}
\end{center}
\end{figure}

Cette carte montre tr\`es clairement l'\'emission des poussi\`eres dans le plan galactique.
En ce qui concerne les fluctuations du CMB, m\^eme ce bolom\`etre 13, le moins bruit\'e, fait appara\^\i tre un niveau de bruit trop important pour les observer directement sur la carte.
L'\'echelle montre, mis \`a part l'incertitude d'\'etalonnage qui peut \^etre importante, que les fluctuations du CMB sont cach\'ees par le bruit (voir la carte du CMB simul\'ee en figure \ref{simusallsky}).

Nous avons \'egalement appliqu\'e la m\'ethode it\'erative COBE aux donn\'ees du vol de Trapani.
Celle-ci se heurte \`a la pr\'esence de syst\'ematiques, et ne parvient pas \`a \'eliminer l'ensemble du bruit corr\'el\'e.
La coupure des basses fr\'equences semble donc n\'ecessaire dans ce cas.

Afin de r\'eduire le bruit par pixel, nous d\'egradons la r\'esolution de la carte en figure \ref{bol13map} jusqu'\`a N$_{side}$ = 8, soit des pixels de 7.3 degr\'es de c\^ot\'e, la r\'esolution angulaire de COBE/DMR.
Nous comparons les cartes d'Archeops et de COBE en figure \ref{arkcob}.
Il semble que le niveau des fluctuations visible dans la carte Archeops corresponde bien \`a celui des anisotropies du CMB.
En revanche, il n'est pas visible de corr\'elation entre les deux cartes (coefficient de corr\'elation sur les pixels proche de 0), ce qui n'est pas tr\`es \'etonnant au vu de la quantit\'e de bruit pr\'esente dans les donn\'ees Archeops mais \'egalement dans les donn\'ees DMR (signal sur bruit $\approx$ 1).

\begin{figure}
\begin{center}
\epsfxsize 10.0 true cm
\leavevmode
\caption[Comparaison Archeops/COBE]{
Carte personnelle non officielle du ciel \`a 143 GHz vu par le bolom\`etre 13 d'Archeops lors du vol de Trapani (haut).
Carte COBE/DMR sur la partie du ciel observ\'ee par Archeops (bas).
Les cartes utilisent des pixels de 7.3$^o$ de c\^ot\'e (HEALPIX 8), afin de r\'eduire le bruit et de comparer la carte Archeops \`a celle de COBE/DMR.
Les parties galactiques des cartes (plus ou moins 20$^o$ en latitude galactique) ont \'et\'e coup\'ees.
Projections Mollweide galactiques centr\'ees sur le centre galactique.}
\label{arkcob}
\end{center}
\end{figure}

 \subsubsection{Donn\'ees Kiruna KS1}

Le premier vol scientifique \`a Kiruna a fourni des donn\'ees de bonne qualit\'e malgr\'e la pr\'esence de nombreux bruits et d'un ``chameau arctique''.
Nous avons fait des cartes de la plupart des 24 bolom\`etres pr\'esents pendant ce vol.
Les bolom\`etres les plus propres de ce vol sont deux canaux \`a 217 GHz: les bolom\`etres 217K04 et 217K06.
Nous avons concentr\'e nos efforts concernant la m\'ethode COBE it\'erative sur ces deux bolom\`etres.
Pr\'ecisons que nous n'avons pas coup\'e de basses fr\'equences avant d'appliquer la m\'ethode it\'erative.
Nous pr\'esentons en figure \ref{217K04opti} la carte obtenue apr\`es convergence pour le bolom\`etre 217K04.
Bien que l'\'etalonnage utilis\'e ne soit que pr\'eliminaire,
nous pouvons dire que l'on observe un effet dip\^olaire sup\'erieur au dip\^ole cosmologique.
De plus, la m\'ethode converge vers une solution qui n'est visiblement pas parfaitement d\'estri\'ee.
Ceci est probablement d\^u \`a la pr\'esence de syst\'ematiques, par ailleurs mises en \'evidence dans le train de donn\'ees.

\begin{figure}
\begin{center}
\epsfxsize 10.0 true cm
\leavevmode
\caption[Carte it\'er\'ee KS1]{Carte personnelle non officielle du ciel vu par le bolom\`etre 217K04 d'Archeops lors du vol KS1.
Cette carte, r\'ealis\'ee avec des pixels de 7$'$ de c\^ot\'e (HEALPIX 512), est obtenue apr\`es convergence de la m\'ethode COBE it\'erative.
N\'eanmoins, la pr\'esence importante de syst\'ematiques fait qu'il reste une quantit\'e importante de stries dans la carte, ainsi qu'un effet dip\^olaire sup\'erieur au dip\^ole cosmologique.
Projection Mollweide galactique centr\'ee sur l'anticentre galactique.}
\label{217K04opti}
\end{center}
\end{figure}

Dans ces conditions, la coupure des basses fr\'equences (approximativement jusqu'\`a la fr\'equence de coupure du bruit 1/f) semble n\'ecessaire.
Lorsque ceci est effectu\'e, le bruit restant dans les cartes est essentiellement d\'epourvu de corr\'elations.
L'application de m\'ethodes optimales de construction de cartes dans ces conditions n'est donc plus tr\`es utile.
Cependant, cette piste est explor\'ee avec succ\`es par certains membres de la collaboration Archeops (D. Yvon, O. Dor\'e {\it et al.}).

Nous avons construit des cartes de fa\c{c}on simple en enlevant une ligne de base, ligne de base construite en lissant le train de donn\'ees sur 500 points, ce qui revient \`a filtrer les basses fr\'equences jusqu'\`a 0.3 Hz.
La fr\'equence de rotation de la nacelle \'etant de 2 tours par minute \`a Kiruna, nous coupons approximativement les grandes \'echelles angulaires jusqu'\`a 30$^o$ (l $\approx$ 10).
Nous moyennons ensuite dans les pixels pour chaque bolom\`etre s\'epar\'ement.
Nous obtenons donc une carte par bolom\`etre utile: six canaux \`a 143 GHz, six \`a 217, cinq \`a 353 (polaris\'es) et deux \`a 545.
En utilisant des \'etalonnages pr\'eliminaires, nous construisons des cartes du ciel par longueur d'onde en moyennant dans les pixels d'une m\^eme carte les diff\'erentes cartes des canaux d'une m\^eme fr\'equence.
Nous pr\'esentons les cartes \`a 143 GHz, 217 GHz et 545 GHz ainsi construites en figure \ref{mapsks1}.

\begin{figure}
\begin{center}
\epsfxsize 10.0 true cm
\leavevmode
\caption[Cartes Archeops KS1]{Cartes personnelles non officielles du ciel vu par Archeops lors de KS1 \`a 143 GHz (haut), 217 GHz (centre) et 545 GHz (bas).
Les cartes sont pr\'esent\'ees \`a basse r\'esolution (pixels de 54$'$ de c\^ot\'e, {\it i.e.} HEALPIX 64) pour bien r\'eduire le bruit.
Notons que l'\'etalonnage utilis\'e pour les canaux \`a 143 GHz n'est probablement pas parfait.
Il y a un art\'efact visible autour de la Galaxie (douves), d\^u au filtrage des basses fr\'equences.
Projections Mollweide galactiques centr\'ees sur l'anticentre galactique.}
\label{mapsks1}
\end{center}
\end{figure}

Bien qu'il reste du bruit dans ces cartes,
il appara\^\i t que les cartes \`a 143 GHz et \`a 217 GHz montrent des fluctuations de forme non stri\'ee en dehors du plan galactique qui ne sont pas visibles sur les cartes \`a 545 GHz.
Ceci est un fort indice permettant de penser que l'on observe bien les fluctuations du CMB, d'autant que le niveau observ\'e de ces fluctuations est celui auquel on s'attend.
Pour l'\'etude des fluctuations du CMB, nous ajoutons les cartes \`a 143 et \`a 217 GHz dans les unit\'es de temp\'erature du CMB ($\ukc$) de sorte \`a construire une carte des fluctuations du CMB la moins bruit\'ee possible.

L'estimation du bruit r\'esiduel dans ces cartes peut \^etre faite de fa\c{c}on simple en prenant la diff\'erence de deux canaux d'une m\^eme fr\'equence, ou d'un panachage de plusieurs canaux; mais ceci suppose que l'\'etalonnage est tr\`es au point.
Dans ce cas, l'absence de fluctuations de type CMB (c'est-\`a-dire gaussiennes aux diff\'erentes \'echelles angulaires, de forme non stri\'ee, et n'apparaissant pas comme la fine granularit\'e du bruit blanc) dans la carte de diff\'erence est un fort indice pour justifier que l'on observe bien les fluctuations du CMB.

 \subsection{S\'eparation de composantes}

Nous avons fait quelques investigations sur une m\'ethode simple de s\'eparation de composantes consistant \`a d\'ecorr\'eler une carte dans une longueur d'onde qui fait appara\^\i tre \`a la fois les fluctuations de temp\'erature du CMB et l'\'emission de la poussi\`ere galactique, d'avec une carte dans une longueur d'onde courte dans laquelle il n'y a que de la poussi\`ere.
Dans le cas d'Archeops, nous avons appliqu\'e cette m\'ethode de d\'ecorr\'elation sur la carte des fluctuations du CMB construite \`a partir des canaux \`a 143 GHz et 217 GHz.
Nous avons retranch\'e \`a cette carte la carte \`a 545 GHz en utilisant un coefficient de corr\'elation empirique ad\'equat pour enlever la Galaxie (900 $\ukc$ par mK$_{RJ}$ \`a 545 GHz).
Le r\'esultat de cette m\'ethode simple (nous ne tenons pas compte d'\'eventuelles variations de couleur dans la Galaxie) est spectaculaire, comme le montre la figure \ref{ks1decor}.

\begin{figure}
\begin{center}
\epsfxsize 10.0 true cm
\leavevmode
\caption[Carte apr\`es extraction de la Galaxie]{Carte personnelle non officielle du ciel vu par Archeops lors de KS1, utilisant des pixels de 54$'$ de c\^ot\'e (HEALPIX 64), en unit\'e de temp\'erature du CMB, r\'ealis\'ee en utilisant les canaux \`a 143 et 217 GHz et en d\'ecorr\'elant la Galaxie d'avec la carte \`a 545 GHz.
L'\'emission galactique est bien \'elimin\'ee de cette carte, bien que l'on n'ait pas tenu compte d'\'eventuelles variations de couleur dans la Galaxie.
Projection Mollweide galactique centr\'ee sur l'anticentre galactique.}
\label{ks1decor}
\end{center}
\end{figure}

Dans le cas du vol KS1, il reste quelques syst\'ematiques sur les cartes (nuages d'ozone atmosph\'erique ?), nous avons donc \'egalement appliqu\'e cette m\'ethode de d\'ecorr\'elation pour \'eliminer ces syst\'ematiques.
En effet, celles-ci sont pr\'esentes \`a toutes les fr\'equences, et elles dominent dans la carte \`a 545 GHz \`a haute latitude galactique.
Nous observons - ce qui n'est pas tr\`es \'etonnant - que ces syst\'ematiques n'ont pas la m\^eme couleur que la Galaxie, autrement dit, le coefficient de d\'ecorr\'elation n'est pas le m\^eme.

\section{ Analyse possible des cartes d'Archeops\label{analarch}}

\`A l'heure o\`u nous \'ecrivons ces lignes, l'analyse des donn\'ees Archeops n'est pas termin\'ee ni publi\'ee.
Les doctorants n'ont donc pas l'autorisation de la collaboration Archeops d'inclure dans leurs m\'emoires des C$_l$ provenant des donn\'ees.
Nous essayons donc de montrer dans cette section une petite partie de l'analyse possible des cartes d'Archeops, en utilisant des simulations.

 \subsection{La Galaxie}

L'ensemble des donn\'ees Archeops (T, KS1, KS2, KS3) couvre une surface tr\`es large du ciel et notamment du plan galactique.
L'int\'er\^et pour la science galactique est tr\`es grand.
Si l'on se r\'ef\`ere \`a la partie galactique de ce m\'emoire (chapitre \ref{pous}), il devient clair que les cartes \`a 550 $\mic$ et 850 $\mic$ d'Archeops peuvent \^etre combin\'ees aux cartes IRAS \`a 60 et 100 $\mic$ pour mesurer la temp\'erature et l'indice spectral des poussi\`eres.
Les cartes Archeops sont particuli\`erement sensibles \`a la poussi\`ere froide, et les cartes millim\'etriques (1.4 et 2.1 mm) sont tr\`es efficaces pour contraindre l'indice spectral des poussi\`eres.
En effet, dans les longueurs d'onde millim\'etriques, l'indice spectral ne peut absolument pas \^etre expliqu\'e par de la poussi\`ere froide m\'elang\'ee \`a d'autres composantes de poussi\`eres, car cette composante froide devrait avoir une temp\'erature de l'ordre de 3 K.
L'effet d'anticorr\'elation que nous mettons en \'evidence en section \ref{anticor} entre la temp\'erature et l'indice spectral des poussi\`eres pourrait donc \^etre bien contraint par les cartes Archeops.

De plus, l'\'etude de la structure de la poussi\`ere froide \`a grande \'echelle dans la Galaxie est une possibilit\'e tr\`es importante d'Archeops, ainsi que la reconstitution de la structure galactique \`a trois dimensions en combinaison avec d'autres donn\'ees \cite{montier02}.

Enfin, la polarisation de sources galactiques est accessible gr\^ace aux canaux polaris\'es \`a 353 GHz.
Ceux-ci sont en cours d'analyse dans la collaboration Archeops.


 \subsection{Spectre de puissance harmonique du CMB}

Nous nous sommes int\'eress\'es \`a l'obtention du spectre de puissance des anisotropies du CMB.
Pour cela, nous avons appliqu\'e une m\'ethode tr\`es simple mais efficace consistant \`a estimer le bruit dans les cartes par diff\'erences entre bolom\`etres, comme expliqu\'e en section \ref{mapmakark}.
Nous obtenons les spectres en C$_l$ des cartes de signal et des cartes de bruit en utilisant l'outil ANAFAST du package HEALPIX ({\it http://www.eso.org/science/healpix} ou G\'orski \etal 1999).
Les spectre en C$_l$ sont des variances qui ont donc le bon go\^ut de pouvoir s'ajouter et se retrancher.
Il est clair que l'information sur le bruit n'est que statistique, et qu'en cons\'equence il faut retrancher au spectre des donn\'ees le spectre du bruit statistiquement moyenn\'e.
En pratique, il s'agit, si l'on ne conna\^\i t pas parfaitement les propri\'et\'es statistiques du bruit par une m\'ethode plus avanc\'ee, d'ajuster le spectre en C$_l$ du bruit par une loi simple (bruit blanc, 1/f, etc).
Un bruit blanc (non corr\'el\'e) se traduit par un spectre en C$_l$ plat, avec des fluctuations statistiques.

\`A l'heure actuelle, nous n'avons pas l'autorisation de la collaboration Archeops d'inclure dans ce m\'emoire des C$_l$ provenant des donn\'ees.
N\'eanmoins, afin de valider la m\'ethode simple que nous avons d\'ecrite pour l'estimation du spectre en C$_l$, nous avons simul\'e des cartes du ciel sur une observation correspondant au vol KS1.
Le CMB simul\'e est celui d\'ecrit en section \ref{simudon}: il est domin\'e par la constante cosmologique, avec $\Omega_\Lambda$ = 0.7, $\Omega_{CDM}$ = 0.25, $\Omega_{bar}$ = 0.05, $H_0$ = 50 km/s/Mpc, et un indice spectral scalaire des fluctuations n \'egal \`a 1.
Nous introduisons un niveau de bruit r\'ealiste par rapport \`a ce que l'on a dans les cartes des vraies donn\'ees.
Dans nos simulations, il s'agit d'un bruit blanc de niveau 1000 $\ukc$ en rms dans les TOI.
Ayant r\'ealis\'e deux cartes de donn\'ees avec du bruit al\'eatoire, nous pouvons faire une carte en moyennant les deux canaux ainsi simul\'es, et une carte de demi-diff\'erence.
Cette carte de demi-diff\'erence ne contient bien entendu plus de signal, mais elle contient le bruit r\'eduit d'un facteur racine de 2 en rms, comme la demi-somme.
En effet, les variances du bruit s'ajoutent (variables al\'eatoires ind\'ependantes), donc le rms cro\^\i t dans la carte somme comme la racine du nombre de cartes co-additionn\'ees.
Dans la carte moyenne, il d\'ecro\^\i t donc comme la racine du nombre de cartes co-additionn\'ees.
La carte de demi-diff\'erence est donc une mesure non biais\'ee du bruit dans la carte de demi-somme.
Ceci vaut aussi dans le cas de bruit autocorr\'el\'e, \`a la seule condition qu'il s'agisse bien de bruits al\'eatoires, ind\'ependants d'un bolom\`etre \`a l'autre.
S'il s'agit de syst\'ematiques corr\'el\'ees entre les canaux, alors il est clair que cette m\'ethode ne sera pas capable d'estimer ces effets.
Pour des bruits \`a tr\`es basse fr\'equence n\'eanmoins, la statistique donn\'ee par la diff\'erence de deux canaux peut ne pas suffire pour caract\'eriser ces bruits.

\begin{figure}
\begin{center}
\epsfxsize 10.0 true cm
\leavevmode
\caption[Simulations de spectres de puissance KS1]{Simulations de spectres de puissance, corrig\'es de la fraction de couverture, obtenus pour la demi-somme de deux canaux (haut, courbe la plus \'elev\'ee en noir), la demi-diff\'erence (le bruit, haut, courbe grise croissante, dans le bas du graphique), le vrai ciel (haut, courbe grise), le spectre corrig\'e (bas) et le spectre pur (th\'eorie sans variance cosmique, bas, courbe grise lisse).
Il sont \'et\'e produits en ne prenant en compte que les pixels en dehors du plan galactique (plus ou moins 20$^o$ en latitude galactique), pour la couverture du ciel de KS1.}
\label{clsimuks1}
\end{center}
\end{figure}

Nous pr\'esentons en figure \ref{clsimuks1} les spectres de puissance corrig\'es de la fraction de couverture obtenus pour la demi-somme (haut, courbe la plus \'elev\'ee en noir), la demi-diff\'erence (le bruit, haut, courbe grise croissante, dans le bas du graphique), le vrai ciel (haut, courbe grise), le spectre corrig\'e (bas) et le spectre pur (th\'eorie sans variance cosmique, bas, courbe grise lisse).
Le spectre corrig\'e a \'et\'e r\'ealis\'e en soustrayant au spectre en C$_l$ des donn\'ees l'ajustement r\'ealis\'e sur la mesure du bruit.
Ces spectres ont \'et\'e produits en ne prenant en compte que les pixels en dehors du plan galactique (plus ou moins 20$^o$ en latitude galactique), pour la couverture du ciel de KS1.
Le r\'esultat obtenu par cette m\'ethode simple d'estimation du bruit et de soustraction des C$_l$ du bruit est spectaculaire.
On observe que le r\'esultat est pr\'ecis malgr\'e le haut niveau de bruit dans les pixels de la carte.
Notons que si les niveaux de bruit ne sont pas exactement les m\^emes pour les deux bolom\`etres utilis\'es, cette m\'ethode fonctionne de la m\^eme fa\c{c}on.
Dans ce cas, le rms du bruit dans la carte moyenn\'ee est:

\begin{equation}
m = {\sqrt{a^2 + b^2} \over 2}
\end{equation}

o\`u a et b sont les rms du bruit pour chaque bolom\`etre.

Puisque cette m\'ethode nous donne une mesure du bruit qui est une r\'ealisation de ce bruit par rapport \`a la moyenne statistique (que l'on ajuste), et que les cartes moyenn\'ees contiennent une autre r\'ealisation du bruit, il est clair que l'erreur sur les C$_l$ corrig\'es est simplement le rms des fluctuations du bruit par rapport \`a sa puissance moyenne statistique.
En moyennant les C$_l$ par bandes, nous pouvons donc obtenir les barres d'erreur par l'\'ecart quadratique moyen des fluctuations du bruit dans la bande divis\'e par la racine du nombre de multip\^oles dans la bande.
Nous pr\'esentons en figure \ref{clsimfin} le r\'esultat sur les simulations du vol KS1 avec les deux bolom\`etres bruit\'es.
Le r\'esultat est tr\`es satisfaisant.
Notons que ces barres d'erreur ne tiennent pas compte de la variance cosmique, le ``bruit'' intrins\`eque \`a notre univers simul\'e, et qu'il est donc normal que, notamment \`a bas l, les barres d'erreur ne soient pas compatibles avec le mod\`ele th\'eorique pur.
Pour les hauts l, on observe un l\'eger effet de diminution de la puissance par rapport \`a la th\'eorie: ceci est d\^u aux effets de pix\'elisation qui commencent \`a se faire sentir, car nous avons utilis\'e des pixels de 13.5$'$ de c\^ot\'e (N$_{side}$ = 256).

\begin{figure}
\begin{center}
\epsfxsize 10.0 true cm
\leavevmode
\caption[Spectres de puissance en bandes]{
Spectres de puissance en C$_l$ de la th\'eorie sans variance cosmique (courbe pleine) et des donn\'ees KS1 simul\'ees moyenn\'ees en bandes, avec leurs barres d'erreur \`a 1 $\sigma$.
Notons que ces barres d'erreur ne tiennent pas compte de la variance cosmique, et qu'il est donc normal que, notamment \`a bas l, les barres d'erreur ne soient pas compatibles avec le mod\`ele th\'eorique pur.
Pour les hauts l, on observe un l\'eger effet de diminution de la puissance par rapport \`a la th\'eorie: ceci est d\^u aux effets de pix\'elisation qui commencent \`a se faire sentir, car nous avons utilis\'e des pixels de 13.5$'$ (N$_{side}$ = 256).
}
\label{clsimfin}
\end{center}
\end{figure}

Bien entendu, ce r\'esultat tr\`es prometteur, en particulier sur les bas l, a \'et\'e produit sans tenir compte d'\'eventuelles syst\'ematiques r\'esiduelles dans les cartes.
Cependant, la m\'ethode d'estimation du bruit et d'extraction des C$_l$ par diff\'erences de deux bolom\`etres est valid\'ee par un tel r\'esultat.

Ce spectre de puissance a \'et\'e obtenu sur deux canaux simul\'es tr\`es bruit\'es (dix fois le niveau de bruit rms suppos\'e des bolom\`etres pour Planck), ce qui montre les possibilit\'es de l'exp\'erience Archeops.
Avec un niveau de bruit nominal sur un nombre important de bolom\`etres, Archeops devrait \^etre capable de contraindre pr\'ecis\'ement un grand domaine de multip\^oles, depuis les bas l (lien avec COBE) jusqu'aux deux premiers pics.

\subsection{Param\`etres cosmologiques}

Le domaine de multip\^oles potentiellement couvert par les mesures d'Archeops permet de contraindre efficacement les param\`etres cosmologiques, notamment $\Omega$ par la position du premier pic, $\Omega_{bar}$ par la hauteur du premier pic mais aussi la position du deuxi\`eme, et l'indice spectral des fluctuations n (bas l).
Ceci est illustr\'e par la figure \ref{cosmodels}, qui montre les variations du spectre de puissance du CMB sans variance cosmique, normalis\'e \`a la mesure de COBE, en fonction des param\`etres cosmologiques $\Omega$, $\Omega_{bar}$ et n, l'indice spectral des fluctuations.

\begin{figure}
\begin{center}
\epsfxsize 10.0 true cm
\leavevmode
\caption[Spectres de puissance de mod\`eles cosmologiques]{Spectres de puissance en C$_l$ sans variance cosmique, normalis\'es \`a la mesure de COBE, en fonction des param\`etres cosmologiques $\Omega$ (haut, mod\`eles sans constante cosmologique), $\Omega_{bar}$ (centre, mod\`eles sans constante cosmologique) et n, l'indice spectral des fluctuations (bas, mod\`eles avec constante cosmologique dominante).
Nous avons produit ces spectres en utilisant le logiciel CMBFAST 4.0 \cite{seljak96}.
}
\label{cosmodels}
\end{center}
\end{figure}

Comme nous le voyons, Archeops a la possibilit\'e d'apporter une contribution essentielle \`a la cosmologie d'aujourd'hui.

\chapter{Conclusion g\'en\'erale}

\vspace*{1em}
{\flushright\it Nunc est bibendum.\\
C'est maintenant qu'il faut boire.\\}
{\flushright Horace\\}
\vspace*{1em}

Ce travail de th\`ese s'est inscrit dans le cadre de l'astrophysique submillim\'etrique et millim\'etrique, en plein d\'eveloppement aujourd'hui, d\'eveloppement qui culminera avec l'analyse des donn\'ees des exp\'eriences Planck et Herschel.
De nombreuses exp\'eriences concernant le continuum d'\'emission dans ce domaine de longueurs d'onde ont produit des r\'esultats dans les derni\`eres ann\'ees ou les produiront bient\^ot, que ce soit dans le domaine de la science galactique ou extragalactique.
Dans ce contexte, notre travail a permis de contribuer au d\'eveloppement des m\'ethodes d'analyse et de la connaissance de l'astrophysique du continuum submillim\'etrique et millim\'etrique.

La construction de cartes occupe une place primordiale dans ce travail.
Nous nous sommes int\'eress\'es au probl\`eme de l'inversion lin\'eaire, dont nous avons \'etudi\'e les solutions optimales.
Notre travail sur les donn\'ees de l'exp\'erience submillim\'etrique ballon PRONAOS nous a permis de d\'evelopper une nouvelle m\'ethode, optimale, de construction de cartes pour cette exp\'erience.
Cette m\'ethode utilise une matrice de Wiener permettant d'inverser le probl\`eme lin\'eaire sur l'ensemble de la carte.
Elle am\'eliore consid\'erablement la qualit\'e des cartes PRONAOS par rapport \`a la m\'ethode pr\'ec\'edemment utilis\'ee.
Nous avons ainsi pu produire des cartes de l'\'emission de la poussi\`ere dans la N\'ebuleuse d'Orion M42 et le complexe M17, sans art\'efacts de reconstruction, et dans lesquelles une grande quantit\'e d'information utile est pr\'eserv\'ee, notamment dans les r\'egions de tr\`es faible intensit\'e o\`u apparaissent souvent des nuages froids.
Le traitement des donn\'ees de M17 a fait l'objet du d\'eveloppement d'une m\'ethode de construction de cartes plus avanc\'ee, dans laquelle le bruit suppos\'e n'est pas ind\'ependant du signal.

L'analyse de ces cartes d'intensit\'e obtenues aux quatre longueurs d'onde de PRONAOS (200, 260, 360 et 580 $\mic$) nous a permis de d\'eduire les param\`etres de l'\'emission des poussi\`eres dans l'ensemble des r\'egions cartographi\'ees.
L'exp\'erience PRONAOS est en effet capable, en compl\'ementarit\'e \'eventuelle avec IRAS, de d\'eriver \`a la fois la temp\'erature et l'indice spectral de la poussi\`ere.
Ceci permet de ne pas faire de suppositions arbitraires sur la valeur de l'un ou l'autre param\`etre.
Nous avons \'etudi\'e les r\'egions de formation d'\'etoiles massives que sont les complexes mol\'eculaires associ\'es aux n\'ebuleuses M42 (Orion) et M17.
La cartographie multi-bandes de ces r\'egions par PRONAOS nous a permis de mettre en \'evidence des variations importantes de la temp\'erature des poussi\`eres (environ de 10 K \`a 100 K) ainsi que de l'indice spectral submillim\'etrique (environ de 1 \`a 2.5).
Nous avons mis en \'evidence des nuages froids dans Orion (11.8, 13.3, 16.9 et 17 K) pr\`es des c{\oe}urs chauds et actifs.
Dans M17, nous avons \'egalement mis en \'evidence des r\'egions froides (14, 17 et 17 K), bien que la r\'esolution spatiale moins bonne aie tendance \`a moyenner l'\'emission des poussi\`eres.
Dans les deux r\'egions \'etudi\'ees, nous mettons en \'evidence pour la premi\`ere fois l'anticorr\'elation entre la temp\'erature et l'indice spectral.
Nous avons r\'ealis\'e des simulations d'ajustements et montr\'e que cet effet avait une r\'eelle signification statistique.
Nous avons \'egalement analys\'e comment la pr\'esence de plusieurs composantes de poussi\`eres sur la ligne de vis\'ee, ayant diff\'erentes temp\'eratures, peut induire un tel effet d'anticorr\'elation.
La conclusion tr\`es nette de ces investigations est que l'effet de m\'elanges de poussi\`eres ne peut pas expliquer l'ensemble de l'anticorr\'elation observ\'ee entre la temp\'erature et l'indice spectral.
Nous privil\'egions une explication fondamentale provenant de la physique microscopique des grains, dont l'effet a \'et\'e mis en \'evidence par des exp\'eriences de laboratoire dans des domaines larges de temp\'erature et de longueur d'onde.
Cet effet va dans le sens de nos observations.
Nous avons pu estimer les densit\'es de colonne dans les r\'egions observ\'ees, en mod\'elisant de fa\c{c}on simple la relation entre la densit\'e de colonne de milieu interstellaire et les param\`etres de l'ajustement r\'ealis\'e sur les donn\'ees PRONAOS.
Nous utilisons pour cela l'opacit\'e \`a 100 $\mic$ provenant de deux mod\`eles de grains, l'un standard, l'autre sp\'ecifique aux c{\oe}urs froids.
Nous d\'erivons \'egalement les densit\'es de colonne d'apr\`es des donn\'ees CO, et il appara\^\i t que les deux estimations sont g\'en\'eralement en bon accord.
Nous d\'erivons des estimations des masses et densit\'es dans les r\'egions \'etudi\'ees, qui font appara\^\i tre l'instabilit\'e gravitationnelle possible de certains nuages froids.

Ce travail de th\`ese s'est \'egalement consacr\'e aux m\'ethodes d'analyse et \`a l'\'etude du rayonnement fossile (CMB).
Nous avons d\'evelopp\'e et test\'e de nouvelles m\'ethodes de construction de cartes pour les donn\'ees concernant le rayonnement fossile, et simul\'e des strat\'egies d'observation d'exp\'eriences ballon et satellite suppos\'ees ou r\'eelles.
Les simulations d'exp\'eriences ballon utilisent une strat\'egie d'observation fond\'ee sur des cercles \`a hauteur constante, ce qui ob\'eit \`a des n\'ecessit\'es techniques simples et permet toutefois une grande vari\'et\'e de types d'observations, suivant l'\'el\'evation du faisceau, le lieu de lancement, le temps de vol, etc.
Nous mettons en \'evidence les diff\'erents avantages de chaque type d'observation, notamment par rapport \`a nos m\'ethodes de construction de cartes.
Nous simulons \'egalement des donn\'ees provenant d'exp\'eriences satellite observant par cercles orthogonaux au plan \'ecliptique.
Dans ce cas, qui est par exemple celui de Planck, il est possible de r\'ealiser diverses observations en faisant varier l'angle d'\'el\'evation du faisceau au-dessus du plan \'ecliptique, et de r\'ealiser \'eventuellement une cartographie compl\`ete du ciel.
Nous avons d\'evelopp\'e des m\'ethodes optimales de construction de cartes pour les exp\'eriences CMB: une m\'ethode de Wiener directe et des m\'ethodes it\'eratives (COBE, Wiener et g\'en\'eralis\'ee).
Ces m\'ethodes permettent de traiter le probl\`eme de construction de cartes de fa\c{c}on optimale pour les larges trains de donn\'ees des exp\'eriences CMB modernes, pour lesquelles l'application des m\'ethodes optimales brutes n'est pas possible.
Les m\'ethodes it\'eratives fonctionnent en construisant peu \`a peu la carte de bruit, en passant du domaine de la carte \`a celui du train de donn\'ees, en filtrant de fa\c{c}on ad\'equate et en repassant dans le domaine de la carte, ceci \`a chaque it\'eration.
Le choix des m\'ethodes COBE ou Wiener d\'epend de l'{\it a priori} sur les propri\'et\'es statistiques du ciel observ\'e.
Nous avons test\'e chacune de ces m\'ethodes sur diff\'erentes simulations d'exp\'eriences ballon et satellite, et il appara\^\i t que les m\'ethodes sont tr\`es efficaces pour reconstruire les cartes.
La comparaison avec des m\'ethodes telles que le moyennage par pixel ou le filtrage simple montre que l\`a o\`u les m\'ethodes simples \'echouent, en laissant le bruit corr\'el\'e ou en \'eliminant le signal, les m\'ethodes optimales obtiennent des r\'esultats tr\`es satisfaisants.
En particulier, les m\'ethodes it\'eratives sont tr\`es pr\'ecises et \'eliminent presque totalement le bruit corr\'el\'e (bruit en 1/f).
Le bruit r\'esiduel dans les cartes finales est constitu\'e dans sa quasi-totalit\'e de bruit blanc moyenn\'e dans les pixels.
Un peu de bruit corr\'el\'e reste n\'eanmoins aux grandes \'echelles angulaires, mais avec un niveau n\'egligeable par rapport au niveau des fluctuations du CMB.
Nous comparons les diff\'erentes strat\'egies d'observation par rapport aux r\'esultats des m\'ethodes de construction de cartes, et nous en d\'eduisons notamment qu'un nombre important de balayages crois\'es n'est pas n\'ecessaire pour construire une carte parfaitement d\'estri\'ee.
En revanche, en pr\'esence de syst\'ematiques p\'eriodiques avec la p\'eriode de rotation (particuli\`erement non stationnaires), les m\'ethodes ne convergent pas vers la bonne solution, ce qui est tout \`a fait normal, mais qui n\'ecessite donc de s'affranchir de tels effets avant d'appliquer une de nos m\'ethodes it\'eratives de construction de cartes optimale.

Notre participation au traitement des donn\'ees de l'exp\'erience ballon Archeops s'est \'egalement concentr\'ee sur la construction de cartes, mais nous avons \'egalement travaill\'e sur l'analyse ult\'erieure.
Nous avons appliqu\'e notre m\'ethode it\'erative COBE ainsi que des m\'ethodes simples pour construire des cartes des vols Trapani et KS1.
Les r\'esultats de la m\'ethode optimale it\'erative sont fauss\'es par la pr\'esence de bruits non stationnaires.
De ce fait, le recouvrement des basses fr\'equences pour ces deux vols semble extr\^emement difficile, malgr\'e les m\'ethodes de construction de cartes optimale d\'evelopp\'ees dans plusieurs \'equipes de la collaboration Archeops.
Par des m\'ethodes simples de filtrage ad\'equat des basses fr\'equences tr\`es bruit\'ees, nous construisons des cartes (personnelles et non officielles) de ces deux vols d'Archeops.
Le vol de Trapani fournit une information galactique riche et fiable, mais peu d'information utile pour la cosmologie.
La corr\'elation des pixels de notre carte Archeops de Trapani avec la carte COBE/DMR n'est pas convaincante, ce qui n'est pas \'etonnant au vu des erreurs de mesure dans ces deux cartes.
Le premier vol \`a Kiruna, en revanche, donne des r\'esultats tr\`es int\'eressants pour la cosmologie ainsi que pour la Galaxie.
Les cartes que nous avons construites de fa\c{c}on simple pour diff\'erentes fr\'equences sugg\`erent fortement que l'on observe bien les fluctuations du CMB.
Nous avons \'egalement d\'eduit des spectres en C$_l$ pour le vol KS1, que nous ne pouvons pas montrer dans ce m\'emoire.
Nous montrons cependant quelques simulations qui sugg\`erent ce qu'Archeops pourrait \^etre capable de faire en ce qui concerne l'estimation du spectre de puissance des anisotropies et la contrainte de certains param\`etres cosmologiques.

L'avenir des observations dans les longueurs d'onde submillim\'etriques et millim\'etriques est tr\`es prometteur, notamment gr\^ace aux futurs satellites Herschel et Planck.
La communaut\'e scientifique qui s'int\'eresse \`a l'Univers froid est en train de d\'ecouvrir de nouvelles propri\'et\'es et de confirmer certaines th\'eories tout en en infirmant d'autres.
Dans le domaine des \'etudes galactiques, la structure du milieu interstellaire et ses propri\'et\'es physico-chimiques commencent \`a peine \`a \^etre contraintes; l'anticorr\'elation entre la temp\'erature et l'indice spectral que nous avons mise en \'evidence gr\^ace aux donn\'ees PRONAOS en est un exemple.
En ce qui concerne la cosmologie observationnelle par l'\'etude des anisotropies du rayonnement fossile, les choses commencent \'egalement \`a se pr\'eciser.
La th\'eorie de l'inflation commence \`a se confirmer, alors que celle des d\'efauts topologiques semble condamn\'ee.
Les pr\'esences de mati\`ere noire non baryonique et d'\'energie du vide sont tr\`es probables aujourd'hui.
L'exp\'erience Archeops devrait contribuer \`a confirmer (ou infirmer) ces d\'ecouvertes.
Il y a fort \`a parier que la contrainte des param\`etres cosmologiques fasse de gros progr\`es dans les dix ann\'ees qui viennent, venant confirmer ou infirmer les th\'eories cosmologiques, et r\'epondre \`a des questions fondamentales.
Les trois derni\`eres ann\'ees ont vu les r\'eponses \`a certaines questions se pr\'eciser fortement: l'Univers semble plat, infini, et rempli de mati\`ere et d'\'energie sombres.
D'autres questions sont pourtant toujours pos\'ees: quelle est la nature de la mati\`ere et de l'\'energie sombres, par exemple.
La mission Planck ne sera pas la fin de la cosmologie.
L'Histoire apprend que de nouvelles questions appara\^\i ssent toujours, contraignant les th\'eories \`a \'evoluer vers une compr\'ehension toujours plus juste (disons plut\^ot toujours moins fausse) de notre monde...

\chapter{Renseignements compl\'ementaires\label{rc}}

\section{ Campagnes, colloques, \'ecoles}

\subsection{Campagnes\label{campagnes}}

J'ai particip\'e aux deux campagnes de l'hiver 2000-2001 pour Archeops, \`a Kiruna, en Laponie su\'edoise, respectivement en novembre-d\'ecembre et en janvier-f\'evrier.
Ces campagnes m'ont permis de participer \`a l'instrumentation attenant \`a mon sujet de th\`ese, notamment sur la connaissance de la structure d'une exp\'erience ballon et de ses d\'etecteurs.
Elles m'ont \'egalement permis de participer au contr\^ole de la mission en vol.

J'ai aussi particip\'e \`a la campagne de lancement de PRONAOS, pour son troisi\`eme vol en septembre 1999, depuis le centre de contr\^ole du CNES \`a Toulouse.

\subsection{Colloques}

J'ai pris part \`a des conf\'erences qui m'ont permis de pr\'esenter mes travaux, de communiquer avec les chercheurs et jeunes chercheurs dans le monde, et de d\'evelopper une vision globale de la recherche dans mon domaine.
Voici l'ensemble des colloques, conf\'erences ou ``ateliers'' auxquels j'ai particip\'e:

- Colloque Rencontres de Moriond ``Energy densities in the Universe'', 22-29 janvier 2000, Les Arcs, Savoie, France.
Participation au colloque.

- Colloque ``From darkness to light'', 3-8 avril 2000, Carg\`ese, Corse, France.
Participation au colloque, pr\'esentation orale concernant Orion, article d'actes Dupac \etal 2001 {\it coll. a}.

- Colloque ``Galactic structure, stars and the interstellar medium'', mai 2000, \`a Grand Teton, \'Etats-Unis.
Pr\'esentation d'un poster concernant Orion.

- Colloque ``Mining the sky'', 31 juillet - 4 ao\^ut 2000, Garching, Allemagne.
Participation au colloque, pr\'esentation d'un poster concernant les m\'ethodes de construction de cartes pour le CMB, article d'actes Dupac \& Giard 2001 {\it coll. b}.

- Colloque ``Infrared and submillimeter space astronomy'' en hommage \`a Guy Serra, 11-13 juin 2001, Toulouse.
Participation au colloque, pr\'esentation d'un poster concernant Orion et M17, article d'actes Dupac \etal 2002 {\it coll. c}.

- ``Planck workshop on image processing'', 4-6 juillet 2001, Pise, Italie.
Participation au colloque, pr\'esentation orale concernant les m\'ethodes de construction de cartes pour Planck.

- Colloque 2K1BC ``Experimental cosmology at millimetre wavelengths'', 9-13 juillet 2001, Breuil-Cervinia, Val d'Aoste, Italie.
Participation au colloque, pr\'esentation d'un poster concernant les m\'ethodes de construction de cartes pour le CMB, article d'actes Dupac 2002 {\it coll. d}.

- Conf\'erence ``Infrared and millimeter waves'' (IRMMW), 10-14 septembre 2001, Toulouse.
Participation au colloque, pr\'esentation orale concernant l'exp\'erience Archeops et le rayonnement fossile, article d'actes Dupac \etal 2002 {\it coll. e}.

\`A ceci s'ajoute les nombreuses r\'eunions et ateliers de travail de la collaboration Archeops.

\subsection{\'Ecoles}

Ma participation \`a deux \'ecoles d'\'et\'e m'a permis d'\'etendre mon exp\'erience \`a des domaines plus larges de l'astronomie.

- \'Ecole d'\'et\'e ``Extragalactic astronomy and cosmology from space'', 18-27 juillet 2000, Alpbach, Tyrol, Autriche.
Cours concernant l'astronomie extragalactique spatiale et d\'eveloppement du projet de mission spatiale SNOOPY (voir section \ref{snoopy}).

- \'Ecole d'\'et\'e IRAM ``Millimeter observing techniques and applications'', 14-21 septembre 2001, Pradollano, Sierra Nevada, Espagne.
Apprentissage de l'observation au radiot\'elescope millim\'etrique de 30 m de l'IRAM, et cours concernant la science attenante.



\section{ Liste des publications\label{publis}}
Cette liste comprend les articles dans lesquels nous apparaissons comme premier
auteur ou co-auteur, dans des journaux (avec rapporteur) ou dans des rapports de conf\'erences.
Les articles de journaux sont
pr\'esent\'es dans leur totalit\'e en annexe \`a la fin du volume.

 \subsection{Articles de journaux (rang A) en premier auteur}
Classement dans l'ordre chronologique:
\vspace*{1em}

{\bf X. Dupac, M. Giard, J.-P. Bernard, J.-M. Lamarre, C. M\'eny, F. Pajot,
  I. Ristorcelli, G. Serra, J.-P. Torre}: {\it Submillimeter mapping and
analysis of cold dust condensations in the Orion M42 star forming
complex}, accept\'e le 9 jan. 2001, \apj, 553, 604-612, juin 2001

arXiv.org: astro-ph/0102407
\vspace*{1em}

{\bf X. Dupac \& M. Giard}: {\it Map-making methods for Cosmic Microwave
Background experiments}, accept\'e le 17 oct. 2001, \mnras, 330 (3), 497-505, mars 2002

arXiv.org: astro-ph/0110407
\vspace*{1em}

{\bf X. Dupac, M. Giard, J.-P. Bernard, N. Boudet, J.-M. Lamarre, C. M\'eny,
  F. Pajot, \'E. Pointecouteau,
  I. Ristorcelli, G. Serra, B. Stepnik, J.-P. Torre}: {\it Submillimeter dust emission of
  the M17 complex with PRONAOS}, accept\'e le 12 juin 2002, \aa, 392, 691-698, sept. 2002

arXiv.org: astro-ph/0206337

 \subsection{Articles de journaux (rang A) en co-auteur}

{\bf A. Beno\^\i t \etal}: {\it Archeops: A High Resolution, Large Sky Coverage Balloon Experiment for Mapping CMB Anisotropies}, \astpart, 17, 101-124, mai 2002

arXiv.org: astro-ph/0106152

 \subsection{Articles de colloques en premier auteur\label{collpaut}}
Classement dans l'ordre chronologique:
\vspace*{1em}

{\bf X. Dupac, M. Giard, J.-P. Bernard, J.-M. Lamarre, F. Pajot,
I. Ristorcelli, G. Serra, J.-P. Torre}: {\it Analysis of ProNaOS
submillimeter maps in the M42 Orion Nebula - Temperature - spectral
index inverse correlation in several regions}, proceeding du colloque
"From Darkness to Light" \`a Carg\`ese, avril 2000, ASP conf. series vol. 243, \'ed. T. Montmerle \& P. Andr\'e, p. 319-324, 2001

arXiv.org: astro-ph/0102104
\vspace*{1em}

{\bf X. Dupac \& M. Giard}: {\it How to make CMB maps from huge timelines with
small computers}, proceeding du colloque "Mining the Sky" \`a Garching,
ao\^ut 2000, Springer ESO Symposia, \'ed. Banday, Zaroubi, Bartelmann, p. 432-434, 2001

arXiv.org: astro-ph/0102102
\vspace*{1em}

{\bf X. Dupac \& la collaboration PRONAOS}: {\it Dust emission in massive star-forming
regions with PRONAOS: the Orion and M17 molecular clouds}, proc. du
colloque G. Serra \`a Toulouse, juin 2001, EAS pub. series vol. 4, \'ed. Giard, Bernard, Klotz et Ristorcelli, p. 263-267, 2002

arXiv.org: astro-ph/0110551
\vspace*{1em}

{\bf X. Dupac}: {\it Iterative map-making methods for Cosmic Microwave Background
data analysis}, proc. du workshop 2K1BC \`a Breuil-Cervinia, Italie, juillet 2001, AIP conf. proceedings vol. 616, \'ed. M. de Petris et M. Gervasi, p. 360-362, 2002

arXiv.org: astro-ph/0109509
\vspace*{1em}

{\bf X. Dupac \& la collaboration Archeops}: {\it Archeops: a large sky coverage
millimeter experiment for mapping Cosmic Microwave Background
anisotropies}, proc. du colloque Infrared and Millimeter Waves (IRMMW)
\`a Toulouse, sept. 2001

[avec comit\'e de lecture et rapporteur]

arXiv.org: astro-ph/0110221

 \subsection{Articles de colloques en co-auteur (liste non exhaustive)\label{collcoaut}}
Classement dans l'ordre chronologique:
\vspace*{1em}

{\bf J.-P. Bernard \etal}: {\it Implications of the PRONAOS observations for the large scale surveys with FIRST},
The Promise of the Herschel Space Observatory, \'ed. G.L. Pilbratt, J. Cernicharo, A.M. Heras, T. Prusti et R. Harris, ESA-SP 460, p. 297, 2001
\vspace*{1em}

{\bf A. Beno\^\i t \& la collaboration Archeops}: {\it Archeops: a balloon experiment for measuring the Cosmic Microwave Background anisotropies}, proc. of the 15th ESA Symposium on European Rocket and Balloon Programmes and Related Research, ESA-SP-471, p. 431, 2001
\vspace*{1em}

{\bf F.-X. D\'esert \& la collaboration Archeops}: {\it Archeops: A CMB anisotropy balloon experiment with a broad range of angular scales}, proc. of the G. Serra Conference Submillimeter Space Astronomy: A Window on the Cold Universe, Toulouse, EAS pub. series vol. 4, \'ed. Giard, Bernard, Klotz et Ristorcelli, p. 219-223, 2002

arXiv.org: astro-ph/0112010
\vspace*{1em}

{\bf C. M\'eny, N. Boudet, J.-P. Bernard, X. Dupac, M. Giard, J.-M. Lamarre, F. Pajot, I. Ristorcelli, G. Serra, B. Stepnik, J.-P. Torre}: {\it PRONAOS submillimeter observation of the Cygnus X IC 1318A nebula}, proc. of the G. Serra Conference Submillimeter Space Astronomy: A Window on the Cold Universe, Toulouse, EAS pub. series vol. 4, \'ed. Giard, Bernard, Klotz et Ristorcelli, p. 127-131, 2002
\vspace*{1em}

{\bf I. Ristorcelli, B. Stepnik, X. Dupac, A. Abergel, J.-P. Bernard, N. Boudet, M. Giard, J.-M. Lamarre, C. M\'eny, F. Pajot, J.-P. Torre, G. Serra}: {\it PRONAOS observations of the interstellar medium: new insights on interstellar dust}, proc. of the G. Serra Conference Submillimeter Space Astronomy: A Window on the Cold Universe, Toulouse, EAS pub. series vol. 4, \'ed. Giard, Bernard, Klotz et Ristorcelli, p. 9-17, 2002
\vspace*{1em}

{\bf A. Beno\^\i t \& la collaboration Archeops}: {\it Archeops: A balloon experiment to measure CMB anisotropies with a broad range of angular sizes}, proc. of the 2K1BC conf.: Experimental Astronomy at Millimeter Wavelengths, Breuil-Cervinia, Val d'Aoste, Italie, juillet 2001, AIP conf. proceedings vol. 616, \'ed. M. de Petris et M. Gervasi, p. 31-38, 2002

arXiv.org: astro-ph/0112012
\vspace*{1em}

{\bf F. Nati \etal}: {\it A simple and reliable star sensor for spinning payloads}, proc. of the 2K1BC conf.: Experimental Astronomy at Millimeter Wavelengths, Breuil-Cervinia, Val d'Aoste, Italie, juillet 2001, AIP conf. proceedings vol. 616, \'ed. M. de Petris et M. Gervasi, p. 52-55, 2002

\vspace*{1em}

{\bf J.-C. Hamilton \& la collaboration Archeops}: {\it Archeops, mapping the CMB sky from large to small angular scales}, proc. of the TAUP 2001 conference, LNGS, Italie, Sept. 2001

arXiv.org: astro-ph/0112009
\vspace*{1em}

{\bf A. Amblard \& la collaboration Archeops}: {\it Archeops: CMB Anisotropies Measurement from Large to Small Angular Scale}, proc. of the Cosmo-01 Workshop, Rovaniemi, Finlande, 29 ao\^ut - 4 sept. 2001

arXiv.org: astro-ph/0112205
\vspace*{1em}



\newpage
\thispagestyle{empty}
\vspace*{5em}
\newpage
\thispagestyle{empty}
\vspace*{5em}
\newpage
\thispagestyle{empty}
\begin{small}

Ce travail de th\`ese s'inscrit dans le d\'eveloppement de l'astrophysique des rayonnements infrarouge lointain et millim\'etrique.
Nous avons travaill\'e sur le traitement et l'analyse de donn\'ees concernant le milieu interstellaire galactique \`a travers l'\'emission thermique des poussi\`eres, et la cosmologie \`a travers l'observation des fluctuations du rayonnement fossile.
Nous nous sommes particuli\`erement int\'eress\'es \`a la construction de cartes optimales par des m\'ethodes d'inversion lin\'eaire.
Ceci nous a permis de d\'evelopper une nouvelle m\'ethode de construction de cartes pour l'exp\'erience ballon submillim\'etrique PRONAOS, fond\'ee sur une matrice d'inversion de Wiener, qui reconstruit la carte de fa\c{c}on globale.
L'analyse des cartes de PRONAOS dans les complexes de formation d'\'etoiles massives que sont Orion et M17 a ensuite permis de d\'ecouvrir les variations importantes des conditions physiques du milieu et des propri\'et\'es des poussi\`eres.
Nous avons notamment mis en \'evidence des condensations froides (T $\approx$ 10 - 20 K) \`a proximit\'e des centres actifs de formation d'\'etoiles.
Certains de ces nuages froids pourraient \^etre gravitationnellement instables.
Il appara\^\i t \'egalement une anticorr\'elation entre la temp\'erature et l'indice spectral submillim\'etrique.
Nos investigations concernant cet effet favorisent des causes li\'ees \`a la physique intrins\`eque des grains.
Nous avons \'egalement d\'evelopp\'e des m\'ethodes optimales de construction de cartes pour les exp\'eriences mesurant les fluctuations du rayonnement fossile.
Nous avons simul\'e diff\'erentes strat\'egies d'observation d'exp\'eriences ballon et satellite (tel que Planck), construit diff\'erents trains de donn\'ees et appliqu\'e nos m\'ethodes de construction de cartes \`a ces donn\'ees simul\'ees.
Les m\'ethodes it\'eratives d\'evelopp\'ees (COBE et Wiener) permettent de reconstruire la carte du ciel avec une grande pr\'ecision, malgr\'e la grande quantit\'e de bruit autocorr\'el\'e pr\'esent dans les trains de donn\'ees.
Nous avons \'egalement particip\'e au traitement et \`a l'analyse des donn\'ees de l'exp\'erience ballon Archeops.
En pr\'esence de bruit important et ayant des caract\'eristiques statistiques complexes, nous avons appliqu\'e des m\'ethodes simples de construction de cartes et d'estimation du spectre de puissance des fluctuations.

\vspace*{1em}

Mots clefs: construction de cartes - ballons - milieu interstellaire - poussi\`ere - cosmologie - rayonnement fossile

\vspace*{2em}

This work takes part of the development of far-infrared and millimeter astrophysics.
We have worked on the data processing and analysis in the fields of the Galactic interstellar medium,
through the dust thermal emission, and cosmology through the observation of the cosmic
microwave background fluctuations.
We have been particularly interested in optimal map-making by inverse linear methods.
We have developed a new map-making method for the balloon-borne submillimeter experiment PRONAOS, based on a Wiener inversion matrix, which allows to globally reconstruct the map.
The analysis of PRONAOS maps in massive star-forming complexes as Orion and M17 allowed us to discover the large variations of the physical conditions and the dust properties.
Notably, we showed cold condensations (T $\approx$ 10 - 20 K) near the active star-forming centers.
Some of these cold clouds could be gravitationally unstable.
Also, we showed an anticorrelation between the temperature and the submillimeter spectral index.
Our investigations concerning this effect favour causes related to the intrinsic physics of the grains.
We have also developed optimal map-making methods for the experiments aiming at measuring the cosmic microwave background fluctuations.
We simulated several different observation strategies for balloon-borne or satellite experiments (such as Planck), constructed timelines and applied our map-making methods to these simulated data.
The iterative methods that we have developed (COBE and Wiener) allow to reconstruct the sky maps very accurately, in spite of the large amount of self-correlated noise present in the timelines.
We have also worked on the data processing and analysis for the Archeops balloon-borne experiment.
In the presence of much noise having complex statistical properties, we have applied simple map-making and power spectrum estimation methods.

\vspace*{1em}

Key words: map-making - balloons - interstellar medium - dust - cosmology - cosmic microwave background

\end{small}




\end{document}